\documentclass{article}
\usepackage{amssymb,amsmath}
\usepackage[dvipdfmx]{graphicx}
\usepackage{bbm}
\usepackage{bm}
\usepackage{caption}
\usepackage{color}
\usepackage{comment}
\usepackage{comment}
\usepackage{enumerate}
\usepackage{ytableau}
\usepackage{here}
\usepackage{subfig}
\DeclareMathOperator*{\Tr}{{\rm Tr}}

\numberwithin{equation}{section}

\begin{document}

\thispagestyle{empty}
\begin{flushright}

\end{flushright}
\vskip1cm
\begin{center}
{\Large \bf Dualities of Corner Configurations\\
\vskip0.3cm 
and 
\vskip0.5cm 
Supersymmetric Indices
}

\vskip1.5cm
Davide Gaiotto\footnote{dgaiotto@perimeterinstitute.ca},
Tadashi Okazaki\footnote{tokazaki@perimeterinstitute.ca}

\bigskip
{\it
Perimeter Institute for Theoretical Physics,\\
Waterloo, Ontario, Canada N2L 2Y5
}

\end{center}

\vskip1cm
\begin{abstract}
We compute supersymmetric indices which count local operators at certain half-BPS interfaces and 
quarter-BPS junctions of interfaces in four-dimensional ${\cal N}=4$ Super Yang-Mills theory. 
We use the indices as very stringent tests of a variety of string theory-inspired conjectures about the action of 
S-duality on such defects. 
\end{abstract}


\newpage
\tableofcontents

\section{Introduction and conclusions}
Superconformal indices \cite{Kinney:2005ej,Romelsberger:2005eg,Romelsberger:2007ec} are a well-established tool to study supersymmetric quantum field theories and their duality properties.
The superconformal index can be used in essentially in any situation where one has two supercharges which square to a combination of the dilatation operator and other symmetry generators of the theory. 
They may be used to ``count'' local operators both in the bulk of a SCFT and at a variety of superconformal defects, in which case they allow one to study properties and dualities of these defects. 
There is a vast literature on the subject, which would be too lengthy to review here. We will refer through the text to the works most relevant for our analysis.  

In this paper we concern ourselves with four-dimensional ${\cal N}=4$ Super Yang-Mills (SYM) theory with $U(N)$ gauge group. 
We are interested in co-dimension one defects (aka boundaries or interfaces) which preserve a three-dimensional ${\cal N}=4$ superconformal algebra,
such as these introduced and studied in \cite{Gaiotto:2008sa,Gaiotto:2008sd,Gaiotto:2008ak}. The local operators at these boundaries can be enumerated by a supersymmetric index 
in a manner analogous to three-dimensional ${\cal N}=4$ SCFTs. Indeed, the three-dimensional superconformal index is well-defined even for 3d ${\cal N}=2$ SCFTs 
\cite{Kim:2009wb,Imamura:2011su,Krattenthaler:2011da} and for half-BPS defects in 4d ${\cal N}=2$ SCFTs \cite{Dimofte:2011py}. 

The superconformal index for some interesting half-BPS interfaces in 4d ${\cal N}=4$ SYM was studied in \cite{Gang:2012ff}. Here we study more examples and we 
go one further step down in dimension, looking at quarter-BPS 2d junctions of half-BPS 3d defects. In particular, we focus on junctions which preserve 
a two-dimensional $\mathcal{N}=(0,4)$ sub-algebra of the bulk supersymmetry algebra, which played an important role in recent work on the Geometric Langlands program
\cite{Gaiotto:2017euk,Creutzig:2017uxh,Frenkel:2018dej} and admit interesting String Theory constructions \cite{Chung:2016pgt,Hanany:2018hlz} 
which result in a precise prescription for the action of S-duality. The associated superconformal index will give us very stringent tests of these conjectural dualitites.

\subsection{Structure}
The structure of the paper is straightforward: we introduce the relevant indices and then write them down explicitly for a long list of pairs (or triples) 
of defects conjecturally related by duality, which should thus have the same index. For some of the examples we give an explicit proof of the equality of the indices, 
for some others we describe a strategy which should allow for a direct proof with some extra work. 
In all cases we compare a large number of terms of the formal power series expansion of the indices. 
We show the several first terms in the $q$-expansions of many examples of indices 
and the orders which they agree up to in Appendix \ref{sec_expansion}. 

\subsection{Open problems}
In this work we confirmed many examples to gain a bottom-up understanding of general identities of indices 
which strongly support the dualities of interfaces/junctions in 4d $\mathcal{N}=4$ SYM theory inspired from string theory. 
However, there are a variety of interesting questions which we leave for future work. In particular:
\begin{itemize}
\item Three-dimensional ${\cal N}=4$ SCFTs can be engineered by RG flow from the compactification of 4d SYM on a segment in the presence of half-BPS boundary conditions and 
interfaces. This is one way to derive many three-dimensional mirror symmetries. Compactifications on a strip or half-strip, involving extra half-BPS boundary conditions and 
interfaces and appropriate 2d junctions, can be used to engineer $\mathcal{N}=(0,4)$ boundary conditions which should have interesting properties under 3d mirror symmetry
and applications to Symplectic Duality \cite{Costello:2018fnz,Costello:2018swh}. It would be interesting to test these dualities with superconformal indices. 
\item Two-dimensional ${\cal N}=(0,4)$ SCFTs can be engineered by RG flow from the compactification of 4d SYM on a square in the presence of half-BPS boundary conditions and 
interfaces. S-duality should lead to interesting dualities for such 2d SCFTs. It would be interesting to test these dualities with superconformal indices. 
\item Our configurations can be enriched further by half-BPS  line defects and surface defects, intersecting our 2d junctions at a point. These enriched configurations 
still admit a superconformal index counting local operators available at that point and will have interesting S-duality properties. Some of the ``Higgsing'' manipulations of 
indices in the bulk of the paper produce some of these decorated indices for free.  
\item The string theory constructions and gauge theory conjectures we employ in this paper have well-studied generalizations involving orthogonal and symplectic gauge groups. 
It would be nice to verify these generalizations at the level of the index. 
\item In this paper we focus on junctions which preserve $\mathcal{N}=(0,4)$ supersymmetry in 2d. Junctions which preserve $\mathcal{N}=(2,2)$ supersymmetry are also 
a possibility and admit brane constructions which can lead to conjectural dualities. It would be nice to explore this further. 
\item It would be interesting to explore dualities for bulk-boundary-corner systems by computing the quarter-indices in other setup with possibly different dimensions and supersymmetries. 
In particular, 4d $\mathcal{N}=2$ bulk, 3d $\mathcal{N}=2$ boundary and 2d $\mathcal{N}=(0,2)$ corner supersymmetric gauge theories should generalize the dualities of half-BPS $(0,2)$  boundary conditions \cite{Gadde:2013wq, Okazaki:2013kaa} in 3d $\mathcal{N}=2$ theories discussed in \cite{Dimofte:2017tpi}.
\item It would be nice to shed light on the holographic dual interpretation of our gauge theoretical bulk-boundary-corner systems by constructing their supergravity solutions, 
as for bulk-boundary systems discussed in \cite{Assel:2011xz, Aharony:2011yc}, and by reproducing the quarter-indices in the large $N$ limit. In particular, our quarter-indices for the Y-junctions should give some generalizations of the MacMahon function which is obtained from the large $N$ limit of the vacuum characters of the Y-algebras \cite{Gaiotto:2017euk}.
\item The operators counted by the superconformal indices should belong to some twisted version of the gauge theory. 
It should thus be possible to ``categorify'' our results to rigorous equalities of certain algebras or modules of local operators. 
\item Our quarter-indices should have rich mathematical properties as they can be viewed as promotions of various characters for VOAs including Kac-Moody superalgebras. In particular, the characters for the affine Kac-Moody Lie superalgebra is known to exhibit intriguing number theoretic properties involving mock modularity \cite{Kac:1994kn, MR1810948, MR3200431, MR2719689, MR2534107}. It would be interesting to examine their transformation and asymptotic properties and to study their associated identities and difference equations. 
\end{itemize}

\section{Indices}
\label{sec_04bc}

\subsection{Definition}
\label{sec_def}
The main actors of this note are supersymmetric indices which count local operators residing on certain co-dimension $2$ defects in four-dimensional ${\cal N}=4$ Supersymmetric Yang-Mills theory, which preserve a two-dimensional $(0,4)$ sub-algebra of the bulk supersymmetry algebra. 

The most general configurations we are interested in involve two-dimensional junctions which lie at the intersection of multiple half-BPS interfaces or boundary conditions. We will denote the corresponding supersymmetric index as a ``quarter-index''
$\mathbb{IV}$. This generalizes simpler indices such as the ``half-index'' $\mathbb{II}$ of boundary or interface local operators and the ``full-index'' $\mathbb{I}$ of bulk local operators. The latter can be also thought of as a specialization of the 
quarter-index to trivial junctions, possibly on a trivial interface. 

The precise definition is 
\begin{align}
\label{INDEX_def}
\mathbb{IV}(t,x;q)
&:={\Tr}_{\mathrm{Op}}(-1)^{F}q^{J+\frac{H+C}{4}}t^{H-C} x^{f}. 
\end{align}
Here the trace is taken over the cohomology of chosen supercharges. $F$ is the fermion number, 
$J$ generates the $Spin(2)$ $\simeq$ $U(1)_{J}$ rotations in the two-dimensional plane on which the local operators are supported. $C$ and $H$ are the Cartan generators of the $SU(2)_{C}$ and $SU(2)_{H}$ R-symmetry groups. 
$f$ stands for the Cartan generators of the flavor symmetry group. 

The choice of fugacity is selected so that the power of $q$ is always strictly positive for a non-trivial local operator, by a unitarity bound. The (quarter)index should be thought as a formal power series in $q$, with coefficients which are Laurent polynomials in the other fugacities. 
\footnote{As in \cite{Costello:2018fnz}, other notational choices $y=t q^{-\frac14}$ may be useful to write neat formulae for indices, but they should always be understood as power series in $q$ for fixed $t$.  
}

\subsubsection{Relations to VOA}
\label{sec_voa}

In some special situations, discussed in \cite{Gaiotto:2017euk,Costello:2018fnz}, $(0,4)$ junctions can be deformed in such a way to be compatible with topological twists of the bulk and boundaries. The junction local operators 
then form a non-trivial and useful vertex operator algebra. At the level of the index, the deformation 
breaks some symmetries and enforces a specialization of $t$ to $q^{\pm \frac14}$, depending on 
the twist using $SU(2)_H$ or $SU(2)_C$. 

After the specialization, the index will coincide with the character of the corresponding VOA. 
In particular, we will make contact with the characters of corner VOA $Y_{L,M,N}[\Psi]$ introduced in \cite{Gaiotto:2017euk}
and boundary VOAs introduced in \cite{Costello:2018fnz}. 

\subsubsection{Localization computation}
\label{sec_localization}

By the state-operator map, the supersymmetric indices can also be thought as equivariant Witten indices 
for appropriate configuration of boundary conditions/interfaces and junctions on a three-sphere: the junctions will run along a great circle of the sphere, while the interfaces/boundary conditions will wrap two-dimensional hemispheres 
ending on that circle. 

These configurations should allow one to compute the index by a localization procedure. We will not pursue that direction 
in detail in this paper, though it will help us explain some interesting phenomena later on. 

In particular, the sphere picture helps motivate certain index manipulations associated to vevs of scalar fields \cite{Gaiotto:2012xa,Gaiotto:2014ina}:
\begin{itemize}
\item There are useful RG flows which are initiated by turning on vevs of local operators which are charged under some $U(1)$ 
flavor symmetry. The vev can preserve some modified R-symmetry which allows a comparison of supersymmetric indices before and after the RG flow. 
The existence of such vevs is associated to certain arrays of poles in the index, whose residues, properly normalized by subtracting off decoupled free fields,
give the index of the IR theory, possibly decorated by certain defects. This manipulation is independent of the specific description of the 
setup, as the poles are an intrinsic property of the theory. 
\item When the system has a gauge theory description, the index is usually written as a contour integral over gauge fugacities. 
Poles in the integrand and their residues may sometimes have a sharp physical interpretation, especially if some FI-like parameter can be turned on, 
enforcing non-trivial vevs of elementary fields carrying gauge charge and thus Higgsing the gauge group. Then the full index can be written 
as a sum of residues which can be interpreted as the index of the Higgsed theory, possibly deformed by dynamical BPS solitons. 
\end{itemize}

We will use these ideas to provide a variety of consistency checks of our proposals throughout the text. 
\subsubsection{Brane construction}
\label{sec_brane}
Our main objective is to compare the supersymmetric indices of gauge theory configurations related by S-duality. 
In order to identify such configurations, it is useful to employ brane constructions. More specifically, we use brane constructions in Type IIB superstring theory \cite{Hanany:1996ie}: stacks of D3-branes
engineer the four-dimensional gauge fields, while a variety of fivebranes engineer co-dimension $1$ boundaries/interfaces, possibly intersecting at co-dimension $2$ junctions.

All branes have in common the $x^{0}, x^{1}$ directions of the 2d junction. The D3-branes wrap the $0126$ directions
and engineer SYM with unitary gauge groups.  
The remaining six directions are split into two groups, $345$ and $789$, which are rotated by the $SO(3)_C \times SO(3)_H$ 
block-diagonal subgroup of the bulk $SO(6)_R$ R-symmetry group. The corresponding triplets of 4d SYM scalar fields are denoted as $Y$ and $X$. 

The junction supersymmetry is compatible with two families of fivebranes: $(p,q)$-fivebranes 
wrapping $345$ and a direction in the $26$ plane with slope $p/q$ 
\footnote{More precisely, the slope is coupling-dependent and equals the argument of $p \tau + q$.}
and $(p,q)$-fivebranes wrapping $789$ and a direction in the $26$ plane with slope $q/p$. 

In particular, we will employ often 
\begin{itemize}
\item NS5-branes extended along $012345$, 
\item D5-branes extended along $012789$, 
\item NS5$'$-branes extended along $016789$, 
\item D5$'$-branes extended along $0123456$
\end{itemize}

Here is a summary of the brane configuration:
\begin{align}
\label{brane1}
\begin{array}{ccccccccccc}
&0&1&2&3&4&5&6&7&8&9\\
\textrm{D3}
&\circ&\circ&\circ&-&-&-&\circ&-&-&- \\
\textrm{NS5}
&\circ&\circ&\circ&\circ&\circ&\circ&-&-&-&- \\
\textrm{D5}
&\circ&\circ&\circ&-&-&-&-&\circ&\circ&\circ \\
\textrm{NS5$'$}
&\circ&\circ&-&-&-&-&\circ&\circ&\circ&\circ \\
\textrm{D5$'$}
&\circ&\circ&-&\circ&\circ&\circ&\circ&-&-&- \\
\end{array}
\end{align}

Depending on the precise geometry of the system, we can engineer a rich variety of 
gauge theory configurations. In particular, we may have:
\begin{enumerate}
\item \textbf{Half-BPS Boundaries and Interfaces in 4d $\mathcal{N}=4$ gauge theories}

These are engineered by configurations of D3-branes wrapping a half-plane, ending on or passing through 
a sequence of NS5- and D5-branes \cite{Gaiotto:2008sa} (or equivalently a sequence of NS5$'$ and D5$'$). 
The general gauge theory configuration can be reconstructed 
with some care from the more elementary interfaces associated with a single fivebrane and a certain number of 
D3-branes on the two sides. 

NS5-branes implement Neumann-like boundary conditions for the gauge theories on the two sides of the interface enriched by 
three-dimensional bi-fundamental matter. D5-branes either add three-dimensional fundamental matter or 
reduce the gauge group at the interface by certain generalizations of Dirichlet boundary conditions, called 
Nahm pole boundary conditions. \cite{Gaiotto:2008sa}. 

We will test some of the expected dualities of boundary conditions and interfaces by computing the appropriate half-indices. 
\footnote{
See \cite{Okazaki:2019ony} for the half-indices of dual interfaces which may involve 3d gauge symmetry. }

\item \textbf{3d $\mathcal{N}=4$ gauge theories}

Configurations of D3-, D5- and NS5-branes where all D3-branes are finite segments in the 
$6$ direction engineer 3d $\mathcal{N}=4$ gauge theories \cite{Hanany:1996ie}
built from vector multiplets and hypermultiplets. 
In a similar manner, finite D3-NS5$'$-D5$'$-brane systems leads to the twisted version of 3d $\mathcal{N}=4$ gauge theory,
where the role of the $SU(2)_{C}$ and $SU(2)_{H}$ R-symmetry groups are swapped. 

The relation between S-duality and mirror symmetries of such 3d gauge theories is well understood. 
We will not explore it further in this paper. 

\item \textbf{Corners in 4d $\mathcal{N}=4$ gauge theories}
Configurations involving fivebranes extending both along the $2$ and the $6$ direction, 
with D3-branes which extend to infinity in both directions, give ``corner'' configurations in 
4ds gauge theory, where two or more boundaries or interfaces intersect at a common 2d junction
preserving $\mathcal{N}=(0,4)$ supersymmetry. If one employs only five-branes extended, say, along the $345$ directions 
one will obtain the junctions whose deformation supports the VOA from \cite{Gaiotto:2017euk}.

We will test some of the expected dualities of such junctions by computing the appropriate quarter-indices.

\item \textbf{Boundaries in 3d $\mathcal{N}=4$ gauge theories}

Configurations involving fivebranes extending both along the $2$ and the $6$ direction, 
with D3-branes which extend to infinity in one direction only will give rise to  
$\mathcal{N}=(0,4)$ supersymmetric boundary conditions in 3d $\mathcal{N}=4$ gauge theory 
as in \cite{Chung:2016pgt, Hanany:2018hlz}. 
These boundary conditions and interfaces are likely to admit the sort of deformations 
which lead to boundary VOAs as in \cite{Costello:2018fnz}. They are also likely to behave in a nice fashion under mirror symmetry. 
The half-indices of $\mathcal{N}=(0,4)$ boundary conditions 
for 3d $\mathcal{N}=4$ Abelian gauge theories have been computed and 
dual boundary conditions have been proposed in \cite{Okazaki:2019bok}. 
We hope to come back to more general boundary conditions and their half-indices. 

\item \textbf{2d $\mathcal{N}=(0,4)$ gauge theories}

Finally, ``brane box'' configurations where all D3-branes are finite in the 2 and 6 directions will engineer two-dimensional 
gauge theories with $\mathcal{N}=(0,4)$ supersymmetry \cite{Hanany:2018hlz}. 
In \cite{Okazaki:2019bok} the simplest $\mathcal{N}=(0,4)$ mirror symmetry between 
Abelian gauge theory and Fermi multiplets has been proposed. 
We leave an exploration of more general theories, their indices and the implications of S-duality. 

\end{enumerate}

These brane systems often have branches of vacua or deformation parameters corresponding to finite or semi-infinite D3-branes
either moving away from the system along some fivebranes or merging and separating from the fivebrane systems in some transverse direction. 
The existence of these geometric deformations is a test of the gauge theory description of the brane systems and offers interesting 
``Higgsing'' manipulations of the supersymmetric indices. 

\subsection{4d $\mathcal{N}=4$ indices}
\label{sec_4dINDEX}

\subsubsection{Indices}
\label{sec_4dindex}
The 4d $\mathcal{N}=4$ SYM theory has $SU(4)_{R}$ R-symmetry. 
We split the adjoint scalar fields transforming as ${\bf 6}$ under the $SU(4)_{R}$ into two scalar fields $X$ and $Y$ 
acted on by $SU(2)_{C}\times SU(2)_{H}$ $\subset$ $SU(4)_{R}$ as $({\bf 1},{\bf 3})$ and $({\bf 3},{\bf 1})$. 
In the brane setup, the scalar fields $X$ and $Y$ describe the motion of D3-branes along 
the $(x^{7}$, $x^{8}$, $x^{9})$ directions and $(x^{3}$, $x^{4}$, $x^{5})$ directions respectively. 
Under the $SU(2)_{C}$ $\times$ $SU(2)_{H}$ the 4d gauginos $\lambda$ transform as $({\bf 2},{\bf 2})$.

The local operators in 4d $\mathcal{N}=4$ SYM theory of gauge group $G$ which contribute to index can have charges 
\begin{align}
\label{4dn4_ch}
\begin{array}{c|cccc}
&\partial_{z}^{n}X&\partial_{z}^{n}Y&\partial_{z}^{n}\lambda&\partial_{z}^{n}\overline{\lambda} \\ \hline
G&\textrm{adj}&\textrm{adj}&\textrm{adj}&\textrm{adj} \\
U(1)_{J}&n&n&n+\frac12&n+\frac12 \\
U(1)_{C}&0&2&+&+ \\
U(1)_{H}&2&0&+&+  \\
\textrm{fugacity}
&q^{n+\frac12}t^{2}s_{\alpha}
&q^{n+\frac12}t^{-2}s_{\alpha}
&-q^{n+1}s_{\alpha}
&-q^{n+1}s_{\alpha} \\
\end{array}
\end{align}

The index for 4d $\mathcal{N}=4$ $U(1)$ gauge theory is 
\begin{align}
\label{4du1_INDEX}
\mathbb{I}^{\textrm{4d $U(1)$}}(t;q)
&=\frac{(q)_{\infty}^{2}}{(q^{\frac12}t^{2};q)_{\infty}(q^{\frac12}t^{-2};q)_{\infty}}\oint \frac{ds}{2\pi is}. 
\end{align}
The denominator comes from the scalar fields $X$ and $Y$ 
and the numerator is contributed from the 4d gauginos $\lambda$ and $\overline{\lambda}$. 

The simplest non-Abelian example is an index for 4d $\mathcal{N}=4$ $U(2)$ gauge theory: 
\begin{align}
\label{4du2_INDEX}
\mathbb{I}^{\textrm{4d $U(2)$}}(t;q)
&=\frac12 \frac{(q)_{\infty}^{4}}
{(q^{\frac12} t^{2};q)_{\infty}^2(q^{\frac12} t^{-2};q)_{\infty}^2}
\oint \frac{ds_{1}}{2\pi is_{1}}\frac{ds_{2}}{2\pi is_{2}}
\nonumber\\
&\frac{\left(\frac{s_{1}}{s_{2}};q\right)_{\infty}
\left(\frac{s_{2}}{s_{1}};q\right)_{\infty}
\left(q\frac{s_{1}}{s_{2}};q\right)_{\infty}
\left(q\frac{s_{2}}{s_{1}};q\right)_{\infty}
}
{
\left(q^{\frac12} t^{2}\frac{s_{1}}{s_{2}};q\right)_{\infty}
\left(q^{\frac12} t^{-2}\frac{s_{1}}{s_{2}};q\right)_{\infty}
\left(q^{\frac12} t^{2}\frac{s_{2}}{s_{1}};q\right)_{\infty}
\left(q^{\frac12} t^{-2}\frac{s_{2}}{s_{1}};q\right)_{\infty}
}. 
\end{align}
Here the integration contour for variables $s_{i}$ is a unit torus $\mathbb{T}^{2}$.

The index for 4d $\mathcal{N}=4$ $U(N)$ gauge theory takes the form
\begin{align}
\label{4duN_INDEX}
\mathbb{I}^{\textrm{4d $U(N)$}}(t;q)
&=\frac{1}{N!} \frac{(q)_{\infty}^{2N}}
{(q^{\frac12} t^{2};q)_{\infty}^N(q^{\frac12} t^{-2};q)_{\infty}^N}
\oint 
\prod_{i=1}^{N}
\frac{ds_{i}}{2\pi is_{i}}
\prod_{i\neq j}
\frac{\left(\frac{s_{i}}{s_{j}};q\right)_{\infty}
\left(q\frac{s_{i}}{s_{j}};q\right)_{\infty}
}
{
\left(q^{\frac12} t^{2}\frac{s_{i}}{s_{j}};q\right)_{\infty}
\left(q^{\frac12} t^{-2}\frac{s_{i}}{s_{j}};q\right)_{\infty}
}
\end{align}
where the integration contour for gauge fugacities $s_{i}$ is a unit torus $\mathbb{T}^{N}$.

\subsubsection{Half-indices}
\label{sec_4dh_index}
The half-BPS boundary conditions in 4d $\mathcal{N}=4$ SYM theory preserve 3d $\mathcal{N}=4$ supersymmetry 
with the R-symmetry group $SU(4)_{R}$ broken down to $SU(2)_{C}$ $\times$ $SU(2)_{H}$. 
They arise for parallel D3-branes ending on a single fivebrane.  
As shown in the brane configuration (\ref{brane1}), 
we consider two types of three-dimensional boundaries/interfaces at $x^2=0$ realized by NS5$'$- and D5$'$-branes 
and those at $x^6=0$ realized by NS5- and D5-branes.

Consider the half-BPS boundary conditions for 4d $\mathcal{N}=4$ Abelian gauge theory with $\theta_{\textrm{YM}}=0$ \cite{Gaiotto:2008sa}. 
The NS5-brane and D5-brane ending on a single D3-brane 
give rise to Neumann b.c. $\mathcal{N}$ and Dirichlet b.c. $\mathcal{D}$ at $x^6=0$ 
for $U(1)$ gauge theory respectively:
\begin{align}
\label{4d_bc1}
\begin{array}{cccc}
\mathcal{N}:& F_{6\mu}|_{\partial}=0,& \partial_{\mu}X|_{\partial}=0,&\partial_{6}Y|_{\partial}=0 \\
\mathcal{D}:& F_{\mu\nu}|_{\partial}=0,& \partial_{6}X|_{\partial}=0,&\partial_{\mu}Y|_{\partial}=0 \\
\end{array}
\qquad \mu,\nu=0,1,2
\end{align}
The NS5$'$-brane and D5$'$-brane ending on a single D3-brane 
leads to Neumann b.c. $\mathcal{N}'$ and Dirichlet b.c. $\mathcal{D}'$ at $x^2=0$ for $U(1)$ gauge theory respectively:
\begin{align}
\label{4d_bc2}
\begin{array}{cccc}
\mathcal{N}':& F_{2\mu}|_{\partial}=0,& \partial_{2}X|_{\partial}=0,&\partial_{\mu}Y|_{\partial}=0 \\
\mathcal{D}':& F_{\mu\nu}|_{\partial}=0,& \partial_{\mu}X|_{\partial}=0,&\partial_{2}Y|_{\partial}=0 \\
\end{array}
\qquad \mu,\nu=0,1,6
\end{align}

The half-indices of Neumann b.c. $\mathcal{N}$ and Dirichlet b.c. $\mathcal{D}'$ 
for 4d $\mathcal{N}=4$ $U(1)$ vector multiplet are given by
\begin{align}
\label{4du1_hINDEX_N1}
\mathbb{II}^{\textrm{4d $U(1)$}}_{\mathcal{N}}(t;q)=
\mathbb{II}^{\textrm{4d $U(1)$}}_{\mathcal{D}'}(t;q)
&=\frac{(q)_{\infty}}{(q^{\frac12}t^{-2};q)_{\infty}}. 
\end{align}
The denominator comes from the scalar fields $Y$ charged under $U(1)_{C}$ 
while the numerator captures a half of the 4d gauginos. 
Similarly, the half-indices of Neumann b.c. $\mathcal{N}'$ and Dirichlet b.c. $\mathcal{D}$ is 
\begin{align}
\label{4du1_hINDEX_N2}
\mathbb{II}^{\textrm{4d $U(1)$}}_{\mathcal{D}}(t;q)=
\mathbb{II}^{\textrm{4d $U(1)$}}_{\mathcal{N}'}(t;q)
&=\frac{(q)_{\infty}}{(q^{\frac12}t^{2};q)_{\infty}}. 
\end{align}
The denominator comes from the scalar fields $X$ charged under $U(1)_{H}$ 
while the numerator describes a half of the 4d gauginos.

The equality of the two half-indices in (\ref{4du1_hINDEX_N1}) agrees with the observation that 
Neumann condition $\mathcal{N}'$ for 4d $\mathcal{N}=4$ $U(1)$ gauge theory is S-dual to 
Dirichlet b.c. $\mathcal{D}$ for 4d $\mathcal{N}=4$ $U(1)$ gauge theory. 

The boundary conditions corresponding to $N$ D3-branes ending on a single NS5-brane (or NS5')
are also Neumann b.c. for the $U(N)$ gauge theory. We can denote them as $\mathcal{N}$ and $\mathcal{N}'$
as in the  Abelian case.

On the other hand, $N$ D3-branes ending on a single D5-brane do {\it not} give rise to Dirichlet boundary conditions, but to 
a modification associated to a ``regular Nahm pole'' \cite{Diaconescu:1996rk,Gaiotto:2008sa}. 
Let us denote the scalar fields by $\vec{X}$ and $\vec{Y}$ to make it explicit that 
they are triplet of $SU(2)_{H}$ and that of $SU(2)_{C}$ respectively. 
When $N$ D3-branes terminate on a single D5-brane or D5$'$-brane, we find the Nahm or Nahm$'$ pole boundary conditions: 
\begin{align}
\label{4d_nahmbc}
\begin{array}{ccccc}
\textrm{Nahm}:& F_{\mu\nu}|_{\partial}=0,& D_{6}\vec{X}+\vec{X}\times \vec{X}|_{\partial}=0,&D_{\mu}\vec{Y}|_{\partial}=0
&\qquad  \mu,\nu=0,1,2 \\
\textrm{Nahm}':& F_{\mu\nu}|_{\partial}=0,& D_{\mu}\vec{X}|_{\partial}=0,&D_{2}\vec{Y}+\vec{Y}\times \vec{Y}|_{\partial}=0
&\qquad \mu,\nu=0,1,6 \\
\end{array}
\end{align}
The Nahm's  equations for the scalar fields $\vec{X}$ and $\vec{Y}$ 
admit singular solutions where $X$ or $Y$ have singularities at $x^6=0$ or at $x^2=0$ respectively:
\begin{align}
\label{4d_nahmbc2}
\vec{X}(x^6)&=\frac{\vec{\mathfrak{t}}}{x^6},& 
\vec{Y}(x^2)&=\frac{\vec{\mathfrak{t}}}{x^2}
\end{align}
where $\vec{\mathfrak{t}}$ $=$ $(\mathfrak{t}_{1},\mathfrak{t}_{2},\mathfrak{t}_{3})$ is a triplet of elements of 
the Lie algebra $\mathfrak{g}$ $=$ $\mathfrak{u}(N)$ satisfying the commutation relation of the Lie algebra $\mathfrak{su}(2)$, 
that is $[\mathfrak{t}_{1},\mathfrak{t}_{2}]=\mathfrak{t}_{3}$ and cyclic permutation thereof. 
Choosing $\vec{\mathfrak{t}}$ specifies a homomorphism of Lie algebras $\rho:$ $\mathfrak{su}(2)$ $\rightarrow$ $\mathfrak{g}$. 

A single D5-brane gives the regular Nahm pole, where $\rho$ maps the fundamental representation of $U(N)$
to the dimension $N$ irreducible representation of $\mathfrak{su}(2)$. Multiple D5-branes can be used to engineer other Nahm poles, including the trivial Nahm pole, aka Dirichlet boundary conditions. That requires $N$ D5-branes. 

Because of the brane construction, S-duality is expected to exchange Neumann and (regular) Nahm pole boundary conditions. 
This is supported by half-index calculations as follows.

As the simplest non-Abelian example, 
a half-index of Neumann b.c. $\mathcal{N}'$ for 4d $\mathcal{N}=4$ $U(2)$ gauge theory is 
\begin{align}
\label{4du2_hINDEX_N1}
\mathbb{II}_{\mathcal{N}'}^{\textrm{4d $U(2)$}}(t;q)
&=
\frac{1}{2}
\frac{(q)_{\infty}^{2}}
{(q^{\frac12} t^{2};q)_{\infty}^{2}}
\oint 
\frac{ds_{1}}{2\pi is_{1}}
\frac{ds_{2}}{2\pi is_{2}}
\frac{\left(
\frac{s_{1}}{s_{2}};q
\right)_{\infty}
\left(
\frac{s_{2}}{s_{1}};q
\right)_{\infty}
}
{
\left(
q^{\frac12} t^{2}\frac{s_{1}}{s_{2}};q
\right)_{\infty}
\left(
q^{\frac12} t^{2}\frac{s_{2}}{s_{1}};q
\right)_{\infty}
}.
\end{align}
The half-index of Dirichlet b.c. $\mathcal{D}$ for $U(2)$ gauge theory is 
\begin{align}
\label{4du2_hINDEX_D1}
\mathbb{II}_{\mathcal{D}}^{\textrm{4d $U(2)$}}(t,s_{i};q)
&=
\frac{(q)_{\infty}^{2} \left(q\frac{s_{1}}{s_{2}};q\right)_{\infty}\left(q\frac{s_{2}}{s_{1}};q\right)_{\infty}}
{(q^{\frac12} t^2;q)_{\infty}^{2}
\left(q^{\frac12} t^{2}\frac{s_{1}}{s_{2}};q\right)_{\infty}
\left(q^{\frac12} t^2 \frac{s_{2}}{s_{1}};q\right)_{\infty}}. 
\end{align}
Here the $s_i$ are fugacities for the $U(2)$  boundary global symmetry which appears at Dirichlet boundary conditions, 
consisting of gauge transformations which are constant at the boundary.

If we evaluate the contour integral in (\ref{4du2_hINDEX_N1}), we find a notably simple answer: 
\begin{align}
\label{4du2_nahm}
\mathbb{II}_{\mathcal{N}'}^{\textrm{4d $U(2)$}}(t;q)
&=
\frac{(q)_{\infty}(q^{\frac32}t^{2};q)_{\infty}}
{(q^{\frac12}t^{2};q)_{\infty}(qt^{4};q)_{\infty}}. 
\end{align}
We would like to identify that with the half-index for Nahm pole boundary conditions 
in 4d $\mathcal{N}=4$ $U(2)$ gauge theory, i.e. we propose that 
\begin{align}
\label{4du2_nahm}
\mathbb{II}_{\textrm{Nahm}}^{\textrm{4d $U(2)$}}(t;q)
&=
\frac{(q)_{\infty}(q^{\frac32}t^{2};q)_{\infty}}
{(q^{\frac12}t^{2};q)_{\infty}(qt^{4};q)_{\infty}}. 
\end{align}

In order to motivate that expression directly from the Nahm pole definition, one would need to do a careful analysis 
of how to define local operators at a singular boundary condition. We will not do so. Instead, we observe that the Nahm pole 
b.c. can also be understood as the result of an RG flow which begins with Dirichlet boundary conditions deformed by 
a regular nilpotent vev for the $X^+ = X^3 + i X^4$ scalar field. The RG flow is expected to produce a regular Nahm pole 
together with some decoupled three-dimensional degrees of freedom, taking the form of a free 3d hypermultiplet with non-standard 
R-symmetry assignement. At the level of the half-index, stripping off the spurious 3d hyper is roughly equivalent to 
multiplying by the index of 3d vector multiplet, cancelling off the contribution of a zeromode in the process. This is analogous to 
the Higgsing procedure in \cite{Gaiotto:2012xa}. 

Concretely, the deformation of the Dirichlet boundary condition forces one to identify the 
boundary global symmetry with the $SU(2)_H$ R-symmetry, identifying the fugacity $s_1/s_2$ with $q^{\frac12} t^2$
and then strip off the index of a 3d hyper (see (\ref{3dthm}) in the next section) with flavor fugacity $x = t q^{\frac14}$. 
This reproduces the expected result. 

As an extra bonus, we can obtain the superconformal index in the presence of ``boundary 't Hooft operators'' 
which can be understood as the result of turning on position-dependent vevs of $X^+$. This corresponds to 
identifying the fugacity $s_1/s_2$ with $q^{\frac12+n} t^2$, leading to something like 
\begin{align}
\label{vortex1}
\mathbb{II}_{\textrm{Nahm};n}^{\textrm{4d $U(2)$}}(t;q)
&=
\frac{(q)_{\infty}^{2} \left(q^{\frac32+n} t^2;q\right)_{\infty}\left(q^{\frac12-n} t^{-2};q\right)_{\infty}}
{(q^{\frac12} t^2;q)_{\infty}^{2}
\left(q^{n+1} t^{4};q\right)_{\infty}
\left(q^{-n} ;q\right)_{\infty}}
\nonumber\\
&=
\frac{(q^{\frac12}t^{-2};q)_{\infty}}{(q^{\frac12} t^{2};q)_{\infty}}
 \frac{(q^{n+1};q)_{\infty} \left(q^{\frac32+n} t^2;q\right)_{\infty}}
{(q^{n+\frac12} t^2;q)_{\infty}
\left(q^{n+1} t^{4};q\right)_{\infty}
} q^{\frac{n}{2}} t^{- 2 n}
\end{align}
though there is some latitude in which factors we do precisely choose to strip off. 

%

On the other hand, 
making use of the $q$-binomial theorem (\ref{q_binomial}), 
we can evaluate the integral (\ref{4du2_hINDEX_N1}) as 
 \footnote{
We  note that the charged version is given by 
 \begin{align}
 \label{4du2_hINDEXsum2}
\mathbb{II}_{\mathcal{N}'_{n}}^{\textrm{4d $U(2)$}}(t;q)
&=
\frac{1}{2}
\frac{(q)_{\infty}^{2}}
{(q^{\frac12} t^{2};q)_{\infty}^{2}}
\oint 
\frac{ds_{1}}{2\pi is_{1}}
\frac{ds_{2}}{2\pi is_{2}} 
s_{1}^n s_{2}^{-n}
\frac{\left(
\frac{s_{1}}{s_{2}};q
\right)_{\infty}
\left(
\frac{s_{2}}{s_{1}};q
\right)_{\infty}
}
{
\left(
q^{\frac12} t^{2}\frac{s_{1}}{s_{2}};q
\right)_{\infty}
\left(
q^{\frac12} t^{2}\frac{s_{2}}{s_{1}};q
\right)_{\infty}
}
\nonumber\\
&=
\frac12 
\frac{(q^{-\frac12} t^{-2};q)_{\infty}^2}
{(q^{\frac12} t^2;q)_{\infty}^2}
\sum_{m=0}^{\infty}
\frac{(q^{1+m};q)_{\infty} (q^{1+m+n};q)_{\infty}}
{(q^{-\frac12+m}t^{-2};q)_{\infty} (q^{-\frac12+m+n}t^{-2};q)_{\infty}}
q^{m+\frac{n}{2}}t^{4m+2n}. 
 \end{align}
 }
\begin{align}
\label{4du2_hINDEXNsum}
\mathbb{II}_{\mathcal{N}'}^{\textrm{4d $U(2)$}}(t;q)
&=\frac12 
\frac{(q^{-\frac12}t^{-2};q)_{\infty}^2}
{(q^{\frac12}t^2 ;q)_{\infty}^2}
\sum_{m=0}^{\infty}\frac{(q^{1+m};q)_{\infty}^2}{(q^{-\frac12+m}t^{-2};q)_{\infty}^2}q^{m}t^{4m}. 
\end{align}
The first term in the sum takes the form of a square of the half-index $\mathbb{II}_{\mathcal{N}'}^{\textrm{4d $U(1)$}}$ 
of the Neumann boundary condition $\mathcal{N}'$ for 4d $\mathcal{N}=4$ $U(1)$ gauge theory. We interpret this sum as associated to a Higgsing 
of the bulk theory from $U(2)$ to $U(1) \times U(1)$.

These half-indices can be generalized to higher rank cases. 
The half-indices of Neumann b.c. $\mathcal{N}'$ for 4d $\mathcal{N}=4$ $U(N)$ gauge theory is 
\begin{align}
\label{4duN_hINDEX_N1}
\mathbb{II}_{\mathcal{N}'}^{\textrm{4d $U(N)$}}(t;q)
&=
\frac{1}{N!}
\frac{(q)_{\infty}^{N}}
{(q^{\frac12} t^{2};q)_{\infty}^{N}}
\oint 
\prod_{i=1}^{N}
\frac{ds_{i}}{2\pi is_{i}}
\prod_{i\neq j}
\frac{\left(
\frac{s_{i}}{s_{j}};q
\right)_{\infty}
}
{
\left(
q^{\frac12} t^{2}\frac{s_{i}}{s_{j}};q
\right)_{\infty}
}.
\end{align}
The half-index of Dirichlet b.c. $\mathcal{D}$ for 4d $\mathcal{N}=4$ $U(N)$ gauge theory is 
\begin{align}
\label{4duN_hINDEX_D1}
\mathbb{II}_{\mathcal{D}}^{\textrm{4d $U(N)$}}(t;q)
&=\frac{(q)_{\infty}^{N} }{(q^{\frac12} t^2;q)_{\infty}^{N}}
\prod_{i\neq j}
\frac{
\left(q\frac{s_{i}}{s_{j}};q\right)_{\infty}}
{\left(q^{\frac12} t^2 \frac{s_{i}}{s_{j}};q\right)_{\infty}}. 
\end{align}
We propose that the half-index for Nahm pole boundary conditions 
in 4d $\mathcal{N}=4$ $U(N)$ gauge theory takes the form 
\begin{align}
\label{4duN_nahm}
\mathbb{II}_{\textrm{Nahm}}^{\textrm{4d $U(N)$}}(t;q)
&=
\prod_{k=1}^{N}
\frac{(q^{\frac{k+1}{2}}t^{2(k-1)};q)_{\infty}}
{(q^{\frac{k}{2}}t^{2k};q)_{\infty}}. 
\end{align}
This results from the Higgsing procedure applied to the Dirichlet half-index, setting $\frac{s_i}{s_{i+1}} = q^{\frac12} t^2$ and 
removing an appropriate collection of 3d free hypers. For some low rank cases we can explicitly check that 
the Nahm pole half-index (\ref{4duN_nahm}) coincides with the Neumann half-index (\ref{4duN_hINDEX_N1}). 
We expect this equality to hold for all $N$. We can write similar expressions for the index in the presence of 
boundary 't Hooft operators. 

Similar expressions for the mirror boundary conditions, 
i.e. $\mathcal{N}$, $\mathcal{D}'$ and Nahm$'$ can be obtained by setting $t\rightarrow t^{-1}$.

\subsubsection{Quarter-indices}
\label{sec_4dq_index}

The gauge theory definition of a junction of two boundary conditions $\mathcal{B}$ and $\mathcal{B}'$ 
will in general require extra data besides the choice of two boundary conditions. After all, extra two-dimensional 
degrees of freedom may be added to the junction. It may also be possible to find non-trivial quarter-BPS singular 
field configurations which define a junction as a disorder defect. 

If some gauge symmetry is preserved at the junction, we must require the cancellation of gauge anomalies there. 
If the boundary conditions are defined by singular field configurations, we must insure that these field configurations can be extended to the junction in a supersymmetric fashion. 

The brane constructions will engineer some specific junctions. We will determine them by a combination of 
string theory lore and judicious guesswork. We will refer to the quarter-index for such junctions as $\mathbb{IV}_{\mathcal{B}\mathcal{B}'}$, indicating the boundary conditions and leaving the choice of junction implicit.

Junctions engineered from NS5 and D5$'$ (or NS5$'$ and D5) intersections appear to exist and have a nice 
gauge theory description for all choices of numbers of D3-branes ending on the fivebrane system. 

Junctions engineered from NS5 and NS5$'$ or from D5 and D5$'$ appear to be somewhat more constrained.
We find a balancing condition on the number of D3-branes ending on such fivebrane systems, 
which can be motivated by linking number conservation in string theory, anomaly cancellation or compatibility of Nahm pole 
boundary conditions in gauge theory. This does not mean necessarily that no such junctions may exist in gauge theory, 
but only that string theory does not provide a simple canonical choice with good duality properties. 

The anomaly cancellation/ Nahm pole constraints do not exist in Abelian gauge theory, 
so we can provide here the quarter-indices for all simple junctions of elementary boundary conditions 
in 4d $\mathcal{N}=4$ $U(1)$ gauge theory:
\begin{align}
\label{qindex_u1nn'}
\mathbb{IV}_{\mathcal{N}\mathcal{N}'}^{\textrm{4d $U(1)$}}(q)
&=(q)_{\infty},
\\
\label{qindex_u1dd'}
\mathbb{IV}_{\mathcal{D}\mathcal{D}'}^{\textrm{4d $U(1)$}}(q)
&=(q)_{\infty},
\\
\label{qindex_u1nd'}
\mathbb{IV}_{\mathcal{N}\mathcal{D}'}^{\textrm{4d $U(1)$}}(t;q)
&=\frac{1}{(q^{\frac12}t^{-2};q)_{\infty}},
\\
\label{qindex_u1n'd}
\mathbb{IV}_{\mathcal{N}'\mathcal{D}}^{\textrm{4d $U(1)$}}(t;q)
&=\frac{1}{(q^{\frac12}t^{2};q)_{\infty}}.
\end{align}
%
In the H-twist limit $t\rightarrow q^{\frac14}$ 
the NS5$'$-D5 quarter-index (\ref{qindex_u1n'd}) becomes the vacuum character for a single $U(1)$ current: 
\begin{align}
\label{qindex_u1H}
\mathbb{IV}_{\mathcal{N}' \mathcal{D}}^{\textrm{4d $U(1)$}}(q)
&=\mathbb{IV}_{\mathcal{N}' \mathcal{D}}^{\textrm{4d $U(1)$}}(t=q^{\frac14};q)
\nonumber\\
&=\frac{1}{(q)_{\infty}}
=\chi_{\mathfrak{u}(1)}(q).
\end{align}
This is indeed the character of the VOA which emerges upon deformation of this system, denoted as $Y_{0,0,1}$ in \cite{Gaiotto:2017euk}. Similarly the quarter-index (\ref{qindex_u1nd'}) reduces to the vacuum character of $U(1)$ current in the C-twist limit $t\rightarrow q^{-\frac14}$.

For $U(N)$ and Neumann/Dirichlet$'$ junctions we propose 
\begin{align}
\label{qindex_nn'}
\mathbb{IV}_{\mathcal{N}\mathcal{D}'}^{\textrm{4d $U(N)$}}(t,s_{i};q)
&=
\frac{1}{(q^{\frac12} t^{-2};q)_{\infty}^{N}}
\prod_{i\neq j}
\frac{1}
{\left(q^{\frac12} t^{-2} \frac{s_{i}}{s_{j}};q\right)_{\infty}},
\\
\label{qindex_n'd}
\mathbb{IV}_{\mathcal{N'}\mathcal{D}}^{\textrm{4d $U(N)$}}(t,s_{i};q)
&=
\frac{1}{(q^{\frac12} t^2;q)_{\infty}^{N}}
\prod_{i\neq j}
\frac{
1}
{\left(q^{\frac12} t^2 \frac{s_{i}}{s_{j}};q\right)_{\infty}}.
\end{align}

For $U(N)$ and Neumann/Nahm$'$ junctions we propose 
\begin{align}
\label{qindex_nn'}
\mathbb{IV}_{\mathcal{N}\textrm{Nahm}'}^{\textrm{4d $U(N)$}}(t;q)
&=
\prod_{k=1}^{N} \frac{1}{(q^{\frac{k}{2}}t^{-2k};q)_{\infty}},
\\
\label{qindex_n'd}
\mathbb{IV}_{\mathcal{N'}\textrm{Nahm}}^{\textrm{4d $U(N)$}}(t;q)
&=
\prod_{k=1}^{N} \frac{1}{(q^{\frac{k}{2}}t^{2k};q)_{\infty}}.
\end{align}

Again, we can reproduce the proposed indices for Neumann$'$/Nahm junctions 
by a Higgsing procedure applied to the Neumann$'$/Dirichlet junctions. For example for $U(2)$ we have 
\begin{align}
\mathbb{IV}_{\mathcal{N'}\mathcal{D}}^{\textrm{4d $U(2)$}}(t,s_{i};q)
&=
\frac{1}{(q^{\frac12} t^2;q)_{\infty}^{2}}
\frac{
1}
{\left(q^{\frac12} t^2 \frac{s_1}{s_2};q\right)_{\infty}\left(q^{\frac12} t^2 \frac{s_2}{s_1};q\right)_{\infty}}
\end{align}
and the pole at $s_1/s_2 = q^{\frac12} t^2$ is associated to the RG flow to 
a  Neumann$'$/Nahm junction. We get the desired result by stripping off a half-index 
with Neumann boundary conditions for the extra decoupled hypermultiplet localized at the Nahm 
boundary.  


%
Taking the H-twist limit of the quarter-index (\ref{qindex_n'd}), 
we obtain the vacuum character of the W-algebra $\mathcal{W}_{N}$,
denoted as $Y_{0,0,N}$ in \cite{Gaiotto:2017euk}
\begin{align}
\label{qindex_uNH}
\mathbb{IV}_{\mathcal{N}'\mathrm{Nahm}}^{\textrm{4d $U(N)$}}(t=q^{\frac14};q)
=\prod_{k=1}^{N}\frac{1}{(q^{k};q)_{\infty}}
=\chi_{\mathcal{W}_{N}}(q)
\end{align}
while the Neumann/Dirichlet$'$ junction gives the character of a Kac-Moody algebra \cite{Creutzig:2017uxh}
\begin{align}
\label{qindex_uNH2}
\mathbb{IV}_{\mathcal{N}'\mathcal{D}}^{\textrm{4d $U(N)$}}(t=q^{\frac14},s_{i};q)
=\frac{1}{(q;q)_{\infty}^{N}}
\prod_{i\neq j}
\frac{
1}
{\left(q \frac{s_{i}}{s_{j}};q\right)_{\infty}}
=\chi_{\mathfrak{\widehat u}(N)}(q).
\end{align}

\subsection{3d $\mathcal{N}=4$ indices}
\label{sec_3dINDEX}

In order to compute the half- and quarter- indices for four-dimensional gauge theories 
with more interesting interfaces and junctions we need to understand the indices and half-indices of 
the three-dimensional matter fields involved in their definition: hypermultiplets and twisted hypermultiplets. 

The 3d $\mathcal{N}=4$ hypermultiplet involves 
a pair of complex scalars $\mathbb{H}$, $\widetilde{\mathbb{H}}$ 
forming a doublet of $SU(2)_{H}$ 
and a pair of complex fermions $\psi_{+}^{\mathbb{H}}$, $\psi_{+}^{\widetilde{\mathbb{H}}}$ 
forming a doublet of $SU(2)_{C}$. 
The 3d $\mathcal{N}=4$ hypermultiplet has charges 
\begin{align}
\label{3dn4_hm_ch}
\begin{array}{c|cccccc}
&\mathbb{H}&\widetilde{\mathbb{H}}&\psi_{+}^{\mathbb{H}}&\psi_{+}^{\widetilde{\mathbb{H}}}&\overline{\psi}_{-}^{\mathbb{H}}&\overline{\psi}_{-}^{\widetilde{\mathbb{H}}} \\ \hline
U(1)_{C}&0&0&-&-&+&+ \\
U(1)_{H}&+&+&0&0&0&0 \\
\end{array}
\end{align}
The 3d $\mathcal{N}=4$ Abelian vector multiplet consists of 
a 3d $U(1)$ gauge field $A_{\mu}^{\textrm{3d}}$, 
real and complex scalars $\sigma$, $\varphi$ forming the $SU(2)_{C}$ triplet, 
and two complex fermions $(\lambda_{\alpha}^{\textrm{3d}}, \eta_{\alpha}^{\textrm{3d}})$. 
They have charges as follows:
\begin{align}
\label{3dn4_vm_ch}
\begin{array}{c|ccccccc}
&A_{\mu}^{\textrm{3d}}&\sigma&\varphi&\lambda_{\pm}^{\textrm{3d}}&\overline{\lambda}_{\pm}^{\textrm{3d}}
&\eta_{\pm}^{\textrm{3d}}&\overline{\eta}_{\pm}^{\textrm{3d}} \\ \hline
U(1)_{C}&0&0&2&+&-&+&- \\
U(1)_{H}&0&0&0&+&-&-&+ \\
\end{array}
\end{align}
Exchanging the $U(1)_{H}$ and $U(1)_{C}$ charges of the hypermultiplet and vector multiplet, 
we obtain the twisted hypermultiplet and twisted vector multiplet respectively.

\subsubsection{Indices}
\label{sec_3dindex}
The index of a 3d $\mathcal{N}=2$ chiral multiplet is
\begin{align}
\label{3dcm}
\mathbb{I}^{\textrm{3d CM}}(x;q)&=
\frac{(qx^{-1};q)_{\infty}}
{(x;q)_{\infty}}.
\end{align}
This counts words made out of derivatives of 
complex scalar and those of fermions included in the 3d $\mathcal{N}=2$ chiral multiplet.

The operators from 3d $\mathcal{N}=4$ hypermultiplet which contribute to index are  
\begin{align}
\label{3dn4_hm_ch2}
\begin{array}{c|cc|cc}
&\partial_{z}^{n}\mathbb{H}&\partial_{z}^{n}\widetilde{\mathbb{H}}&
\partial_{z}^{n}\overline{\psi}_{-}^{\mathbb{H}}&\partial_{z}^{n}\overline{\psi}_{-}^{\widetilde{\mathbb{H}}} \\ \hline
U(1)_{f}&+&-&+&- \\
U(1)_{J}&n&n&n+\frac12&n+\frac12 \\ 
U(1)_{C}&0&0&+&+ \\
U(1)_{H}&+&+&0&0  \\
\textrm{fugacity}&q^{n+\frac14}tx&q^{n+\frac14}tx^{-1}
&-q^{n+\frac34}t^{-1}x&-q^{n+\frac34}t^{-1}x^{-1} \\
\end{array}
\end{align}
The index for 3d $\mathcal{N}=4$ hypermultiplet is 
\begin{align}
\label{3dhm}
\mathbb{I}^{\textrm{3d HM}}(t,x;q)
&=\frac{(q^{\frac34}t^{-1}x;q)_{\infty}(q^{\frac34}t^{-1}x^{-1};q)_{\infty}}
{(q^{\frac14}t x;q)_{\infty}(q^{\frac14}tx^{-1};q)_{\infty}}. 
\end{align}
It has an expansion
\begin{align}
\label{thm_series}
\mathbb{I}^{\textrm{3d HM}}(t,x;q)
&=\sum_{n=0}^{\infty}\sum_{k=0}^{n}
\frac{(q^{\frac12}t^{-2};q)_{k} (q^{\frac12} t^{-2};q)_{n-k}}
{(q)_{k}(q)_{n-k}}
q^{\frac{n}{4}}t^{n}x^{n-2k}. 
\end{align}
In the H-twist limit $t\rightarrow q^{\frac14}$, 
the index (\ref{3dhm}) becomes 1. 
This reflects the fact that 
free hypermultiplet has no Coulomb branch local operators surviving in the H-twist.

Similarly, the operators of twisted hypermultiplet which contribute to index are
\begin{align}
\label{3dn4_thm_ch}
\begin{array}{c|cc|cc}
&\partial_{z}^{n}\mathbb{T}&\partial_{z}^{n}\widetilde{\mathbb{T}}&
\partial_{z}^{n}\overline{\widetilde{\psi}}_{-}^{\mathbb{T}}&\partial_{z}^{n}\overline{\widetilde{\psi}}_{-}^{\widetilde{\mathbb{T}}} \\ \hline
U(1)_{f}&+&-&+&- \\
U(1)_{J}&n&n&n+\frac12&n+\frac12 \\ 
U(1)_{C}&+&+&0&0 \\
U(1)_{H}&0&0&+&+  \\
\textrm{fugacity}&q^{n+\frac14}t^{-1}x&q^{n+\frac14}t^{-1}x^{-1}&-q^{n+\frac34}tx&-q^{n+\frac34}tx^{-1} \\
\end{array}
\end{align}
The index for 3d $\mathcal{N}=4$ twisted hypermultiplet is
\begin{align}
\label{3dthm}
\mathbb{I}^{\textrm{3d tHM}}(t,x;q)
&=\frac{(q^{\frac34}tx;q)_{\infty}(q^{\frac34}tx^{-1};q)_{\infty}}
{(q^{\frac14}t^{-1}x;q)_{\infty}(q^{\frac14}t^{-1}x^{-1};q)_{\infty}}.
\end{align}
This can be obtained from the hypermultiplet index (\ref{3dhm}) 
by setting $t\rightarrow t^{-1}$. 
It has an expansion
\begin{align}
\label{thm_series}
\mathbb{I}^{\textrm{3d tHM}}(t,x;q)
&=\sum_{n=0}^{\infty}\sum_{k=0}^{n}
\frac{(q^{\frac12}t^2;q)_{k} (q^{\frac12} t^2;q)_{n-k}}
{(q)_{k}(q)_{n-k}}
q^{\frac{n}{4}}t^{-n}x^{n-2k}. 
\end{align}


For completeness, we also describe some properties of 3d gauge multiplets. 

The charges of operators in 3d $\mathcal{N}=4$ vector multiplet 
which contribute to the index are 
\begin{align}
\label{3dn4_vm_ch2}
\begin{array}{c|cc|cc}
&D_{z}^{n}(\sigma+i\rho)&D_{z}^{n}\varphi&
D_{z}^{n}\overline{\lambda}_{-}^{\textrm{3d}}&D_{z}^{n}\overline{\eta}_{-}^{\textrm{3d}} \\ \hline
G&\textrm{adj}&\textrm{adj}&\textrm{adj}&\textrm{adj} \\
U(1)_{J}&n&n&n+\frac12&n+\frac12 \\ 
U(1)_{C}&0&2&-&- \\
U(1)_{H}&0&0&-&+  \\
\textrm{fugacity}
&q^{n}s_{\alpha}
&q^{n+\frac12}t^{-2}s_{\alpha}
&-q^{n}s_{\alpha}
&-q^{n+\frac12}t^{2}s_{\alpha} \\
\end{array}
\end{align}
The index for 3d $\mathcal{N}=4$ $U(1)$ vector multiplet is 
\begin{align}
\label{3du1vm}
\mathbb{I}^{\textrm{3d $U(1)$}}(t;q)
&=\frac{(q^{\frac12} t^{2};q)_{\infty}}
{(q^{\frac12}t^{-2};q)_{\infty}}\oint \frac{ds}{2\pi is}.
\end{align}
%
The index for 3d $\mathcal{N}=4$ $U(N)$ vector multiplet is 
\begin{align}
\label{3duNvm}
\mathbb{I}^{\textrm{3d $U(N)$}}(t;q)
&=
\frac{1}{N!}\frac{(q^{\frac12}t^2;q)_{\infty}^{N}}
{(q^{\frac12}t^{-2};q)_{\infty}^{N}}
\oint\prod_{i=1}^{N}\frac{ds_{i}}{2\pi is_{i}}
\prod_{i\neq j}
\left(1-\frac{s_{i}}{s_{j}}\right)
\frac{
\left(q^{\frac12}t^2 \frac{s_{i}}{s_{j}};q\right)_{\infty}
}{
\left(q^{\frac12}t^{-2} \frac{s_{i}}{s_{j}};q\right)_{\infty}
}.
\end{align}
%

As a consistency check, notice that the trivial interface in 4d gauge theory can be obtained starting from two Dirichlet boundary conditions and gauging the diagonal boundary global symmetry. Correspondingly, the full index (\ref{4duN_INDEX}) of 4d $\mathcal{N}=4$ $U(N)$ gauge theory 
can be recovered from two copies of the Dirichlet half-index (\ref{4duN_hINDEX_D1}) by gauging with the measure for a 3d $U(N)$ gauge theory:
\begin{align}
\label{4dindexgauging}
\mathbb{I}^{\textrm{4d $U(N)$}}
&=
\underbrace{
\frac{1}{N!}\frac{(q^{\frac12}t^2;q)_{\infty}^{N}}
{(q^{\frac12}t^{-2};q)_{\infty}^{N}}\oint \prod_{i=1}^{N} \frac{ds_{i}}{2\pi is_{i}}
\prod_{i\neq j}\left(1-\frac{s_{i}}{s_{j}}\right)
\frac{\left(q^{\frac12}t^2\frac{s_{i}}{s_{j}};q\right)_{\infty}}
{\left(q^{\frac12}t^{-2}\frac{s_{i}}{s_{j}};q\right)_{\infty}}
}_{\mathbb{I}^{\textrm{3d $U(N)$}}}
\nonumber\\
&\times 
\underbrace{
\frac{(q)_{\infty}^{N}}{(q^{\frac12}t^2;q)_{\infty}^{N}}
\prod_{i\neq j}\frac{\left(q\frac{s_{i}}{s_{j}};q\right)_{\infty}}
{\left(q^{\frac12}t^{2}\frac{s_{i}}{s_{j}};q\right)_{\infty}}
}_{\mathbb{II}_{\mathcal{D}}^{\textrm{4d $U(N)$}}}
\cdot 
\underbrace{
\frac{(q)_{\infty}^{N}}{(q^{\frac12}t^2;q)_{\infty}^{N}}
\prod_{i\neq j}\frac{\left(q\frac{s_{i}}{s_{j}};q\right)_{\infty}}
{\left(q^{\frac12}t^{2}\frac{s_{i}}{s_{j}};q\right)_{\infty}}
}_{\mathbb{II}_{\mathcal{D}}^{\textrm{4d $U(N)$}}}. 
\end{align}

Similarly, Neumann boundary conditions can be obtained by gauging the global symmetry of Dirichlet boundary conditions.
Correspondingly, the half-index of Neumann b.c. $\mathcal{N}$ for 4d $\mathcal{N}=4$ $U(N)$ gauge theory can be recovered 
from the Dirichlet half-index (\ref{4duN_hINDEX_D1}) by gauging with the measure for a 3d $U(N)$ gauge theory:
\begin{align}
\label{4dhindexgauging}
\mathbb{II}_{\mathcal{N}}^{\textrm{4d $U(N)$}}
&=
\underbrace{
\frac{1}{N!}\frac{(q^{\frac12}t^2;q)_{\infty}^{N}}
{(q^{\frac12}t^{-2};q)_{\infty}^{N}}\oint \prod_{i=1}^{N} \frac{ds_{i}}{2\pi is_{i}}
\prod_{i\neq j}\left(1-\frac{s_{i}}{s_{j}}\right)
\frac{\left(q^{\frac12}t^2\frac{s_{i}}{s_{j}};q\right)_{\infty}}
{\left(q^{\frac12}t^{-2}\frac{s_{i}}{s_{j}};q\right)_{\infty}}
}_{\mathbb{I}^{\textrm{3d $U(N)$}}}
\nonumber\\
&\times 
\underbrace{
\frac{(q)_{\infty}^{N}}{(q^{\frac12}t^2;q)_{\infty}^{N}}
\prod_{i\neq j}\frac{\left(q\frac{s_{i}}{s_{j}};q\right)_{\infty}}
{\left(q^{\frac12}t^{2}\frac{s_{i}}{s_{j}};q\right)_{\infty}}
}_{\mathbb{II}_{\mathcal{D}}^{\textrm{4d $U(N)$}}}. 
\end{align}

Notice that a similar prescription can be used to give us the half-index of general enriched Neumann boundary conditions, by 3d gauging the 
product of the half-index for Dirichlet boundary conditions and the 3d index for whatever extra 3d degrees of freedom we want to include at the boundary. Observe that the index for 3d gauge theories would generically include contributions from monopole operators. 
Here these contributions drop out because the half-index for Dirichlet boundary conditions vanishes in the presence of a 
monopole background \cite{Dimofte:2011py}.

The charges of operators in 3d $\mathcal{N}=4$ twisted vector multiplet are given by
\begin{align}
\label{3dn4_tvm_ch}
\begin{array}{c|cc|cc}
&D_{z}^{n}(\widetilde{\sigma}+i\widetilde{\rho})&D_{z}^{n}\widetilde{\varphi}&
D_{z}^{n}\overline{\widetilde{\lambda}}_{-}^{\textrm{3d}}&D_{z}^{n}\overline{\widetilde{\eta}}_{-}^{\textrm{3d}} \\ \hline
G&\textrm{adj}&\textrm{adj}&\textrm{adj}&\textrm{adj} \\
U(1)_{J}&n&n&n+\frac12&n+\frac12 \\ 
U(1)_{C}&0&0&+&+ \\
U(1)_{H}&0&2&+&-  \\
\textrm{fugacity}
&q^{n}s_{\alpha}
&q^{n+\frac12}t^{2}s_{\alpha}
&-q^{n}s_{\alpha}
&-q^{n+\frac12}t^{-2}s_{\alpha} \\
\end{array}
\end{align}
The index of 3d $\mathcal{N}=4$ twisted vector multiplet is obtained 
by setting $t$ $\rightarrow$ $t^{-1}$ for the index of 3d $\mathcal{N}=4$ vector multiplet.

\subsubsection{$(0,4)$ half-indices of 3d matter multiplet}
\label{sec_04half_matter}
Consider now $\mathcal{N}=(0,4)$ supersymmetric boundary conditions for 3d $\mathcal{N}=4$ hypermultiplet. 
The simplest, free examples of boundary conditions are 
Neumann b.c. $N'$ and Dirichlet b.c. $D'$ respectively:
\begin{align}
\label{3dhm_bc}
\begin{array}{ccc}
N':& \partial_{2} \mathbb{H}|_{\partial}=0,& \partial_{2}\widetilde{\mathbb{H}}|_{\partial}=0, \\
D':& \partial_{\mu} \mathbb{H}|_{\partial}=0,& \partial_{\mu}\widetilde{\mathbb{H}}|_{\partial}=0, \\
\end{array}
\qquad \mu=0,1
\end{align}
We will conjecturally encounter these boundary conditions at simple brane junctions.

The half-index of Neumann b.c. for 3d $\mathcal{N}=4$ hypermultiplet is 
\begin{align}
\label{hindex_HM_N}
\mathbb{II}_{N}^{\textrm{3d HM}}(t,x;q)
&=\frac{1}{(q^{\frac14} tx;q)_{\infty}(q^{\frac14} tx^{-1};q)_{\infty}}. 
\end{align}
%
The Neumann b.c. for 3d $\mathcal{N}=4$ hypermultiplet can be deformed to be compatible with the H-twist (mirror Rozanski-Witten)
and the deformed boundary condition supports the VOA Sb of symplectic bosons $X(z)$ and $Y(z)$ with OPE
\cite{Gaiotto:2017euk, Costello:2018fnz} 
\begin{align}
\label{sb_ope}
X(z)Y(w)&\sim \frac{1}{z-w}
\end{align}
and conformal dimension $1/2$. 
In the H-twist limit $t\rightarrow q^{\frac14}$ of the half-index (\ref{hindex_HM_N}) 
we obtain the vacuum character for the symplectic boson VOA Sb:
\begin{align}
\label{hindex_HM_N_H}
\mathbb{II}_{N^{(H)}}^{\textrm{3d HM}}(x;q)
&=\mathbb{II}_{N}^{\textrm{HM}}(t=q^{\frac14},x;q)\nonumber\\
&=\frac{1}{(q^{\frac12}x;q)_{\infty}(q^{\frac12}x^{-1};q)_{\infty}}
=\chi_{\mathrm{Sb}}(x;q). 
\end{align}
%
The half-index of Dirichlet b.c. for 3d $\mathcal{N}=4$ hypermultiplet is 
\begin{align}
\label{hindex_HM_D}
\mathbb{II}_{D}^{\textrm{3d HM}}(t,x;q)
&=(q^{\frac34}t^{-1}x;q)_{\infty}
(q^{\frac34}t^{-1}x^{-1};q)_{\infty}. 
\end{align}
It can be expanded as 
\begin{align}
\label{dirthm_sum1}
\mathbb{II}_{D}^{\textrm{3d HM}}(t,x;q)
&=
\frac{1}{(q)_{\infty}^2}
\sum_{n=0}^{\infty}\sum_{k=0}^{n}
(q^{1+k};q)_{\infty}
(q^{1+n-k};q)_{\infty} 
(-1)^n 
x^{-n+2k} t^{-n}
q^{\frac{n^2-n}{2}+k(k-n)+\frac{3n}{4}}. 
\end{align}
%
The Dirichlet b.c. for 3d $\mathcal{N}=4$ hypermultiplet can be deformed to be compatible with the C-twist (Rozanski-Witten)
and the deformed boundary condition supports the VOA Fc of fermionic currents $x(z)$ and $y(z)$ with OPE
\begin{align}
\label{fc_ope}
x(z)y(w)&\sim \frac{1}{(z-w)^2}
\end{align}
and conformal dimension $1$. 
This can be equivalently defined as $\mathfrak{psu}(1|1)$ Kac-Moody VOA. 
In the C-twist limit $t\rightarrow q^{-\frac14}$ of the half-index (\ref{hindex_HM_D}) 
we obtain the vacuum character for the fermionic current VOA Fc:
\begin{align}
\label{hindex_HM_D_C}
\mathbb{II}_{D^{(C)}}^{\textrm{3d HM}}(x;q)&=
\mathbb{II}_{D}^{\textrm{HM}}(t=q^{-\frac14},x;q)\nonumber\\
&=(qx;q)_{\infty}(qx^{-1};q)_{\infty}
=\chi_{Fc}(x;q).
\end{align}

The half-index of Neumann b.c. for 3d $\mathcal{N}=4$ twisted hypermultiplet is 
\begin{align}
\label{hindex_tHM_N}
\mathbb{II}_{N}^{\textrm{3d tHM}}(t,x;q)
&=\frac{1}{(q^{\frac14} t^{-1}x;q)_{\infty}
(q^{\frac14} t^{-1}x^{-1};q)_{\infty}}
\end{align}
and the half-index for Dirichlet b.c. for 3d $\mathcal{N}=4$ twisted hypermultiplet is 
\begin{align}
\label{hindex_tHM_D}
\mathbb{II}_{D}^{\textrm{3d tHM}}(t,x;q)
&=(q^{\frac34}tx;q)_{\infty}(q^{\frac34}tx^{-1};q)_{\infty}. 
\end{align}
By setting $t\rightarrow t^{-1}$, 
the half-indices of 3d twisted hypermultiplets convert into those of 3d hypermultplets and vice versa. 
Therefore the C-twist limit $t\rightarrow q^{-\frac14}$ of Dirichlet b.c. for 3d twisted hyper leads to the vacuum character (\ref{hindex_HM_D}) 
of symplectic boson VOA Sb while the H-twist limit $t\rightarrow q^{\frac14}$ of Neumann b.c. for 3d twisted hyper reproduces
the vacuum character (\ref{hindex_HM_D_C}) of fermionic current VOA Fc.

\subsubsection{$(0,4)$ Neumann b.c. for 3d vector multiplet}
\label{sec_hindex_vec_N}

For completeness, we also describe some properties of 3d gauge multiplets. 

The 3d $\mathcal{N}=4$ $U(1)$ vector multiplet admits two types of $\mathcal{N}=(0,4)$ supersymmetric boundary conditions, 
i.e. Neumann boundary condition $\mathcal{N}'$ and Dirichlet boundary condition $\mathcal{D}'$
\begin{align}
\label{3du1vm_bc1}
\begin{array}{cccc}
\mathcal{N}':& F_{2\mu}|_{\partial}=0,& D_{\mu}\sigma|_{\partial}=0,& D_{\mu}\varphi|_{\partial}=0,  \\
\mathcal{D}':& F_{\mu\nu}|_{\partial}=0,& D_{2}\sigma|_{\partial}=0,& D_{2}\varphi|_{\partial}=0, \\
\end{array}
\qquad \mu,\nu=0,1
\end{align}

Let us discuss the half-index of $\mathcal{N}=(0,4)$ Neumann boundary condition $\mathcal{N}'$ for the 3d $\mathcal{N}=4$ vector multiplet. 
\footnote{
The half-index of $\mathcal{N}=(0,2)$ Neumann b.c. for 3d $\mathcal{N}=2$ gauge multiplets is studied in 
\cite{Gadde:2013wq, Gadde:2013sca, Yoshida:2014ssa, Dimofte:2017tpi}.
}

The half-index for 3d $\mathcal{N}=4$ $U(1)$ vector multiplet obeying $\mathcal{N}=(0,4)$ Neumann boundary condition $\mathcal{N}$ is 
\begin{align}
\label{hindex_u1_N}
\mathbb{II}_{\mathcal{N}'}^{\textrm{3d $U(1)$}}(t;q)
&=(q)_{\infty}(q^{\frac12}t^2;q)_{\infty}\oint \frac{ds}{2\pi is}. 
\end{align}

The half-index of Neumann boundary condition $\mathcal{N}'$ for 3d $\mathcal{N}=4$ $U(N)$ vector multiplet takes the form
\begin{align}
\label{hindex_vm_N}
\mathbb{II}_{\mathcal{N}'}^{\textrm{3d $U(N)$}}(t;q)
&=
\frac{1}{N!}
\left[(q)_{\infty}
(q^{\frac12}t^{2};q)_{\infty}
\right]^{N}
\oint \prod_{i=1}^{N}
\frac{ds_{i}}{2\pi s_{i}}
\prod_{i\neq j}
\left(\frac{s_{i}}{s_{j}};q\right)_{\infty}
\left(q^{\frac12}t^{2}\frac{s_{i}}{s_{j}};q\right)_{\infty}. 
\end{align}
In this paper we focus on the case where the integration contour can be taken as a unit torus $\mathbb{T}^{N}$. 
However, it is important to note that 
when we have enriched Neumann boundary conditions including 2d bosonic matter fields, 
one has to carefully choose the integration contour of Neumann half-index of 3d vector multiplet in (\ref{hindex_u1_N}) and (\ref{hindex_vm_N}). 
In this note we only encounter situations where auxiliary 
2d matter consists of Fermi multiplets only and thus does not contribute any interesting poles to the integrand. 

The main challenge in computing the half-index of 3d Dirichlet boundary conditions is that it includes contributions from boundary monopole operators, which are not present, instead, in the half-index of Neumann boundary conditions. \footnote{But may be present for enriched Neumann b.c. if bosonic boundary matter is present.} We will not need to consider such monopole contributions in the 4d gauge theory examples 
considered in this paper, though. 

In a sector of zero monopole charge, the half-index of $\mathcal{N}=(0,4)$ Dirichlet b.c. for 3d $\mathcal{N}=4$ $U(1)$ vector multiplet takes the form  
\begin{align}
\label{hindex_3du1vmD0}
\mathbb{II}_{\mathcal{D}'}^{\textrm{3d $U(1)$}}(t;q)
&=\frac{1}{(q)_{\infty}(q^{\frac12}t^{-2};q)_{\infty}}.
\end{align}

\subsection{2d $\mathcal{N}=(0,4)$ indices}
\label{sec_2d04_index}
The index of 2d theory is identified with the flavored elliptic genus 
\cite{Gadde:2013wq, Gadde:2013ftv, Benini:2013nda, Benini:2013xpa, Gadde:2015tra, Putrov:2015jpa, KIm:2016foj, Kim:2018gak}. 

In the rest of the paper we will only need the index of Fermi multiplets. The index of $\mathcal{N}=(0,2)$ Fermi multiplet of charge $+1$ under $U(1)_{x}$ flavor symmetry and $U(1)_{R}$ R-charge $1$ is given by
\begin{align}
\label{fm_index}
F(x)=(x;q)_{\infty} (qx^{-1};q)_{\infty}. 
\end{align}
This counts the local operators as a pair of left-moving fermions $\gamma_{-}$, $\overline{\gamma}_{-}$
and their derivatives. 
The index of $\mathcal{N}=(0,2)$ Fermi multiplet of charge $+1$ under $U(1)_{x}$ flavor symmetry and $U(1)_{R}$ R-charge $0$ has an expansion
\begin{align}
\label{fm_series}
F(q^{\frac12}x;q)
&=\sum_{m=0}^{\infty}\sum_{n=0}^{\infty}
\frac{(-1)^{m+n} q^{\frac{m^{2}+n^{2}}{2}}}
{(q;q)_{m}(q;q)_{n}}x^{m-n}. 
\end{align}

\subsection{(Boundary) Anomalies}
\label{sec_bdyanomaly}
$\mathcal{N}=(0,4)$ supersymmetric boundary conditions for 3d $\mathcal{N}=4$ can involve anomalies. 
We review the results in \cite{Dimofte:2017tpi}, 
where the boundary anomalies for $\mathcal{N}=(0,2)$ supersymmetric b.c. were examined. \footnote{
We also refer the reader to the results in \cite{Hanany:2018hlz} 
where the gauge anomalies of $\mathcal{N}=(0,4)$ supersymmetric b.c. are specified by linking numbers in brane construction.} 

Let ${\bf f}$ be the field strength for the $U(1)$ symmetry which rotates fermions with charge $1$. 
The anomaly polynomial $\mathcal{I}$ is given by \cite{Dimofte:2017tpi}
\begin{align}
\label{AN_bdy1}
\begin{array}{c|c}
\textrm{anomaly contribution}&\textrm{anomaly polynomial $\mathcal{I}$} \\ \hline
\textrm{2d left-handed chiral fermion}&{\bf f}^2 \\  
\textrm{2d right-handed chiral fermion}&-{\bf f}^2 \\  
\textrm{3d $U(1)$ Chern-Simons coupling of level $k$}&k{\bf f}^2 \\  
\textrm{3d fermion with b.c. $\psi_{+}|_{\partial}=0$}&\frac12{\bf f}^2 \\  
\textrm{3d fermion with b.c. $\psi_{-}|_{\partial}=0$}&-\frac12{\bf f}^2 \\  
\end{array}
\end{align}
In particular, the boundary anomaly of 3d fermions equals half of the anomaly of a 2d fermion with the same 
charges as the component of the 3d fermion which survives at the boundary.

Let ${\bf f}$ be the field strength of a simple compact group $G$ 
under which fermions transform as irreducible unitary representation ${\bf R}$. 
The contribution to the anomaly polynomial is 
\begin{align}
\label{AN_bdy2}
\begin{array}{c|c}
\textrm{anomaly contribution}&\textrm{anomaly polynomial $\mathcal{I}$} \\ \hline
\textrm{2d left-handed chiral fermion}&T_{\bf R}\Tr({\bf f}^2) \\  
\textrm{2d right-handed chiral fermion}&-T_{\bf R}\Tr({\bf f}^2) \\  
\textrm{3d $U(1)$ Chern-Simons coupling of level $k$}&k\Tr ({\bf f}^2) \\  
\textrm{3d fermion with b.c. $\psi_{+}|_{\partial}=0$}&\frac12T_{\bf R}\Tr({\bf f}^2) \\  
\textrm{3d fermion with b.c. $\psi_{-}|_{\partial}=0$}&-\frac12T_{\bf R}\Tr({\bf f}^2) \\  
\end{array}
\end{align}
Here $T_{\bf R}$ is the quadratic index for representation $\bf R$ 
of the symmetry group $G$ which is defined by a sum over length-square of weights $\lambda$:
\begin{align}
\label{AN_bdy3}
T_{\bm R}&:=\frac{1}{\mathrm{rank}(G)} \sum_{\lambda\in \bf R}\|{\lambda}\|^2
\end{align}
where the length-square $\|{\alpha}\|$ of long roots $\alpha$ is 2. 
For example, 
for $G=SU(N)$ the quadratic index is $T_{\square}=1$ and $T_{\mathrm{adjoint}}=2N$. 
When $G$ is not simple, 
the anomaly should be fixed by additionally taking into account 
the consistency of Abelian anomaly for the maximal torus of $G$.

For example, let us consider the 3d $\mathcal{N}=2$ chiral multiplets of 
$U(1)_{R}$ R-charge $\rho$ in $U(N_{c})$ gauge theory with $SU(N_{f})$ global symmetry. 
Let ${\bf s}$, ${\bf r}$ and ${\bf x}$ be the 
$U(N_{c})$, $U(1)_{R}$ and $SU(N_{f})$ curvatures respectively. 
The boundary anomaly polynomials for these chiral multiplets are given as follows \cite{Dimofte:2017tpi}:
\begin{align}
\label{AN_3dn2cm}
\begin{array}{c|c|c|c}
U(N_c)&U(1)_{R}&SU(N_{f})&\textrm{anomaly polynomial $\mathcal{I}$} \\ \hline
\textrm{adj}&\rho&{\bf 1}&
\pm \left[N_{c}\Tr ({\bf s}^{2})-(\Tr {\bf s})^{2}+\frac{N_{c}^{2}}{2} 
\left((\rho-1){\bf r}\right)^{2}\right] \\  
\square&\rho&{\bf 1}
&\pm \left[\frac12 \Tr({\bf s}^{2})+(\Tr {\bf s})\cdot (\rho-1){\bf r}+\frac{N_{c}}{2}\left((\rho-1){\bf r}\right)^{2}\right] \\  
\overline{\square}&\rho&{\bf 1}
&\pm \left[\frac12 \Tr({\bf s}^{2})-(\Tr {\bf s})\cdot (\rho-1){\bf r}+\frac{N_{c}}{2}\left((\rho-1){\bf r}\right)^{2}\right] \\  
\square&\rho&{\bf N_{f}}
&\pm\left[
\frac12 N _{c}\Tr ({\bf x}^{2})+
\frac12 N_{f}\Tr({\bf s}^{2})
+N_{f}(\Tr {\bf s})\cdot (\rho-1){\bf r}
+\frac{N_{c}N_{f}}{2}\left((\rho-1){\bf r}\right)^{2}
\right]  \\  
\overline{\square}&\rho&{\overline{\bf N_{f}}}
&\pm\left[
\frac12 N _{c}\Tr ({\bf x}^{2})+
\frac12 N_{f}\Tr({\bf s}^{2})
-N_{f}(\Tr {\bf s})\cdot (\rho-1){\bf r}
+\frac{N_{c}N_{f}}{2}\left((\rho-1){\bf r}\right)^{2}
\right]  \\  
\end{array}
\end{align}
where the $+$ and $-$ sings correspond to 
the Dirichlet b.c. and Neumann b.c. respectively.

Now we would like to consider the boundary anomalies for 3d $\mathcal{N}=4$ gauge theories obeying 
$\mathcal{N}=(0,4)$ supersymmetric boundary conditions. 
Unlike $\mathcal{N}=(0,2)$ boundary conditions, 
as the R-symmetry $SU(2)_{C}$ $\times$ $SU(2)_{H}$ is non-Abelian, it will not mix with any flavor symmetries. 
The boundary anomaly polynomial for 
3d $\mathcal{N}=4$  $U(N_{c})$ SQCD with $N_{f}$ fundamental hypermultiplets obeying 
$(\mathcal{N}, N)$ boundary condition is 
\begin{align}
\label{ncnf_AN}
\mathcal{I}_{(\mathcal{N},N)}^{(N_{c})-[N_{f}]}
&=
\underbrace{
2N_{c} \Tr ({\bf s}^{2})-2(\Tr {\bf s})^{2}
}_{\textrm{$\mathcal{N}$ of $U(N_{c})$ gaugino}}
+
\underbrace{
2(\Tr {\bf s})\cdot {\bf z}}_{\textrm{FI term}}
\underbrace{
-N_{c} \Tr({\bf x}^{2})
-N_{f} \Tr ({\bf s}^{2})
}_{\textrm{$N$ of hyper}}
\nonumber\\
&=(2N_{c}-N_{f})\Tr ({\bf s}^{2})
-N_{c}\Tr ({\bf x}^{2})
+2(\Tr {\bf s})\cdot \left[
-(\Tr {\bf s})+ {\bf z}
\right]
\end{align}
where ${\bf z}$ is the topological $U(1)_{t}$ field strength. 
The boundary anomaly polynomial for 3d $\mathcal{N}=4$ $\prod_{i}^{n} U(N_{i})$ 
linear quiver gauge theory with bi-fundamental hypermultiplets 
obeying $(\mathcal{N},N)$ boundary conditions is 
\begin{align}
\label{quiver_AN}
\mathcal{I}^{(N_{1})-(N_{2})- \cdots (N_{n})}_{(\mathcal{N}, N)}
&=
\sum_{i=1}^{n}
\underbrace{
\left[
2N_{i}\Tr ({\bf s}_{i}^{2})
-2(\Tr {\bf s}_{i})^{2}
+2(\Tr {\bf s}_{i})\cdot {\bf z}_{i}
\right]
}_{\textrm{$\mathcal{N}$ of $U(N)_{i}$ gaugino $+$ FI}}
\nonumber\\
&+\sum_{i=1}^{n-1}
\underbrace{
\left[
-N_{i+1}\Tr ({\bf s}_{i}^{2})
-N_{i}\Tr ({\bf s}_{i+1}^{2})
-\left(
N_{i+1}\Tr {\bf s}_{i}-N_{i}\Tr {\bf s}_{i+1}
\right)^{2}
\right]
}_{\textrm{$N$ of bi-fundamental hypers}}
\nonumber\\
&=(2N_{1}-N_{2})\Tr ({\bf s}_{1}^{2})
+(\Tr {\bf s}_{1})\cdot \left[
-(N_{2}^{2}+2)(\Tr {\bf s}_{1})+N_{1}N_{2}\Tr {\bf s}_{2}+2{\bf z}_{1}
\right]
\nonumber\\
&+\sum_{i=2}^{n-1}(2N_{i}-N_{i-1}-N_{i+1})\Tr ({\bf s}_{i}^{2})
\nonumber\\
&+(\Tr {\bf s}_{i})\cdot \left[
-(N_{i-1}^{2}+N_{i+1}^{2}+2)(\Tr {\bf s}_{i})+N_{i}N_{i-1}\Tr {\bf s}_{i-1}
+N_{i}N_{i+1}\Tr {\bf s}_{i+1}+2{\bf z}_{i}
\right]
\nonumber\\
&+(2N_{n}-N_{n-1})\Tr {\bf s}_{n}^2+(\Tr {\bf s}_{n})\cdot 
\left[
-(N_{n-1}^{2}+2)(\Tr {\bf s}_{n})
+N_{n}N_{n-1}\Tr {\bf s}_{n-1}+2{\bf z}_{n}
\right]. 
\end{align}

Similar anomalies exist for junctions of enriched Neumann boundary conditions and interfaces. They will play an important role in the study of
NS5-NS5$'$ junctions.

\section{Half-indices of interfaces in $\mathcal{N}=4$ SYM and S-dualities}
\label{sec_3bdy}

We will now describe the gauge interfaces associated to a single D5 or NS5$'$ with semi-infinite D3-branes on both sides of the fivebrane.
These will be interfaces between 4d $U(N)$ and $U(M)$ gauge theories. We will compute the half-indices and check the identities 
required by S-duality \cite{Gaiotto:2008sa, Gaiotto:2008sd, Gaiotto:2008ak}. 
\footnote{
See \cite{Okazaki:2019ony} for more general half-indices of dual interfaces in 4d $\mathcal{N}=4$ SYM theory. 
}

\subsection{$U(N)|U(N)$ interfaces}
\label{sec_3dn|n}
\begin{figure}
\begin{center}
\includegraphics[width=8.5cm]{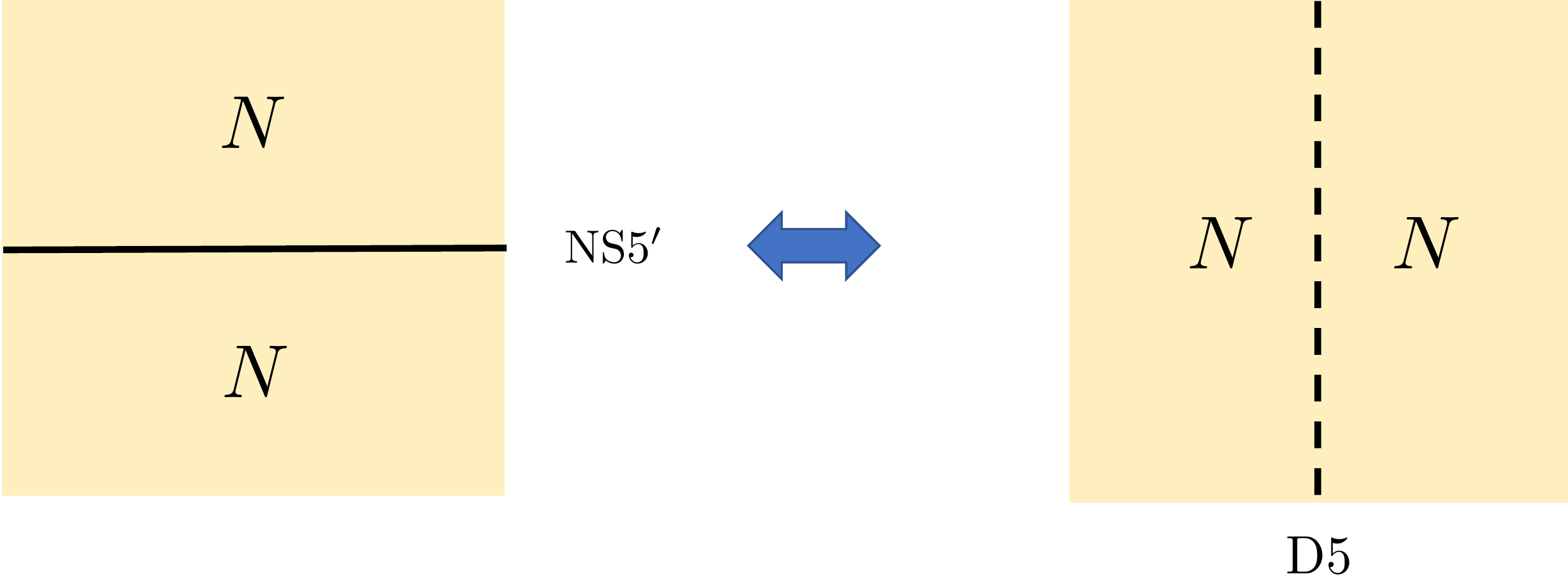}
\caption{
3d interface of NS5$'$-type and of D5-type with equal numbers of D3-branes 
where the horizontal and vertical directions are $x^6$ and $x^2$. 
The NS5$'$-type interface includes 3d $\mathcal{N}=4$ bi-fundamental twisted hypermultiplet 
while the D5-type interface has 3d $\mathcal{N}=4$ fundamental hypermultiplets that come from D3-D5 strings. }
\label{fig4dunun}
\end{center}
\end{figure}

We start with the case of equal numbers of D3-branes ending on the two sides of fivebranes (see Figure \ref{fig4dunun}). 
%
%
%
%
%

\subsubsection{$U(1)|U(1)$}
\label{sec_3du1u1}
The simplest example is the 3d interface for 4d $\mathcal{N}=4$ Abelian gauge theories. 
The NS5$'$-type half-BPS interface would be described by 4d $\mathcal{N}=4$ $U(1)$ gauge theory 
coupled to 3d $\mathcal{N}=4$ bi-fundamental twisted hypermultiplet at $x^2=0$. 

The half-index of NS5$'$-type interface between two 4d $\mathcal{N}=4$ $U(1)$ gauge theories is 
\begin{align}
\label{4du1u1_hindex}
\mathbb{II}_{\mathcal{N}'}^{\textrm{4d $U(1)|U(1)$}}
&=
\underbrace{
\frac{(q)_{\infty}}
{(q^{\frac12} t^2;q)_{\infty}}
\oint \frac{ds_{1}}{2\pi i s_{1}}
}_{\mathbb{II}_{\mathcal{N}'}^{\textrm{4d $U(1)$}}}
\underbrace{
\frac{(q)_{\infty}}
{(q^{\frac12} t^2;q)_{\infty}}
\oint \frac{ds_{2}}{2\pi i s_{2}}
}_
{\mathbb{II}_{\mathcal{N}'}^{\textrm{4d $U(1)$}}}
\underbrace{
\frac{
\left(q^{\frac34} t\frac{s_{1}}{s_{2}};q\right)_{\infty}
\left(q^{\frac34} t\frac{s_{2}}{s_{1}};q\right)_{\infty}
}
{
\left(q^{\frac14}t^{-1}\frac{s_{1}}{s_{2}};q\right)_{\infty}
\left(q^{\frac14}t^{-1}\frac{s_{2}}{s_{1}};q\right)_{\infty}
}
}_{\mathbb{I}^{\textrm{3d tHM}}
\left(\frac{s_{1}}{s_{2}}\right)}. 
\end{align}
The S-dual of the NS5$'$ interface is the D5-type interface. 
The D5-brane interface breaks the $U(1)$ $\times$ $U(1)$ gauge symmetry down to a diagonal $U(1)$
and couples it to a fundamental hypermultiplet. In other words, we have a 4d $\mathcal{N}=4$ 
$U(1)$ gauge theory defined on the full spacetime and coupled to a 3d $\mathcal{N}=4$ charged hypermultiplet living at $x^2=0$. 

The corresponding index is
\begin{align}
\label{4du1u1_hindex2}
\mathbb{II}_{\mathcal{D}}^{\textrm{4d $U(1)|U(1)$}}
&=
\underbrace{
\frac{(q)_{\infty}^{2}}
{(q^{\frac12} t^2;q)_{\infty}
(q^{\frac12}t^{-2};q)_{\infty}}
\oint \frac{ds}{2\pi is}
}_{\mathbb{I}^{\textrm{4d $U(1)$}}}
\underbrace{
\frac{
(q^{\frac34} t^{-1}s;q)_{\infty}
(q^{\frac34} t^{-1}s^{-1};q)_{\infty}
}
{
(q^{\frac14} ts;q)_{\infty}
(q^{\frac14} ts^{-1};q)_{\infty}
}
}_{\mathbb{I}^{\textrm{3d HM}}(s)}.  
\end{align}
We will momentarily demonstrate that it coincides with the NS5$'$ index. 

Notice that the interface has two interesting deformations. One corresponds to the D3-brane separating from the 
fivebrane. The other to the D3-brane breaking in two halves which separate from each other in a direction parallel to the fivebrane. 
In the D5-brane description, the fundamental hyper gets a vev when the D3-brane is split in two. In the NS5$'$ description 
the bi-fundamental field gets a vev when the D3-brane separates from the fivebrane.

 These vevs can be enforced by turning on boundary FI parameters for the 
bulk gauge fields, which shift the relative values of the centra of mass degrees of freedom on the two sides of the interface. 
We thus expect to be able to give a physical interpretation to the expansion of the half-index into a sum of residues of poles in either 
of the two descriptions above. 

Firstly, we can evaluate the integral in (\ref{4du1u1_hindex}) as the sum of residues at poles $\frac{s_{1}}{s_{2}}$ $=$ $q^{\frac14+m}t^{-1}$ 
for the bi-fundamental twisted hypermultiplet:
\begin{align}
\label{4du1u1_sum1}
\mathbb{II}_{\mathcal{N}'}^{\textrm{4d $U(1)|U(1)$}}
&=
\frac{(q^{\frac12} t^{-2};q)_{\infty}}{(q^{\frac12}t^2;q)_{\infty}}
\sum_{m=0}^{\infty}\frac{(q^{1+m};q)_{\infty}^{2}}{(q^{\frac12+m}t^{-2};q)_{\infty}^2}q^{\frac{m}{2}}t^{2m}.
\end{align}
This would correspond to the IR description 
via the Higgsing procedure of giving a vev to the  bi-funamental twisted hypermultiplet at the defect, which removes the fivebrane from a single D3-brane. 
Consequently, the $U(1)\times U(1)$ gauge symmetry is broken to $U(1)$ 
and a whole 4d $\mathcal{N}=4$ $U(1)$ gauge theory remains. 
In fact, the sum (\ref{4du1u1_sum1}) begins with the full-index $\mathbb{I}^{\textrm{4d $U(1)$}}$ of 4d $\mathcal{N}=4$ $U(1)$ gauge theory. 

On the other hand, we observe that the half-index in the (\ref{4du1u1_hindex2}) formulation can be expressed as the sum of residues at $s = q^{\frac14 + n} t$:
\begin{align}
\label{4du1u1_hindex3}
\mathbb{II}_{\mathcal{D}}^{\textrm{4d $U(1)|U(1)$}}
=\sum_{n=0}^{\infty}
\frac{(q^{1+n};q)^{2}_{\infty}}
{(q^{\frac12 +n}t^2;q)_{\infty}^2}
q^{\frac{n}{2}}t^{-2n}. 
\end{align}
The first term in the sum is clearly the square of $\mathbb{II}^{\textrm{4d $U(1)$}}_{\mathcal{D}}$, compatible with the IR description as 
a product of two Dirichlet boundary conditions for the gauge fields on the two sides of the interface. The other terms should have a tentative interpretation as
contributions from configuration of $n$ BPS vortices. In the IR, the boundary vortices take the appearance of boundary 't Hooft lines 
deforming the Dirichlet boundary conditions. In an Abelian theory the boundary 't Hooft lines do not change the half-index,
so the $n$ dependence must come from some Witten index of the vortex moduli space. It would be interesting to explore this point further.

By applying the $q$-binomial theorem (\ref{q_binomial}) to the D5 half-index (\ref{4du1u1_hindex2}), 
we can alternatively evaluate the integral (\ref{4du1u1_hindex2}) as
\begin{align}
\label{4du1u1_sum2}
\mathbb{II}_{\mathcal{D}}^{\textrm{4d $U(1)|U(1)$}}
&=
\frac{(q)_{\infty}^2}{(q^{\frac12}t^2;q)_{\infty}(q^{\frac12}t^{-2};q)_{\infty}}
\oint \frac{ds}{2\pi is}
\sum_{m,n=0}^{\infty}\frac{(q^{\frac12}t^{-2};q)_{m}}{(q)_{m}}
\frac{(q^{\frac12}t^{-2};q)_{n}}{(q)_{n}} q^{\frac{m+n}{4}}t^{m+n}s^{m-n}
\nonumber\\
&=
\frac{(q^{\frac12} t^{-2};q)_{\infty}}{(q^{\frac12}t^2;q)_{\infty}}
\sum_{m=0}^{\infty}\frac{(q^{1+m};q)_{\infty}^{2}}{(q^{\frac12+m}t^{-2};q)_{\infty}^2}q^{\frac{m}{2}}t^{2m}. 
\end{align}
This agrees with the residue sum (\ref{4du1u1_sum1}) for the bi-fundamental twisted hypermultiplet in the NS5$'$-type interface.

\subsubsection{$U(2)|U(2)$}
\label{sec_3du2u2}
Now consider the simplest non-Abelian example, that is 
the 3d interface between $U(2)$ and $U(2)$. 
The NS5$'$-type interface is expected to be described by 
4d $\mathcal{N}=4$ $U(2)$ gauge theory coupled to 3d $\mathcal{N}=4$ bi-fundamental twisted hypermultiplet living at $x^2=0$. 

The half-index for the NS5$'$-type interface between two $U(2)$ gauge theories is 
\begin{align}
\label{4du2u2_hindex}
\mathbb{II}_{\mathcal{N}'}^{\textrm{4d $U(2)|U(2)$}}
&=
\underbrace{
\frac{1}{2}
\frac{(q)_{\infty}^{2}}
{(q^{\frac12} t^2;q)_{\infty}^2}
\oint \frac{ds_{1}}{2\pi is_{1}}\frac{ds_{2}}{2\pi is_{2}}
\frac{
\left(\frac{s_{1}}{s_{2}};q\right)_{\infty} 
\left(\frac{s_{2}}{s_{1}};q\right)_{\infty}
}
{
\left(q^{\frac12} t^2 \frac{s_{1}}{s_{2}};q\right)_{\infty}
\left(q^{\frac12} t^2 \frac{s_{2}}{s_{1}};q\right)_{\infty}
}
}_{\mathbb{II}_{\mathcal{N}'}^{\textrm{4d $U(2)$}}}
\nonumber\\
&\times 
\underbrace{
\frac{1}{2}
\frac{(q)_{\infty}^{2}}
{(q^{\frac12} t^2;q)_{\infty}^2}
\oint \frac{ds_{3}}{2\pi is_{3}}\frac{ds_{4}}{2\pi is_{4}}
\frac{
\left(\frac{s_{3}}{s_{4}};q\right)_{\infty} 
\left(\frac{s_{4}}{s_{3}};q\right)_{\infty}
}
{
\left(q^{\frac12} t^2 \frac{s_{3}}{s_{4}};q\right)_{\infty}
\left(q^{\frac12} t^2 \frac{s_{4}}{s_{3}};q\right)_{\infty}
}
}_{\mathbb{II}_{\mathcal{N}'}^{\textrm{4d $U(2)$}}}
\nonumber\\
&\times 
\prod_{i=1}^{2}\prod_{k=3}^{4}
\underbrace{
\frac{
\left(q^{\frac34} t \frac{s_{i}}{s_{k}};q\right)_{\infty}
\left(q^{\frac34} t \frac{s_{k}}{s_{i}};q\right)_{\infty}
}
{
\left(q^{\frac14} t^{-1} \frac{s_{i}}{s_{k}};q\right)_{\infty}
\left(q^{\frac14} t^{-1} \frac{s_{k}}{s_{i}};q\right)_{\infty}
}
}_{\mathbb{I}^{\textrm{3d tHM}}\left(\frac{s_{i}}{s_{k}}\right)}
. 
\end{align}
The S-dual configuration is the D5-type interface between two 4d $\mathcal{N}=4$ $U(2)$ gauge theories. 
The interface has 3d $\mathcal{N}=4$ hypermultiplets transforming as the fundamental representation 
under the $U(2)$ gauge symmetry, which comes from D3-D5 strings. 
The half-index of D5-type interface is given by
\begin{align}
\label{4du2u2_hindex2}
\mathbb{II}_{\mathcal{D}}^{\textrm{4d $U(2)|U(2)$}}
&=
\underbrace{
\frac12 \frac{(q)_{\infty}^{4}}
{(q^{\frac12} t^2;q)_{\infty}^2 (q^{\frac12} t^{-2};q)_{\infty}^2}
\oint 
\prod_{i=1}^{2}
\frac{ds_{i}}{2\pi is_{i}}
\prod_{i\neq j}
\frac{
\left(\frac{s_{i}}{s_{j}};q\right)_{\infty}
\left(q \frac{s_{i}}{s_{j}};q\right)_{\infty}
}
{
\left(q^{\frac12} t^2 \frac{s_{i}}{s_{j}};q\right)_{\infty}
\left(q^{\frac12} t^{-2} \frac{s_{i}}{s_{j}};q\right)_{\infty}
}
}_{\mathbb{I}^{\textrm{4d $U(2)$}}}
\nonumber\\
&\times 
\prod_{i=1}^{2}
\underbrace{
\frac{
\left(q^{\frac34} t^{-1} s_{i};q\right)_{\infty}
\left(q^{\frac34} t^{-1} s_{i}^{-1};q\right)_{\infty}
}
{
\left(q^{\frac14} t s_{i};q\right)_{\infty}
\left(q^{\frac14} t s_{i}^{-1};q\right)_{\infty}
}
}_{\mathbb{I}^{\textrm{3d HM}}(s_{i})}. 
\end{align}

Experimentally, we find that the half-indices (\ref{4du2u2_hindex}) and (\ref{4du2u2_hindex2}) coincide.

Furthermore, the latter contour integral can be expressed as a sum over residues 
\begin{align}
\label{4du2u2_hindex3}
\mathbb{II}_{\mathcal{N}'}^{\textrm{4d $U(2)|U(2)$}}&=
\mathbb{II}_{\mathcal{D}}^{\textrm{4d $U(2)|U(2)$}}
\nonumber\\
&=\sum_{n=0}^{\infty}\sum_{m=0}^{\infty}
\frac{
(q^{1+n};q)_{\infty}^2 (q^{\frac32+n+m}t^2;q)_{\infty}^2
}
{
(q^{\frac12+n}t^2;q)_{\infty}^2 (q^{1+n+m} t^4;q)_{\infty}^2
}
q^{\frac{2n+m}{2}} t^{-2(2n+m)}
\end{align}
associated to the branch of vacua where both D3-branes break and the left pair is translated with respect to the right pair 
along the fivebrane. Notice that the low energy description of that system involves Nahm pole boundary conditions for both $U(2)$ gauge theories,
which is the first term in the sum. The other terms should arise from contributions of BPS vortices, which take the appearance of 
boundary 't Hooft lines in the IR theory. It would be nice to push the comparison further.  

It would be also nice to express the (\ref{4du2u2_hindex}) index in a manner which corresponds to giving a vev to
the bi-fundamental hypers. Finally, if would be nice to reproduce the same expansion by some judicious manipulation of the other contour integral.
This would be a reasonable strategy to prove the equality of (\ref{4du2u2_hindex}) and (\ref{4du2u2_hindex2}).

\subsubsection{$U(3)|U(3)$}
\label{sec_3du3u3}
Furthermore we can analyze three D3-branes on the two sides of a single fivebrane. 
The half-index for the NS5$'$-type junction 
between two $U(3)$ gauge theories is given by
\begin{align}
\label{4du3u3_hindex}
\mathbb{II}_{\mathcal{N}'}^{\textrm{4d $U(3)|U(3)$}}
&=
\underbrace{
\frac{1}{3!}
\frac{(q)_{\infty}^{3}}
{(q^{\frac12} t^2;q)_{\infty}^3}
\oint 
\prod_{i=1}^3
\frac{ds_{i}}{2\pi is_{i}}
\prod_{i\neq j}
\frac{
\left(\frac{s_{i}}{s_{j}};q\right)_{\infty} 
}
{
\left(q^{\frac12} t^2 \frac{s_{i}}{s_{j}};q\right)_{\infty}
}
}_{\mathbb{II}_{\mathcal{N}'}^{\textrm{4d $U(3)$}}}
\nonumber\\
&\times 
\underbrace{
\frac{1}{3!}
\frac{(q)_{\infty}^{3}}
{(q^{\frac12} t^2;q)_{\infty}^3}
\oint 
\prod_{i=4}^6
\frac{ds_{i}}{2\pi is_{i}}
\prod_{i\neq j}
\frac{
\left(\frac{s_{i}}{s_{j}};q\right)_{\infty} 
}
{
\left(q^{\frac12} t^2 \frac{s_{i}}{s_{j}};q\right)_{\infty}
}
}_{\mathbb{II}_{\mathcal{N}'}^{\textrm{4d $U(3)$}}}
\nonumber\\
&\times 
\prod_{i=1}^{3}\prod_{k=4}^{6}
\underbrace{
\frac{
\left(q^{\frac34} t \frac{s_{i}}{s_{k}};q\right)_{\infty}
\left(q^{\frac34} t \frac{s_{k}}{s_{i}};q\right)_{\infty}
}
{
\left(q^{\frac14} t^{-1} \frac{s_{i}}{s_{k}};q\right)_{\infty}
\left(q^{\frac14} t^{-1} \frac{s_{k}}{s_{i}};q\right)_{\infty}
}
}_{\mathbb{I}^{\textrm{3d tHM}}\left(\frac{s_{i}}{s_{k}}\right)}
. 
\end{align}
The mirror configuration is a D5-type domain wall interpolating between two 4d $\mathcal{N}=4$ $U(3)$ gauge theories. 
The gauge group is reduced from $U(3)$ $\times$ $U(3)$ to $U(3)$. 
The domain wall includes hyper multiplets transforming as the fundamental representation under the $U(3)$ gauge group. 
The half-index of the D5-type interface is
\begin{align}
\label{4du3u3_hindex2}
\mathbb{II}_{\mathcal{D}}^{\textrm{4d $U(3)|U(3)$}}
&=
\underbrace{
\frac{1}{3!} \frac{(q)_{\infty}^{6}}
{(q^{\frac12} t^2;q)_{\infty}^3 (q^{\frac12} t^{-2};q)_{\infty}^3}
\oint 
\prod_{i=1}^{3}
\frac{ds_{i}}{2\pi is_{i}}
\prod_{i\neq j}
\frac{
\left(\frac{s_{i}}{s_{j}};q\right)_{\infty}
\left(q \frac{s_{i}}{s_{j}};q\right)_{\infty}
}
{
\left(q^{\frac12} t^2 \frac{s_{i}}{s_{j}};q\right)_{\infty}
\left(q^{\frac12} t^{-2} \frac{s_{i}}{s_{j}};q\right)_{\infty}
}
}_{\mathbb{I}^{\textrm{4d $U(3)$}}}
\nonumber\\
&\times 
\prod_{i=1}^{3}
\underbrace{
\frac{
\left(q^{\frac34} t^{-1} s_{i};q\right)_{\infty}
\left(q^{\frac34} t^{-1} s_{i}^{-1};q\right)_{\infty}
}
{
\left(q^{\frac14} t s_{i};q\right)_{\infty}
\left(q^{\frac14} t s_{i}^{-1};q\right)_{\infty}
}
}_{\mathbb{I}^{\textrm{3d HM}}(s_{i})}. 
\end{align}
Experimentally, this coincides with the half-index (\ref{4du3u3_hindex}). 

The half-indices (\ref{4du3u3_hindex}) and (\ref{4du3u3_hindex2}) can be expressed (again experimentally) as 
\begin{align}
\label{4du3u3_hindex3}
\mathbb{II}_{\mathcal{N}'}^{\textrm{4d $U(3)|U(3)$}}&=
\mathbb{II}_{\mathcal{D}}^{\textrm{4d $U(3)|U(3)$}}
\nonumber\\
&=
\sum_{n=0}^{\infty}\sum_{m=0}^{\infty}\sum_{l=0}^{\infty}
\frac{
(q^{1+n};q)_{\infty}^2
(q^{\frac32+n+m}t^2;q)_{\infty}^2
(q^{2+n+m+l} t^4;q)_{\infty}^2
}
{
(q^{\frac12+n}t^2;q)_{\infty}^2 
(q^{1+n+m} t^4;q)_{\infty}^2
(q^{\frac32+n+m+l}t^6;q)_{\infty}^2
}
q^{\frac{3n+2m+l}{2}} 
t^{-2(3n+2m+l)}. 
\end{align}
We associate these expressions to the branch of vacua where all D3-branes break and the left set is translated with respect to the right set 
along the fivebrane. Notice that the low energy description of that system involves Nahm pole boundary conditions,
whose half-index is quite visible as the first term in the sum. 

Again, it would be also nice to find an expansion corresponding to the bi-fundamental hypers getting a vev, and to prove the various identities proposed in this section.

\subsubsection{$U(N)|U(N)$}
\label{sec_3duNuN}

The half-index for the NS5$'$-type junction between $U(N)$ and $U(N)$ gauge theories is 
\begin{align}
\label{4duNuN_hindex}
\mathbb{II}_{\mathcal{N}'}^{\textrm{4d $U(N)|U(N)$}}
&=
\underbrace{
\frac{1}{N!}
\frac{(q)_{\infty}^{N}}
{(q^{\frac12} t^2;q)_{\infty}^N}
\oint 
\prod_{i=1}^N
\frac{ds_{i}}{2\pi is_{i}}
\prod_{i\neq j}
\frac{
\left(\frac{s_{i}}{s_{j}};q\right)_{\infty} 
}
{
\left(q^{\frac12} t^2 \frac{s_{i}}{s_{j}};q\right)_{\infty}
}
}_{\mathbb{II}_{\mathcal{N}'}^{\textrm{4d $U(N)$}}}
\nonumber\\
&\times 
\underbrace{
\frac{1}{N!}
\frac{(q)_{\infty}^{N}}
{(q^{\frac12} t^2;q)_{\infty}^N}
\oint 
\prod_{i=N+1}^{2N}
\frac{ds_{i}}{2\pi is_{i}}
\prod_{i\neq j}
\frac{
\left(\frac{s_{i}}{s_{j}};q\right)_{\infty} 
}
{
\left(q^{\frac12} t^2 \frac{s_{i}}{s_{j}};q\right)_{\infty}
}
}_{\mathbb{II}_{\mathcal{N}'}^{\textrm{4d $U(N)$}}}
\nonumber\\
&\times 
\prod_{i=1}^{N}\prod_{k=N+1}^{2N}
\underbrace{
\frac{
\left(q^{\frac34} t \frac{s_{i}}{s_{k}};q\right)_{\infty}
\left(q^{\frac34} t \frac{s_{k}}{s_{i}};q\right)_{\infty}
}
{
\left(q^{\frac14} t^{-1} \frac{s_{i}}{s_{k}};q\right)_{\infty}
\left(q^{\frac14} t^{-1} \frac{s_{k}}{s_{i}};q\right)_{\infty}
}
}_{\mathbb{I}^{\textrm{3d tHM}}\left(\frac{s_{i}}{s_{k}}\right)}
. 
\end{align}
Under S-duality the NS5$'$-type interface maps to the D5-type interface. 
The D5-brane breaks the $U(N)$ $\times$ $U(N)$ gauge symmetry down into a diagonal $U(N)$ 
so that the four-dimensional $U(N)$ gauge fields couple to 3d $\mathcal{N}=4$ fundamental hypermultiplets. 
The half-index (\ref{4duNuN_hindex}) will coincide with the half-index of the D5-type interface which takes the form: 
\begin{align}
\label{4duNuN_hindex2}
\mathbb{II}_{\mathcal{D}}^{\textrm{4d $U(N)|U(N)$}}
&=
\underbrace{
\frac{1}{N!} \frac{(q)_{\infty}^{2N}}
{(q^{\frac12} t^2;q)_{\infty}^N (q^{\frac12} t^{-2};q)_{\infty}^N}
\oint 
\prod_{i=1}^{N}
\frac{ds_{i}}{2\pi is_{i}}
\prod_{i\neq j}
\frac{
\left(\frac{s_{i}}{s_{j}};q\right)_{\infty}
\left(q \frac{s_{i}}{s_{j}};q\right)_{\infty}
}
{
\left(q^{\frac12} t^2 \frac{s_{i}}{s_{j}};q\right)_{\infty}
\left(q^{\frac12} t^{-2} \frac{s_{i}}{s_{j}};q\right)_{\infty}
}
}_{\mathbb{I}^{\textrm{4d $U(N)$}}}
\nonumber\\
&\times 
\prod_{i=1}^{N}
\underbrace{
\frac{
\left(q^{\frac34} t^{-1} s_{i};q\right)_{\infty}
\left(q^{\frac34} t^{-1} s_{i}^{-1};q\right)_{\infty}
}
{
\left(q^{\frac14} t s_{i};q\right)_{\infty}
\left(q^{\frac14} t s_{i}^{-1};q\right)_{\infty}
}
}_{\mathbb{I}^{\textrm{3d HM}}(s_{i})}. 
\end{align}
Experimentally, we will have the following identity between 
the NS5$'$ interface half-indedx (\ref{4duNuN_hindex}) and the D5-type interface half-index (\ref{4duNuN_hindex2}):
\begin{align}
\label{4duNuN_hindex3}
&\mathbb{II}_{\mathcal{N}'}^{\textrm{4d $U(N)|U(N)$}}
=
\mathbb{II}_{\mathcal{D}}^{\textrm{4d $U(N)|U(N)$}}
\nonumber\\
&=
\sum_{n_{1},\cdots, n_{N}=0}^{\infty}
\prod_{i=1}^{N}
\frac{
\left(q^{1+\frac{i-1}{2}+\sum_{k=1}^{i}n_{k}}t^{2(i-1)};q\right)_{\infty}^2
}
{
\left(q^{\frac12+\frac{i-1}{2}+\sum_{k=1}^{i}n_{k}}t^{2+2(i-1)};q\right)_{\infty}^2
}
q^{\frac{\sum_{k=1}^{N}(N+1-k)n_{k}}{2}} 
t^{-2\sum_{k=1}^{N}(N+1-k)n_{k}}
. 
\end{align}
We associate these expressions to the branch of vacua where all D3-branes break and the left set is translated with respect to the right set 
along the fivebrane. Notice that the low energy description of that system involves Nahm pole boundary conditions,
whose half-index is quite visible as the first in the sum. 

Again, it would be also nice to find an expansion corresponding to the bi-fundamental hypers getting a vev, and to prove the various identities proposed in this section.

\subsection{$U(N)|U(M)$ interfaces}
\label{sec_3dn|m}
\begin{figure}
\begin{center}
\includegraphics[width=8.5cm]{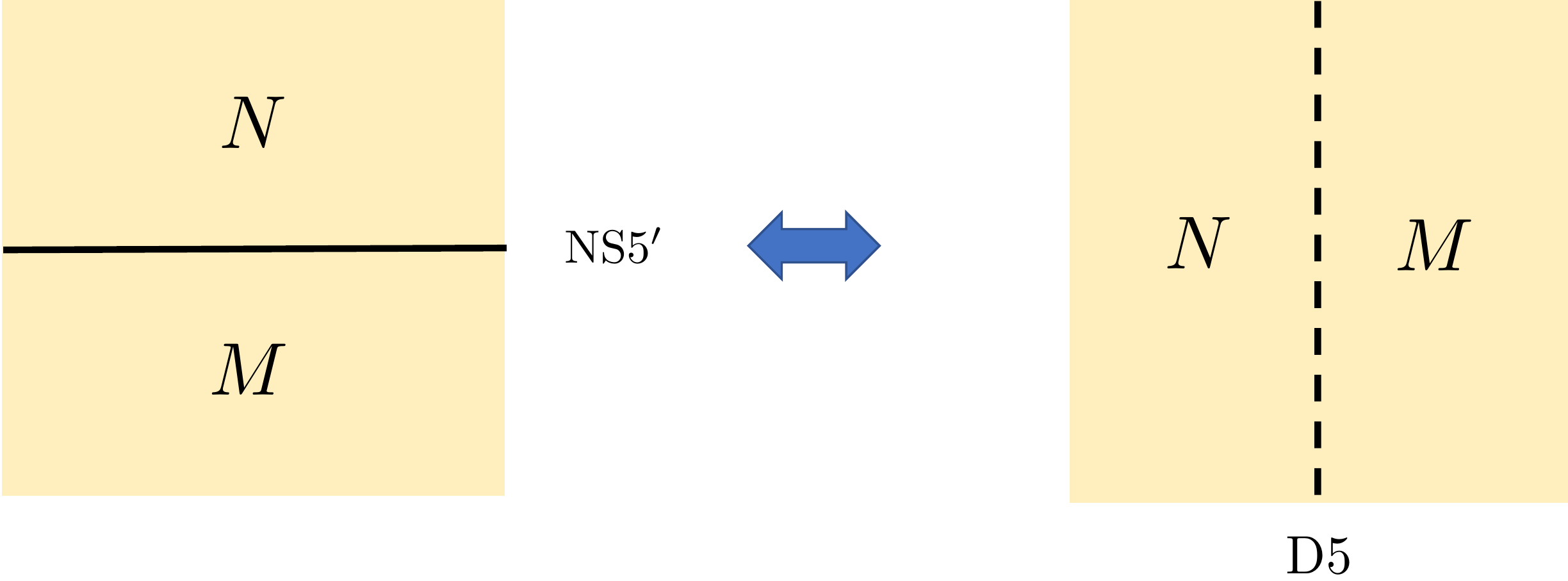}
\caption{3d interface of NS5$'$-type and of D5-type with different numbers of D3-branes. 
For D5-type interface there is no 3d hypermultiplet at the interface. 
When the difference of numbers of D3-branes is larger than one, 
the defect introduces the Nahm pole. }
\label{fig4dunum}
\end{center}
\end{figure}

Next, we can look at the setup with unequal numbers of D3-branes ending on the two sides of fivebranes 
(see Figure \ref{fig4dunum}). 
%
%
%
%
%

\subsubsection{$U(2)|U(1)$}
\label{sec_3du2u1}
The simplest example is the 3d interface between 4d $\mathcal{N}=4$ $U(2)$ and $U(1)$ gauge theories. 
The physics of NS5$'$-type interface can be described by 
4d $\mathcal{N}=4$ $U(2)$ gauge theory for $x^2>0$, 
4d $\mathcal{N}=4$ $U(1)$ gauge theory for $x^2<0$ 
and 3d bi-fundamental twisted hypermultiplets at $x^2=0$ transforming in the representation 
$({\bf 2},-)$ $\oplus$ $(\overline{{\bf 2}},+)$ of the gauge group $U(2)$ $\times$ $U(1)$.

The half-index of NS5$'$-type interface between $U(2)$ and $U(1)$ gauge theories 
involving bi-fundamental twisted hypermultiplets is 
\begin{align}
\label{4du2u1_hindex}
\mathbb{II}_{\mathcal{N}'}^{\textrm{4d $U(2)|U(1)$}}
&=
\underbrace{
\frac{1}{2!}
\frac{(q)_{\infty}^{2}}
{(q^{\frac12} t^{2};q)_{\infty}^{2}}
\oint \frac{ds_{1}}{2\pi is_{1}}
\frac{ds_{2}}{2\pi is_{2}}
\frac{
\left(\frac{s_{1}}{s_{2}};q\right)_{\infty}
\left(\frac{s_{2}}{s_{1}};q\right)_{\infty}
}
{
\left(q^{\frac12} t^2\frac{s_{1}}{s_{2}}\right)_{\infty}
\left(q^{\frac12} t^2\frac{s_{2}}{s_{1}}\right)_{\infty}
}
}_{\mathbb{II}_{\mathcal{N}'}^{\textrm{4d $U(2)$}}}
\underbrace{
\frac{(q)_{\infty}}{(q^{\frac12} t^2;q)_{\infty}}
\oint \frac{ds_{3}}{2\pi is_{3}}
}_{\mathbb{II}_{\mathcal{N}'}^{\textrm{4d $U(1)$}}}
\nonumber\\
&
\underbrace{
\frac
{
\left(q^{\frac34} t \frac{s_{1}}{s_{3}};q\right)_{\infty}
\left(q^{\frac34} t \frac{s_{3}}{s_{1}};q\right)_{\infty}
}
{
\left(q^{\frac14} t^{-1} \frac{s_{1}}{s_{3}};q\right)_{\infty}
\left(q^{\frac14} t^{-1} \frac{s_{3}}{s_{1}};q\right)_{\infty}
}
}
_{\mathbb{I}^{\textrm{3d tHM}}\left(\frac{s_{1}}{s_{3}}\right)}
\underbrace{
\frac
{
\left(q^{\frac34} t \frac{s_{2}}{s_{3}};q\right)_{\infty}
\left(q^{\frac34} t \frac{s_{3}}{s_{2}};q\right)_{\infty}
}
{
\left(q^{\frac14} t^{-1} \frac{s_{2}}{s_{3}};q\right)_{\infty}
\left(q^{\frac14} t^{-1} \frac{s_{3}}{s_{2}};q\right)_{\infty}
}
}
_{\mathbb{I}^{\textrm{3d tHM}}\left(\frac{s_{2}}{s_{3}}\right)}.
\end{align}
The S-dual configuration is realized as a single D5-brane 
on which two D3-branes in $x^6<0$ and a single D3-brane in $x^6>0$ terminate. 
In the gauge theory, 
the $U(2)$ 4d gauge group for $x^6>0$ is reduced to a $U(1)$ subgroup at the interface (generated by the ${1\,0 \choose 0 \, 0}$ generator) 
and identified with the $U(1)$ 4d gauge group for $x^6<0$. The commuting $U(1)$ subgroup (generated by the ${0\,0 \choose 0 \, 1}$ generator) 
becomes a global symmetry at the interface.

In contrast to the case with equal numbers of D3-branes, 
there is no 3d $\mathcal{N}=4$ hypermultiplet at the interface. 
On the other hand, in $x^6>0 $ there are 4d gauginos and scalar fields of the $U(2)$ gauge theory 
that do not belong to the surviving $U(1)$ gauge theory. 
They may contribute to the index. 

In fact, we can check experimentally that the half-index (\ref{4du2u1_hindex}) matches with the 
following half-index of the S-dual D5-type interface: 
\begin{align}
\label{4du2u1_hindex2}
\mathbb{II}_{\mathcal{D}}^{\textrm{4d $U(1)|U(2)$}}
&=
\underbrace{
\frac{(q)_{\infty}^{2}}
{(q^{\frac12} t^2;q)_{\infty}
(q^{\frac12} t^{-2};q)_{\infty}}
\oint \frac{ds}{2\pi is}
}_{\mathbb{I}^{\textrm{4d $U(1)$}}}
\underbrace{
\frac{(q)_{\infty}}
{(q^{\frac12} t^2;q)_{\infty}}
}_{\mathbb{II}_{\mathcal{D}}^{\textrm{4d $U(1)$}}}
\frac{(qs;q)_{\infty}(qs^{-1};q)_{\infty}}
{(q^{\frac12} t^2 s;q)_{\infty}(q^{\frac12} t^2 s^{-1};q)_{\infty}}. 
\end{align}
The integrand in (\ref{4du2u1_hindex2}) would correspond to 
the 4d gauginos and scalar fields of the $U(2)$ gauge theory 
that do not belong to the surviving $U(1)$ gauge theory in $x^6>0$. 

The half-index in the integral form (\ref{4du2u1_hindex2}) can be written as a sum over residues again
\begin{align}
\label{4du2u1_hindex3}
\mathbb{II}_{\mathcal{D}}^{\textrm{4d $U(1)|U(2)$}}
&=
\underbrace{
\frac{(q)_{\infty}}{(q^{\frac12} t^2;q)_{\infty}}
}_{\mathbb{II}_{\mathcal{D}}^{\textrm{4d $U(1)$}}}
\sum_{n=0}^{\infty}
\frac{(q^{1+n};q)_{\infty} (q^{\frac32 +n} t^2;q)_{\infty}}
{(q^{\frac12 +n} t^2;q)_{\infty} (q^{1+n} t^4;q)_{\infty}}
q^{\frac{n}{2}}t^{-2n}. 
\end{align}
We associate these expressions to the branch of vacua where all D3-branes break and the left set is translated with respect to the right set 
along the fivebrane. Notice that the low energy description of that system involves Nahm pole boundary conditions for the $U(2)$ gauge fields and Dirichlet 
for the $U(1)$,whose half-index is the first term in the sum. 

The half-indices (\ref{4du2u1_hindex}) and (\ref{4du2u1_hindex2}) can be alternatively expanded as 
\begin{align}
\label{4du2u1_hindex4}
\mathbb{II}_{\mathcal{N}'}^{\textrm{4d $U(2)|U(1)$}}
&=\mathbb{II}_{\mathcal{D}}^{\textrm{4d $U(1)|U(2)$}}
\nonumber\\
&=
\frac{(q)_{\infty} (q^{\frac12}t^{-2};q)_{\infty}}
{(q^{\frac12}t^2;q)_{\infty}^2}\sum_{n=0}^{\infty}
\frac{(q^{1+n};q)_{\infty}^2}{(q^{\frac12+n}t^{-2};q)_{\infty}^2}
q^n t^{4n}. 
\end{align}
The sum has the first term as 
the product of the index $\mathbb{I}^{\textrm{4d $U(1)$}}$ of 4d $\mathcal{N}=4$ $U(1)$ gauge theory and 
the half-index $\mathbb{II}_{\mathcal{N}'}^{\textrm{4d $U(1)$}}$ of the Neumann b.c. $\mathcal{N}'$ for 
4d $\mathcal{N}=4$ $U(1)$ gauge theory. 
The expansion (\ref{4du2u1_hindex4}) would be associated to the Higgsing process of giving vevs to the bi-fundamental twisted hypermultiplet. 
In the brane configuration this is realized by detaching a single D3-brane from the NS5$'$-brane, 
attaching another semi-infinite D3-brane to the NS5$'$-brane and separating them.

\subsubsection{$U(3)|U(1)$}
\label{sec_3du3u1}
Let us consider the 3d interface between $U(3)$ and $U(1)$ gauge theories. 
When the difference between the numbers of D3-branes is larger than one, 
the D5-brane interface involves Nahm poles \cite{Gaiotto:2008sa}.

The NS5$'$-type interface includes 4d $\mathcal{N}=4$ $U(3)$ gauge theory in $x^2>0$, 
4d $\mathcal{N}=4$ $U(1)$ gauge theory in $x^2<0$ 
and 3d $\mathcal{N}=4$ bi-fundamental hypermultiplet transforming as 
$({\bf 3},-)$ $\oplus$ $(\overline{{\bf 3}},+)$ under the $U(3)$ $\times$ $U(1)$ gauge group. 
The half-index for NS5$'$-type interface between $U(3)$ gauge theory and $U(1)$ gauge theory is 
\begin{align}
\label{4du3u1_hindex}
\mathbb{II}_{\mathcal{N}'}^{\textrm{4d $U(3)|U(1)$}}
&=
\underbrace{
\frac{1}{3!}\frac{(q)_{\infty}^{3}}
{(q^{\frac12} t^2;q)_{\infty}^{3}}
\oint \prod_{i=1}^{3}
\frac{ds_{i}}{2\pi is_{i}}
\prod_{i\neq j}
\frac{\left(\frac{s_{i}}{s_{j}};q\right)_{\infty}}
{\left(q^{\frac12} t^2 \frac{s_{i}}{s_{j}};q\right)_{\infty}}
}_{\mathbb{II}_{\mathcal{N}'}^{\textrm{4d $U(3)$}}}
\underbrace{
\frac{(q)_{\infty}}
{(q^{\frac12} t^2 ;q)_{\infty}}
\oint \frac{ds_{4}}{2\pi is_{4}}
}_{\mathbb{II}_{\mathcal{N}'}^{\textrm{4d $U(1)$}}}
\nonumber\\
&
\prod_{i=1}^{3}
\underbrace{
\frac{
\left(q^{\frac34} t\frac{s_{i}}{s_{4}};q\right)_{\infty}
\left(q^{\frac34} t\frac{s_{4}}{s_{i}};q\right)_{\infty}
}
{
\left(q^{\frac14} t^{-1}\frac{s_{i}}{s_{4}};q\right)_{\infty}
\left(q^{\frac14} t^{-1}\frac{s_{4}}{s_{i}};q\right)_{\infty}
}}_{\mathbb{I}^{\textrm{3d tHM}}\left(\frac{s_{i}}{s_{4}}\right)}. 
\end{align}

The S-dual D5-type interface has three D3-branes in $x^6<0$ and a single D3-brane in $x^6>0$. 
The D5-brane defect will break the 4d gauge group $U(3)$ in $x^6<0$ down to  a block-diagonal $U(1)$
identified with the gauge group for $x^6>0$. The commuting $U(2)$ block supports a Nahm pole configuration
as in (\ref{4d_nahmbc2}) with a pole associated to an embedding $\rho:$ $\mathfrak{su}(2)$ $\rightarrow$ $\mathfrak{u}(2)$.  
There is no 3d $\mathcal{N}=4$ hypermultiplet at the defect. 
The diagonal $U(1)$ in the $U(2)$ subgroup survives as a global symmetry at the interface. 

Experimentally, 
the NS5-type interface half-index (\ref{4du3u1_hindex}) coincides with 
\begin{align}
\label{4du3u1_hindex2}
\mathbb{II}_{\mathcal{D}}^{\textrm{4d $U(1)|U(3)$}}
&=
\underbrace{
\frac{(q)_{\infty}^{2}}
{(q^{\frac12} t^2;q)_{\infty}
(q^{\frac12} t^{-2};q)_{\infty}}
\oint \frac{ds}{2\pi is}
}_{\mathbb{I}^{\textrm{4d $U(1)$}}}
\underbrace{
\frac{(q)_{\infty}(q^{\frac32} t^2;q)_{\infty}}
{(q^{\frac12} t^2;q)_{\infty}(q t^4;q)_{\infty}}
}_{\mathbb{II}_{\textrm{Nahm}}^{\textrm{4d $U(2)$}}}
\frac{(q^{\frac54} ts;q)_{\infty}(q^{\frac54}ts^{-1};q)_{\infty}}
{(q^{\frac34} t^3 s;q)_{\infty}(q^{\frac34} t^3 s^{-1};q)_{\infty}}. 
\end{align}
This should be viewed as the half-index of D5-type interface between $U(3)$ and $U(1)$ gauge theories. 
The integrand in (\ref{4du3u1_hindex2}) would correspond to 
the 4d gauginos and scalar fields which are not contained in the whole 4d $U(1)$ gauge theory.

The half-indices (\ref{4du3u1_hindex}) and (\ref{4du3u1_hindex2}) admit the expansion 
\begin{align}
\label{4du3u1_hindex3}
\mathbb{II}_{\mathcal{N}'}^{\textrm{4d $U(3)|U(1)$}}
&=\mathbb{II}_{\mathcal{D}}^{\textrm{4d $U(1)|U(3)$}}
\nonumber\\
&=
\underbrace{
\frac{(q)_{\infty}(q^{\frac32} t^2;q)_{\infty}}
{(q^{\frac12} t^2;q)_{\infty}(q t^4;q)_{\infty}}
}_{\mathbb{II}_{\textrm{Nahm}}^{\textrm{4d $U(2)$}}}
\sum_{n=0}^{\infty}
\frac{(q^{1+n};q)_{\infty} (q^{2+n}t^4;q)_{\infty}}
{(q^{\frac12+n}t^2;q)_{\infty} (q^{\frac32+n}t^6;q)_{\infty}}
q^{\frac{n}{2}}t^{-2n}. 
\end{align}
This expansion corresponds to the Higgsing procedure 
giving vevs to the fundamental hypers, 
which decomposes the whole D3-branes into 
a stack of three semi-infinite D3-branes in $x^6<0$ and a single semi-infinite D3-brane in $x^6>0$ 
as its first term takes the form of the product of the half-indices $\mathbb{II}_{\textrm{Nahm}}^{\textrm{4d $U(3)$}}$ 
and $\mathbb{II}_{\mathcal{D}}^{\textrm{4d $U(1)$}}$. 


The half-indices (\ref{4du3u1_hindex}) and (\ref{4du3u1_hindex2}) have another expansion
\begin{align}
\label{4du3u1_hindex4}
\mathbb{II}_{\mathcal{N}'}^{\textrm{4d $U(3)|U(1)$}}
&=\mathbb{II}_{\mathcal{D}}^{\textrm{4d $U(1)|U(3)$}}
\nonumber\\
&=\frac{(q)_{\infty}(q^{\frac12}t^{-2};q)_{\infty} (q^{\frac32}t^2;q)_{\infty}}
{(q^{\frac12}t^2;q)_{\infty}^2 (qt^4;q)_{\infty}}
\sum_{n=0}^{\infty}
\frac{(q^{1+n};q)_{\infty}^2}{(q^{\frac12+n}t^{-2};q)_{\infty}^2}q^{\frac{3n}{2}}t^{6n}. 
\end{align}
The first term in the sum is the product of the 4d $U(1)$ index $\mathbb{I}^{\textrm{4d $U(1)$}}$ 
and the $U(2)$ Nahm index $\mathbb{II}_{\textrm{Nahm}}^{\textrm{4d $U(2)$}}$. 
As the $U(2)$ Nahm index is equivalent to the $U(2)$ Neumann index $\mathbb{II}_{\mathcal{N}'}^{\textrm{4d $U(2)$}}$, 
the sum (\ref{4du3u1_hindex4}) will correspond to the separation of the NS5$'$-brane 
attaching two semi-infinite D3-branes in from a single D3-brane.

\subsubsection{$U(4)|U(1)$}
\label{sec_3du4u1}
To gain more insight, we can look at another concrete example before tackling the general case. 

Consider the NS5$'$-type interface between four D3-branes in $x^2>0$ and a single D3-brane in $x^2<0$. 
It is described by 4d $\mathcal{N}=4$ $U(4)$ gauge theory in $x^2>0$, 
4d $\mathcal{N}=4$ $U(1)$ gauge theory in $x^2<0$ and 3d $\mathcal{N}=4$ bi-fundamental hypermultiplets. 
The half-index is given by
\begin{align}
\label{4du4u1_hindex}
\mathbb{II}_{\mathcal{N}'}^{\textrm{4d $U(4)|U(1)$}}
&=
\underbrace{
\frac{1}{4!} 
\frac{(q)_{\infty}^4}{(q^{\frac12} t^2;q)_{\infty}^4} 
\oint \prod_{i=1}^4 \frac{ds_{i}}{2\pi is_{i}} \prod_{i\neq j} 
\frac{\left(\frac{s_{i}}{s_{j}};q\right)_{\infty}}
{\left(q^{\frac12} t^2 \frac{s_{i}}{s_{j}};q\right)_{\infty}}
}_{\mathbb{II}_{\mathcal{N}'}^{\textrm{4d $U(4)$}}}
\underbrace{
\frac{(q)_{\infty}}
{(q^{\frac12} t^2;q)_{\infty}}
\oint \frac{ds_{5}}{2\pi is_{5}}
}_{\mathbb{II}_{\mathcal{N}'}^{\textrm{4d $U(1)$}}}
\nonumber\\
&\prod_{i=1}^{4}
\underbrace{
\frac{
\left(q^{\frac34} t\frac{s_{i}}{s_{5}};q\right)_{\infty}
\left(q^{\frac34} t\frac{s_{5}}{s_{i}};q\right)_{\infty}
}
{
\left(q^{\frac14} t^{-1}\frac{s_{i}}{s_{5}};q\right)_{\infty}
\left(q^{\frac14} t^{-1}\frac{s_{5}}{s_{i}};q\right)_{\infty}
}}_{\mathbb{I}^{\textrm{3d tHM}}\left(\frac{s_{i}}{s_{5}}\right)}. 
\end{align}

The S-dual D5-type interface will break the $U(4)$ gauge symmetry in $x^6<0$ down to the $U(1)$ gauge symmetry in $x^6>0$. 
There is no 3d $\mathcal{N}=4$ hypermultiplet at $x^6=0$. 
In this case the 4d scalar fields have a singular behavior which is controlled by the Nahm pole boundary condition 
with a pole associated to an embedding $\rho:$ $\mathfrak{su}(2)$ $\rightarrow$ $\mathfrak{u}(3)$. 
The diagonal $U(1)$ in the $U(3)$ subgroup survives as a global symmetry at the interface.

We can check that the NS5-type half-index (\ref{4du4u1_hindex}) coincides with 
the following half-index for D5-type interface between $U(1)$ and $U(4)$ gauge theories:
\begin{align}
\label{4du4u1_hindex2}
\mathbb{II}_{\mathcal{D}}^{\textrm{4d $U(1)|U(4)$}}
&=
\underbrace{
\frac{(q)_{\infty}^{2}}
{(q^{\frac12} t^2;q)_{\infty}
(q^{\frac12} t^{-2};q)_{\infty}}
\oint \frac{ds}{2\pi is}
}_{\mathbb{I}^{\textrm{4d $U(1)$}}}
\underbrace{
\frac{(q)_{\infty}(q^{\frac32} t^2;q)_{\infty}(q^2 t^4;q)_{\infty}}
{(q^{\frac12} t^2;q)_{\infty}(q t^4;q)_{\infty}(q^{\frac32} t^6;q)_{\infty}}
}_{\mathbb{II}_{\textrm{Nahm}}^{\textrm{4d $U(3)$}}}
\frac{(q^{\frac32} t^2s;q)_{\infty}(q^{\frac32}t^2s^{-1};q)_{\infty}}
{(q t^4 s;q)_{\infty}(q t^4 s^{-1};q)_{\infty}}. 
\end{align}
The half-indices (\ref{4du4u1_hindex}) and (\ref{4du4u1_hindex2}) admit the expansion
\begin{align}
\label{4du4u1_hindex3}
\mathbb{II}_{\mathcal{N}'}^{\textrm{4d $U(4)|U(1)$}}
&=\mathbb{II}_{\mathcal{D}}^{\textrm{4d $U(1)|U(4)$}}
\nonumber\\
&=
\underbrace{
\frac{(q)_{\infty}(q^{\frac32}t^2;q)_{\infty} (q^2 t^4;q)_{\infty}}
{(q^{\frac12} t^2;q)_{\infty}(q t^4;q)_{\infty} (q^{\frac32} t^6;q)_{\infty}}
}_{\mathbb{II}_{\textrm{Nahm}}^{\textrm{4d $U(3)$}}}
\sum_{n=0}^{\infty}
\frac
{(q^{1+n};q)_{\infty} (q^{\frac52+n} t^6;q)_{\infty}}
{(q^{\frac12+n}t^2;q)_{\infty}(q^{2+n}t^8;q)_{\infty}}
q^{\frac{n }{2}} t^{-2n}. 
\end{align}
This residue sum corresponds to the Higgsing procedure by giving vevs to 
the fundamental hypers. As its first term is the product of two half-indices 
$\mathbb{II}_{\textrm{Nahm}}^{\textrm{4d $U(4)$}}$ 
and $\mathbb{II}_{\mathcal{D}}^{\textrm{4d $U(1)$}}$, 
this is physically realized as the splitting of D3-branes into two halves in $x^6<0$ and in $x^6>0$.


Another expansion of the half-indices (\ref{4du4u1_hindex}) and (\ref{4du4u1_hindex2}) takes the form
\begin{align}
\label{4du4u1_hindex4}
\mathbb{II}_{\mathcal{N}'}^{\textrm{4d $U(4)|U(1)$}}
&=\mathbb{II}_{\mathcal{D}}^{\textrm{4d $U(1)|U(4)$}}
\nonumber\\
&=
\frac{(q)_{\infty} (q^{\frac12} t^{-2};q)_{\infty} (q^{\frac32}t^2;q)_{\infty} (q^2t^{4};q)_{\infty}}
{(q^{\frac12}t^2;q)_{\infty}^2 (qt^4;q)_{\infty} (q^{\frac32}t^6;q)_{\infty}}
\sum_{n=0}^{\infty}
\frac{(q^{1+n};q)_{\infty}^2}{(q^{\frac12+n}t^{-2};q)_{\infty}^2}q^{2n} t^{8n}. 
\end{align}
The sum begins with the product of 4d $U(1)$ full index $\mathbb{I}^{\textrm{4d $U(1)$}}$ 
and the half-index $\mathbb{II}_{\textrm{Nahm}}^{\textrm{4d $U(3)$}}$ for the Nahm pole boundary condition of rank 3. 
This would be associated to the Higgsing procedure of giving vevs to the twisted hypermultiplet, 
which removes three semi-infinite D3-branes ending on the NS5$'$-brane 
from a single D3-brane.

\subsubsection{$U(N)|U(1)$}
\label{sec_3duNu1}
For the interface between 4d $\mathcal{N}=4$ $U(N)$ gauge theory and Abelian gauge theory, 
the half-index for the NS5$'$-type interface takes the form
\begin{align}
\label{4duNu1_hindex}
\mathbb{II}_{\mathcal{N}'}^{\textrm{4d $U(N)|U(1)$}}
&=
\underbrace{
\frac{1}{N!} 
\frac{(q)_{\infty}^N}{(q^{\frac12} t^2;q)_{\infty}^N} 
\oint \prod_{i=1}^N \frac{ds_{i}}{2\pi is_{i}} \prod_{i\neq j} 
\frac{\left(\frac{s_{i}}{s_{j}};q\right)_{\infty}}
{\left(q^{\frac12} t^2 \frac{s_{i}}{s_{j}};q\right)_{\infty}}
}_{\mathbb{II}_{\mathcal{N}'}^{\textrm{4d $U(4)$}}}
\underbrace{
\frac{(q)_{\infty}}
{(q^{\frac12} t^2;q)_{\infty}}
\oint \frac{ds_{N+1}}{2\pi is_{N+1}}
}_{\mathbb{II}_{\mathcal{N}'}^{\textrm{4d $U(1)$}}}
\nonumber\\
&\times \prod_{i=1}^{N}
\underbrace{
\frac{
\left(q^{\frac34} t\frac{s_{i}}{s_{N+1}};q\right)_{\infty}
\left(q^{\frac34} t\frac{s_{N+1}}{s_{i}};q\right)_{\infty}
}
{
\left(q^{\frac14} t^{-1}\frac{s_{i}}{s_{N+1}};q\right)_{\infty}
\left(q^{\frac14} t^{-1}\frac{s_{N+1}}{s_{i}};q\right)_{\infty}
}}_{\mathbb{I}^{\textrm{3d tHM}}\left(\frac{s_{i}}{s_{N+1}}\right)}. 
\end{align}
This will match with the half-index for the D5-type interface between $U(1)$ and $U(N)$ gauge theories:
\begin{align}
\label{4duNu1_hindex2}
\mathbb{II}_{\mathcal{D}}^{\textrm{4d $U(1)|U(N)$}}
&=
\underbrace{
\frac{(q)_{\infty}^{2}}
{(q^{\frac12} t^2;q)_{\infty}
(q^{\frac12} t^{-2};q)_{\infty}}
\oint \frac{ds}{2\pi is}
}_{\mathbb{I}^{\textrm{4d $U(1)$}}}
\underbrace{
\prod_{k=1}^{N-1}
\frac{(q^{\frac{k+1}{2}}t^{2(k-1)};q)_{\infty}}
{(q^{\frac{k}{2}}t^{2k};q)_{\infty}}
}_{\mathbb{II}_{\textrm{Nahm}}^{\textrm{4d $U(N-1)$}}}
\nonumber\\
&\times 
\frac{
(q^{\frac34+\frac{N-1}{4}} t^{-1+(N-1)}s;q)_{\infty}
(q^{\frac34+\frac{N-1}{4}} t^{-1+(N-1)}s^{-1};q)_{\infty}}
{(q^{\frac14+\frac{N-1}{4}} t^{1+(N-1)} s;q)_{\infty}
(q^{\frac14+\frac{N-1}{4}} t^{1+(N-1)} s^{-1};q)_{\infty}}. 
\end{align}
The integrand is the expected contributions from the 4d $U(N)$ gauge theory fields 
controlled by the Nahm pole boundary conditions associated to the embedding 
$\rho:$ $\mathfrak{su}(2)$ $\rightarrow$ $\mathfrak{u}(N-1)$. 

The half-indices (\ref{4du4u1_hindex}) and (\ref{4du4u1_hindex2}) have an expansion 
\begin{align}
\label{4duNu1_hindex3}
\mathbb{II}_{\mathcal{N}'}^{\textrm{4d $U(N)|U(1)$}}
&=\mathbb{II}_{\mathcal{D}}^{\textrm{4d $U(1)|U(N)$}}
\nonumber\\
&=
\underbrace{
\prod_{k=1}^{N-1}
\frac{(q^{\frac{k+1}{2}}t^{2(k-1)};q)_{\infty}}
{(q^{\frac{k}{2}}t^{2k};q)_{\infty}}
}_{\mathbb{II}_{\textrm{Nahm}}^{\textrm{4d $U(N-1)$}}}
\sum_{n=0}^{\infty}
\frac
{(q^{1+n};q)_{\infty} (q^{\frac{N+1}{2}+n} t^{2(N-1)};q)_{\infty}}
{(q^{\frac12+n}t^2;q)_{\infty}(q^{\frac{N}{2}+n}t^{2N};q)_{\infty}}
q^{\frac{n}{2}} t^{-2n}. 
\end{align}
This is obtained by picking up the residues of at $s=q^{\frac{N}{4}+n}t^{N}$ in the D5 interface index (\ref{4duNu1_hindex2}). 
As the residue sum starts with the product of the two half-indices 
$\mathbb{II}_{\textrm{Nahm}}^{\textrm{4d $U(N)$}}$ and 
$\mathbb{II}_{\mathcal{D}}^{\textrm{4d $U(1)$}}$, 
the expansion (\ref{4duNu1_hindex3}) is associated to the Higgsing process invoking a decomposition of the D3-branes into 
a single D3-brane in $x^6<0$ and $N$ D3-branes in $x^6>0$.


The half-indices (\ref{4duNu1_hindex}) and (\ref{4duNu1_hindex2}) also have an expansion 
\begin{align}
\label{4duNu1_hindex4}
\mathbb{II}_{\mathcal{N}'}^{\textrm{4d $U(N)|U(1)$}}
&=\mathbb{II}_{\mathcal{D}}^{\textrm{4d $U(1)|U(N)$}}
\nonumber\\
&=\frac{(q^{\frac12}t^{-2};q)_{\infty}}
{(q^{\frac12}t^2;q)_{\infty}}\prod_{k=1}^{N-1}\frac{(q^{\frac{k+1}{2}}t^{2(k-1)};q)_{\infty}}
{(q^{\frac{k}{2}}t^{2k};q)_{\infty}}
\sum_{n=0}^{\infty} \frac{(q^{1+n};q)_{\infty}^2}
{(q^{\frac12+n}t^{-2};q)_{\infty}^2} q^{\frac{Nn}{2}} t^{2Nn}. 
\end{align}
The first term in the sum is the product of the 4d $U(1)$ index $\mathbb{I}^{\textrm{4d $U(1)$}}$ 
and the half-index $\mathbb{II}_{\textrm{Nahm}}^{\textrm{4d $U(N-1)$}}$ for the Nahm pole boundary condition of rank $(N-1)$. 
This expansion for the half-index would be interpreted as 
the Higgsing process of giving vevs to the bi-fundamental twisted hypermultiplet, 
which attaches the $(N-1)$ semi-infinite D3-branes to the NS5$'$-brane 
and then separates from the remaining D3-brane.

\begin{figure}
\begin{center}
\includegraphics[width=12.5cm]{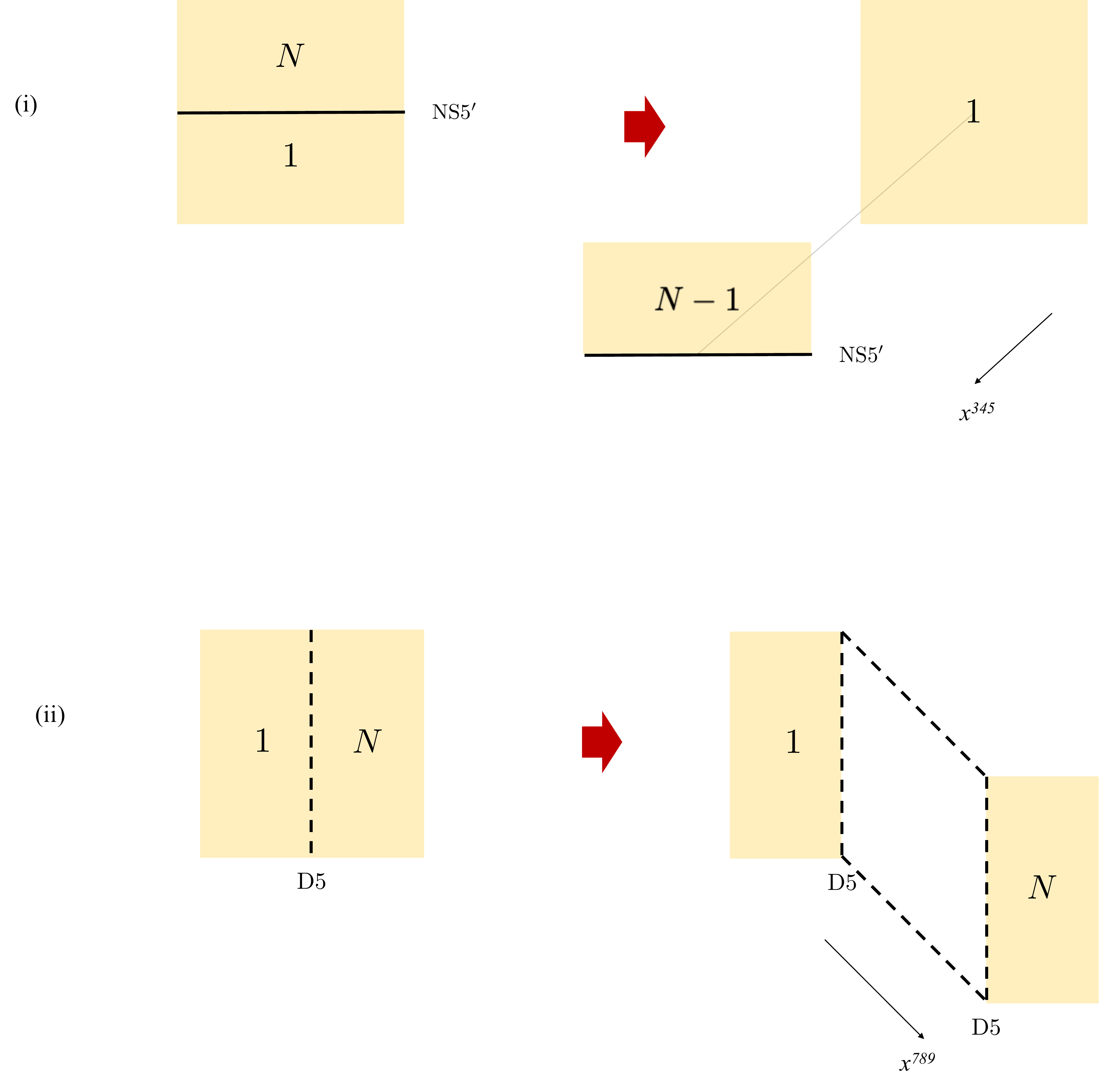}
\caption{Higgsing procedures of the $U(N)|U(1)$ interfaces in 4d $\mathcal{N}=4$ SYM theories corresponding to the deformations of Type IIB brane configurations. (i) Separation of the NS5$'$-brane which breaks gauge symmetry from $U(N)\times U(1)$ to $U(N-1)$ $\times$ $U(1)$. This corresponds to giving vevs to the 3d $\mathcal{N}=4$ bi-fundamental twisted hypermultiplets. (ii) Splitting of D3-branes along the D5-brane leading to a pair of $\mathcal{D}$ and $\textrm{Nahm}/\mathcal{D}$ b.c. for 4d $\mathcal{N}=4$ $U(1)$ and $U(N)$ gauge theories respectively. This corresponds to giving vevs to the 3d $\mathcal{N}=4$ fundamental hypermultiplets. }
\label{fighiggsing}
\end{center}
\end{figure}

The Higgsing processes corresponding to the two deformations in the brane configurations for the $U(N)|U(1)$ 3d interfaces are illustrated in Figure \ref{fighiggsing}. 
%
%
%
%
%

\subsubsection{$U(2)|U(3)$}
\label{sec_3du2u3}
As the last example, let us consider the interface between two non-Abelian gauge theories of 
gauge group $U(2)$ and $U(3)$. 
For the D5-type interface, this admits the non-Abelian gauge symmetry in four-dimensions. 

The NS5$'$ interface involves 
the 4d $\mathcal{N}=4$ $U(2)$ gauge theory in $x^2>0 $, 
4d $\mathcal{N}=4$ $U(3)$ gauge theory in $x^2<0$ 
and 3d $\mathcal{N}=4$ hypermultiplets transforming under 
the gauge group $U(2)\times U(3)$ as $({\bf 2},\overline{{\bf 3}})$ $\oplus$ 
$(\overline{{\bf 2}},{\bf 3})$. 

The half-index for NS5$'$-type interface between $U(2)$ and $U(3)$ gauge theories is 
\begin{align}
\label{4du2u3_hindex}
\mathbb{II}_{\mathcal{N}'}^{\textrm{4d $U(2)|U(3)$}}
&=
\underbrace{
\frac{1}{3!}\frac{(q)_{\infty}^3}{(q^{\frac12} t^2;q)_{\infty}^3}
\oint \prod_{i=1}^{3}\frac{ds_{i}}{2\pi is_{i}}
\prod_{i\neq j}\frac{\left(\frac{s_{i}}{s_{j}};q\right)_{\infty}}
{\left(
q^{\frac12} t^2\frac{s_{i}}{s_{j}};q
\right)_{\infty}}
}_{\mathbb{II}_{\mathcal{N}'}^{\textrm{4d $U(3)$}}}
\underbrace{
\frac{1}{2!}\frac{(q)_{\infty}^2}{(q^{\frac12} t^2;q)_{\infty}^2}
\oint \prod_{i=4}^{5}\frac{ds_{i}}{2\pi is_{i}}
\prod_{i\neq j}\frac{\left(\frac{s_{i}}{s_{j}};q\right)_{\infty}}
{\left(
q^{\frac12} t^2\frac{s_{i}}{s_{j}};q
\right)_{\infty}}
}_{\mathbb{II}_{\mathcal{N}'}^{\textrm{4d $U(2)$}}}
\nonumber\\
&\times 
\prod_{i=1}^{3}\prod_{k=4}^{5}
\underbrace{
\frac{
\left(q^{\frac34}t \frac{s_{i}}{s_{k}};q\right)_{\infty}
\left(q^{\frac34}t \frac{s_{k}}{s_{i}};q\right)_{\infty}
}
{
\left(q^{\frac14}t^{-1} \frac{s_{i}}{s_{k}};q\right)_{\infty}
\left(q^{\frac14}t^{-1} \frac{s_{k}}{s_{i}};q\right)_{\infty}
}
}_{\mathbb{I}^{\textrm{3d tHM}}\left(\frac{s_{i}}{s_{k}}\right)}. 
\end{align}

For the D5-type interface the gauge group $U(3)$ in $x^6>0$ would be broken down to a $U(2)$ block-diagonal subgroup to be identified with the  
$U(2)$ gauge symmetry in $x^6<0$. Another block-diagonal $U(1)$ subgroup survives as a global symmetry at the interface. 
The defect has no 3d $\mathcal{N}=4$ hypermultiplet. 
As the difference of the numbers of D3-branes is one, 
there is no Nahm pole at $x^6=0$. 

We can check experimentally that 
the half-index (\ref{4du2u3_hindex}) coincides with the following half-index for D5-type interface between $U(2)$ and $U(3)$ gauge theories:
\begin{align}
\label{4du2u3_hindex2}
\mathbb{II}_{\mathcal{D}}^{\textrm{4d $U(3)|U(2)$}}
&=
\underbrace{
\frac{1}{2}
\frac{(q)_{\infty}^4}
{(q^{\frac12} t^2;q)_{\infty}^2 (q^{\frac12}t^{-2};q)_{\infty}^2}
\oint \prod_{i=1}^{2}
\frac{ds_{i}}{2\pi is_{i}}
\prod_{i\neq j}
\frac{
\left(\frac{s_{i}}{s_{j}};q\right)_{\infty}
\left(q \frac{s_{i}}{s_{j}};q\right)_{\infty}
}
{
\left(q^{\frac12} t^2 \frac{s_{i}}{s_{j}};q\right)_{\infty}
\left(q^{\frac12} t^{-2} \frac{s_{i}}{s_{j}};q\right)_{\infty}
}
}_{\mathbb{I}^{\textrm{4d $U(2)$}}}
\nonumber\\
&\times 
\underbrace{
\frac{(q)_{\infty}}{(q^{\frac12} t^2;q)_{\infty}}
}_{\mathbb{II}_{\mathcal{D}}^{\textrm{4d $U(1)$}}}
\cdot 
\prod_{i=1}^{2}
\frac{
(qs_{i};q)_{\infty}
(qs_{i}^{-1};q)_{\infty}}
{
(q^{\frac12} t^2 s_{i};q)_{\infty}
(q^{\frac12} t^2 s_{i}^{-1};q)_{\infty}
}.
\end{align}
The integrand has the same form as (\ref{4du2u1_hindex2}) 
where the difference of the numbers of D3-branes is one. 
Hence it would capture the 4d gauginos and scalar fields in $x^6>0$ of the original $U(3)$ gauge theory 
which are not part of the $U(2)$ gauge theory. 

The half-indices (\ref{4du2u3_hindex}) and (\ref{4du2u3_hindex2}) can be expressed experimentally as
\begin{align}
\label{4du2u3_hindex3}
&\mathbb{II}_{\mathcal{N}'}^{\textrm{4d $U(2)|U(3)$}}=
\mathbb{II}_{\mathcal{D}}^{\textrm{4d $U(3)|U(2)$}}
\nonumber\\
&=
\underbrace{
\frac{(q)_{\infty}}{(q^{\frac12} t^2;q)_{\infty}}
}_{\mathbb{II}_{\mathcal{D}}^{\textrm{4d $U(1)$}}}
\sum_{n=0}^{\infty}\sum_{m=0}^{\infty}
\frac{
(q^{1+n};q)_{\infty}
(q^{\frac32+n}t^2;q)_{\infty}}
{(q^{\frac12+n}t^2;q)_{\infty}
(q^{1+n}t^4;q)_{\infty}}
\cdot 
\frac{
(q^{\frac32+n+m}t^2;q)_{\infty}
(q^{2+n+m} t^4;q)_{\infty}}
{(q^{1+n+m}t^{4};q)_{\infty} (q^{\frac32+n+m}t^6;q)_{\infty}}
q^{\frac{2n+m}{2}} t^{-2(2n+m)}. 
\end{align}

It would be nice to find a similar expansion for the bi-fundamental Higgsing, and to prove all these conjectural equalities.

\subsubsection{$U(N)|U(M)$}
\label{sec_3duNuM}
Now we would like to propose the generalization of the results so far. We already discussed the $N=M$ case above, 
so we can focus on $N < M$ without loss of generality. 

For the NS5$'$-type interface between $N$ D3-branes in $x^2>0$ and $M$ D3-branes in $x^2<0$, 
we have 4d $\mathcal{N}=4$ $U(N)$ gauge theory for $x^2>0$, 
4d $\mathcal{N}=4$ $U(M)$ gauge theory for $x^2<0$ 
and 3d $\mathcal{N}=4$ hypermultiplets 
transforming as $({\bf N},\overline{\bf M})$ $\oplus$ $(\overline{{\bf N}},{\bf M})$ 
under the gauge group $U(N)$ $\times$ $U(M)$. 

The half-index for NS5$'$-type interface between $U(N)$ and $U(M)$ gauge theories is 
\begin{align}
\label{4duNuM_hindex}
\mathbb{II}_{\mathcal{N}'}^{\textrm{4d $U(N)|U(M)$}}
&=
\underbrace{
\frac{1}{N!}\frac{(q)_{\infty}^N}{(q^{\frac12} t^2;q)_{\infty}^N}
\oint \prod_{i=1}^{N}\frac{ds_{i}}{2\pi is_{i}}
\prod_{i\neq j}\frac{\left(\frac{s_{i}}{s_{j}};q\right)_{\infty}}
{\left(
q^{\frac12} t^2\frac{s_{i}}{s_{j}};q
\right)_{\infty}}
}_{\mathbb{II}_{\mathcal{N}'}^{\textrm{4d $U(N)$}}}
\underbrace{
\frac{1}{M!}\frac{(q)_{\infty}^M}{(q^{\frac12} t^2;q)_{\infty}^M}
\oint \prod_{i=N+1}^{N+M}\frac{ds_{i}}{2\pi is_{i}}
\prod_{i\neq j}\frac{\left(\frac{s_{i}}{s_{j}};q\right)_{\infty}}
{\left(
q^{\frac12} t^2\frac{s_{i}}{s_{j}};q
\right)_{\infty}}
}_{\mathbb{II}_{\mathcal{N}'}^{\textrm{4d $U(M)$}}}
\nonumber\\
&\times 
\prod_{i=1}^{N}\prod_{k=N+1}^{N+M}
\underbrace{
\frac{
\left(q^{\frac34}t \frac{s_{i}}{s_{k}};q\right)_{\infty}
\left(q^{\frac34}t \frac{s_{k}}{s_{i}};q\right)_{\infty}
}
{
\left(q^{\frac14}t^{-1} \frac{s_{i}}{s_{k}};q\right)_{\infty}
\left(q^{\frac14}t^{-1} \frac{s_{k}}{s_{i}};q\right)_{\infty}
}
}_{\mathbb{I}^{\textrm{3d tHM}}\left(\frac{s_{i}}{s_{k}}\right)}. 
\end{align}

The S-dual configuration is a single D5-brane at $x^6=0$, 
$N$ D3-branes in $x^6<0$ and $M$ D3-branes in $x^6>0$. 
The D5-brane break down 4d gauge group from $U(N)$ $\times$ $U(M)$ to $U(N)$ at the interface.
If $|N-M|>1$ there is a regular Nahm pole  in the block-diagonal $U(M-N)$ sugbroup of $U(M)$.
The diagonal $U(1)$ in $U(M-N)$ survives as a global symmetry at the interface.
There are no hypermultiplets at the defect. 

The half-index for D5-type interface between $U(N)$ and $U(M)$ gauge theories will take the form:
\begin{align}
\label{4duNuM_hindex2}
&\mathbb{II}_{\mathcal{D}}^{\textrm{4d $U(N)|U(M)$}}
\nonumber\\
&=
\underbrace{
\frac{1}{N!}
\frac{(q)_{\infty}^{2N}}
{(q^{\frac12} t^2;q)_{\infty}^{N} (q^{\frac12}t^{-2};q)_{\infty}^{N}}
\oint \prod_{i=1}^{N}
\frac{ds_{i}}{2\pi is_{i}}
\prod_{i\neq j}
\frac{
\left(\frac{s_{i}}{s_{j}};q\right)_{\infty}
\left(q \frac{s_{i}}{s_{j}};q\right)_{\infty}
}
{
\left(q^{\frac12} t^2 \frac{s_{i}}{s_{j}};q\right)_{\infty}
\left(q^{\frac12} t^{-2} \frac{s_{i}}{s_{j}};q\right)_{\infty}
}
}_{\mathbb{I}^{\textrm{4d $U(N)$}}}
\nonumber\\
&\times 
\underbrace{
\prod_{k=1}^{M-N}
\frac{(q^{\frac{k+1}{2}}t^{2(k-1)};q)_{\infty}}
{(q^{\frac{k}{2}}t^{2k};q)_{\infty}}
}_{\mathbb{II}_{\textrm{Nahm}}^{\textrm{4d $U(M-N)$}}}
\cdot 
\prod_{i=1}^{N}
\frac{
\left(q^{\frac{3}{4}+\frac{M-N}{4}}t^{-1+M-N}s_{i};q\right)_{\infty}
\left(q^{\frac{3}{4}+\frac{M-N}{4}}t^{-1+M-N}s_{i}^{-1};q\right)_{\infty}
}
{
\left(q^{\frac14+\frac{M-N}{4}}t^{1+M-N}s_{i};q\right)_{\infty}
\left(q^{\frac14+\frac{M-N}{4}}t^{1+M-N}s_{i}^{-1};q\right)_{\infty}
}. 
\end{align}
We expect that the half-index (\ref{4duNuM_hindex}) of the NS5$'$-type interface 
and the half-index (\ref{4duNuM_hindex2}) of the D5-type interface coincide. 

It would be nice to derive some appropriate Higgsing expansion.

\section{NS5$'$-D5 junctions}
\label{sec_ndjunction}
When the intersecting fivebranes are NS5$'$ and D5, we find no constraints on the numbers of D3-branes in the four quadrants. 
We label the NS5$'$-D5 junction by the numbers of D3-branes as 
$\left(\begin{smallmatrix}
N&M\\L&K\\\end{smallmatrix}\right)$ 
in such a way that the NS5$'$-brane splits four elements into two rows 
$\left(\begin{smallmatrix}N&M\\\end{smallmatrix}\right)$ and 
$\left(\begin{smallmatrix}L&K\\\end{smallmatrix}\right)$ 
while the D5-brane separates them into two columns 
$\left(\begin{smallmatrix}N\\L\\\end{smallmatrix}\right)$ and 
$\left(\begin{smallmatrix}M\\K\\\end{smallmatrix}\right)$. 

The NS5$'$-brane will impose Neumann boundary conditions to the gauge groups on the two sides, 
together with bi-fundamental twisted hypers. The D5-brane will either support fundamental hypers or 
impose appropriate reductions of the gauge groups and Nahm poles. 

We conjecture that at the junction the fundamental hypers, if present, will have Neumann boundary conditions. 
The bi-fundamental twisted hypers, instead, will be glued across the junction in a manner analogous to 
the respective gauge groups. Extra Fermi multiplets may be needed in order to cancel anomalies. 
The charge assignment of hypers, twisted hypers and Fermi's will allow the introduction of extra cubic 
couplings required to preserve $\mathcal{N}=(0,4)$ SUSY \cite{Tong:2014cha}.

The Fermi multiplets and cubic couplings also play an important role in enforcing the correct spaces of vacua and 
deformations for the gauge theory setup. In particular, it must be possible to find deformations and vacua which correspond 
to D3-brane pieces moving along the fivebranes, or recombining and moving away from some fivebranes,
possibly in conjunction with the fivebranes moving away from each other. When possible, the corresponding Higgsing manipulations of the 
quarter-indices must match the expected IR physics. 

Finally, as for the Y-junctions we study later on, these junctions are expected to admit a deformation supporting a vertex operator algebra
as in \cite{Gaiotto:2017euk}. We will describe the relation between the appropriate specialization of the quarter-indices and 
appropriate constructions of the vertex algebras as BRST reductions of simpler algebras. 

\subsection{${N \, N \choose 0 \, 0}$ and ${0 \, N \choose 0 \, N}$}
\label{sec_2dndNN00}
\subsubsection{${1 \, 1 \choose 0 \, 0}$ and ${0 \, 1 \choose 0 \, 1}$}
Consider the 
$\left(\begin{smallmatrix}
1&1\\0&0\\\end{smallmatrix}\right)$ 
NS5$'$-D5 junction. Both the $U(1)$ gauge theory and the extra fundamental 3d hyper associated to the D5-brane
should have Neumann boundary conditions at the NS5$'$-brane. 
According to the brane box analysis in \cite{Hanany:2018hlz} and gauge anomaly cancellation, 
there should be an additional charged Fermi multiplet at the intersection of NS5$'$-brane and D5-brane. 
This Fermi multiplet will be also charged under an extra $U(1)_x$ flavor symmetry with fugacity $x$.
Notice that because of mixed anomalies, $U(1)_x$ is also identified with the ``topological'' symmetries present at the Neumann b.c., with currents $*F_\partial$ and the $U(1)$ global symmetry of the fundamental hyper. 
In the brane setup, this flavor symmetry is identified with a diagonal combination of the $U(1)$ 
gauge symmetries on the fivebrane worldvolumes. 

We thus obtain the following quarter-index for the $\left(\begin{smallmatrix}
1&1\\0&0\\\end{smallmatrix}\right)$ NS5$'$-D5 junction:
\begin{align}
\label{nd1100}
\mathbb{IV}_{\mathcal{N}'\mathcal{D}}^{\left(\begin{smallmatrix}
1&1\\
0&0\\
\end{smallmatrix}\right)}&=
\underbrace{
\frac{(q)_{\infty}}
{(q^{\frac12}t^2;q)_{\infty}}\oint \frac{ds}{2\pi is}
}_{\mathbb{II}_{\mathcal{N}'}^{\textrm{4d $U(1)$}}}
\underbrace{
\frac{(q^{\frac12} sx;q)_{\infty}(q^{\frac12}s^{-1}x^{-1};q)_{\infty}}
{(q^{\frac14}ts;q)_{\infty}(q^{\frac14} ts^{-1};q)_{\infty}}
}_{\mathbb{II}_{N}^{\textrm{3d HM}}(s)\cdot F(q^{\frac12} sx)}.
\end{align}

The S-dual configuration is 
the $\left(\begin{smallmatrix}
0&1\\0&1\\\end{smallmatrix}\right)$ NS5$'$-D5 junction. 
The D5-brane completely breaks the two $U(1)$ gauge groups, imposing Dirichlet b.c. to both gauge fields and to the 
3d $\mathcal{N}=4$ twisted hypermultiplet associated to the NS5$'$-brane.
The boundary global symmetries for the gauge Dirichlet boundary conditions only act on the boundary 
values of the twisted hypermultiplet. We identify that with the $U(1)_x$ action in the dual picture, and with the worldvolume 
gauge symmetry theory of the fivebranes.  

Then the quarter-index of the $\left(\begin{smallmatrix}
0&1\\0&1\\\end{smallmatrix}\right)$ NS5$'$-D5 junction takes the form 
\begin{align}
\label{nd0101}
\mathbb{IV}_{\mathcal{N}'\mathcal{D}}^{\left(\begin{smallmatrix}
0&1\\
0&1\\
\end{smallmatrix}\right)}&=
\underbrace{
\frac{1}{(q^{\frac12} t^2;q)_{\infty}}
}_{\mathbb{IV}_{\mathcal{N}'\mathcal{D}}^{\textrm{4d $U(1)$}}}
\underbrace{
\frac{1}{(q^{\frac12} t^2;q)_{\infty}}
}_{\mathbb{IV}_{\mathcal{N}'\mathcal{D}}^{\textrm{4d $U(1)$}}}
\underbrace{
(q^{\frac34} tx;q)_{\infty}(q^{\frac34}tx^{-1};q)_{\infty}
}_{\mathbb{II}_{D}^{\textrm{3d tHM}}}. 
\end{align}
As expected, the quarter-indices (\ref{nd1100}) and (\ref{nd0101}) coincide. 

A few observations are in order. First of all, the index has no poles as a function of the $x$ fugacity. There are no gauge-invariant bosonic operators charged under 
$U(1)_x$ which can get a vev. 

Next, we can expand the contour integral in $\mathbb{IV}_{\mathcal{N}'\mathcal{D}}^{\left(\begin{smallmatrix}
1&1\\
0&0\\
\end{smallmatrix}\right)}$ as a sum over residues at $s = q^{\frac14 + n} t$
\begin{align}
\label{nd1100a}
\mathbb{IV}_{\mathcal{N}'\mathcal{D}}^{\left(\begin{smallmatrix}
1&1\\
0&0\\
\end{smallmatrix}\right)}&=
\frac{1}{(q)_{\infty} (q^{\frac12}t^2;q)_{\infty}}
(q^{\frac34}tx;q)_{\infty}(q^{\frac14}t^{-1}x^{-1};q)_{\infty}
\sum_{n=0}^{\infty}
\frac{(q^{1+n};q)_{\infty}}
{(q^{\frac12+n}t^2;q)_{\infty}}q^{\frac{n}{4}}t^{-n}x^{-n}. 
\end{align}
This is associated to a relative separation of the 
two D3-brane quadrants enforced by a vev of the fundamental hypermultiplet. 
The sum begins with the square of the 
$\mathbb{IV}_{\mathcal{N}'\mathcal{D}}^{\left(\begin{smallmatrix}
1&0\\
0&0\\
\end{smallmatrix}\right)}$ quarter-index as well as the Fermi index $F(q^{\frac34}tx)$. 
Applying the $q$-binomial theorem (\ref{q_binomial}), 
we can see that (\ref{nd1100a}) is equal to the quarter-index (\ref{nd0101}). 

The quarter-index of the $\left(\begin{smallmatrix}
0&1\\0&1\\\end{smallmatrix}\right)$ NS5$'$-D5 junction also has an expansion 
\begin{align}
\label{nd0101_sum}
\mathbb{IV}_{\mathcal{N}'\mathcal{D}}^{\left(\begin{smallmatrix}
0&1\\0&1\\\end{smallmatrix}\right)}
&=
\frac{1}
{(q^{\frac12} t^2;q)_{\infty}^2}
\frac{1}{(q)_{\infty}^2}
\sum_{n=0}^{\infty}\sum_{k=0}^{n}
(q^{1+k};q)_{\infty}
(q^{1+n-k};q)_{\infty} 
(-1)^{n}
x^{2k-n}
t^{n}
q^{\frac{n^2}{2}+\frac{n}{4}+k(k-n)}. 
\end{align}
Unlike the residue sum (\ref{nd1100a}), 
the sum (\ref{nd0101_sum}) has the first term as just the square of the $\mathbb{IV}_{\mathcal{N}'\mathcal{D}}^{\left(\begin{smallmatrix}
1&0\\
0&0\\
\end{smallmatrix}\right)}$ quarter-index. We are not sure about the significance of that fact.

Furthermore, we can evaluate the integral (\ref{nd1100}) as 
\begin{align}
\label{nd1100b}
\mathbb{IV}_{\mathcal{N}'\mathcal{D}}^{\left(\begin{smallmatrix}
1&1\\
0&0\\
\end{smallmatrix}\right)}&=
\frac{(q^{\frac14} t^{-1}x;q)_{\infty}(q^{\frac14}t^{-1}x^{-1};q)_{\infty}}
{(q)_{\infty}(q^{\frac12}t^2;q)_{\infty}}
\sum_{m=0}^{\infty}
\frac{(q^{1+m};q)_{\infty}^2}
{(q^{\frac14+m}t^{-1}x;q)_{\infty} (q^{\frac14+m}t^{-1}x^{-1};q)_{\infty}}
q^{\frac{m}{2}}t^{2m} 
\end{align}
by using the $q$-binomial theorem (\ref{q_binomial}). 
In contrast to the sum of residue at $s = q^{\frac14 + n} t$, the sum (\ref{nd1100b}) starts with the half-index $\mathbb{II}_{\mathcal{N}'}^{\textrm{4d $U(1)$}}$ of the Neumann boundary condition $\mathcal{N}'$ 
for 4d $\mathcal{N}=4$ $U(1)$ gauge theory. It should correspond to the two quarter D3-branes merging and separating from the 
D5-brane.

The specialization $t \to q^{\frac14}$ in the quarter-indices gives characters of the $\mathfrak{\widehat u}(1|1)$ Kac-Moody algebra,
either directly or by the description as $\mathfrak{\widehat u}(1)$-BRST reduction of $ \mathfrak{\widehat u}(1) \times \mathrm{Sb} \times \mathrm{Fc} \times \mathfrak{\widehat u}(1)$.

Finally, we can try to add a Wilson line $\mathcal{W}_{n}$ of charge $n$. 
It will end at the junction on a local operator of gauge charge $-n$. 
The quarter-index for such local operators tentatively takes the form
\begin{align}
\label{nd1100wil}
\mathbb{IV}_{\mathcal{N}'\mathcal{D}+\mathcal{W}_{n}}^{\left(\begin{smallmatrix}
1&1\\0&0\\\end{smallmatrix}\right)}
&=
\frac{(q)_{\infty}}{(q^{\frac12}t^2;q)_{\infty}}
\oint \frac{ds}{2\pi is}
\frac{
(q^{\frac12}sx;q)_{\infty}
(q^{\frac12}s^{-1}x^{-1};q)_{\infty}
}
{(q^{\frac14}ts;q)_{\infty}
(q^{\frac14}ts^{-1};q)_{\infty}}s^{n}. 
\end{align}

By picking up the residues at poles $s=q^{\frac14+m}t$ 
and using the $q$-binomial theorem (\ref{q_binomial}), 
the quarter-index (\ref{nd1100wil}) is evaluated as 
\begin{align}
\label{nd1100wil_sum1}
\mathbb{IV}_{\mathcal{N}'\mathcal{D}+\mathcal{W}_{n}}^{\left(\begin{smallmatrix}
1&1\\0&0\\\end{smallmatrix}\right)}
&=
\frac{
(q^{\frac34} tx;q)_{\infty} 
(q^{\frac14}t^{-1}x^{-1};q)_{\infty}
}
{(q^{\frac12} t^2;q)_{\infty}^2 }
\sum_{m=0}^{\infty}
\frac{(q^{\frac12}t^2;q)_{m}}
{(q)_{m}}
q^{\frac{m}{4}+nm+\frac{n}{4}}
t^{-m+n}
x^{-m}
\nonumber\\
&=
\frac{1}{(q^{\frac12}t^2;q)_{\infty}^2}
(q^{\frac34}tx;q)_{\infty}
(q^{\frac34+n}tx^{-1};q)_{\infty}
\frac{
(q^{\frac14}t^{-1}x^{-1};q)_{\infty}
}
{
(q^{\frac14+n}t^{-1}x^{-1};q)_{\infty}
}
q^{\frac{n}{4}}t^n
\nonumber\\
&=
\frac{
(q^{\frac34+n}tx^{-1};q)_{\infty}
}
{
(q^{\frac14+n}t^{-1}x^{-1};q)_{\infty}
}
\frac{
(q^{\frac14}t^{-1}x^{-1};q)_{\infty}
}
{
(q^{\frac34}tx^{-1};q)_{\infty}
}
q^{\frac{n}{4}}t^n
\cdot
\mathbb{IV}_{\mathcal{N}'\mathcal{D}}^{\left(\begin{smallmatrix}
0&1\\0&1\\\end{smallmatrix}\right)}. 
\end{align}
which should have an interpretation in terms of endpoints of a 't Hooft operator of charge $n$. 
We defer discussions of the dualities involving the line operators to future work.

\subsubsection{${2 \, 2 \choose 0 \, 0}$ and ${0 \, 2 \choose 0 \, 2}$}
For the 
$\left(\begin{smallmatrix}
2&2\\0&0\\\end{smallmatrix}\right)$ 
NS5$'$-D5 junction, 
there are two semi-infinite D3-branes in $x^2>0$ which terminate on the NS5$'$-brane. 
This corresponds to the 4d $U(2)$ gauge theory with Neumann boundary condition $\mathcal{N}'$. 
There is the 3d $\mathcal{N}=4$ fundamental hypermultiplet arising from the D5-brane. 
It should obey the Neumann boundary condition $N'$. 
Similarly to the Abelian case, anomaly cancellation requires fundamental Fermi multiplets as well. 

The quarter-index for the $\left(\begin{smallmatrix}
2&2\\0&0\\\end{smallmatrix}\right)$ NS5$'$-D5 junction takes the form 
\begin{align}
\label{nd2200}
\mathbb{IV}_{\mathcal{N}'\mathcal{D}}^{\left(\begin{smallmatrix}
2&2\\0&0\\\end{smallmatrix}\right)}
&=
\underbrace{
\frac12 \frac{(q)_{\infty}^2}
{(q^{\frac12} t^2;q)_{\infty}^2}
\oint \frac{ds_{1}}{2\pi is_{1}}\frac{ds_{2}}{2\pi is_{2}}
\frac{
\left(\frac{s_{1}}{s_{2}};q\right)_{\infty}
\left(\frac{s_{2}}{s_{1}};q\right)_{\infty}}
{
\left(q^{\frac12} t^2\frac{s_{1}}{s_{2}};q\right)_{\infty}
\left(q^{\frac12} t^2\frac{s_{2}}{s_{1}};q\right)_{\infty}
}
}_{\mathbb{II}_{\mathcal{N}'}^{\textrm{4d $U(2)$}}}
\nonumber\\
&\times 
\underbrace{
\frac{(q^{\frac12}s_{1}x;q)_{\infty} (q^{\frac12} s_{1}^{-1}x^{-1};q)_{\infty}\cdot 
(q^{\frac12}s_{2}x;q)_{\infty} (q^{\frac12} s_{2}^{-1}x^{-1};q)_{\infty}}
{
(q^{\frac14} t s_{1};q)_{\infty} (q^{\frac14} t s_{1}^{-1};q)_{\infty}\cdot 
(q^{\frac14} t s_{2};q)_{\infty} (q^{\frac14} t s_{2}^{-1};q)_{\infty}
}
}_{\mathbb{II}_{N}^{\textrm{3d HM}}(s_{1})\cdot \mathbb{II}_{N}^{\textrm{3d HM}}(s_{2})\cdot F(q^{\frac12} s_{1}x)\cdot F(q^{\frac12} s_{2}x)}. 
\end{align}

The S-dual configuration is the $\left(\begin{smallmatrix}
0&2\\0&2\\\end{smallmatrix}\right)$ NS5$'$-D5 junction. 
Now the difference of numbers of D3-branes across the D5-brane is two
and the two $U(2)$ gauge fields must have regular Nahm pole boundary conditions there. 

We need to understand the effect of the Nahm pole boundary conditions on the bi-fundamental twisted hypermultiplet.
Rather than trying to identify the correct ``disorder'' definition of the boundary condition, i.e. the 
correct singular solution of the joint equations of motion for the gauge field and twisted hypermultiplet, 
we can tentatively define the system as an RG flow from Dirichlet boundary conditions for gauge fields and 3d twisted hypers,
triggered by deforming the gauge fields b.c. by a nilpotent expectation value for $X^+$, removing by hand eventual 
decoupled fields. 

In any case, we can check that 
the quarter-index (\ref{nd2200}) of the 
$\left(\begin{smallmatrix}
2&2\\0&0\\\end{smallmatrix}\right)$ 
NS5$'$-D5 junction 
turns out to coincide with the following expression, which is a very natural proposal for the quarter-index of 
the  $\left(\begin{smallmatrix} 0&2\\0&2\\\end{smallmatrix}\right)$ NS5$'$-D5 junction:
\begin{align}
\label{nd0202}
\mathbb{IV}_{\mathcal{N}'\mathcal{D}}^{\left(\begin{smallmatrix}
0&2\\0&2\\\end{smallmatrix}\right)}
&=
\underbrace{
\frac{1}
{
(q^{\frac12} t^2;q)_{\infty} 
(q t^4;q)_{\infty} 
}
}_{\mathbb{IV}_{\mathcal{N}'\textrm{Nahm}}^{\textrm{4d $U(2)$}}}
\underbrace{
\frac{1}
{
(q^{\frac12} t^2;q)_{\infty} 
(q t^4;q)_{\infty} 
}
}_{\mathbb{IV}_{\mathcal{N}'\textrm{Nahm}}^{\textrm{4d $U(2)$}}}
(q^{\frac34} tx;q)_{\infty}
(q^{\frac34} tx^{-1};q)_{\infty}
\cdot 
(q^{\frac54} t^3x;q)_{\infty}
(q^{\frac54} t^3x^{-1};q)_{\infty}. 
\end{align}
We recognize the gauge field contributions we found in section \ref{sec_3du3u1} in the denominator, 
while the numerator can clearly arise from the same Higgsing procedure applied to the bi-fundamental twisted hypermultiplet. 

More precisely, before we subtract off the decoupled free fields originating from the twisted hypermultiplet, the numerator would be 
\begin{equation}
(q^{\frac14} t^{-1} x;q)_{\infty}
(q^{\frac14} t^{-1} x^{-1};q)_{\infty}
\cdot 
(q^{\frac34} tx;q)^2_{\infty}
(q^{\frac34} tx^{-1};q)^2_{\infty}
\cdot 
(q^{\frac54} t^3x;q)_{\infty}
(q^{\frac54} t^3x^{-1};q)_{\infty}. 
\end{equation}
The factors $(q^{\frac14} t^{-1} x;q)_{\infty}(q^{\frac34} tx^{-1};q)_{\infty}$ and $(q^{\frac14} t^{-1} x^{-1};q)_{\infty} (q^{\frac34} tx;q)_{\infty}$
we strip off are indices of 2d Fermi multiplets with fugacities $q^{\frac14} t^{-1} x^{\pm 1}$. 

The quarter-index of the $\left(\begin{smallmatrix}
0&2\\0&2\\\end{smallmatrix}\right)$ NS5$'$-D5 junction can be expanded as
\begin{align}
\label{nd0202_sum}
\mathbb{IV}_{\mathcal{N}'\mathcal{D}}^{\left(\begin{smallmatrix}
0&2\\0&2\\\end{smallmatrix}\right)}
&=
\frac{1}{
(q^{\frac12}t^2;q)_{\infty}^2 
(q t^4;q)_{\infty}^2
}
\frac{1}{(q)_{\infty}^4}
\sum_{n=0}^{\infty}
\sum_{m=0}^{\infty}
\sum_{k=0}^{n}\sum_{l=0}^{m}
(q^{1+k};q)_{\infty}
(q^{1+l};q)_{\infty}
(q^{1+n-k};q)_{\infty}
(q^{1+m-l};q)_{\infty}
\nonumber\\
&\times 
(-1)^{n+m}
x^{-n-m+2k+2l}
t^{n+3m}
q^{\frac{n^2+m^2-n-m}{2}
+k(k-n)+l(l-m)+\frac{n+m}{2}+\frac{n+3m}{4}}. 
\end{align}
The expansion starts from the square of the $\mathbb{IV}_{\mathcal{N}'\mathcal{D}}^{\left(\begin{smallmatrix}
2&0\\
0&0\\
\end{smallmatrix}\right)}$ quarter-index and should correspond to giving a vev to the fundamental hypermultiplet. 

Another expansion from which we may try to draw a lesson is 
\begin{align}
\label{nd2200expand}
\mathbb{IV}_{\mathcal{N}'\mathcal{D}}^{\left(\begin{smallmatrix}
2&2\\0&0\\\end{smallmatrix}\right)}
&=
\mathbb{IV}_{\mathcal{N}'\mathcal{D}}^{\left(\begin{smallmatrix}
0&2\\0&2\\\end{smallmatrix}\right)}
\nonumber\\
&=
\frac{1}{2}
\frac{(q^{-\frac12}t^{-2};q)_{\infty}^2 (q^{\frac34}tx;q)_{\infty}^2 (q^{\frac14}t^{-1}x^{-1};q)_{\infty}^2}
{(q)_{\infty}^2 (q^{\frac12}t^2;q)_{\infty}^4}\nonumber\\
&\times 
\sum_{n=0}^{\infty}\sum_{k=0}^{n }
\frac{(q^{\frac34+n-2k}tx^{-1};q)_{\infty} (q^{\frac34+2k-n}tx^{-1};q)_{\infty} (q^{1+k};q)_{\infty} (q^{1+n-k};q)_{\infty}}
{(q^{\frac14+n-2k}t^{-1}x^{-1};q)_{\infty} (q^{\frac14+2k-n}t^{-1}x^{-1};q)_{\infty} (q^{-\frac12+k}t^{-2};q)_{\infty} (q^{-\frac12+n-k}t^{-2};q)_{\infty}} 
q^{\frac{n}{2}} t^{2n}. 
\end{align}
The first term is the product of 
the fourth power of the quarter-index $\mathbb{IV}_{\mathcal{N}'\mathcal{D}}^{\textrm{4d $U(1)$}}$ and 
the square of the half-index $\mathbb{II}_{D}^{\textrm{3d tHM}}(x)$ of the Dirichlet boundary condition $D$ for the twisted hypermultiplet. 
This may have a Higgsing interpretation.

The specialization $t \to q^{\frac14}$ in the quarter-indices gives characters of the Drinfeld-Sokolov reduction of the 
$\mathfrak{\widehat u}(2|2)$ Kac-Moody algebra by the regular $\mathfrak{su}(2)$ embedding in both $\mathfrak{\widehat u}(2)$ 
sub-algebras, either directly or by the description as $\mathfrak{\widehat u}(2)$-BRST reduction of $\mathfrak{\widehat u}(2) \times \mathrm{Sb}^2 \times \mathrm{Fc}^2 \times \mathfrak{\widehat u}(2)$.
The DS reduction involves a certain collection of fermionic and bosonic ghosts which remove Kac-Moody generators whose dimension 
would have been non-positive. These have precisely the effect of stripping off the denominator and numerator factors above. 

\subsubsection{${3 \, 3 \choose 0 \, 0}$ and ${0 \, 3 \choose 0 \, 3}$}
To get more insight, let us proceed to the 
$\left(\begin{smallmatrix}
3&3\\0&0\\\end{smallmatrix}\right)$ 
NS5$'$-D5 junction. 
Three semi-infinite D3-branes in $x^2>0$ ending on the NS5$'$-brane 
give the Neumann boundary condition $\mathcal{N}'$ for 4d $\mathcal{N}=4$ $U(3)$ gauge theory. 
The D5-brane ending on NS5$'$ brane give the 3d $\mathcal{N}=4$ fundamental hypermultiplet 
obeying the Neumann boundary condition $N$ together with the fundamental Fermi multiplets. 

Then the quarter-index for the $\left(\begin{smallmatrix}
3&3\\0&0\\\end{smallmatrix}\right)$ NS5$'$-D5 junction is given by 
\begin{align}
\label{nd3300}
\mathbb{IV}_{\mathcal{N}'\mathcal{D}}^{\left(\begin{smallmatrix}
3&3\\0&0\\\end{smallmatrix}\right)}
&=
\underbrace{
\frac{1}{3!}\frac{(q)_{\infty}^{3}}
{(q^{\frac12} t^2;q)_{\infty}^{3}}
\oint \prod_{i=1}^{3}\frac{ds_{i}}{2\pi is_{i}}
\prod_{i\neq j}\frac{
\left(\frac{s_{i}}{s_{j}};q\right)_{\infty}
}
{
\left(
q^{\frac12} t^2 \frac{s_{i}}{s_{j}};q
\right)_{\infty}
}
}_{\mathbb{II}_{\mathcal{N}'}^{\textrm{4d $U(N)$}}}
\prod_{i=1}^{3}
\underbrace{
\frac{(q^{\frac12}s_{i}x;q)_{\infty} (q^{\frac12}s_{i}^{-1}x^{-1};q)_{\infty}}
{(q^{\frac14}t s_{i};q)_{\infty}(q^{\frac14}t s_{i}^{-1};q)_{\infty}}
}_{\mathbb{II}_{N}^{\textrm{3d HM}}(s_{i})\cdot F(q^{\frac12}s_{i}x)}. 
\end{align}

It turns out that 
the quarter-index (\ref{nd3300}) or (\ref{nd0303_sum}) is equal to
\begin{align}
\label{nd0303}
\mathbb{IV}_{\mathcal{N}'\mathcal{D}}^{\left(\begin{smallmatrix}
0&3\\0&3\\\end{smallmatrix}\right)}
&=
\underbrace{
\frac{1}
{
(q^{\frac12}t^2;q)_{\infty}
(q t^4;q)_{\infty}
(q^{\frac32}t^6;q)_{\infty}
}
}_{\mathbb{IV}_{\mathcal{N}'\textrm{Nahm}}^{\textrm{4d $U(3)$}}}
\underbrace{
\frac{1}
{
(q^{\frac12}t^2;q)_{\infty}
(q t^4;q)_{\infty}
(q^{\frac32}t^6;q)_{\infty}
}
}_{\mathbb{IV}_{\mathcal{N}'\textrm{Nahm}}^{\textrm{4d $U(3)$}}}
\nonumber\\
&\times 
(q^{\frac34}tx;q)_{\infty} 
(q^{\frac34}tx^{-1};q)_{\infty} \cdot 
(q^{\frac54}t^3x;q)_{\infty} 
(q^{\frac54}t^3x^{-1};q)_{\infty} \cdot 
(q^{\frac74}t^5x;q)_{\infty} 
(q^{\frac74}t^5x^{-1};q)_{\infty}
\end{align}
which is very reasonable for the junction involving two regular Nahm poles 
for the $U(3)$ gauge groups and related boundary condition for the bi-fundamental twisted hypers. 

Again, to match with the Higgsing of a Dirichlet quarter-index to the Nahm pole index we strip off numerator factors
 $(q^{-\frac14} t^{-1} x^\pm;q)_{\infty}(q^{\frac54} tx^{\mp};q)_{\infty}$ and $(q^{\frac14} t^{-1} x^{\pm};q)^2_{\infty} (q^{\frac34} t x^{\mp} ;q)^2_{\infty}$
which are indices of 2d Fermi multiplets.

The quarter-index (\ref{nd3300}) can be expanded as 
\begin{align}
\label{nd0303_sum}
\mathbb{IV}_{\mathcal{N}'\mathcal{D}}^{\left(\begin{smallmatrix}
0&3\\0&3\\\end{smallmatrix}\right)}
&=
\frac{1}
{(q^{\frac12} t^2;q)_{\infty}^2 (qt^{4};q)_{\infty}^2 (q^{\frac32}t^6;q)_{\infty}^2}
\frac{1}{(q)_{\infty}^6}
\sum_{n_{1}=0}^{\infty}
\sum_{n_{2}=0}^{\infty}
\sum_{n_{3}=0}^{\infty}
\sum_{k_{1}=0}^{n_{1}}
\sum_{k_{2}=0}^{n_{2}}
\sum_{k_{3}=0}^{n_{3}}
\nonumber\\
&\times 
(q^{1+k_{1}};q)_{\infty}
(q^{1+n_{1}-k_{1}};q)_{\infty}
(q^{1+k_{2}};q)_{\infty}
(q^{1+n_{2}-k_{2}};q)_{\infty}
(q^{1+k_{3}};q)_{\infty}
(q^{1+n_{3}-k_{3}};q)_{\infty}
\nonumber\\
&\times 
(-1)^{n_{1}+n_{2}+n_{3}}
x^{2k_{1}+2k_{2}+2k_{3}-n_{1}-n_{2}-n_{3}}
t^{n_{1}+3n_{2}+5n_{3}}
\nonumber\\
&\times 
q^{\frac{n_{1}^2+n_{2}^2+n_{3}^2}{2}+\frac{n_{1}+3n_{2}+5n_{3}}{4}+k_{1}(k_{1}-n_{1})+k_{2}(k_{2}-n_{2})+k_{3}(k_{3}-n_{3})}. 
\end{align}
The first term in the sum takes the form of the square of the $\mathbb{IV}_{\mathcal{N}'\mathcal{D}}^{\left(\begin{smallmatrix}
3&0\\
0&0\\
\end{smallmatrix}\right)}$ quarter-index and it may be possible to interprete this sum in terms of a Higgsing  process  
giving a vev to the fundamental hypermultiplet.

\subsubsection{${N \, N \choose 0 \, 0}$ and ${0 \, N \choose 0 \, N}$}
Now we would like to propose the generalization for the $\left(\begin{smallmatrix}
N&N\\0&0\\\end{smallmatrix}\right)$ NS5$'$-D5 junction. Again, all fields get Neumann b.c. at the NS5$'$ brane and we add a fundamental Fermi multiplet at the junction.

Therefore the quarter-index 
for the $\left(\begin{smallmatrix}
N&N\\0&0\\\end{smallmatrix}\right)$ NS5$'$-D5 junction takes the form 
\begin{align}
\label{ndNN00}
\mathbb{IV}_{\mathcal{N}'\mathcal{D}}^{\left(\begin{smallmatrix}
N&N\\0&0\\\end{smallmatrix}\right)}
&=
\underbrace{
\frac{1}{N!}\frac{(q)_{\infty}^{N}}
{(q^{\frac12} t^2;q)_{\infty}^{N}}
\oint \prod_{i=1}^{N}\frac{ds_{i}}{2\pi is_{i}}
\prod_{i\neq j}\frac{
\left(\frac{s_{i}}{s_{j}};q\right)_{\infty}
}
{
\left(
q^{\frac12} t^2 \frac{s_{i}}{s_{j}};q
\right)_{\infty}
}
}_{\mathbb{II}_{\mathcal{N}'}^{\textrm{4d $U(N)$}}}
\prod_{i=1}^{N}
\underbrace{
\frac{(q^{\frac12}s_{i}x;q)_{\infty} (q^{\frac12}s_{i}^{-1}x^{-1};q)_{\infty}}
{(q^{\frac14}t s_{i};q)_{\infty}(q^{\frac14}t s_{i}^{-1};q)_{\infty}}
}_{\mathbb{II}_{N}^{\textrm{3d HM}}(s_{i})\cdot F(q^{\frac12}s_{i}x)}. 
\end{align}

In a manner analogous to the previous examples, 
we expect that the quarter-index for the $\left(\begin{smallmatrix}
N&N\\0&0\\\end{smallmatrix}\right)$ NS5$'$-D5 junction 
will coincide with the following quarter-index for 
the $\left(\begin{smallmatrix}
0&N\\0&N\\\end{smallmatrix}\right)$ NS5$'$-D5 junction:
\begin{align}
\label{nd0N0N}
\mathbb{IV}_{\mathcal{N}'\mathcal{D}}^{\left(\begin{smallmatrix}
0&N\\0&N\\\end{smallmatrix}\right)}
&=
\underbrace{
\prod_{k=1}^{N}\frac{1}{(q^{\frac{k}{2}} t^{2k};q)_{\infty}}
}_{\mathbb{IV}_{\mathcal{N}'\textrm{Nahm}/\mathcal{D}}^{\textrm{4d $U(N)$}}}
\underbrace{
\prod_{k=1}^{N}\frac{1}{(q^{\frac{k}{2}} t^{2k};q)_{\infty}}
}_{\mathbb{IV}_{\mathcal{N}'\textrm{Nahm}/\mathcal{D}}^{\textrm{4d $U(N)$}}}
\prod_{k=1}^{N}
\left(
q^{\frac34+\frac{k-1}{2}}t^{1+2(k-1)}x;q
\right)_{\infty}
\left(
q^{\frac34+\frac{k-1}{2}}t^{1+2(k-1)}x^{-1};q
\right)_{\infty}. 
\end{align}

The quarter-index (\ref{ndNN00}) of 
the $\left(\begin{smallmatrix}
0&N\\0&N\\\end{smallmatrix}\right)$ NS5$'$-D5 junction has an expansion
\begin{align}
\label{nd0N0N_sum}
\mathbb{IV}_{\mathcal{N}'\mathcal{D}}^{\left(\begin{smallmatrix}
0&N\\0&N\\\end{smallmatrix}\right)}
&=
\prod_{k=1}^{N}
\frac{1}{(q^{\frac{k}{2}}t^{2k};q)_{\infty}^2}
\frac{1}{(q)_{\infty}^{2N}}
\prod_{i=1}^{N}
\sum_{n_{i}=0}^{\infty}
\sum_{k_{i}=0}^{n_{i}}
(q^{1+k_{i}};q)_{\infty}
(q^{1+n_{i}-k_{i}};q)_{\infty}
\nonumber\\
&\times 
(-1)^{n_{i}}
x^{2k_{i}-n_{i}}
t^{(2i-1)n_{i}}
q^{\frac{n_{i}(n_{i}-1)}{2}
+k_{i}(k_{i}-n_{i})
+\frac{n_{i}}{4}
+\frac{in_{i}}{2}
}. 
\end{align}
The sum begins with the square of the $\mathbb{IV}_{\mathcal{N}'\mathcal{D}}^{\left(\begin{smallmatrix}
N&0\\
0&0\\
\end{smallmatrix}\right)}$ quarter-index and may be associated to giving a vev to the fundamental hypermultiplet.

\subsection{${N \, M \choose 0 \, 0}$ and ${0 \, N \choose 0 \, M}$}
\label{sec_2dndNM00}
\subsubsection{${1 \, 2 \choose 0 \, 0}$ and ${0 \, 1 \choose 0 \, 2}$}

Now consider the case with different numbers of D3-branes across the NS5$'$-brane or the D5-brane. 

We start with the $\left(\begin{smallmatrix}
1&2\\0&0\\\end{smallmatrix}\right)$ NS5$'$-D5 junction. 

We now have a 4d gauge theory with a gauge group which is reduced from $U(2)$ to $U(1)$ across the D5 
interface. We assign Neumann boundary conditions at the NS5$'$ boundary and add a single Fermi multiplet 
charged under $U(1)$ at the junction for anomaly cancellation or based on the brane box analysis \cite{Hanany:2018hlz}.

Then the quarter-index for the $\left(\begin{smallmatrix}
1&2\\0&0\\\end{smallmatrix}\right)$ NS5$'$-D5 junction is given by
\begin{align}
\label{nd1200}
\mathbb{IV}_{\mathcal{N}'\mathcal{D}}^{\left(\begin{smallmatrix}
1&2\\0&0\\\end{smallmatrix}\right)}
&=
\underbrace{
\frac{(q)_{\infty}}
{(q^{\frac12} t^2;q)_{\infty}}\oint \frac{ds}{2\pi is}
}_{\mathbb{II}_{\mathcal{N}'}^{\textrm{4d $U(1)$}}}
\underbrace{
\frac{1}
{(q^{\frac12} t^2;q)_{\infty}}
}_{\mathbb{IV}_{\mathcal{N}'\mathcal{D}}^{\textrm{4d $U(1)$}}}
\frac{(q^{\frac12} sx;q)_{\infty} (q^{\frac12}s^{-1}x^{-1};q)_{\infty}}
{(q^{\frac12} t^2 s;q)_{\infty} (q^{\frac12} t^2 s^{-1};q)_{\infty}}. 
\end{align}
This can be computed as the sum over residues at $s=q^{\frac12+m}t^2$
\begin{align}
\label{nd1200sum}
\mathbb{IV}_{\mathcal{N}'\mathcal{D}}^{\left(\begin{smallmatrix}
1&2\\0&0\\\end{smallmatrix}\right)}
&=\frac{1}{(q)_{\infty}(q^{\frac12} t^2;q)_{\infty}^2} 
(q t^2x;q)_{\infty} (t^{-2}x^{-1};q)_{\infty}
\sum_{m=0}^{\infty} 
\frac{(q^{1+m};q)_{\infty}}
{(q^{1+m} t^4;q)_{\infty}} t^{-2m}x^{-m}.
\end{align}
The first term includes the product of the quarter-indices 
$\mathbb{IV}_{\mathcal{N}'\mathcal{D}}^{\left(\begin{smallmatrix}
1&0\\0&0\\\end{smallmatrix}\right)}$ and 
$\mathbb{IV}_{\mathcal{N}'\mathcal{D}}^{\left(\begin{smallmatrix}
0&2\\0&0\\\end{smallmatrix}\right)}$ as well as the Fermi index $F(q t^2 x)$. 
We associate the residue sum (\ref{nd1200sum}) to the  
splitting of the NS5$'$-D5 junction as 
$\left(\begin{smallmatrix}
1&2\\0&0\\\end{smallmatrix}\right)$ $\rightarrow$ $\left(\begin{smallmatrix}
1&0\\0&0\\\end{smallmatrix}\right)$ $\oplus$ 
$\left(\begin{smallmatrix}
0&2\\0&0\\\end{smallmatrix}\right)$. 

In addition, the integral can be expanded in terms of the $q$-binomial theorem (\ref{q_binomial}) as
\begin{align}
\label{nd1200sum2}
\mathbb{IV}_{\mathcal{N}'\mathcal{D}}^{\left(\begin{smallmatrix}
1&2\\0&0\\\end{smallmatrix}\right)}
&=
\frac{1}{(q)_{\infty} (q^{\frac12}t^2;q)_{\infty}^2} 
(t^{-2}x;q)_{\infty} (t^{-2}x^{-1};q)_{\infty}
\sum_{m=0}^{\infty}
\frac{(q^{1+m};q)_{\infty}^2}
{(q^{m}t^{-2}x;q)_{\infty} (q^{m}t^{-2}x^{-1};q)_{\infty}} 
q^{m} t^{4m}. 
\end{align}
This sum has the first term as the product of 
the half-index $\mathbb{II}_{\mathcal{N}'}^{\textrm{4d $U(1)$}}$ 
and the quarter-index $\mathbb{IV}_{\mathcal{N}'\mathcal{D}}^{\textrm{4d $U(1)$}}$ for 4d $\mathcal{N}=4$ gauge theory. 
The associated deformation of brane configuration is 
the decomposition of the NS5$'$-D5 junction as 
$\left(\begin{smallmatrix}
1&2\\0&0\\\end{smallmatrix}\right)$ $\rightarrow$ $\left(\begin{smallmatrix}
1&1\\0&0\\\end{smallmatrix}\right)$ $\oplus$ 
$\left(\begin{smallmatrix}
0&1\\0&0\\\end{smallmatrix}\right)$. 

After the usual specialization of $t$ this reduces to the character of a $\mathfrak{u}(1)$-BRST reduction of 
$\mathfrak{\widehat u}(2) \times \mathfrak{\widehat u}(1) \times \mathrm{Ff}$.  

The S-dual is the $\left(\begin{smallmatrix}
0&1\\0&2\\\end{smallmatrix}\right)$ NS5$'$-D5 junction. 
We have the usual bi-fundamental twisted hyper at the NS5$'$ interface
and we impose Dirichlet boundary condition for the $U(1)$ gauge fields and regular Nahm pole 
for the $U(2)$ gauge fields at the D5 boundary. 

In fact, the quarter-index (\ref{nd1200sum}) for the $\left(\begin{smallmatrix}
1&2\\0&0\\\end{smallmatrix}\right)$ NS5$'$-D5 junction
turns out to coincides with the following 
quarter-index for the $\left(\begin{smallmatrix}
0&1\\0&2\\\end{smallmatrix}\right)$ NS5$'$-D5 junction:
\begin{align}
\label{nd0102}
\mathbb{IV}_{\mathcal{N}'\mathcal{D}}^{\left(\begin{smallmatrix}
0&1\\0&2\\\end{smallmatrix}\right)}
&=
\underbrace{
\frac{1}{(q^{\frac12} t^2;q)_{\infty}(qt^4;q)_{\infty}}
}_{\mathbb{IV}_{\mathcal{N}'\textrm{Nahm}}^{\textrm{4d $U(2)$}}}
\underbrace{
\frac{1}{(q^{\frac12} t^2;q)_{\infty}}
}_{\mathbb{IV}_{\mathcal{N}'\mathcal{D}}^{\textrm{4d $U(1)$}}}
(q t^2x;q)_{\infty} (q t^2x^{-1};q)_{\infty}. 
\end{align}
This can be justified as we did before, as the effect of the Nahm pole 
modification of the Dirichlet b.c. indices of the gauge theories and twisted hypermultiplet,
where we strip off the Fermi index $(q^{\frac12} x;q)_{\infty} (q^{\frac12}  x^{-1};q)_{\infty}$
from the numerator. 

After the usual specialization of $t$ this reduces to the character of the standard DS reduction of
$\mathfrak{\widehat u}(2|1)$.

\subsubsection{${1 \, 3 \choose 0 \, 0}$ and ${0 \, 1 \choose 0 \, 3}$}
The next simplest case is the $\left(\begin{smallmatrix}
1&3\\0&0\\\end{smallmatrix}\right)$ NS5$'$-D5 junction. 

Now the $U(3)$ gauge group is reduced to $U(1)$ across the D5 interface by a rank $2$ Nahm pole. 
The gauge fields are given a uniform Neumann b.c. at the NS5$'$ boundary condition. We again need a 
single extra Fermi multiplet charged under the $U(1)$ gauge group. 
 
The quarter-index for the $\left(\begin{smallmatrix}
1&3\\0&0\\\end{smallmatrix}\right)$ NS5$'$-D5 junction can thus be expressed as
\begin{align}
\label{nd1300}
\mathbb{IV}_{\mathcal{N}'\mathcal{D}}^{\left(\begin{smallmatrix}
1&3\\0&0\\\end{smallmatrix}\right)}
&=
\underbrace{
\frac{(q)_{\infty}}
{(q^{\frac12} t^2;q)_{\infty}}\oint \frac{ds}{2\pi is}
}_{\mathbb{II}_{\mathcal{N}'}^{\textrm{4d $U(1)$}}}
\underbrace{
\frac{1}
{(q^{\frac12} t^2;q)_{\infty} (q t^4;q)_{\infty}}
}_{\mathbb{IV}_{\mathcal{N}'\mathrm{Nahm}}^{\textrm{4d $U(2)$}}}
\frac{(q^{\frac12} sx;q)_{\infty} (q^{\frac12}s^{-1}x^{-1};q)_{\infty}}
{(q^{\frac34} t^3 s;q)_{\infty} (q^{\frac34} t^3 s^{-1};q)_{\infty}}. 
\end{align}
Evaluating the integral as the sum over residues at $s=q^{\frac34+m}t^3$, we find 
\begin{align}
\label{nd1300sum}
\mathbb{IV}_{\mathcal{N}'\mathcal{D}}^{\left(\begin{smallmatrix}
1&3\\0&0\\\end{smallmatrix}\right)}
&=
\frac{
(q^{\frac54} t^3x;q)_{\infty} (q^{-\frac14}t^{-3}x^{-1};q)_{\infty}
}{(q)_{\infty} (q^{\frac12}t^2;q)_{\infty}^2 (qt^4;q)_{\infty}}
\sum_{m=0}^{\infty}\frac{(q^{1+m};q)_{\infty}}
{(q^{\frac32+m}t^6;q)_{\infty}} 
q^{-\frac{m}{4}}t^{-3m} x^{-m}
\end{align}
The first term in the sum is the product of the quarter-indices 
$\mathbb{IV}_{\mathcal{N}'\mathcal{D}}^{\left(\begin{smallmatrix}
1&0\\0&0\\\end{smallmatrix}\right)}$ and 
$\mathbb{IV}_{\mathcal{N}'\mathcal{D}}^{\left(\begin{smallmatrix}
0&3\\0&0\\\end{smallmatrix}\right)}$ as well as the Fermi index $F(q^{\frac54} t^3 x)$. 
Physically, the residue sum (\ref{nd1300sum}) describes the deformation of the 
NS5$'$-D5 junction as 
$\left(\begin{smallmatrix}
1&3\\0&0\\\end{smallmatrix}\right)$ $\rightarrow$ $\left(\begin{smallmatrix}
1&0\\0&0\\\end{smallmatrix}\right)$ $\oplus$ 
$\left(\begin{smallmatrix}
0&3\\0&0\\\end{smallmatrix}\right)$. 

Besides, we have another expansion of the quarter-index for the $\left(\begin{smallmatrix}
1&3\\0&0\\\end{smallmatrix}\right)$ NS5$'$-D5 junction:
\begin{align}
\label{nd1300sum2}
\mathbb{IV}_{\mathcal{N}'\mathcal{D}}^{\left(\begin{smallmatrix}
1&3\\0&0\\\end{smallmatrix}\right)}
&=
\frac{
(q^{-\frac14}t^{-3}x;q)_{\infty} (q^{-\frac14}t^{-3}x^{-1};q)_{\infty}
}
{
(q)_{\infty} (q^{\frac12}t^2;q)_{\infty}^2 (qt^4;q)_{\infty}
}
\sum_{m=0}^{\infty} 
\frac{
(q^{1+m};q)_{\infty}^2
}
{
(q^{-\frac14+m}t^{-3}x;q)_{\infty} 
(q^{-\frac14+m}t^{-3}x^{-1};q)_{\infty}
}q^{\frac{3m}{2}}t^{6m}. 
\end{align}
The expansion (\ref{nd1300sum2}) begins with the product of 
the half-index $\mathbb{II}_{\mathcal{N}'}^{\textrm{4d $U(1)$}}$ 
and the quarter-index $\mathbb{IV}_{\mathcal{N}'\textrm{Nahm}}^{\textrm{4d $U(2)$}}$ for 4d $\mathcal{N}=4$ gauge theory. 
This should be associated to a Higgsing procedure splitting the 
NS5$'$-D5 junction as 
$\left(\begin{smallmatrix}
1&3\\0&0\\\end{smallmatrix}\right)$ $\rightarrow$ $\left(\begin{smallmatrix}
1&1\\0&0\\\end{smallmatrix}\right)$ $\oplus$ 
$\left(\begin{smallmatrix}
0&2\\0&0\\\end{smallmatrix}\right)$.

For the S-dual $\left(\begin{smallmatrix}
0&1\\0&3\\\end{smallmatrix}\right)$ NS5$'$-D5 junction, 
we have again the bi-fundamental NS5$'$ interface based on a combination of Dirichlet 
b.c. for the $U(1)$ gauge field and regular Nahm pole for the $U(3)$ gauge group. 

We propose that the quarter-index (\ref{nd1300}) for the $\left(\begin{smallmatrix}
1&3\\0&0\\\end{smallmatrix}\right)$ NS5$'$-D5 junction
is equal to the following 
quarter-index for the $\left(\begin{smallmatrix}
0&1\\0&3\\\end{smallmatrix}\right)$ NS5$'$-D5 junction:
\begin{align}
\label{nd0103}
\mathbb{IV}_{\mathcal{N}'\mathcal{D}}^{\left(\begin{smallmatrix}
0&1\\0&3\\\end{smallmatrix}\right)}
&=
\underbrace{
\frac{1}{(q^{\frac12} t^2;q)_{\infty}(qt^4;q)_{\infty}(q^{\frac32} t^6;q)_{\infty}}
}_{\mathbb{IV}_{\mathcal{N}'\textrm{Nahm}}^{\textrm{4d $U(3)$}}}
\underbrace{
\frac{1}{(q^{\frac12} t^2;q)_{\infty}}
}_{\mathbb{IV}_{\mathcal{N}'\mathcal{D}}^{\textrm{4d $U(1)$}}}
(q^{\frac54}t^3x;q)_{\infty} 
(q^{\frac54}t^3x^{-1};q)_{\infty}. 
\end{align}
We can verify experimentally that the expressions 
(\ref{nd1300}), (\ref{nd1300sum}), (\ref{nd1300sum2}) and (\ref{nd0103}) give the same answer.

\subsubsection{${1 \, N \choose 0 \, 0}$ and ${0 \, 1 \choose 0 \, N}$}

We are led to the generalization to the junction with 
a single D3-brane and $N$ D3-branes filled across fivebrane. 

The quarter-index for the $\left(\begin{smallmatrix}
1&N\\0&0\\\end{smallmatrix}\right)$ NS5$'$-D5 junction can be written as 
\begin{align}
\label{nd1n00}
\mathbb{IV}_{\mathcal{N}'\mathcal{D}}^{\left(\begin{smallmatrix}
1&N\\0&0\\\end{smallmatrix}\right)}
&=
\underbrace{
\frac{(q)_{\infty}}
{(q^{\frac12} t^2;q)_{\infty}}\oint \frac{ds}{2\pi is}
}_{\mathbb{II}_{\mathcal{N}'}^{\textrm{4d $U(1)$}}}
\underbrace{
\prod_{k=1}^{N-1}\frac{1}{(q^{\frac{k}{2}}t^{2k};q)_{\infty}}
}_{\mathbb{IV}_{\mathcal{N}'\mathcal{D}/\textrm{Nahm}}^{\textrm{4d $U(N-1)$}}}
\frac{(q^{\frac12}sx;q)_{\infty} (q^{\frac12}s^{-1} x^{-1};q)_{\infty}}
{(q^{\frac14+\frac{N-1}{4}}t^{1+(N-1)}s;q)_{\infty}
q^{\frac14+\frac{N-1}{4}}t^{1+(N-1)}s^{-1};q)_{\infty}}
\end{align}
where the denominator in the integrand captures the bosonic operators 
which will originate from the broken $U(N)$ gauge theory 
while the numerator describes the charged Fermi multiplet which cancels the $U(1)$ gauge anomaly.

\begin{figure}
\begin{center}
\includegraphics[width=13.5cm]{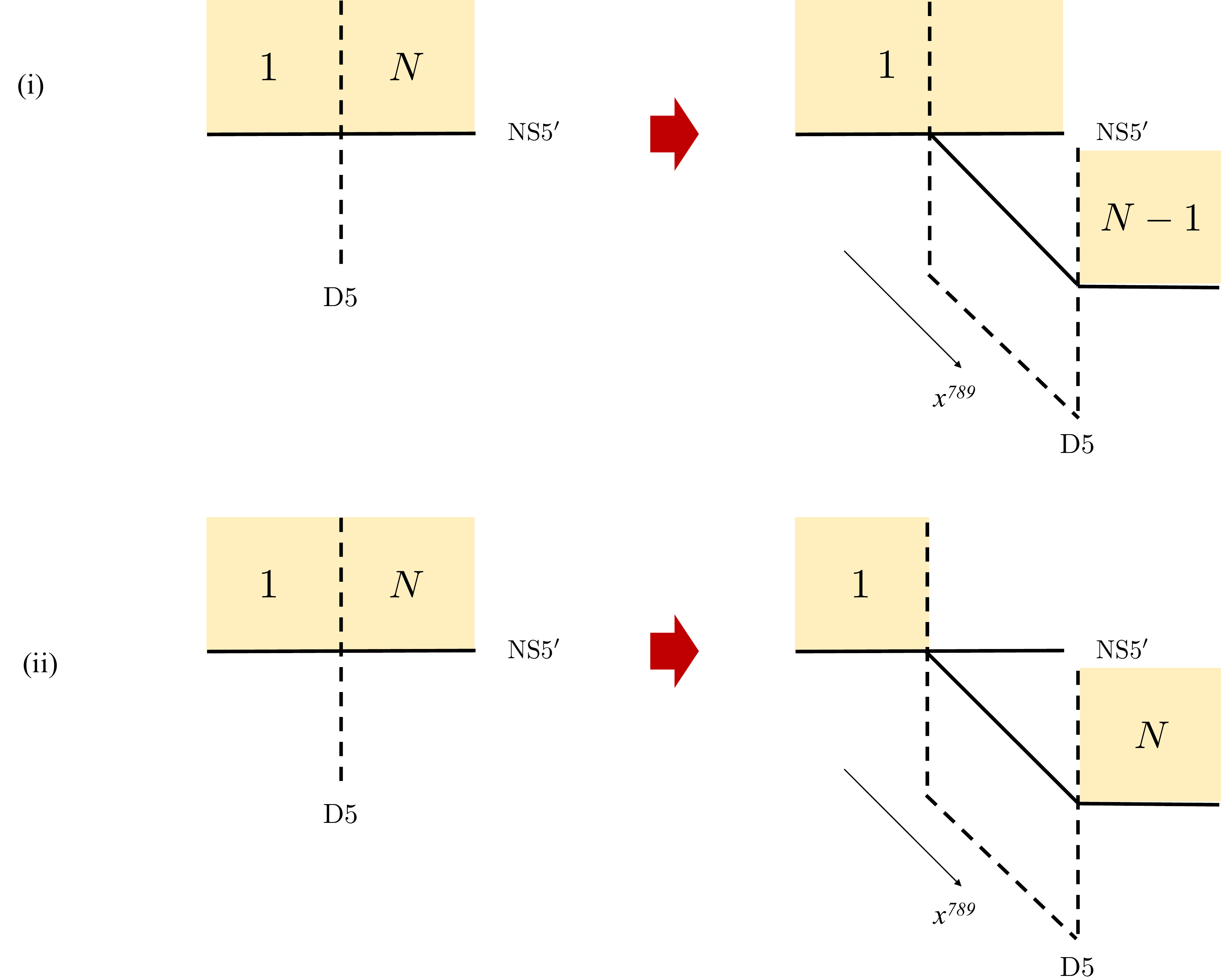}
\caption{Higgsing procedures corresponding to the deformations of the ${1 \, N \choose 0 \, 0}$ NS5$'$-D5 junction in Type IIB brane configurations. 
(i) ${1 \, N \choose 0 \, 0}$ $\rightarrow$ ${1 \, 1 \choose 0 \, 0}\oplus {\ 0\ \, N-1 \choose 0\ \ \ \ \, 0}$ (ii) ${1 \, N \choose 0 \, 0}$ $\rightarrow$ ${1 \, 0 \choose 0 \, 0}\oplus 
{0 \, N \choose 0 \, 0}$. }
\label{fighiggsing2}
\end{center}
\end{figure}

The quarter-index (\ref{nd1n00}) can be computed as 
the sum of residues at poles $s=q^{\frac{N}{4}+n}t^{N}$ 
\begin{align}
\label{nd1n00sum1}
\mathbb{IV}_{\mathcal{N}'\mathcal{D}}^{\left(\begin{smallmatrix}
1&N\\0&0\\\end{smallmatrix}\right)}
&=
\prod_{k=1}^{N}\frac{1}{(q^{\frac{k}{2}} t^{2k};q)_{\infty}}
 \frac{1}{(q^{\frac12}t^{2};q)_{\infty}}
 (q^{\frac{N}{4}+\frac12}t^{N}x;q)_{\infty}
 (q^{-\frac{N}{4}+\frac12}t^{-N}x^{-1};q)_{\infty}
 \nonumber\\
 &\times 
 \sum_{n=0}^{\infty}\frac{(q^{\frac{N}{2}}t^{2N};q)_{n}}{(q)_{n}}
 q^{-\frac{Nn}{4}+\frac{n}{2}}t^{-Nn}x^{-n}. 
\end{align}
The sum in (\ref{nd1n00sum1}) has the first term which takes the form of the product of the quarter-indices 
$\mathbb{IV}_{\mathcal{N}'\mathcal{D}}^{\left(\begin{smallmatrix}
1&0\\0&0\\\end{smallmatrix}\right)}$ and 
$\mathbb{IV}_{\mathcal{N}'\mathcal{D}}^{\left(\begin{smallmatrix}
0&N\\0&0\\\end{smallmatrix}\right)}$ as well as the Fermi index $F(q^{\frac{N}{4}+\frac12} t^N x)$. 
In the brane configuration the D3-branes are divided from 
$\left(\begin{smallmatrix}
1&N\\0&0\\\end{smallmatrix}\right)$ to $\left(\begin{smallmatrix}
1&0\\0&0\\\end{smallmatrix}\right)$ $\oplus$ $\left(\begin{smallmatrix}
0&N\\0&0\\\end{smallmatrix}\right)$ as shown in Figure \ref{fighiggsing2}.

The integral in (\ref{nd1n00}) can also be computed by expanding the integrand in terms of the $q$-binomial theorem (\ref{q_binomial}). 
We find that 
\begin{align}
\label{nd1n00sum2}
\mathbb{IV}_{\mathcal{N}'\mathcal{D}}^{\left(\begin{smallmatrix}
1&N\\0&0\\\end{smallmatrix}\right)}
&=\frac{1}{(q)_{\infty}(q^{\frac12}t^2;q)_{\infty}}
\prod_{k=1}^{N-1}\frac{1}{(q^{\frac{k}{2}}t^{2k};q)_{\infty}}
(q^{\frac12-\frac{N}{4}} t^{-N}x^{\pm};q)_{\infty}
\sum_{m=0}^{\infty}
\frac{
(q^{1+m};q)_{\infty}^2
}
{
(q^{\frac12-\frac{N}{4}+m}t^{-N}x^{\pm};q)_{\infty}
}
q^{\frac{Nm}{2}}t^{2Nm}. 
\end{align}
Unlike the sum in (\ref{nd1n00sum1}), 
the first term in (\ref{nd1n00sum2}) is the product of 
the half-index $\mathbb{II}_{\mathcal{N}'}^{\textrm{4d $U(1)$}}$ 
and the quarter-index $\mathbb{IV}_{\mathcal{N}'\mathcal{D}/\textrm{Nahm}}^{\textrm{4d $U(N-1)$}}$ for 4d $\mathcal{N}=4$ gauge theory. 
This sum is associated to the deformation of brane configuration where the D3-branes are divided from 
$\left(\begin{smallmatrix}
1&N\\0&0\\\end{smallmatrix}\right)$ to $\left(\begin{smallmatrix}
1&1\\0&0\\\end{smallmatrix}\right)$ $\oplus$ $\left(\begin{smallmatrix}
0&N-1\\0&0\\\end{smallmatrix}\right)$, as illustrated in Figure \ref{fighiggsing2}. 

The S-dual junction is the $\left(\begin{smallmatrix}
0&1\\0&N\\\end{smallmatrix}\right)$ NS5$'$-D5 junction. 
We expect that its quarter-index reads
\begin{align}
\label{nd010N}
\mathbb{IV}_{\mathcal{N}'\mathcal{D}}^{\left(\begin{smallmatrix}
0&1\\0&N\\\end{smallmatrix}\right)}
&=
\underbrace{
\prod_{k=1}^{N}
\frac{1}{(q^{\frac{k}{2}}t^{2k};q)_{\infty}}
}_{\mathbb{IV}_{\mathcal{N}'\mathcal{D}/\textrm{Nahm}}^{\textrm{4d $U(N)$}}}
\underbrace{
\frac{1}{(q^{\frac12} t^2;q)_{\infty}}
}_{\mathbb{IV}_{\mathcal{N}'\mathcal{D}}^{\textrm{4d $U(1)$}}}
(q^{\frac34+\frac{N-1}{4}} t^{1+(N-1)}x;q)_{\infty}
(q^{\frac34+\frac{N-1}{4}} t^{1+(N-1)}x^{-1};q)_{\infty}. 
\end{align}
In fact,  we can show that 
the residue sum (\ref{nd1n00sum1}) is equal to the dual quarter-index (\ref{nd010N}) 
by employing the $q$-binomial theorem (\ref{q_binomial}).

Introducing a Wilson line $\mathcal{W}_{n}$ of charge $n$ to the $\left(\begin{smallmatrix}
1&N\\0&0\\\end{smallmatrix}\right)$ NS5$'$-D5 junction, 
the quarter-index takes the form 
\begin{align}
\label{nd1N00wil}
\mathbb{IV}_{\mathcal{N}'\mathcal{D}+\mathcal{W}_{n}}^{\left(\begin{smallmatrix}
1&N\\0&0\\\end{smallmatrix}\right)}
&=
\frac{(q)_{\infty}}{(q^{\frac12}t^2;q)_{\infty}}
\oint \frac{ds}{2\pi is}
\prod_{k=1}^{N-1}
\frac{1}{(q^{\frac{k}{2}} t^{2k};q)_{\infty}}
\frac{
(q^{\frac12}sx;q)_{\infty}
(q^{\frac12}s^{-1}x^{-1};q)_{\infty}
}
{
(q^{\frac{N}{4}}t^{N}s;q)_{\infty}
(q^{\frac{N}{4}}t^{N}s^{-1};q)_{\infty}
}s^{n}. 
\end{align}

By taking the sum over residues at poles $s=q^{\frac{N}{4}+m}t^{N}$, we get
\begin{align}
\label{nd1N00wil_sum}
\mathbb{IV}_{\mathcal{N}'\mathcal{D}+\mathcal{W}_{n}}^{\left(\begin{smallmatrix}
1&N\\0&0\\\end{smallmatrix}\right)}
&=
\frac{1}
{
(q^{\frac12} t^2;q)_{\infty}
}
\prod_{k=1}^{N}
\frac{1}{(q^{\frac{k}{2}}t^{2k};q)_{\infty}}
(q^{\frac12+\frac{N}{4}}t^{N}x;q)_{\infty}
(q^{\frac12-\frac{N}{4}}t^{-N}x^{-1};q)_{\infty}
\nonumber\\
&\times 
\sum_{m=0}^{\infty}
\frac{
(q^{\frac{N}{2}}t^{2N};q)_{m}
}
{
(q)_{m}
}
q^{\frac{m}{2}-\frac{Nm}{4}+nm+\frac{Nn}{4}}
t^{-Nm+Nn}
x^{-m}
\nonumber\\
&=
\frac{1}{(q^{\frac12} t^2;q)_{\infty}}
\prod_{k=1}^{N}\frac{1}{(q^{\frac{k}{2}} t^{2k};q)_{\infty}}
%
(q^{\frac12+\frac{N}{4}}t^{N}x;q)_{\infty}
(q^{\frac12+\frac{N}{4}+n}t^{N}x^{-1};q)_{\infty}
\frac{
(q^{\frac12-\frac{N}{4}}t^{-N}x^{-1};q)_{\infty}
}
{
(q^{\frac12-\frac{N}{4}+n }t^{-N}x^{-1};q)_{\infty}
}
q^{\frac{Nn}{4}}t^{Nn}
\nonumber\\
&=
\frac{
(q^{\frac12+\frac{N}{4}+n}t^{N}x^{-1};q)_{\infty}
}
{
(q^{\frac12+\frac{N}{4}}t^{N}x^{-1};q)_{\infty}
}
\frac{
(q^{\frac12-\frac{N}{4}+n}t^{-N}x^{-1};q)_{\infty}
}
{
(q^{\frac12-\frac{N}{4}+n}t^{-N}x^{-1};q)_{\infty}
}
q^{\frac{Nn}{4}}t^{Nn}
\cdot
\mathbb{IV}_{\mathcal{N}'\mathcal{D}}^{\left(\begin{smallmatrix}
0&1\\0&N\\\end{smallmatrix}\right)}
\end{align}
where we have used the $q$-binomial theorem (\ref{q_binomial}). 
We leave the analysis of duality involving the line operators for future work.

\subsubsection{${2 \, 3 \choose 0 \, 0}$ and ${0 \, 2 \choose 0 \, 3}$}
Now consider the NS5$'$-D5 junction between two non-Abelian gauge groups.

A simple example is the $\left(\begin{smallmatrix}
2&3\\0&0\\\end{smallmatrix}\right)$ NS5$'$-D5 junction.
We describe it in the usual manner as a reduction from $U(3)$ to $U(2)$ across the 
D5 interface and Neumann b.c. at the NS5$'$ junction, with an extra fundamental Fermi multiplet
for $U(2)$ to cancel the anomaly. 

Then the quarter-index for the $\left(\begin{smallmatrix}
2&3\\0&0\\\end{smallmatrix}\right)$ NS5$'$-D5 junction is written as
\begin{align}
\label{nd2300}
\mathbb{IV}_{\mathcal{N}'\mathcal{D}}^{\left(\begin{smallmatrix}
2&3\\0&0\\\end{smallmatrix}\right)}
&=
\underbrace{
\frac12 \frac{(q)_{\infty}^2}{(q^{\frac12} t^2;q)_{\infty}^2}
\oint 
\frac{ds_{1}}{2\pi is_{1}}
\frac{ds_{2}}{2\pi is_{2}}
\frac
{
\left(\frac{s_{1}}{s_{2}};q\right)_{\infty}
\left(\frac{s_{2}}{s_{1}};q\right)_{\infty}
}
{
\left(q^{\frac12} t^2\frac{s_{1}}{s_{2}};q\right)_{\infty}
\left(q^{\frac12} t^2\frac{s_{2}}{s_{1}};q\right)_{\infty}
}
}_{\mathbb{II}_{\mathcal{N}'}^{\textrm{4d $U(2)$}}}
\nonumber\\
&\times 
\underbrace{
\frac{1}{(q^{\frac12}t^2;q)_{\infty}}
}_{\mathbb{IV}_{\mathcal{N}'\mathcal{D}}^{\textrm{4d $U(1)$}}}
\frac{
(q^{\frac12}s_{1}x;q)_{\infty}
(q^{\frac12}s_{1}^{-1}x^{-1};q)_{\infty}
\cdot 
(q^{\frac12}s_{2}x;q)_{\infty}
(q^{\frac12}s_{2}^{-1}x^{-1};q)_{\infty}
}
{
(q^{\frac12} t^2s_{1}x;q)_{\infty}
(q^{\frac12} t^2s_{1}^{-1}x^{-1};q)_{\infty}
\cdot 
(q^{\frac12} t^2s_{2}x;q)_{\infty}
(q^{\frac12} t^2s_{2}^{-1}x^{-1};q)_{\infty}
}. 
\end{align}

The S-dual configuration is the $\left(\begin{smallmatrix}
0&2\\0&3\\\end{smallmatrix}\right)$ NS5$'$-D5 junction. 

In fact we find that 
the quarter-index (\ref{nd2300}) for the $\left(\begin{smallmatrix}
0&2\\0&3\\\end{smallmatrix}\right)$ NS5$'$-D5 junction matches with  
\begin{align}
\label{nd0203}
\mathbb{IV}_{\mathcal{N}'\mathcal{D}}^{\left(\begin{smallmatrix}
0&2\\0&3\\\end{smallmatrix}\right)}
&=
\underbrace{
\frac{1}
{(q^{\frac12}t^2;q)_{\infty} (q t^4;q)_{\infty} (q^{\frac32} t^6;q)_{\infty}}
}_{\mathbb{IV}_{\mathcal{N}'\textrm{Nahm}}^{\textrm{4d $U(3)$}}}
\underbrace{
\frac{1}{
(q^{\frac12}t^2;q)_{\infty} 
(q t^4;q)_{\infty}}
}_{\mathbb{IV}_{\mathcal{N}'\textrm{Nahm}}^{\textrm{4d $U(2)$}}}
\nonumber\\
&\times 
(qt^2 x;q)_{\infty} (q t^2 x^{-1}:q)_{\infty}
\cdot 
(q^{\frac32}t^4 x;q)_{\infty} (q^{\frac32}t^4 x^{-1};q)_{\infty}
\end{align}
which should be explained in terms of the regular Nahm poles for $U(3)$ and $U(2)$ gauge fields 
and related boundary condition for the twisted bi-fundamental hypermultiplet.

\subsubsection{${2 \, 4 \choose 0 \, 0}$ and ${0 \, 2 \choose 0 \, 4}$}
As a next example, consider the $\left(\begin{smallmatrix}
2&4\\0&0\\\end{smallmatrix}\right)$ NS5$'$-D5 junction. 

The quarter-index for the $\left(\begin{smallmatrix}
2&4\\0&0\\\end{smallmatrix}\right)$ NS5$'$-D5 junction is calculated as 
\begin{align}
\label{nd2400}
\mathbb{IV}_{\mathcal{N}'\mathcal{D}}^{\left(\begin{smallmatrix}
2&4\\0&0\\\end{smallmatrix}\right)}
&=
\underbrace{
\frac12 \frac{(q)_{\infty}^2}{(q^{\frac12} t^2;q)_{\infty}^2}
\oint 
\frac{ds_{1}}{2\pi is_{1}}
\frac{ds_{2}}{2\pi is_{2}}
\frac
{
\left(\frac{s_{1}}{s_{2}};q\right)_{\infty}
\left(\frac{s_{2}}{s_{1}};q\right)_{\infty}
}
{
\left(q^{\frac12} t^2\frac{s_{1}}{s_{2}};q\right)_{\infty}
\left(q^{\frac12} t^2\frac{s_{2}}{s_{1}};q\right)_{\infty}
}
}_{\mathbb{II}_{\mathcal{N}'}^{\textrm{4d $U(2)$}}}
\nonumber\\
&\times 
\underbrace{
\frac{1}{(q^{\frac12}t^2;q)_{\infty} (qt^4;q)_{\infty}}
}_{\mathbb{IV}_{\mathcal{N}'\textrm{Nahm}}^{\textrm{4d $U(2)$}}}
\frac{
(q^{\frac12}s_{1}x;q)_{\infty}
(q^{\frac12}s_{1}^{-1}x^{-1};q)_{\infty}
\cdot 
(q^{\frac12}s_{2}x;q)_{\infty}
(q^{\frac12}s_{2}^{-1}x^{-1};q)_{\infty}
}
{
(q^{\frac34} t^3s_{1};q)_{\infty}
(q^{\frac34} t^3s_{1}^{-1};q)_{\infty}
\cdot 
(q^{\frac34} t^3s_{2};q)_{\infty}
(q^{\frac34} t^3s_{2}^{-1};q)_{\infty}
}. 
\end{align}

For the S-dual $\left(\begin{smallmatrix}
0&2\\0&4\\\end{smallmatrix}\right)$ NS5$'$-D5 junction, 
we can check that 
the quarter-index (\ref{nd2400}) for the $\left(\begin{smallmatrix}
2&4\\0&0\\\end{smallmatrix}\right)$ NS5$'$-D5 junction coincides with 
the following quarter-index for the $\left(\begin{smallmatrix}
0&2\\0&4\\\end{smallmatrix}\right)$ NS5$'$-D5 junction
\begin{align}
\label{nd0204}
\mathbb{IV}_{\mathcal{N}'\mathcal{D}}^{\left(\begin{smallmatrix}
0&2\\0&4\\\end{smallmatrix}\right)}
&=
\underbrace{
\frac{1}
{(q^{\frac12}t^2;q)_{\infty} 
(qt^{4};q)_{\infty}
(q^{\frac32}t^6;q)_{\infty}
(q^2 t^8;q)_{\infty}
}
}_{\mathbb{IV}_{\mathcal{N}'\textrm{Nahm}}^{\textrm{4d $U(4)$}}}
\underbrace{
\frac{1}
{(q^{\frac12} t^2;q)_{\infty}
(qt^4;q)_{\infty}}
}_{\mathbb{IV}_{\mathcal{N}'\textrm{Nahm}}^{\textrm{4d $U(2)$}}}
\nonumber\\
&\times 
(q^{\frac54}t^3x;q)_{\infty}
(q^{\frac54}t^3x^{-1};q)_{\infty}
(q^{\frac74}t^5x;q)_{\infty}
(q^{\frac74}t^5x^{-1};q)_{\infty}. 
\end{align}

\subsubsection{${3 \, 4 \choose 0 \, 0}$ and ${0 \, 3 \choose 0 \, 4}$}

The quarter-index for the $\left(\begin{smallmatrix}
3&4\\0&0\\\end{smallmatrix}\right)$ NS5$'$-D5 junction is 
\begin{align}
\label{nd3400}
\mathbb{IV}_{\mathcal{N}'\mathcal{D}}^{\left(\begin{smallmatrix}
3&4\\0&0\\\end{smallmatrix}\right)}
&=
\underbrace{
\frac{1}{3!} \frac{(q)_{\infty}^3}{(q^{\frac12} t^2;q)_{\infty}^3}
\oint 
\prod_{i=1}^{3}
\frac{ds_{i}}{2\pi is_{i}}
\prod_{i\neq j}
\frac
{
\left(\frac{s_{i}}{s_{j}};q\right)_{\infty}
}
{
\left(q^{\frac12} t^2\frac{s_{i}}{s_{j}};q\right)_{\infty}
}
}_{\mathbb{II}_{\mathcal{N}'}^{\textrm{4d $U(3)$}}}
\underbrace{
\frac{1}{(q^{\frac12}t^2;q)_{\infty}}
}_{\mathbb{IV}_{\mathcal{N}'\mathcal{D}}^{\textrm{4d $U(1)$}}}
\prod_{i=1}^{3}
\frac{
(q^{\frac12}s_{i}x;q)_{\infty}
(q^{\frac12}s_{i}^{-1}x^{-1};q)_{\infty}
}
{
(q^{\frac12} t^2s_{i};q)_{\infty}
(q^{\frac12} t^2s_{i}^{-1};q)_{\infty}
}. 
\end{align}

We find that 
the quarter-index for the $\left(\begin{smallmatrix}
3&4\\0&0\\\end{smallmatrix}\right)$ NS5$'$-D5 junction coincides with 
the following the quarter-index for the $\left(\begin{smallmatrix}
0&3\\0&4\\\end{smallmatrix}\right)$ NS5$'$-D5 junction:
\begin{align}
\label{nd0304}
\mathbb{IV}_{\mathcal{N}'\mathcal{D}}^{\left(\begin{smallmatrix}
0&3\\0&4\\\end{smallmatrix}\right)}
&=
\underbrace{
\frac{1}
{
(q^{\frac12}t^2;q)_{\infty}
(q t^4;q)_{\infty}
(q^{\frac32}t^6;q)_{\infty}
(q^2 t^8;q)_{\infty}
}
}_{\mathbb{IV}_{\mathcal{N}'\textrm{Nahm}}^{\textrm{4d $U(4)$}}}
\underbrace{
\frac{1}{
(q^{\frac12} t^2;q)_{\infty}
(q t^4;q)_{\infty}
(q^{\frac32} t^6;q)_{\infty}
}}_{\mathbb{IV}_{\mathcal{N}'\textrm{Nahm}}^{\textrm{4d $U(3)$}}}
\nonumber\\
&\times 
(qt^2x;q)_{\infty}(qt^2x^{-1};q)_{\infty}
\cdot
(q^{\frac32}t^4x;q)_{\infty}(q^{\frac32}t^4x^{-1};q)_{\infty}
\cdot 
(q^2 t^6x;q)_{\infty}(q^2t^6x^{-1};q)_{\infty}. 
\end{align}

\subsubsection{${N \, M \choose 0 \, 0}$ and ${0 \, N \choose 0 \, M}$}
Now we would like to propose the generalization 
of the duality between the $\left(\begin{smallmatrix}
N&M\\0&0\\\end{smallmatrix}\right)$ NS5$'$-D5 junction 
and the $\left(\begin{smallmatrix}
0&N\\0&M\\\end{smallmatrix}\right)$ NS5$'$-D5 junction with $N< M$. 

The $\left(\begin{smallmatrix}
N&M\\0&0\\\end{smallmatrix}\right)$ NS5$'$-D5 junction 
can be realized by adding the NS5$'$-brane 
to the D5-type interface of $U(N)|U(M)$. 
The initial $U(N)$ $\times$ $U(M)$ gauge symmetry is broken to $U(N)$. 
Consequently, the whole $U(N)$ gauge symmetry is kept in half-space $x^2>0$ 
and the broken part of the original $U(M)$ gauge theory 
gives contributions to the index. 
As the D5-type interface of $U(N)|U(M)$ further ends on the NS5$'$-brane, 
the 4d $\mathcal{N}=4$ $U(N)$ gauge theory satisfies the 
Neumann boundary condition $\mathcal{N}'$. 
In addition, the bosonic degrees of freedom in the broken part of the original $U(M)$ gauge theory 
remain at the junction and they transform as fundamental representation under the $U(N)$ gauge symmetry. 
The junction also has the fundamental Fermi multiplet 
that has fivebrane charge and cancels the boundary gauge anomaly.

The resulting quarter-index for the $\left(\begin{smallmatrix}
N&M\\0&0\\\end{smallmatrix}\right)$ NS5$'$-D5 junction takes the form
\begin{align}
\label{ndNM00}
\mathbb{IV}_{\mathcal{N}'\mathcal{D}}^{\left(\begin{smallmatrix}
N&M\\0&0\\\end{smallmatrix}\right)}
&=
\underbrace{
\frac{1}{N!}\frac{(q)_{\infty}^{N}}
{(q^{\frac12}t^2;q)_{\infty}^{N}}
\oint \prod_{i=1}^{N}
\frac{ds_{i}}{2\pi is_{i}}
\prod_{i\neq j}
\frac{
\left(
\frac{s_{i}}{s_{j}};q
\right)_{\infty}
}
{
\left(
q^{\frac12} t^2 \frac{s_{i}}{s_{j}};q
\right)_{\infty}
}
}_{\mathbb{II}_{\mathcal{N}'}^{\textrm{4d $U(N)$}}}
\nonumber\\
&\times 
\underbrace{
\prod_{k=1}^{M-N}
\frac{1}{(q^{\frac{k}{2}}t^{2k};q)_{\infty}}
}_{\mathbb{IV}_{\mathcal{N}'\mathcal{D}/\textrm{Nahm}}^{\textrm{4d $U(M-N)$}}}
\prod_{i=1}^{N}
\frac{
(q^{\frac12}s_{i}x;q)_{\infty}
(q^{\frac12}s_{i}^{-1}x^{-1};q)_{\infty}
}
{
\left(q^{\frac14+\frac{M-N}{4}}t^{1+M-N}s_{i};q\right)_{\infty}
\left(q^{\frac14+\frac{M-N}{4}}t^{1+M-N}s_{i}^{-1};q\right)_{\infty}
}. 
\end{align}

For the $\left(\begin{smallmatrix}
0&N\\0&M\\\end{smallmatrix}\right)$ NS5$'$-D5 junction, the D5-brane breaks the gauge symmetry. 
The local operators from 
4d $\mathcal{N}=4$ $U(N)$ and $U(M)$ gauge theories at the corner 
with boundary conditions $\mathcal{N}'$ and $\mathcal{D}$ live at the junction. 
In addition, 
the junction contains the 3d $\mathcal{N}=4$ twisted hypermultiplets 
arising from D3-D3 strings across the NS5$'$-brane. 
They may receive boundary conditions which are different from 
the regular Dirichlet boundary conditions, because of the Nahm poles.

\begin{figure}
\begin{center}
\includegraphics[width=10cm]{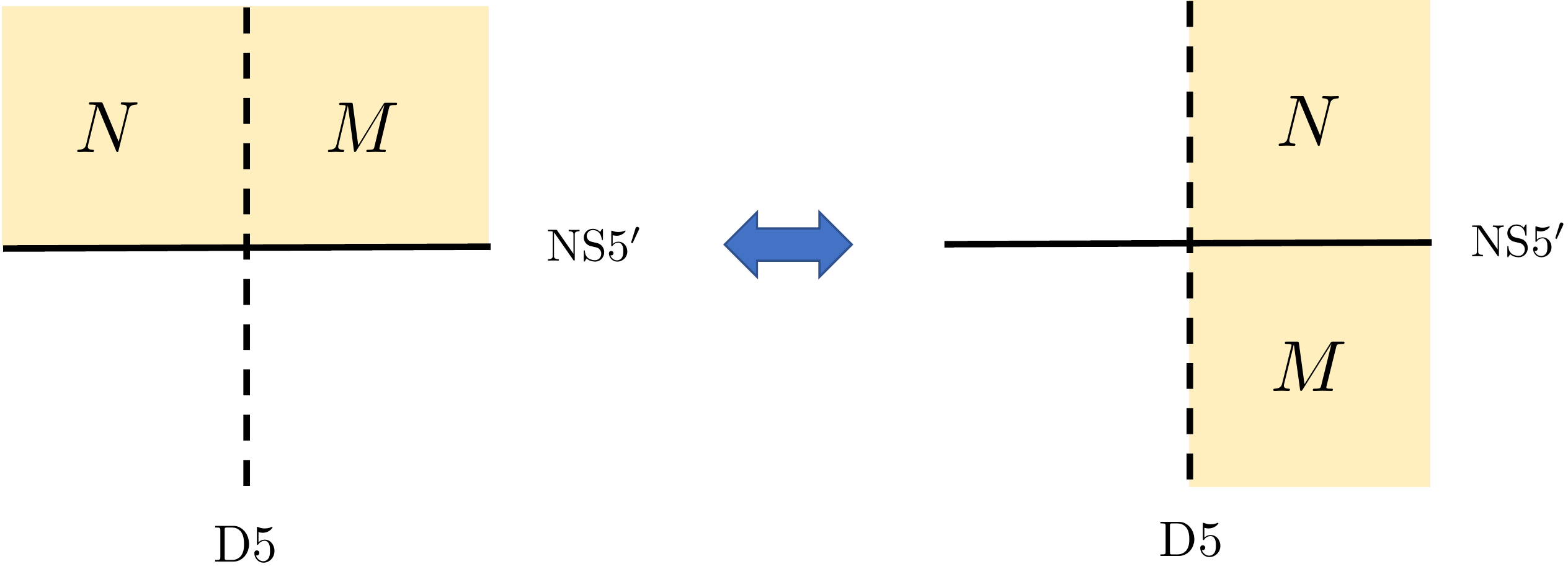}
\caption{The ${N \, M \choose 0 \, 0}$ and ${0 \, N \choose 0 \, M}$ NS5$'$-D5 junction.}
\label{figndjunction}
\end{center}
\end{figure}

We expect that 
the quarter-index for the $\left(\begin{smallmatrix}
0&N\\0&M\\\end{smallmatrix}\right)$ NS5$'$-D5 junction is given by 
\begin{align}
\label{nd0N0M}
\mathbb{IV}_{\mathcal{N}'\mathcal{D}}^{\left(\begin{smallmatrix}
0&N\\0&M\\\end{smallmatrix}\right)}
&=
\underbrace{
\prod_{k=1}^{M}
\frac{1}{(q^{\frac{k}{2}}t^{2k};q)_{\infty}}
}_{\mathbb{IV}_{\mathcal{N}'\mathcal{D}/\textrm{Nahm}}^{\textrm{4d $U(M)$}}}
\underbrace{
\prod_{l=1}^{N}
\frac{1}{(q^{\frac{l}{2}}t^{2l};q)_{\infty}}
}_{\mathbb{IV}_{\mathcal{N}'\mathcal{D}/\textrm{Nahm}}^{\textrm{4d $U(N)$}}}
\prod_{i=1}^{\min (N,M)}
\left(
q^{\frac{1+|N-M|}{4}+\frac{i}{2}}t^{|N-M|+2i-1}x^{\pm};q\right)_{\infty}
\end{align}
and this coincides with the quarter-index (\ref{ndNM00}) for the $\left(\begin{smallmatrix}
N&M\\0&0\\\end{smallmatrix}\right)$ NS5$'$-D5 junction. 
The brane configuration is illustrated in Figure \ref{figndjunction}. 

In order to understand the numerator factor, start from the $N \times M$ matrix of twisted hypers with Dirichlet boundary conditions and shift the 
fugacities according to the two regular Nahm poles. The factors which have non-positive powers of $t$ pair up with factors which have non-negative powers of $t$, 
to give Fermi multiplet indices which we strip off. After specialization of the $t$ fugacity this gives the standard regular DS reduction of the $\mathfrak{\widehat{u}}(N|M)$ Kac-Moody algebra. 

Again the quarter-index of the $\left(\begin{smallmatrix}
0&N\\0&M\\\end{smallmatrix}\right)$ NS5$'$-D5 junction has the following expansion:
\begin{align}
\label{nd0N0M_sum}
\mathbb{IV}_{\mathcal{N}'\mathcal{D}}^{\left(\begin{smallmatrix}
0&N\\0&M\\\end{smallmatrix}\right)}
&=\prod_{k=1}^{N}
\frac{1}{(q^{\frac{k}{2}}t^{2k};q)_{\infty}}
\prod_{l=1}^{M}
\frac{1}{(q^{\frac{k}{2}}t^{2k};q)_{\infty}}
\nonumber\\
&\times
\frac{1}{(q)_{\infty}^{2 \min (N,M)}}
\prod_{i=1}^{\min(N,M)}
\sum_{n_{i}=0}^{\infty}
\sum_{k_{i}=0}^{n_{i}}
(q^{1+k_{i}};q)_{\infty}
(q^{1+n_{i}-k_{i}};q)_{\infty}
\nonumber\\
&\times 
(-1)^{n_{i}}
x^{2k_{i}-n_{i}}
t^{(|N-M|+2i-1)n_{i}}
q^{\frac{n_{i}(n_{i}-1)}{2}+k_{i}(k_{i}-n_{i})
+\frac{(1+|N-M|)n_{i}}{4}+\frac{in_{i}}{2}}.
\end{align}
We see that the first term in the sum is identified with 
the product of the quarter-indices $\mathbb{IV}_{\mathcal{N}'\mathcal{D}}^{\left(\begin{smallmatrix}
N&0\\
0&0\\
\end{smallmatrix}\right)}$ and 
$\mathbb{IV}_{\mathcal{N}'\mathcal{D}}^{\left(\begin{smallmatrix}
0&M\\
0&0\\
\end{smallmatrix}\right)}$ 
which will be realized by giving a vev to the fundamental hypermultiplet.

%
%
%
%
%

\subsection{${N \, N \choose N \, N}$}
\label{sec_2dndNNNN}
Now we would like to study the NS5$'$-D5 junction 
with all quadrants occupied by the equal numbers of D3-branes (see Figure \ref{figndNNNN}). 
This type of junctions is self S-dual so that 
there is no non-trivial identity without additional line defects.

\subsubsection{${1 \, 1 \choose 1 \, 1}$}
Let us start with the $\left(\begin{smallmatrix}
1&1\\1&1\\\end{smallmatrix}\right)$ NS5$'$-D5 junction. 
We have two 4d $\mathcal{N}=4$ $U(1)$ gauge theories living respectively in $x^2>0$ and in $x^2<0$. 
For each of the $U(1)$ gauge theories, 
there is the 3d $\mathcal{N}=4$ charged hypermultiplet at the D5 interface. 
The NS5$'$-brane requires the Neumann boundary condition $\mathcal{N}'$ 
for the 4d $\mathcal{N}=4$ $U(1)$ gauge theories, 
and the Neumann boundary condition 
for the 3d $\mathcal{N}=4$ charged hypermultiplets. 
In addition, 
there are charged Fermi multiplets which cancel the boundary gauge anomalies 
and the 3d $\mathcal{N}=4$ bi-fundamental twisted hypermultiplet 
arising from the D3-D3 strings across the NS5$'$-brane. The Fermi multiplet fugacities are such as to allow 
cubic fermionic superpotential  couplings to the hypers and twisted hypers. 

We then obtain the quarter-index for the $\left(\begin{smallmatrix}
1&1\\1&1\\\end{smallmatrix}\right)$ NS5$'$-D5 junction
\begin{align}
\label{nd1111}
\mathbb{IV}_{\mathcal{N}'\mathcal{D}}^{\left(\begin{smallmatrix}
1&1\\1&1\\\end{smallmatrix}\right)}
&=
\underbrace{
\frac{(q)_{\infty}}{(q^{\frac12}t^2;q)_{\infty}}\oint \frac{ds_{1}}{2\pi is_{1}}
}_{\mathbb{II}_{\mathcal{N}'}^{\textrm{4d $U(1)$}}}
\underbrace{
\frac{(q)_{\infty}}{(q^{\frac12}t^2;q)_{\infty}}\oint \frac{ds_{2}}{2\pi is_{2}}
}_{\mathbb{II}_{\mathcal{N}'}^{\textrm{4d $U(1)$}}}
\nonumber\\
&\times 
\underbrace{
\frac{
\left(q^{\frac12}s_{2}x;q\right)_{\infty}
\left(q^{\frac12}s_{2}^{-1}x^{-1};q\right)_{\infty}
}{
(q^{\frac14}ts_{1}x;q)_{\infty}
(q^{\frac14}ts_{1}^{-1}x^{-1};q)_{\infty}
}
}_{\mathbb{II}_{N}^{\textrm{3d HM}}(s_{1}x)\cdot F(q^{\frac12}s_{2}x)}
\underbrace{
\frac{
\left(q^{\frac12}s_{1}x^{-1};q\right)_{\infty}
\left(q^{\frac12}s_{1}^{-1}x;q\right)_{\infty}
}{
(q^{\frac14}ts_{2}x^{-1};q)_{\infty}
(q^{\frac14}ts_{2}^{-1}x;q)_{\infty}
}
}_{\mathbb{II}_{N}^{\textrm{3d HM}}(s_{2}x^{-1})\cdot F(q^{\frac12}s_{1}x^{-1})}
\nonumber\\
&
\times 
\underbrace{
\frac{
\left( q^{\frac34}t \frac{s_{1}}{s_{2}};q\right)_{\infty}
\left( q^{\frac34}t \frac{s_{2}}{s_{1}};q\right)_{\infty}
}
{
\left( q^{\frac14}t^{-1} \frac{s_{1}}{s_{2}};q\right)_{\infty}
\left( q^{\frac14}t^{-1} \frac{s_{2}}{s_{1}};q\right)_{\infty}
}
}_{\mathbb{I}^{\textrm{3d tHM}}\left(\frac{s_{1}}{s_{2}}\right)}. 
\end{align}

A few observations are in order. First of all, we now have potential poles in the $x$ fugacity. They can potentially appear from 
situations where two poles in the integrand pinch the contour of integration. The corresponding gauge-invariant operators would combine a 
bi-fundamental twisted hypermultiplet with the two fundamental hypers. These potential poles, though, are precisely cancelled by the 
Fermi multiplet zeroes because of our choice of Fermi multiplet fugacities. 

Physically, the cubic fermionic superpotential couplings between Fermi's, hypers and twisted hypers constrain the 
potential vevs of the scalar fields. For example, if the twisted hypers get a vev it means that the D3-branes on the two sides of the 
NS5$'$ interface are glued together and separated from the NS5$'$-brane. 
That means the fundamental hypers on the two sides of the 
NS5$'$ interface should also be glued together instead of having Neumann b.c. This is enforced by the Fermi couplings. Similarly, if the fundamental hypers are given a vev, breaking the D3-branes across the D5 interface, then the twisted hypers should effectively be split across the D5 interface as well 
and receive Dirichlet b.c. there.  Again, this is enforced by the Fermi couplings.

\begin{figure}
\begin{center}
\includegraphics[width=10.5cm]{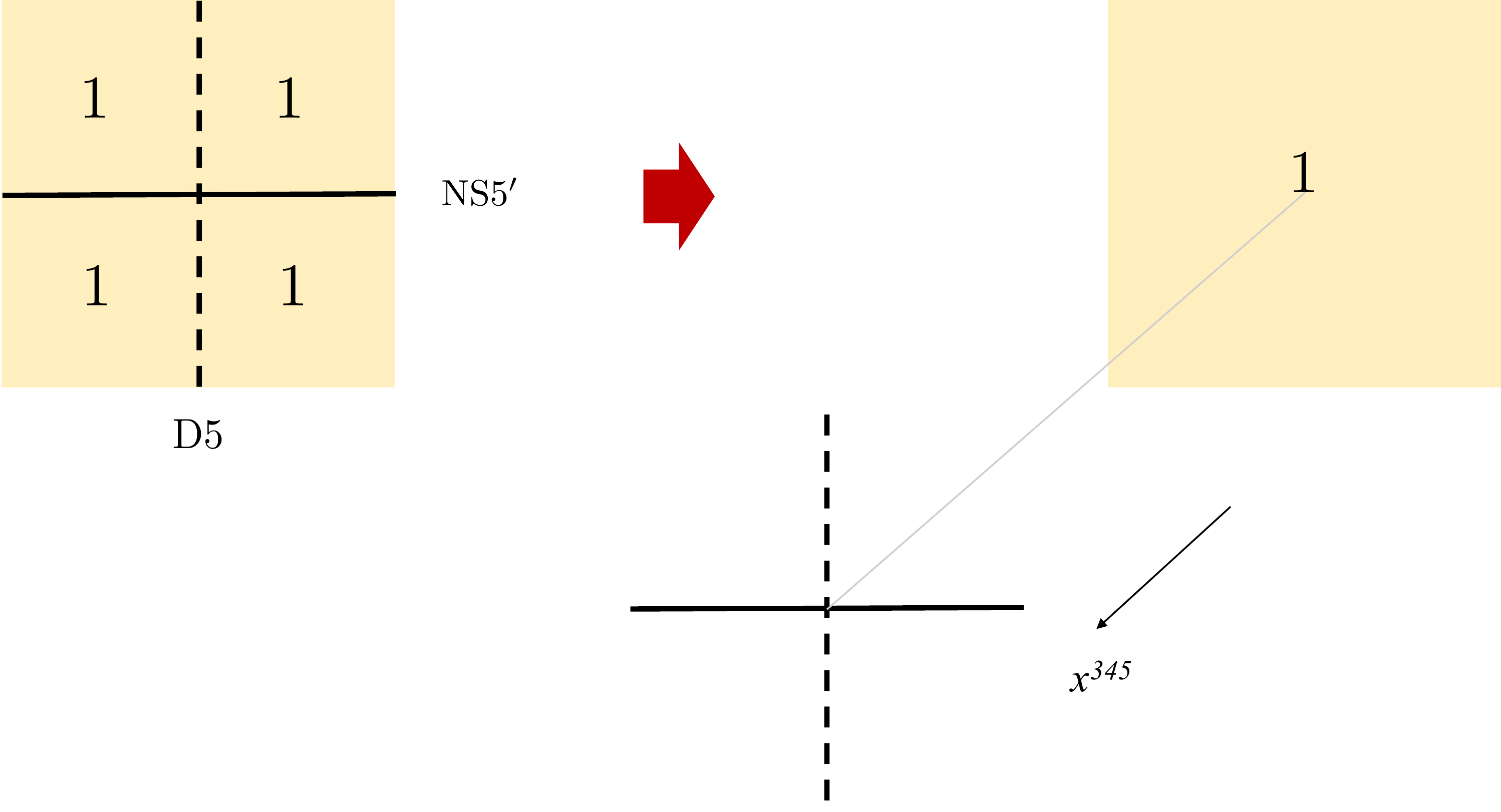}
\caption{Higgsing process of the ${1 \, 1 \choose 1 \, 1}$ NS5$'$-D5 junction which separates the fivebranes from the D3-brane.}
\label{fighiggsing3b}
\end{center}
\end{figure}

We can express the quarter-index (\ref{nd1111}) as
\begin{align}
\label{nd1111sum0}
\mathbb{IV}_{\mathcal{N}'\mathcal{D}}^{
\left(
\begin{smallmatrix}
1&1\\
1&1\\
\end{smallmatrix}
\right)}
&=
\frac{(q^{\frac12}t^{-2};q)_{\infty}}
{(q)_{\infty}^4 (q^{\frac12}t^2;q)_{\infty}}
(q^{\frac14}t^{-1}x^{\pm 2};q)_{\infty}^2
\sum_{n=0}^{\infty}\sum_{m=0}^{\infty}\sum_{k=0}^{\infty}\sum_{l=0}^{\infty}
\nonumber\\
&\times 
\frac{(q^{1+n};q)_{\infty} (q^{1+k};q)_{\infty} 
(q^{1+m};q)_{\infty} (q^{1+m+n-k};q)_{\infty} 
(q^{1+l};q)_{\infty}^2
}{
(q^{\frac14+n}t^{-1}x^{-2};q)_{\infty} 
(q^{\frac14+k}t^{-1}x^2;q)_{\infty} 
(q^{\frac14+m}t^{-1}x^{2};q)_{\infty} 
(q^{\frac14+m+n-k}t^{-1}x^{-2};q)_{\infty}
(q^{\frac12+l}t^{-2};q)_{\infty}^2
}
\nonumber\\
&\times 
q^{\frac{3n}{4}-\frac{k}{4}+\frac{m}{2}+\frac{l}{2}+(n-k)l}
t^{n+k+2m+2l}x^{2n-2k}
\end{align}
This is obtained by picking up the residues at twisted hypermultiplet poles 
$\frac{s_{1}}{s_{2}}$ $=$ $q^{\frac14+l}t^{-1}$. 
The first term in the sum is the 4d $U(1)$ index $\mathbb{I}^{\textrm{4d $U(1)$}}$. 
We expect this expression to be associated to a Higgsing process detaching the fivebranes from the D3-brane (see Figure \ref{fighiggsing3b}). 

\begin{figure}
\begin{center}
\includegraphics[width=10.5cm]{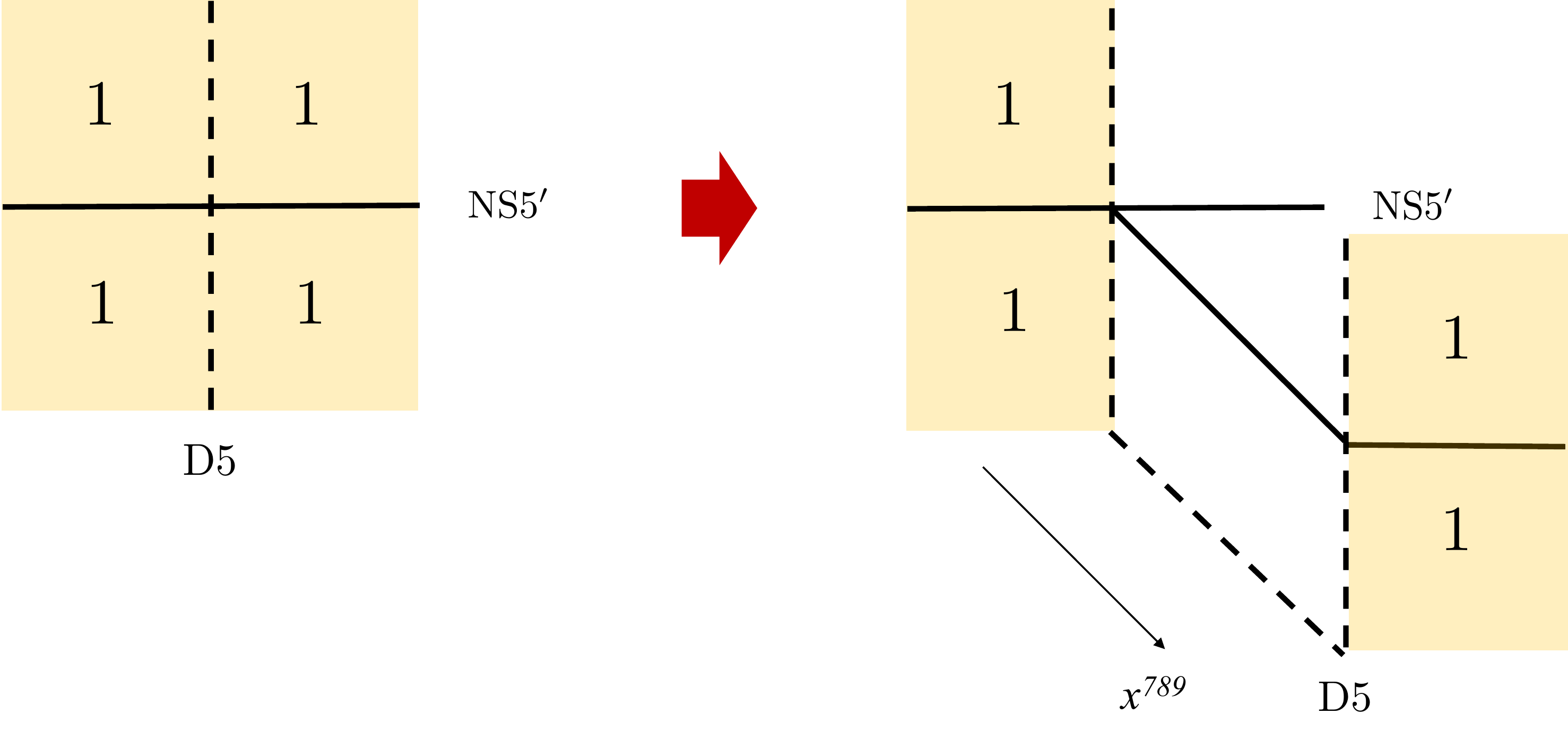}
\caption{Higgsing procedure corresponding to the deformations of the ${1 \, 1 \choose 1 \, 1}$ NS5$'$-D5 junction, 
which results in a pair of ${1 \, 0 \choose 1 \, 0}$ and ${0 \, 1 \choose 0 \, 1}$ NS5$'$-D5 junctions. 
The associated expansion of the quarter-index begins with the fourth power of the quarter-index 
$\mathbb{IV}_{\mathcal{N}'\mathcal{D}}^{\textrm{4d $U(1)$}}$ and the square of the half-index $\mathbb{II}_{D}^{\textrm{3d tHM}}(x^2)$. }
\label{fighiggsing3}
\end{center}
\end{figure}

\begin{figure}
\begin{center}
\includegraphics[width=4cm]{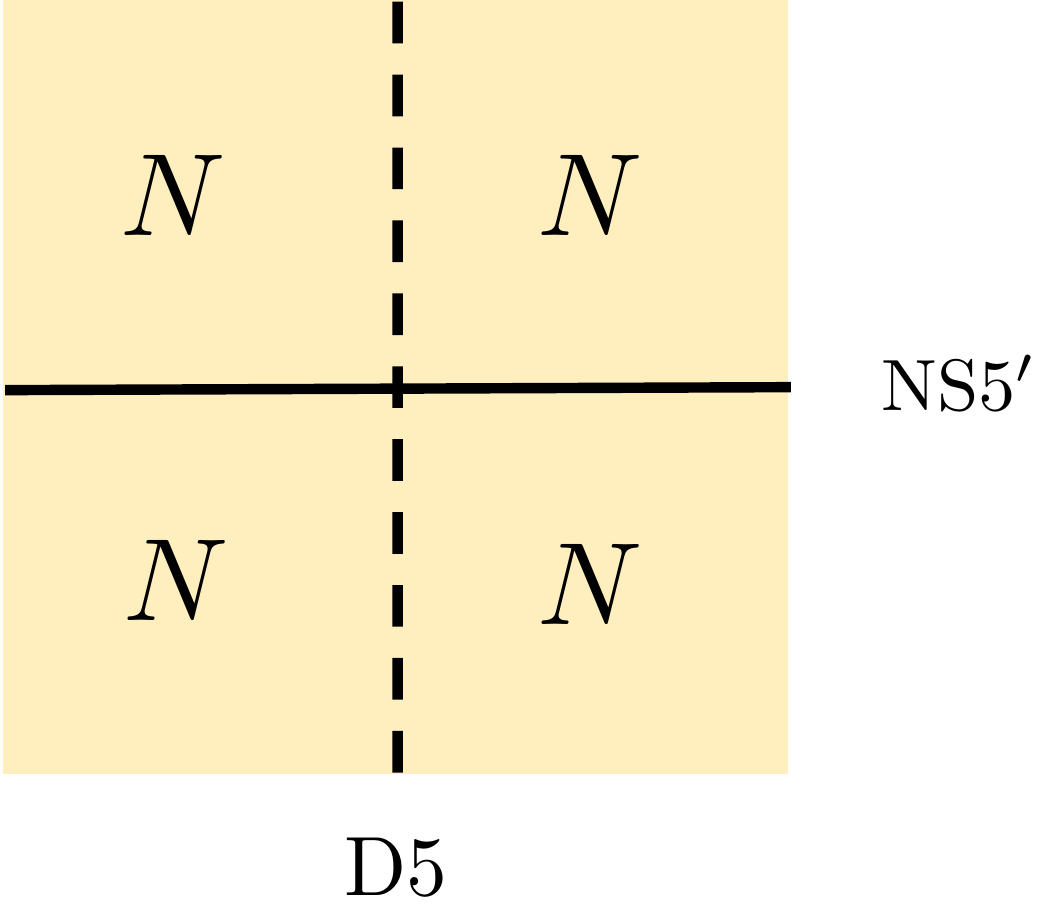}
\caption{The self S-dual ${N \, N \choose N \, N}$ NS5$'$-D5 junction.}
\label{figndNNNN}
\end{center}
\end{figure}

Making use of of the summation formula (\ref{thm_series}) for the full-index of 3d $\mathcal{N}=4$ twisted hypermultiplet, 
we can expand the quarter-index (\ref{nd1111}) as 
\begin{align}
\label{nd1111sum}
\mathbb{IV}_{\mathcal{N}'\mathcal{D}}^{\left(\begin{smallmatrix}
1&1\\1&1\\\end{smallmatrix}\right)}
&=\frac{1}{(q)_{\infty}^2 (q^{\frac12}t^2;q)_{\infty}^2} 
\underbrace{
(q^{\frac14}t^{-1}x^{2};q)_{\infty} 
(q^{\frac34}tx^{-2};q)_{\infty}
}_{F(q^{\frac14}t^{-1}x^2)}
\underbrace{
(q^{\frac14}t^{-1}x^{-2};q)_{\infty}
(q^{\frac34}tx^{2};q)_{\infty}
}_{F(q^{\frac14}t^{-1}x^{-2})}
\nonumber\\
&\times 
\sum_{n=0}^{\infty}
\sum_{k=0}^{n}
\frac{(q^{1+k};q)_{\infty} (q^{1+n-k};q)_{\infty}}
{(q^{\frac12+k}t^2;q)_{\infty} (q^{\frac12+n-k}t^2;q)_{\infty}}
\frac{(q^{\frac34+n-2k}tx^2;q)_{\infty}}
{(q^{\frac14+2k-n}t^{-1}x^{-2};q)_{\infty}}
\frac{(q^{\frac34+2k-n}tx^{-2};q)_{\infty}}
{(q^{\frac14+n-2k}t^{-1}x^{2};q)_{\infty}}
q^{\frac{n}{4}}t^{-n}x^{4k-2n}. 
\end{align}
This expansion would be compatible with the IR description associated to the Higgsing procedure 
of giving vevs to the fundamental hypers. 
The first term in the sum consists of the fourth power of the quarter-index 
$\mathbb{IV}_{\mathcal{N}'\mathcal{D}}^{\textrm{4d $U(1)$}}$ and the square of the half-index $\mathbb{II}_{D}^{\textrm{3d tHM}}(x^2)$ of the Dirichlet boundary condition $D$ for 3d $\mathcal{N}=4$ twisted hypermultiplet, 
which indicates that a single D3-brane decompose into a pair of semi-infinite D3-branes in $x^6<0$ and $x^6>0$ along the fivebrane, as depicted in Figure \ref{fighiggsing3}. 

Upon specialization of the $t$ fugacity, the contour integral expression of the index matches the naive character of an $\mathfrak{\widehat{u}}(1|1)$-BRST reduction of 
$\mathfrak{\widehat{u}}(1|1) \times \mathfrak{\widehat{u}}(1|1) \times \mathrm{Ff}^{(1|1) \times (1|1)}$: a combination of a complex fermion and a pair of symplectic bosons
gives a vertex algebra $\mathrm{Ff}^{(1|1)}$ which has an $\mathfrak{\widehat{u}}(1|1)$ sub-algebra with level $\pm 1$ depending on the choice of Grassmann parity
of the fields; the BRST complex involves two copies of this algebra, with opposite levels, which we denote as $\mathrm{Ff}^{(1|1) \times (1|1)}$.
We say ``naive character'' because after specialization of $t$ we have dangerous numerator factors, associated to zeromodes of the super-ghosts. 
The calculation of the character of the VOA requires some form of regularization, and we claim that taking the $t$ specialization {\it after} computing the contour integral is
a good choice of regularization.

\subsubsection{${N \, N \choose N \, N}$}
For the $\left(\begin{smallmatrix}
N&N\\N&N\\\end{smallmatrix}\right)$ NS5$'$-D5 junction, 
there are two 4d $\mathcal{N}=4$ $U(N)$ gauge theories living in $x^2>0$ and in $x^2<0$. 
Each of the $U(N)$ gauge theories couples to 
the 3d $\mathcal{N}=4$ fundamental hypermultiplet at the D5-brane defect. 
At the NS5$'$-brane interface, 
the 4d $\mathcal{N}=4$ $U(N)$ gauge theories receive  Neumann boundary condition $\mathcal{N}'$ 
and the 3d $\mathcal{N}=4$ fundamental hypermultiplets receive Neumann boundary condition $N'$. 
The setup also includes fundamental Fermi multiplets which cancel boundary gauge anomaly \cite{Hanany:2018hlz} 
and the 3d $\mathcal{N}=4$ bi-fundamental twisted hypermultiplet corresponding to the D3-D3 strings across the NS5$'$-brane. 

We find the following quarter-index for 
the $\left(\begin{smallmatrix}
N&N\\N&N\\\end{smallmatrix}\right)$ NS5$'$-D5 junction:
\begin{align}
\label{ndNNNN}
\mathbb{IV}_{\mathcal{N}'\mathcal{D}}^{\left(\begin{smallmatrix}
N&N\\N&N\\\end{smallmatrix}\right)}
&=
\underbrace{
\frac{1}{N!}\frac{(q)_{\infty}^{N}}
{(q^{\frac12} t^2;q)_{\infty}^{N}}
\oint \prod_{i=1}^{N}\frac{ds_{i}}{2\pi is_{i}}
\prod_{i\neq j}\frac{
\left(\frac{s_{i}}{s_{j}};q\right)_{\infty}
}
{
\left(q^{\frac12} t^2 \frac{s_{i}}{s_{j}};q\right)_{\infty}
}
}_{\mathbb{II}_{\mathcal{N}'}^{\textrm{4d $U(N)$}}}
\underbrace{
\frac{1}{N!}\frac{(q)_{\infty}^{N}}
{(q^{\frac12} t^2;q)_{\infty}^{N}}
\oint \prod_{i=N+1}^{2N}\frac{ds_{i}}{2\pi is_{i}}
\prod_{i\neq j}\frac{
\left(\frac{s_{i}}{s_{j}};q\right)_{\infty}
}
{
\left(q^{\frac12} t^2 \frac{s_{i}}{s_{j}};q\right)_{\infty}
}
}_{\mathbb{II}_{\mathcal{N}'}^{\textrm{4d $U(N)$}}}
\nonumber\\
&\times 
\prod_{i=1}^{N}
\underbrace{
\frac{
\left(q^{\frac12}s_{i}x;q\right)_{\infty}
\left(q^{\frac12}s_{i}^{-1}x^{-1};q\right)_{\infty}
}{
(q^{\frac14}ts_{i}x;q)_{\infty}
(q^{\frac14}ts_{i}^{-1}x^{-1};q)_{\infty}
}
}_{\mathbb{II}_{N}^{\textrm{3d HM}}(s_{i}x)\cdot F(q^{\frac12}s_{i}x)}
\prod_{i=N+1}^{2N}
\underbrace{
\frac{
\left(q^{\frac12}s_{i}x^{-1};q\right)_{\infty}
\left(q^{\frac12}s_{i}^{-1}x;q\right)_{\infty}
}{
(q^{\frac14}ts_{i}x^{-1};q)_{\infty}
(q^{\frac14}ts_{i}^{-1}x;q)_{\infty}
}
}_{\mathbb{II}_{N}^{\textrm{3d HM}}(s_{i}x^{-1})\cdot F(q^{\frac12}s_{i}x^{-1})}
\nonumber\\
&
\times 
\prod_{i=1}^{N}\prod_{k=N+1}^{2N}
\underbrace{
\frac{
\left( q^{\frac34}t \frac{s_{i}}{s_{k}};q\right)_{\infty}
\left( q^{\frac34}t \frac{s_{k}}{s_{i}};q\right)_{\infty}
}
{
\left( q^{\frac14}t^{-1} \frac{s_{i}}{s_{k}};q\right)_{\infty}
\left( q^{\frac14}t^{-1} \frac{s_{k}}{s_{i}};q\right)_{\infty}
}
}_{\mathbb{I}^{\textrm{3d tHM}}\left(\frac{s_{i}}{s_{k}}\right)}. 
\end{align}

%
%
%
%
%

\subsection{${N \, M \choose L\ \, 0}$}
\label{sec_2dndNML0}
Consider the $\left(\begin{smallmatrix}
N&M\\ L&0\\\end{smallmatrix}\right)$ NS5$'$-D5 junction 
with three out of four quadrants filled by D3-branes. 

\subsubsection{${1 \, 1 \choose 2 \, 0}$ and ${2 \, 1 \choose 0 \, 1}$}
Let us begin with the $\left(\begin{smallmatrix}
1&1\\ 2&0\\\end{smallmatrix}\right)$ NS5$'$-D5 junction. 
In the half-space $x^2>0$, 
there is a 4d $\mathcal{N}=4$ $U(1)$ gauge theory with 
the Neumann boundary condition $\mathcal{N}'$ at $x^2=0$ 
and the 3d $\mathcal{N}=4$ charged hypermultiplet with 
the Neumann boundary condition $N'$. 
The 4d $\mathcal{N}=4$ $U(2)$ gauge theory in the lower left quadrant 
obeys the Neumann boundary condition $\mathcal{N}'$ 
and the Nahm pole boundary condition 
specified by an embedding $\rho:$ $\mathfrak{su}(2)$ $\rightarrow$ $\mathfrak{u}(2)$. 
There are also local operators from D3-D3 string across the NS5$'$-brane 
as the 3d $\mathcal{N}=4$ twisted hypermultiplet obeying the Nahm pole boundary condition.

The quarter-index for the 
$\left(\begin{smallmatrix}
1&1\\ 2&0\\\end{smallmatrix}\right)$ NS5$'$-D5 junction reads
\begin{align}
\label{nd1120}
\mathbb{IV}_{\mathcal{N}'\mathcal{D}}^{\left(\begin{smallmatrix}
1&1\\2&0\\\end{smallmatrix}\right)}
&=
\underbrace{
\frac{(q)_{\infty}}{(q^{\frac12} t^2;q)_{\infty}}\oint \frac{ds}{2\pi is}
}_{\mathbb{II}_{\mathcal{N}'}^{\textrm{4d $U(1)$}}}
\underbrace{
\frac{1}{\left(q^{\frac14}ts;q\right)_{\infty}\left(q^{\frac14}ts^{-1};q\right)_{\infty}}
}_{\mathbb{II}_{N}^{\textrm{3d HM}}\left(s\right)}
(qt^2 s^{\pm}x^{\mp};q)_{\infty}
\underbrace{
\frac{1}{(q^{\frac12} t^2;q)_{\infty} (qt^4;q)_{\infty}}
}_{\mathbb{IV}_{\mathcal{N}'\textrm{Nahm}}^{\textrm{4d $U(2)$}}}. 
\end{align}

Picking up poles from the fundamental hypermultiplet at $s=q^{\frac14+m}t$, 
we find that 
\begin{align}
\label{nd1120sum1}
\mathbb{IV}_{\mathcal{N}'\mathcal{D}}^{\left(\begin{smallmatrix}
1&1\\2&0\\\end{smallmatrix}\right)}
&=
\frac{(q^{\frac34}tx;q)_{\infty} (q^{\frac14}t^{-1}x^{-1};q)_{\infty}}
{(q)_{\infty} (q^{\frac12}t^2;q)_{\infty}^2 (qt^4;q)_{\infty}}
\sum_{m=0}^{\infty}
\frac{(q^{1+m};q)_{\infty} (q^{\frac54+m}t^3x^{-1};q)_{\infty}}
{(q^{\frac12+m}t^2;q)_{\infty} (q^{\frac14+m}t^{-1}x^{-1};q)_{\infty}}
q^{\frac{3m}{4}}t^{m}x^{m}. 
\end{align}
The first term in the series (\ref{nd1120sum1}) is the product of three quarter-indices 
$\mathbb{IV}^{\textrm{4d $U(1)$}}_{\mathcal{N}'\mathcal{D}}$, 
$\mathbb{IV}^{\textrm{4d $U(1)$}}_{\mathcal{N}'\mathcal{D}}$ 
and 
$\mathbb{IV}^{\textrm{4d $U(2)$}}_{\mathcal{N}'\textrm{Nahm}}$ 
corresponding to the three separate corners together with 
the extra fermionic contributions 
$(q^{\frac34}tx;q)_{\infty}$ $(q^{\frac54}t^3x^{-1};q)_{\infty}$ living at the junction. Presumably, this 
expansion is associated to a Higgsing procedure separating these three corners. 

Alternatively, we can expand the integral in (\ref{nd1120}) as
\begin{align}
\label{nd1120sum2}
\mathbb{IV}_{\mathcal{N}'\mathcal{D}}^{\left(\begin{smallmatrix}
1&1\\2&0\\\end{smallmatrix}\right)}
&=\frac{(q^{\frac34}tx;q)_{\infty} (q^{\frac34}tx^{-1};q)_{\infty}}
{(q)_{\infty} (q^{\frac12}t^2;q)_{\infty}^2 (qt^{4};q)_{\infty}}
\sum_{n=0}^{\infty} 
\frac{(q^{1+n};q)_{\infty}^2}{(q^{\frac34+n}tx;q)_{\infty} (q^{\frac34+n}tx^{-1};q)_{\infty}}
q^{\frac{n}{2}} t^{2n}. 
\end{align}
The sum starts from the product of 
the half-index $\mathbb{II}_{\mathcal{N}'}^{\textrm{4d $U(1)$}}$ of Neumann b.c. $\mathcal{N}'$ for 4d $\mathcal{N}=4$ $U(1)$ gauge theory 
and the quarter-index $\mathbb{IV}_{\mathcal{N}'\textrm{Nahm}}^{\textrm{4d $U(2)$}}$. It should also have a Higgsing interpretation.

The S-dual $\left(\begin{smallmatrix}
2&1\\ 0&1\\\end{smallmatrix}\right)$ NS5$'$-D5 junction also has 
a 4d $\mathcal{N}=4$ $U(1)$ gauge theory in $x^2>0$ with 
the Neumann boundary condition $\mathcal{N}'$ 
while a $U(2)$ gauge symmetry in the upper right corner is broken to $U(1)$ at the interface. 
There is no 3d $\mathcal{N}=4$ hypermultiplet 
since the number of D3-branes jumps when crossing the D5-brane. 
The 4d $\mathcal{N}=4$ $U(1)$ gauge theory in the lower right corner
satisfies the Neumann boundary condition $\mathcal{N}'$ and Dirichlet boundary condition $\mathcal{D}$. 
In addition, 
there are contributions from the 3d $\mathcal{N}=4$ twisted hypermultiplet 
obeying the Dirichlet boundary condition $D$. 

We obtain the quarter-index for the 
$\left(\begin{smallmatrix}
2&1\\ 0&1\\\end{smallmatrix}\right)$ NS5$'$-D5 junction
\begin{align}
\label{nd2101}
\mathbb{IV}_{\mathcal{N}'\mathcal{D}}^{\left(\begin{smallmatrix}
2&1\\0&1\\\end{smallmatrix}\right)}
&=
\underbrace{
\frac{(q)_{\infty}}{(q^{\frac12} t^2;q)_{\infty}}\oint \frac{ds}{2\pi is}
}_{\mathbb{II}_{\mathcal{N}'}^{\textrm{4d $U(1)$}}}
\underbrace{
\frac{1}{(q^{\frac12} t^2;q)_{\infty}}
}_{\mathbb{IV}_{\mathcal{N}'\mathcal{D}}^{\textrm{4d $U(1)$}}}
\frac{1}{(q^{\frac12}t^2 s^{\pm}x^{\mp};q)_{\infty}}
\underbrace{
(q^{\frac34}t s^{\pm};q)_{\infty}
}_{\mathbb{II}_{D}^{\textrm{3d tHM}}(s)}
\underbrace{
\frac{1}{(q^{\frac12}t^2;q)_{\infty}}
}_{\mathbb{IV}_{\mathcal{N}'\mathcal{D}}^{\textrm{4d $U(1)$}}}. 
\end{align}
We see that the quarter-index (\ref{nd1120}) 
for the $\left(\begin{smallmatrix}
1&1\\ 2&0\\\end{smallmatrix}\right)$ NS5$'$-D5 junction
agrees with the quarter-index (\ref{nd2101}) for the $\left(\begin{smallmatrix}
2&1\\ 0&1\\\end{smallmatrix}\right)$ NS5$'$-D5 junction. 

The contour integral in (\ref{nd2101}) can be evaluated by 
taking the sum of residues at poles $s=q^{\frac12+m}t^2$
\begin{align}
\label{nd2101sum1}
\mathbb{IV}_{\mathcal{N}'\mathcal{D}}^{\left(\begin{smallmatrix}
2&1\\0&1\\\end{smallmatrix}\right)}
&=
\frac{(q^{\frac14}t^{-1}x^{-1};q)_{\infty} (q^{\frac34}tx;q)_{\infty}}
{(q)_{\infty} (q^{\frac12}t^2;q)_{\infty}}
\sum_{m=0}^{\infty}
\frac{(q^{1+m};q)_{\infty} (q^{\frac54+m}t^3x;q)_{\infty}}
{(q^{1+m}t^4;q)_{\infty} (q^{\frac34+m}tx;q)_{\infty}}
q^{\frac{m}{4}}t^{-m}x^{-m}.
\end{align}
This begins with the product of the three quarter-indices 
$\mathbb{IV}_{\mathcal{N}'\mathcal{D}}^{\textrm{4d $U(1)$}}$, 
$\mathbb{IV}_{\mathcal{N}'\mathcal{D}}^{\textrm{4d $U(1)$}}$, 
$\mathbb{IV}_{\mathcal{N}'\textrm{Nahm}}^{\textrm{4d $U(2)$}}$  for three corners 
and the extra fermionic contributions 
$(q^{\frac14}t^{-1}x^{-1};q)_{\infty}$ 
$(q^{\frac54}t^3 x;q)_{\infty}$ appearing at the junction. It should also have a Higgsing interpretation.

On the other hand, we have another expression of 
the quarter-index (\ref{nd2101}) as
\begin{align}
\label{nd2101sum2}
\mathbb{IV}_{\mathcal{N}'\mathcal{D}}^{\left(\begin{smallmatrix}
2&1\\0&1\\\end{smallmatrix}\right)}
&=\frac{(q^{\frac14}t^{-1}x;q)_{\infty} (q^{\frac14}t^{-1}x^{-1};q)_{\infty}}
{(q)_{\infty} (q^{\frac12}t^2;q)_{\infty}^3}
\sum_{n=0}^{\infty}
\frac{(q^{1+n};q)_{\infty}^2}{(q^{\frac14+n}t^{-1}x;q)_{\infty} (q^{\frac14+n}t^{-1}x^{-1})}q^{n}t^{4n}. 
\end{align}
In contrast to the expansion (\ref{nd2101sum1}), 
this sum has the first term as 
the product of the half-index $\mathbb{II}_{\mathcal{N}'}^{\textrm{4d $U(1)$}}$ 
and the square of the quarter-index $\mathbb{IV}_{\mathcal{N}'\mathcal{D}}^{\textrm{4d $U(1)$}}$. It should also have a Higgsing interpretation.

\subsubsection{${1 \, 1 \choose N \, 0}$ and ${N \, 1 \choose 0 \, 1}$}
Now consider the $\left(\begin{smallmatrix}
1&1\\ N&0\\\end{smallmatrix}\right)$ NS5$'$-D5 junction which may include a Wilson line $\mathcal{W}_{n}$ of charge $n$. 
\footnote{We can take $n=0$ when the Wilson line is turned off. } 
The quarter-index reads
\begin{align}
\label{nd11N0wil}
\mathbb{IV}_{\mathcal{N}'\mathcal{D}+\mathcal{W}_{n}}^{\left(\begin{smallmatrix}
1&1\\N&0\\\end{smallmatrix}\right)}
&=
\underbrace{
\frac{(q)_{\infty}}{(q^{\frac12}t^2;q)_{\infty}}
\oint \frac{ds}{2\pi is}
}_{\mathbb{II}_{\mathcal{N}'}^{\textrm{4d $U(1)$}}}
\frac{1}
{
(q^{\frac14}ts^{\pm};q)_{\infty}
}
(q^{\frac12+\frac{N}{4}}t^{N}s^{\pm}x^{\pm};q)_{\infty}
\underbrace{
\prod_{k=1}^{N}\frac{1}{(q^{\frac{k}{2}} t^{2k};q)_{\infty}} 
}_{\mathbb{IV}_{\mathcal{N}'\mathcal{D}/\textrm{Nahm}}}
s^{n}. 
\end{align}
Evaluating the integral by taking the sum of residues at poles $s=q^{\frac14+m}t$, we obtain 
\begin{align}
\label{nd11N0wil_sum1}
\mathbb{IV}_{\mathcal{N}'\mathcal{D}+\mathcal{W}_{n}}^{\left(\begin{smallmatrix}
1&1\\N&0\\\end{smallmatrix}\right)}
&=
\frac{1}{(q)_{\infty}(q^{\frac12}t^2;q)_{\infty}}
\prod_{k=1}^{N}\frac{1}{(q^{\frac{k}{2}} t^{2k};q)_{\infty}} 
(q^{\frac34-\frac{N}{4}}t^{-N+1}x;q)_{\infty}
(q^{\frac14+\frac{N}{4}}t^{N-1}x^{-1};q)_{\infty}
\nonumber\\
&\times 
\sum_{m=0}^{\infty}
\frac{
(q^{1+m};q)_{\infty}
(q^{\frac34+\frac{N}{4}+m}t^{N+1};q)_{\infty}
}
{
(q^{\frac12+m}t^2;q)_{\infty}
(q^{\frac34-\frac{N}{4}+m}t^{-N+1};q)_{\infty}
}
q^{\frac{m}{4}+\frac{Nm}{4}+nm+\frac{n}{4}}
t^{(N-1)m+n}. 
\end{align}
The first term in the sum includes the product of quarter-indices 
$\mathbb{IV}_{\mathcal{N}'\mathcal{D}}^{\left(\begin{smallmatrix}
1&0\\0&0\\\end{smallmatrix}\right)}$, 
$\mathbb{IV}_{\mathcal{N}'\mathcal{D}}^{\left(\begin{smallmatrix}
0&1\\0&0\\\end{smallmatrix}\right)}$, and 
$\mathbb{IV}_{\mathcal{N}'\mathcal{D}}^{\left(\begin{smallmatrix}
0&0\\N&0\\\end{smallmatrix}\right)}$ along with the extra fermionic contributions $(q^{\frac34+\frac{N}{4}}t^{N+1}x;q)_{\infty}$ $(q^{\frac14+\frac{N}{4}}t^{N-1}x^{-1})_{\infty}$ which do not behave as the $\mathcal{N}=(0,2)$ Fermi multiplet. 
The associated Higgsing procedure is the splitting of the $\left(\begin{smallmatrix}
1&1\\ N&0\\\end{smallmatrix}\right)$ NS5$'$-D5 junction into 
the three pieces; 
$\left(\begin{smallmatrix}
1&0\\ 0&0\\\end{smallmatrix}\right)$, 
$\left(\begin{smallmatrix}
0&1\\ 0&0\\\end{smallmatrix}\right)$, 
and 
$\left(\begin{smallmatrix}
0&0\\ N&0\\\end{smallmatrix}\right)$ NS5$'$-D5 junctions.

We can also expand the integral in (\ref{nd11N0wil}) as
\begin{align}
\label{nd11N0wil_sum2}
\mathbb{IV}_{\mathcal{N}'\mathcal{D}+\mathcal{W}_{n}}^{\left(\begin{smallmatrix}
1&1\\N&0\\\end{smallmatrix}\right)}
&=
\frac{(q^{\frac14+\frac{N}{4}}t^{N-1}x^{\pm};q)_{\infty}}
{(q)_{\infty} (q^{\frac12}t^2;q)_{\infty}}
\prod_{k=1}^{N}\frac{1}{(q^{\frac{k}{2}}t^{2k};q)_{\infty}}
\sum_{m=0}^{\infty} 
\frac{(q^{1+m};q)_{\infty}}
{(q^{\frac14+\frac{N}{4}+m}t^{N-1}x^{\pm};q)_{\infty}}
q^{\frac{m}{2}-\frac{n}{4}}t^{2m-n}.
\end{align}
Unlike the residue sum (\ref{nd11N0wil_sum1}), 
the first term for $n=0$ in the expression (\ref{nd11N0wil_sum2}) is the product of 
the half-index $\mathbb{II}_{\mathcal{N}'}^{\textrm{4d $U(1)$}}$ of the Neumann b.c. $\mathcal{N}'$ 
for 4d $\mathcal{N}=4$ $U(1)$ gauge theory and the quarter-index $\mathbb{IV}_{\mathcal{N}'\mathcal{D}/\textrm{Nahm}}^{\textrm{4d $U(N)$}}$. 
The sum (\ref{nd11N0wil_sum2}) will correspond to the decomposition $\left(\begin{smallmatrix}
1&1\\ N&0\\\end{smallmatrix}\right)$ 
$\rightarrow$ 
$\left(\begin{smallmatrix}
1&1\\ 0&0\\\end{smallmatrix}\right)$ 
$\oplus$ 
$\left(\begin{smallmatrix}
0&0\\ N&0\\\end{smallmatrix}\right)$ 
of the NS5$'$-D5 junction.

Next consider the S-dual $\left(\begin{smallmatrix}
N&1\\ 0&1\\\end{smallmatrix}\right)$ NS5$'$-D5 junction with a Wilson line $\mathcal{W}_{n}$ of charge $n$. 
We can express the quarter-index as the contour integral
\begin{align}
\label{ndN101wil}
\mathbb{IV}_{\mathcal{N}'\mathcal{D}+\mathcal{W}_{n}}^{\left(\begin{smallmatrix}
N&1\\0&1\\\end{smallmatrix}\right)}
&=
\frac{(q)_{\infty}}{(q^{\frac12}t^2;q)_{\infty}}
\oint \frac{ds}{2\pi is} 
\prod_{k=1}^{N-1}
\frac{1}{(q^{\frac{k}{2}}t^{2k};q)_{\infty}}
\frac{1}
{
(q^{\frac{N}{4}} t^{N}s^{\pm}x^{\mp};q)_{\infty}
}
(q^{\frac34}t s^{\pm};q)_{\infty}
\frac{1}{(q^{\frac12}t^2;q)_{\infty}}
s^{n}. 
\end{align}
It is calculated as the sum of residues at $s=q^{\frac{N}{4}+m}t^{N}x$
\begin{align}
\label{ndN101wil_sum1}
\mathbb{IV}_{\mathcal{N}'\mathcal{D}+\mathcal{W}_{n}}^{\left(\begin{smallmatrix}
N&1\\0&1\\\end{smallmatrix}\right)}
&=
\frac{1}{(q)_{\infty}(q^{\frac12} t^2;q)_{\infty}^2}
\prod_{k=1}^{N-1}
\frac{1}{(q^{\frac{k}{2}}t^{2k};q)_{\infty}}
(q^{\frac14+\frac{N}{4}}t^{N-1};q)_{\infty} 
(q^{\frac34-\frac{N}{4}}t^{1-N};q)_{\infty} 
\nonumber\\
&\times 
\sum_{m=0}^{\infty}
\frac{
(q^{1+m};q)_{\infty}
(q^{\frac{N}{4}+\frac34+m}t^{N+1};q)_{\infty}
}
{
(q^{\frac{N}{2}+m}t^{2N};q)_{\infty}
(q^{\frac{1}{4}+\frac{N}{4}+m}t^{N-1};q)_{\infty}
}
q^{\frac{3m}{4}-\frac{Nm}{4}+nm+\frac{Nn}{4}} 
t^{(1-N)m+Nn}. 
\end{align}
The residue sum (\ref{ndN101wil_sum1}) begins with 
the product of three quarter-indices 
$\mathbb{IV}_{\mathcal{N}'\mathcal{D}}^{\textrm{4d $U(1)$}}$, 
$\mathbb{IV}_{\mathcal{N}'\mathcal{D}}^{\textrm{4d $U(1)$}}$ 
and 
$\mathbb{IV}_{\mathcal{N}'\mathcal{D}/\textrm{Nahm}}^{\textrm{4d $U(N)$}}$ 
as well as the extra fermionic factors 
$(q^{\frac34-\frac{N}{4}}t^{1-N};q)_{\infty}$ 
$(q^{\frac{N}{4}+\frac34}t^{N+1};q)_{\infty}$. 
This is compatible with the Higgsing process as the division of the 
of the NS5$'$-D5 junction: 
$\left(\begin{smallmatrix}
N&1\\ 0&1\\\end{smallmatrix}\right)$ 
$\rightarrow$ 
$\left(\begin{smallmatrix}
N&0\\ 0&0\\\end{smallmatrix}\right)$ 
$\oplus$ 
$\left(\begin{smallmatrix}
0&1\\ 0&0\\\end{smallmatrix}\right)$ 
$\oplus$ 
$\left(\begin{smallmatrix}
0&0\\ 0&1\\\end{smallmatrix}\right)$.

The quarter-index (\ref{ndN101wil}) can be also expressed as 
\begin{align}
\label{ndN101wil_sum2}
\mathbb{IV}_{\mathcal{N}'\mathcal{D}+\mathcal{W}_{n}}^{\left(\begin{smallmatrix}
N&1\\0&1\\\end{smallmatrix}\right)}
&=
\frac{(q^{\frac34-\frac{N}{4}}t^{1-N}x^{\pm};q)_{\infty}}
{(q)_{\infty} (q^{\frac12}t^2;q)_{\infty}^2}
\prod_{k=1}^{N-1}
\frac{1}{(q^{\frac{k}{2}}t^{2k};q)_{\infty}}
\sum_{m=0}^{\infty}
\frac{(q^{1+m};q)_{\infty}^2}
{(q^{\frac34-\frac{N}{4}+m} t^{1-N}x^{\pm};q)_{\infty}} 
q^{\frac{Nm}{2}-\frac{Nn}{4}} 
t^{2Nm-Nn}x^{-n}.
\end{align}
The first term in the sum can be factorized as 
the product of the half-index $\mathbb{II}_{\mathcal{N}'}^{\textrm{4d $U(1)$}}$ 
of the Neumann b.c. $\mathcal{N}'$ for 4d $U(1)$ gauge theory, 
the quarter-indices $\mathbb{IV}_{\mathcal{N}'\mathcal{D}}^{\textrm{4d $U(1)$}}$ and 
$\mathbb{IV}_{\mathcal{N}'\mathcal{D}/\textrm{Nahm}}^{\textrm{4d $U(N-1)$}}$. 
Thus the sum is associated to the Higgsing procedure which leads to the splitting 
$\left(\begin{smallmatrix}
N&1\\ 0&1\\\end{smallmatrix}\right)$ 
$\rightarrow$ 
$\left(\begin{smallmatrix}
1&1\\ 0&0\\\end{smallmatrix}\right)$ 
$\oplus$ 
$\left(\begin{smallmatrix}
N-1&0\\ 0&0\\\end{smallmatrix}\right)$ 
$\oplus$ 
$\left(\begin{smallmatrix}
0&0\\ 0&1\\\end{smallmatrix}\right)$ 
of the NS5$'$-D5-junction.

In the absence of the Wilson lines, i.e. $n=0$, 
the quarter-indices (\ref{nd11N0wil}) and (\ref{ndN101wil}) coincide, as in 
the equivalence between (\ref{nd1120}) and (\ref{nd2101}). 
This amounts to a transformation formula for a basic hypergeometric series:
\begin{align}
\label{nd11N0wid}
&\frac{1}{(q^{\frac{N}{2}}t^{2N};q)_{\infty}}
\sum_{m=0}^{\infty}
\frac{
\left(q^{1+m};q\right)_{\infty}
\left(q^{\frac34+\frac{N}{4}+m}t^{N+1};q\right)_{\infty}
}
{
\left(q^{\frac12+m}t^2;q\right)_{\infty}
\left(q^{1-\left(\frac14+\frac{N}{4}\right)+m}t^{-(N-1)};q\right)_{\infty}
}
q^{\frac14+\frac{N}{4}}
t^{(N-1)m}
\nonumber\\
&=
\frac{1}{(q^{\frac12}t^2;q)_{\infty}}
\sum_{m=0}^{\infty}
\frac{
\left(q^{1+m};q\right)_{\infty}
\left(q^{\frac34+\frac{N}{4}+m}t^{N+1};q\right)_{\infty}
}
{
\left(q^{\frac{N}{2}+m} t^{2N};q\right)_{\infty}
\left(q^{\frac14+\frac{N}{4}+m} t^{N-1};q \right)_{\infty}
}
q^{\left[ 1- \left(\frac14+\frac{N}{4}\right) \right]}
t^{-(N-1)m}. 
\end{align}
In the presence of the Wilson lines, i.e. $n\neq 0$, 
the quarter-indices (\ref{nd11N0wil}) and (\ref{ndN101wil}) do not coincide,
of course, as S-duality would match them to indices involving a 't Hooft line. 
Again we leave the duality involving the line operators for future work.

\subsubsection{${2 \, 1 \choose 3 \, 0}$ and ${3 \, 2 \choose 0 \, 1}$}
Next consider the $\left(\begin{smallmatrix}
2&1\\ 3&0\\\end{smallmatrix}\right)$ NS5$'$-D5 junction. 
It contains a 4d $\mathcal{N}=4$ $U(1)$ gauge theory with 
the Neumann boundary condition $\mathcal{N}'$ at $x^2=0$ 
and bosonic local operators coming from the broken $U(2)$ gauge theory. 
The 4d $\mathcal{N}=4$ $U(3)$ gauge theory in the lower left quadrant 
obeys the Neumann boundary condition $\mathcal{N}'$ 
and the Nahm pole boundary condition 
characterized by a homomorphism $\rho:$ $\mathfrak{su}(2)$ $\rightarrow$ $\mathfrak{u}(3)$. 
There are also local operators from the 3d $\mathcal{N}=4$ twisted hypermultiplet 
obeying the Nahm pole boundary condition. 

We find the quarter-index for the 
$\left(\begin{smallmatrix}
2&1\\ 3&0\\\end{smallmatrix}\right)$ NS5$'$-D5 junction
\begin{align}
\label{nd2130}
\mathbb{IV}_{\mathcal{N}'\mathcal{D}}^{\left(\begin{smallmatrix}
2&1\\3&0\\\end{smallmatrix}\right)}
&=
\underbrace{
\frac{(q)_{\infty}}{(q^{\frac12} t^2;q)_{\infty}}\oint \frac{ds}{2\pi is}
}_{\mathbb{II}_{\mathcal{N}'}^{\textrm{4d $U(1)$}}}
\underbrace{
\frac{1}{(q^{\frac12} t^2;q)_{\infty}}
}_{\mathbb{IV}_{\mathcal{N}'\mathcal{D}}^{\textrm{4d $U(1)$}}}
\frac{1}{(q^{\frac12}t^2 s^{\pm}x_{1}^{\mp};q)_{\infty}}
(q^{\frac54}t^3 s^{\pm}x_{2}^{\mp};q)_{\infty}
\nonumber\\
&\times 
\underbrace{
\frac{1}{(q^{\frac12}t^2;q)_{\infty} (qt^4;q)_{\infty} (q^{\frac32} t^6;q)_{\infty}}
}_{\mathbb{IV}_{\mathcal{N}'\textrm{Nahm}}^{\textrm{4d $U(3)$}}}
(q^{\frac54}t^3x_{1}^{\pm}x_{2}^{\mp};q)_{\infty}. 
\end{align}
Here the same contributions as in (\ref{nd0103}) appear in the last line, 
which we found for the $\left(\begin{smallmatrix}
1&0\\ 3&0\\\end{smallmatrix}\right)$ NS5$'$-D5 junction. 

The quarter-index (\ref{nd2130}) can be calculated as the sum of residues at poles 
$s=q^{\frac12+m}t^2 x_{1}$
\begin{align}
\label{nd2130sum1}
\mathbb{IV}_{\mathcal{N}'\mathcal{D}}^{\left(\begin{smallmatrix}
2&1\\3&0\\\end{smallmatrix}\right)}
&=
\frac{(q^{\frac54}t^3 x_{1}^{\pm}x_{2}^{\mp};q)_{\infty} 
\left(q^{\frac34}t \frac{x_{2}}{x_{1}};q\right)_{\infty}
\left(q^{\frac14}t \frac{x_{1}}{x_{2}};q\right)_{\infty}}
{(q)_{\infty} (q^{\frac12}t^2;q)_{\infty}^3 (qt^4;q)_{\infty} (q^{\frac32}t^6;q)_{\infty}}
\nonumber\\
&\times 
\sum_{m=0}^{\infty}
\frac{
(q^{1+m};q)_{\infty} \left(q^{\frac74+m}t^5 \frac{x_{1}}{x_{2}};q\right)_{\infty}
}
{
(q^{1+m}t^{4};q)_{\infty}\left(q^{\frac14+m}t^{-1}\frac{x_{1}}{x_{2}};q\right)_{\infty}
}
q^{\frac{3m}{4}} t^{m} x_{1}^{-m} x_{2}^{m}. 
\end{align}
The sum in (\ref{nd2130}) starts from 
the product of three quarter-indices 
$\mathbb{IV}_{\mathcal{N}'\mathcal{D}}^{\textrm{4d $U(1)$}}$, 
$\mathbb{IV}_{\mathcal{N}'\textrm{Nahm}}^{\textrm{4d $U(2)$}}$, 
$\mathbb{IV}_{\mathcal{N}'\textrm{Nahm}}^{\textrm{4d $U(3)$}}$ 
and the extra fermionic factors 
$(q^{\frac{5}{4}}t^3x_{1}^{\pm}x_{2}^{\mp};q)_{\infty}$ 
$\left(q^{\frac{7}{4}}t^{5}\frac{x_{1}}{x_{2}};q\right)_{\infty}$ 
$\left(q^{\frac{3}{4}}t\frac{x_{2}}{x_{1}};q\right)_{\infty}$. 

We can also expand the quarter-index (\ref{nd2130}) as
\begin{align}
\label{nd2130sum2}
\mathbb{IV}_{\mathcal{N}'\mathcal{D}}^{\left(\begin{smallmatrix}
2&1\\3&0\\\end{smallmatrix}\right)}
&=
\frac{(q^{\frac54}t^3 x_{1}^{\pm}x_{2}^{\mp};q)_{\infty} (q^{\frac34}tx_{1}^{\pm}x_{2}^{\mp};q)_{\infty}}
{(q)_{\infty} (q^{\frac12}t^2;q)_{\infty}^3 (qt^{4};q)_{\infty} (q^{\frac32}t^6;q)_{\infty}}
\sum_{m=0}^{\infty}
\frac{(q^{1+m};q)_{\infty}^2}
{(q^{\frac34+m}tx_{1}^{\pm}x_{2}^{\mp};q)_{\infty}}q^{m}t^{4m}. 
\end{align}
The first term in the sum takes the form of the product of 
the half-index $\mathbb{II}_{\mathcal{N}'}^{\textrm{4d $U(1)$}}$ of the Neumann b.c. $\mathcal{N}'$ 
for 4d $\mathcal{N}=4$ $U(1)$ gauge theory surviving in $x^2>0$ 
and two quarter-indices $\mathbb{IV}_{\mathcal{N}'\mathcal{D}}^{\textrm{4d $U(1)$}}$, 
$\mathbb{IV}_{\mathcal{N}'\textrm{Nahm}}^{\textrm{4d $U(3)$}}$ 
as well as the factor $(q^{\frac54}t^3x_{1}^{\pm}x_{2}^{\mp};q)_{\infty}$ associated to the Nahm pole of rank 3 for 3d $\mathcal{N}=4$ twisted hypermultiplet.

The 
$\left(\begin{smallmatrix}
2&1\\ 3&0\\\end{smallmatrix}\right)$ NS5$'$-D5 junction 
turns into the 
$\left(\begin{smallmatrix}
3&2\\ 0&1\\\end{smallmatrix}\right)$ NS5$'$-D5 junction under S-duality. 
The $\left(\begin{smallmatrix}
3&2\\ 0&1\\\end{smallmatrix}\right)$ NS5$'$-D5 junction has 
a 4d $\mathcal{N}=4$ $U(2)$ SYM theory in $x^2>0$ obeying the Neumann boundary condition $\mathcal{N}'$ 
and bosonic local operators appear from the broken $U(3)$ gauge theory. 
The $U(1)$ gauge symmetry in the lower right quadrant is broken 
and the additional 3d $\mathcal{N}=4$ twisted hypermultiplet should receive the Dirichlet boundary condition $D$. 
 
We can compute the quarter-index for the 
$\left(\begin{smallmatrix}
3&2\\ 0&1\\\end{smallmatrix}\right)$ NS5$'$-D5 junction as
\begin{align}
\label{nd3201}
\mathbb{IV}_{\mathcal{N}'\mathcal{D}}^{\left(\begin{smallmatrix}
3&2\\0&1\\\end{smallmatrix}\right)}
&=
\underbrace{
\frac12 
\frac{(q)_{\infty}^2}{(q^{\frac12} t^2;q)_{\infty}^2} 
\oint \prod_{i=1}^2 \frac{ds_{i}}{2\pi is_{i}}
\prod_{i\neq j}\frac{\left(\frac{s_{i}}{s_{j}};q\right)_{\infty}}
{
\left(q^{\frac12} t^2 \frac{s_{i}}{s_{j}};q\right)_{\infty}
}
}_{\mathbb{II}_{\mathcal{N}'}^{\textrm{4d $U(2)$}}}
\underbrace{
\frac{1}{(q^{\frac12} t^2;q)_{\infty}}
}_{\mathbb{IV}_{\mathcal{N}'\mathcal{D}}^{\textrm{4d $U(1)$}}}
\prod_{i=1}^2 
\frac{1}{(q^{\frac12}t^2 s_{i}^{\pm}x_{2}^{\mp};q)_{\infty}}
\underbrace{
(q^{\frac34}t s_{i}^{\pm}x_{1}^{\mp};q)_{\infty}
}_{\mathbb{II}_{D}^{\textrm{3d tHM}}(s_{i}x_{1}^{-1})}
\nonumber\\
&\times 
\underbrace{
\frac{1}{(q^{\frac12}t^2;q)_{\infty}}
}_{\mathbb{IV}_{\mathcal{N}'\mathcal{D}}^{\textrm{4d $U(1)$}}}. 
\end{align}
This coincides with the quarter-index (\ref{nd2130}) 
for the $\left(\begin{smallmatrix}
2&1\\ 3&0\\\end{smallmatrix}\right)$ 
NS5$'$-D5 junction.

\subsubsection{${3 \, 1 \choose 2 \, 0}$ and ${2 \, 3 \choose 0 \, 1}$}

An inequivalent D3-branes configuration gives the 
$\left(\begin{smallmatrix}
3&1\\ 2&0\\\end{smallmatrix}\right)$ NS5$'$-D5 junction. 
It includes a 4d $\mathcal{N}=4$ $U(1)$ gauge theory in $x^2>0$ with the Neumann boundary condition $\mathcal{N}'$ 
and bosonic local operators originating from the broken $U(3)$ gauge theory. 
The 4d $\mathcal{N}=4$ $U(2)$ gauge theory in the lower left quadrant 
satisfies the Neumann boundary condition $\mathcal{N}'$ 
and Nahm pole boundary condition associated with a homomorphism 
$\rho:$ $\mathfrak{su}(2)$ $\rightarrow$ $\mathfrak{u}(2)$. 
In addition, there are contributions from the Nahm pole boundary condition for 
the 3d $\mathcal{N}=4$ twisted hypermultiplet 
and those from the neutral fields which follow from the 
$\left(\begin{smallmatrix}
2&0\\ 2&0\\\end{smallmatrix}\right)$ NS5$'$-D5 junction. 
 
The quarter-index for the 
$\left(\begin{smallmatrix}
3&1\\ 2&0\\\end{smallmatrix}\right)$ NS5$'$-D5 junction is 
\begin{align}
\label{nd3120}
\mathbb{IV}_{\mathcal{N}'\mathcal{D}}^{\left(\begin{smallmatrix}
3&1\\2&0\\\end{smallmatrix}\right)}
&=
\underbrace{
\frac{(q)_{\infty}}{(q^{\frac12} t^2;q)_{\infty}}\oint \frac{ds}{2\pi is}
}_{\mathbb{II}_{\mathcal{N}'}^{\textrm{4d $U(1)$}}}
\underbrace{
\frac{1}{(q^{\frac12} t^2;q)_{\infty} (qt^4;q)_{\infty}}
}_{\mathbb{IV}_{\mathcal{N}'\textrm{Nahm}}^{\textrm{4d $U(2)$}}}
\frac{1}{(q^{\frac34}t^3 s^{\pm}x_{1}^{\mp};q)_{\infty}}
(q t^2 s^{\pm}x_{2}^{\mp};q)_{\infty}
\nonumber\\
&\times 
\underbrace{
\frac{1}{(q^{\frac12}t^2;q)_{\infty} (qt^4;q)_{\infty}}
}_{\mathbb{IV}_{\mathcal{N}'\textrm{Nahm}}^{\textrm{4d $U(2)$}}}
(q^{\frac34}tx_{1}^{\pm}x_{2}^{\mp};q)_{\infty}
(q^{\frac54}t^3x_{1}^{\pm}x_{2}^{\mp};q)_{\infty}. 
\end{align}
The contributions in the last line describe 
the neutral local operators appearing in the 
$\left(\begin{smallmatrix}
2&0\\ 2&0\\\end{smallmatrix}\right)$ NS5$'$-D5 junction, 
which we obtained in (\ref{nd0202}). 
As we cannot get the 
$\left(\begin{smallmatrix}
3&1\\ 2&0\\\end{smallmatrix}\right)$ NS5$'$-D5 junction 
by S-duality from the 
$\left(\begin{smallmatrix}
2&1\\ 3&0\\\end{smallmatrix}\right)$ 
and 
$\left(\begin{smallmatrix}
3&2\\ 0&1\\\end{smallmatrix}\right)$ 
NS5$'$-D5 junction, 
(\ref{nd3120}) is different from the quarter-indices 
(\ref{nd2130}) and (\ref{nd3201}). 

By considering the residues at poles $s=q^{\frac34+m}t^3 x_{1}$, 
we can evaluate the contour integral in (\ref{nd3120}) and find that
\begin{align}
\label{nd3120sum1}
\mathbb{IV}_{\mathcal{N}'\mathcal{D}}^{\left(\begin{smallmatrix}
3&1\\2&0\\\end{smallmatrix}\right)}
&=\frac{(q^{\frac34}t x_{1}^{\pm}x_{2}^{\mp};q)_{\infty} (q^{\frac54}t^3 x_{1}^{\pm} x_{2}^{\mp};q)_{\infty} 
\left(q^{\frac14}t^{-1}\frac{x_{1}}{x_{2}};q\right)_{\infty} \left(q^{\frac34}t \frac{x_{1}}{x_{2}};q\right)_{\infty}}
{(q)_{\infty} (q^{\frac12}t^2;q)_{\infty}^3 (qt^{4};q)_{\infty}^2}
\nonumber\\
&\times 
\sum_{m=0}^{\infty}
\frac{
(q^{1+m};q)_{\infty} \left(q^{\frac74+m}t^5 \frac{x_{1}}{x_{2}};q\right)_{\infty}
}
{(q^{\frac32+m}t^6;q)_{\infty} \left(q^{\frac34+m}t \frac{x_{1}}{x_{2}}\right)_{\infty}} q^{\frac{m}{4}} t^{-m} x_{1}^{-m}x_{2}^{m}
\end{align}
The first term in the sum is the product of three quarter-indices 
$\mathbb{IV}_{\mathcal{N}'\mathcal{D}}^{\textrm{4d $U(1)$}}$, 
$\mathbb{IV}_{\mathcal{N}'\textrm{Nahm}}^{\textrm{4d $U(2)$}}$, 
$\mathbb{IV}_{\mathcal{N}'\textrm{Nahm}}^{\textrm{4d $U(3)$}}$,  
corresponding to the three corners of 
the 4d $\mathcal{N}=4$ $U(1)$, $U(2)$ and $U(3)$ gauge theoreis 
and the extra fermionic factors $\left(q^{\frac14}t^{-1}\frac{x_{1}}{x_{2}};q\right)_{\infty}$ 
$\left(q^{\frac74}t^5 \frac{x_{1}}{x_{2}};q\right)_{\infty}$. 

The quarter-index (\ref{nd3120}) can be also expressed as 
\begin{align}
\label{nd3120sum2}
\mathbb{IV}_{\mathcal{N}'\mathcal{D}}^{\left(\begin{smallmatrix}
3&1\\2&0\\\end{smallmatrix}\right)}
&=\frac{(q^{\frac14}t^{-1}x_{1}^{\pm}x_{2}^{\mp})_{\infty} (q^{\frac34}tx_{1}^{\pm}x_{2}^{\mp};q)_{\infty} 
(q^{\frac54}t^3 x_{1}^{\pm}x_{2}^{\mp})_{\infty}}{(q)_{\infty} (q^{\frac12}t^2;q)_{\infty}^3 (qt^{4};q)_{\infty}^2}
\sum_{m=0}^{\infty} 
\frac{(q^{1+m};q)_{\infty}^2}{(q^{\frac14+m}t^{-1}x_{1}^{\pm}x_{2}^{\mp};q)_{\infty}} 
q^{\frac{3 m}{2}} t^{6m}. 
\end{align}
The sum begins with the product of 
the half-index $\mathbb{II}_{\mathcal{N}'}^{\textrm{4d $U(1)$}}$ of the Neumann b.c. $\mathcal{N}'$ 
for 4d $\mathcal{N}=4$ $U(1)$ gauge theory living in $x^6>0$ 
and the square of the quarter-index $\mathbb{IV}_{\mathcal{N}'\textrm{Nahm}}^{\textrm{4d $U(2)$}}$ 
and the extra fermionic contributions 
$(q^{\frac34}tx_{1}^{\pm}x_{2}^{\mp};q)_{\infty}$ 
$(q^{\frac54}t^3x_{1}^{\pm}x_{2}^{\mp};q)_{\infty}$.

The action of S-duality turns the 
$\left(\begin{smallmatrix}
3&1\\ 2&0\\\end{smallmatrix}\right)$ NS5$'$-D5 junction
into the 
$\left(\begin{smallmatrix}
2&3\\ 0&1\\\end{smallmatrix}\right)$ NS5$'$-D5 junction. 
There is a 4d $\mathcal{N}=4$ $U(2)$ gauge theory in $x^2>0$ with the Neumann boundary condition $\mathcal{N}'$ 
and bosonic local operators appear from the broken $U(3)$ gauge theory. 
The 4d $\mathcal{N}=4$ $U(1)$ gauge theory in the lower right space  
obeys the Neumann boundary condition $\mathcal{N}'$ 
and Dirichlet boundary condition $\mathcal{D}$. 
In addition, we have the contributions from the Dirichlet boundary condition $D$ for 
the 3d $\mathcal{N}=4$ twisted hypermultiplet 
and those from the neutral fields which follow from the 
$\left(\begin{smallmatrix}
0&1\\ 0&1\\\end{smallmatrix}\right)$ NS5$'$-D5 junction.

We get the quarter-index for the 
$\left(\begin{smallmatrix}
2&3\\ 0&1\\\end{smallmatrix}\right)$ NS5$'$-D5 junction 
\begin{align}
\label{nd2301}
\mathbb{IV}_{\mathcal{N}'\mathcal{D}}^{\left(\begin{smallmatrix}
2&3\\0&1\\\end{smallmatrix}\right)}
&=
\underbrace{
\frac12 
\frac{(q)_{\infty}^2}{(q^{\frac12} t^2;q)_{\infty}^2} 
\oint \prod_{i=1}^2 \frac{ds_{i}}{2\pi is_{i}}
\prod_{i\neq j}\frac{\left(\frac{s_{i}}{s_{j}};q\right)_{\infty}}
{
\left(q^{\frac12} t^2 \frac{s_{i}}{s_{j}};q\right)_{\infty}
}
}_{\mathbb{II}_{\mathcal{N}'}^{\textrm{4d $U(2)$}}}
\underbrace{
\frac{1}{(q^{\frac12} t^2;q)_{\infty}}
}_{\mathbb{IV}_{\mathcal{N}'\mathcal{D}}^{\textrm{4d $U(1)$}}}
\prod_{i=1}^2 
\frac{1}{(q^{\frac12}t^2 s_{i}^{\pm}x_{2}^{\mp};q)_{\infty}}
\underbrace{
(q^{\frac34}t s_{i}^{\pm}x_{1}^{\mp};q)_{\infty}
}_{\mathbb{II}_{D}^{\textrm{3d tHM}}(s_{i}x_{1}^{-1})}
\nonumber\\
&\times 
\underbrace{
\frac{1}{(q^{\frac12}t^2;q)_{\infty}}
}_{\mathbb{IV}_{\mathcal{N}'\mathcal{D}}^{\textrm{4d $U(1)$}}}
(q^{\frac34}tx_{1}^{\pm}x_{2}^{\mp};q)_{\infty}
\end{align}
where the factors in the last line 
include the contributions from the broken gauge theory 
associated to the 
$\left(\begin{smallmatrix}
0&1\\ 0&1\\\end{smallmatrix}\right)$ NS5$'$-D5 junction, 
which takes the same form as in (\ref{nd0101}). 
The quarter-index (\ref{nd2301}) matches with the quarter-index (\ref{nd3120}) 
for the $\left(\begin{smallmatrix}
3&1\\ 2&0 \\\end{smallmatrix}\right)$ NS5$'$-D5 junction.

\begin{figure}
\begin{center}
\includegraphics[width=4cm]{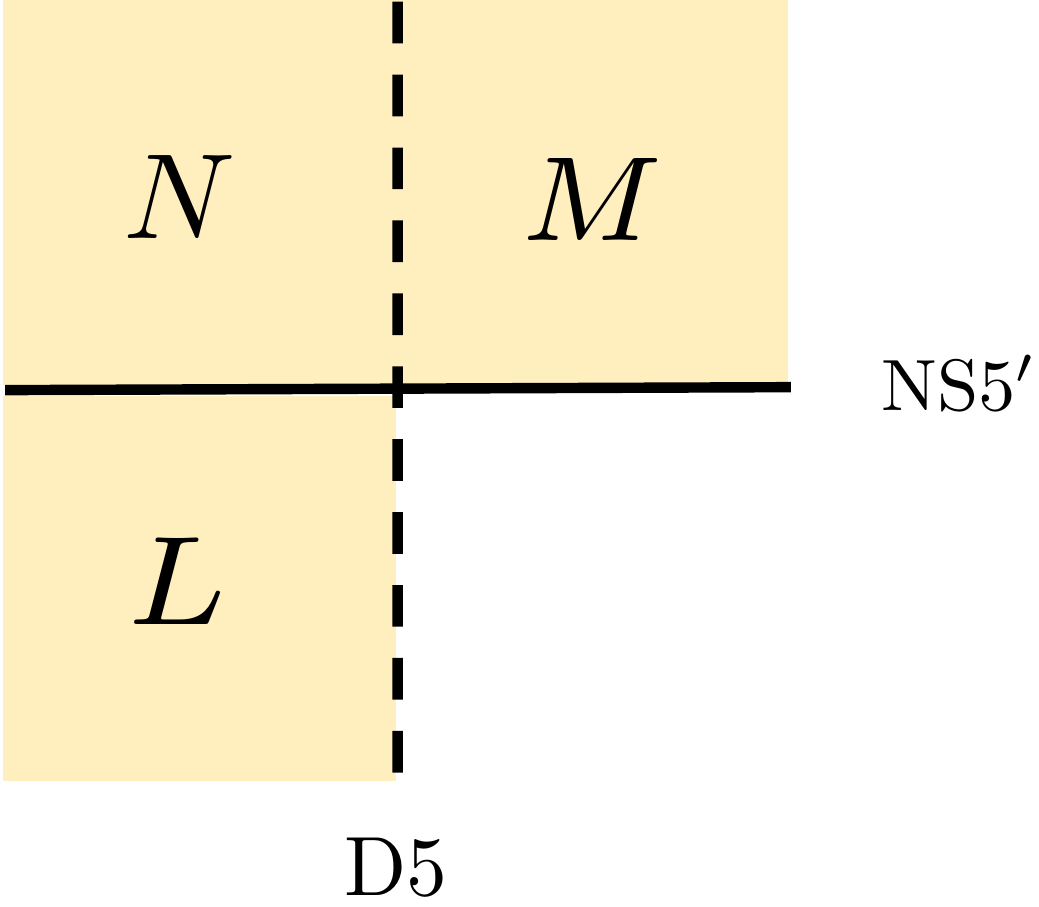}
\caption{The ${N \, M \choose L \, 0}$ NS5$'$-D5 junction.}
\label{figndNML0}
\end{center}
\end{figure}

The brane picture is illustrated in Figure \ref{figndNML0}. 
%
%
%
%
%

\subsection{${N \, M \choose L \, K}$}
\label{sec_2dndNMLK}
Finally, let us discuss the most general NS5$'$-D5 junction, 
i.e. $\left(\begin{smallmatrix}
N&M\\ L&K\\\end{smallmatrix}\right)$ NS5$'$-D5 junction. 

\subsubsection{${1 \, 2 \choose 1 \, 2}$ and ${1 \, 1 \choose 2 \, 2}$}

As an example, consider the $\left(\begin{smallmatrix}
1&2\\ 1&2\\\end{smallmatrix}\right)$ NS5$'$-D5 junction. 
Both 4d $\mathcal{N}=4$ $U(1)$ gauge theories in $x^2>0$ and in $x^2<0$ 
obey the Neumann boundary condition $\mathcal{N}'$. 
In addition, there will be the 3d $\mathcal{N}=4$ bi-fundamental twisted hyprmultiplet,
as well as the contributions from the broken gauge symmetry 
which are associated to 
the $\left(\begin{smallmatrix}
0&1\\ 0&1\\\end{smallmatrix}\right)$ 
NS5$'$-D5 junctions. 

The quarter-index for the $\left(\begin{smallmatrix}
1&2\\ 1&2\\\end{smallmatrix}\right)$ NS5$'$-D5 junction is written as 
\begin{align}
\label{nd1212}
\mathbb{IV}_{\mathcal{N}'\mathcal{D}}^{\left(\begin{smallmatrix}
1&2\\1&2\\\end{smallmatrix}\right)}
&=
\underbrace{
\frac{(q)_{\infty}}{(q^{\frac12}t^2;q)_{\infty}}\oint \frac{ds_{1}}{2\pi is_{1}}
}_{\mathbb{II}_{\mathcal{N}'}^{\textrm{4d $U(1)$}}}
\underbrace{
\frac{1}{(q^{\frac12}t^2;q)_{\infty}}
}_{\mathbb{IV}_{\mathcal{N}'\mathcal{D}}^{\textrm{4d $U(1)$}}}
\frac{1}{(q^{\frac12} t^2 s_{1}^{\pm}x_{1}^{\mp};q)_{\infty}}
\underbrace{
(q^{\frac34} t s_{1}^{\pm}x_{2}^{\mp};q)_{\infty}
}_{\mathbb{II}_{D}^{\textrm{3d tHM}}(s_{1}x_{2}^{-1})}
\nonumber\\
&\times 
\underbrace{
\frac{(q)_{\infty}}{(q^{\frac12}t^2;q)_{\infty}}\oint \frac{ds_{2}}{2\pi is_{2}}
}_{\mathbb{II}_{\mathcal{N}'}^{\textrm{4d $U(1)$}}}
\underbrace{
\frac{1}{(q^{\frac12}t^2;q)_{\infty}}
}_{\mathbb{IV}_{\mathcal{N}'\mathcal{D}}^{\textrm{4d $U(1)$}}}
\frac{1}{(q^{\frac12} t^2 s_{2}^{\pm}x_{2}^{\mp};q)_{\infty}}
\underbrace{
(q^{\frac34} t  s_{2}^{\pm}x_{1}^{\mp};q)_{\infty}
}_{\mathbb{II}_{D}^{\textrm{3d tHM}}(s_{2}x_{1}^{-1})}
\nonumber\\
&\times 
\underbrace{
\frac{
\left(q^{\frac34}t \frac{s_{1}}{s_{2}};q\right)_{\infty}
\left(q^{\frac34}t \frac{s_{2}}{s_{1}};q\right)_{\infty}
}
{
\left(q^{\frac14}t^{-1} \frac{s_{1}}{s_{2}};q\right)_{\infty}
\left(q^{\frac14}t^{-1} \frac{s_{2}}{s_{1}};q\right)_{\infty}
}
}_{\mathbb{I}^{\textrm{3d tHM}}\left(\frac{s_{1}}{s_{2}}\right)}
(q^{\frac34}t x_{1}^{\pm}x_{2}^{\mp};q)_{\infty}. 
\end{align}
The factors in the last line describe 
the additional 3d $\mathcal{N}=4$ bi-fundamentaql twisted hypermultiplet 
and the contributions from the $\left(\begin{smallmatrix}
0&1\\ 0&1\\\end{smallmatrix}\right)$ NS5$'$-D5 junction 
which we obtained in (\ref{nd0101}).

The S-dual is the $\left(\begin{smallmatrix}
1&1\\ 2&2\\\end{smallmatrix}\right)$ NS5$'$-D5 junction. 
This can be constructed by combining 
the $\left(\begin{smallmatrix}
1&1\\ 0&0\\\end{smallmatrix}\right)$ 
and $\left(\begin{smallmatrix}
0&0\\ 2&2\\\end{smallmatrix}\right)$ NS5$'$-D5 junctions. 
The whole gauge symmetry is $U(1)$ for $x^2>0$ and $U(2)$ for $x^2<0$. 
Since these decomposed junctions have equal numbers of D3-branes across the D5-brane, 
they include 3d $\mathcal{N}=4$ fundamental hypermultiplts together with fundamental Fermi multiplets. 
The 3d $\mathcal{N}=4$ bi-fundamental twisted hypermultiplet across the NS5$'$-brane 
will contribute to the index. 

The quarter-index for the S-dual $\left(\begin{smallmatrix}
1&1\\ 2&2\\\end{smallmatrix}\right)$ NS5$'$-D5 junction is 
\begin{align}
\label{nd1122}
\mathbb{IV}_{\mathcal{N}'\mathcal{D}}^{\left(\begin{smallmatrix}
1&1\\2&2\\\end{smallmatrix}\right)}
&=
\underbrace{
\frac{(q)_{\infty}}{(q^{\frac12}t^2;q)_{\infty}}\oint \frac{ds_{1}}{2\pi is_{1}}
}_{\mathbb{II}_{\mathcal{N}'}^{\textrm{4d $U(1)$}}}
\underbrace{
\frac{
(q^{\frac12} s_{1}x_{1}^{-1};q)_{\infty} (q^{\frac12} s_{1}^{-1}x_{1};q)_{\infty}}
{(q^{\frac14}t  s_{1}x_{2}^{-1};q)_{\infty} (q^{\frac14}t s_{1}^{-1}x_{2};q)_{\infty}}
}_{\mathbb{II}_{N}^{\textrm{3d HM}}(s_{1}x_{2}^{-1})\cdot F(q^{\frac12}s_{1}x_{1}^{-1})}
\nonumber\\
&\times 
\underbrace{
\frac12 \frac{(q)_{\infty}^{2}}{(q^{\frac12} t^2;q)_{\infty}^{2}} 
\oint \prod_{i=2}^{3}\frac{ds_{i}}{2\pi is_{i}} \prod_{i\neq j}\frac{\left(\frac{s_{i}}{s_{j}};q\right)_{\infty}}{\left(q^{\frac12}t^2 \frac{s_{i}}{s_{j}};q\right)_{\infty}}
}_{\mathbb{II}_{\mathcal{N}'}^{\textrm{4d $U(2)$}}}
\underbrace{
\frac{
(q^{\frac12} s_{i}x_{2}^{-1};q)_{\infty} (q^{\frac12}s_{i}^{-1}x_{2};q)_{\infty}
}
{(q^{\frac14}t s_{i}x_{1}^{-1};q)_{\infty}(q^{\frac14}t s_{i}^{-1}x_{1};q)_{\infty}
}
}_{\mathbb{II}_{N}^{\textrm{3d HM}}(s_{i}x_{1}^{-1})\cdot F(q^{\frac12} s_{i}x_{2}^{-1})}
\nonumber\\
&\times 
\prod_{i=2}^{3}
\underbrace{
\frac{
\left( q^{\frac34} t\frac{s_{1}}{s_{i}};q\right)_{\infty}
\left( q^{\frac34} t\frac{s_{i}}{s_{1}};q\right)_{\infty}
}
{
\left( q^{\frac14} t^{-1}\frac{s_{1}}{s_{i}};q\right)_{\infty}
\left( q^{\frac14} t^{-1}\frac{s_{i}}{s_{1}};q\right)_{\infty}
}
}_{\mathbb{I}^{\textrm{3d tHM}}\left(\frac{s_{1}}{s_{i}}\right)}
\end{align}
where the factors in the last line is the additional 3d $\mathcal{N}=4$ bi-fundamental twisted hypermultiplet. 
In fact the quarter-indices (\ref{nd1212}) and (\ref{nd1122}) coincide.

\subsubsection{${1 \, 1 \choose 2 \, 3}$ and ${2 \, 1 \choose 3 \, 1}$}

The next example is the $\left(\begin{smallmatrix}
1&1\\ 2&3\\\end{smallmatrix}\right)$ NS5$'$-D5 junction. 
The 3d $\mathcal{N}=4$ charged hypermultiplet satisfies the Neumann boundary condition $N'$. 
The gauge anomaly is canceled by the contributions from the Dirichlet boundary condition $D$ 
for the 3d $\mathcal{N}=4$ charged twisted hypermultiplet arising from D3-D3 string across the NS5$'$-brane. 
In $x^2<0$, 
the $U(3)$ gauge symmetry in the lower right corner is broken to $U(2)$ 
and there remains a 4d $\mathcal{N}=4$ $U(2)$ SYM theory. 
Instead of the 3d $\mathcal{N}=4$ hypermultiplet, 
there are bosonic local operators from the broken $U(3)$ gauge theory 
together with the fundamental Fermi multiplet that cancel the gauge anomaly. 
Besides, there is the 3d $\mathcal{N}=4$ twisted hypermultiplet 
transforming as bi-fundamental representation under the $U(1)$ $\times$ $U(2)$ gauge symmetry.

The quarter-index for the $\left(\begin{smallmatrix}
1&1\\ 2&3\\\end{smallmatrix}\right)$ NS5$'$-D5 junction is computed as
\begin{align}
\label{nd1123}
\mathbb{IV}_{\mathcal{N}'\mathcal{D}}^{\left(\begin{smallmatrix}
1&1\\2&3\\\end{smallmatrix}\right)}
&=
\underbrace{
\frac{(q)_{\infty}}{(q^{\frac12}t^2;q)_{\infty}}\oint \frac{ds_{1}}{2\pi is_{1}}
}_{\mathbb{II}_{\mathcal{N}'}^{\textrm{4d $U(1)$}}}
\underbrace{
\frac{1}{(q^{\frac14} t s_{1}^{\pm}x_{1}^{\mp};q)_{\infty}}
}_{\mathbb{II}_{N}^{\textrm{3d HM}}(s_{1}x_{1}^{-1})}
\underbrace{
(q^{\frac34}t s_{1}^{\pm}x_{2}^{\mp};q)_{\infty}
}_{\mathbb{II}_{D}^{\textrm{3d tHM}}(s_{1}x_{2}^{-1})}
\nonumber\\
&\times 
\underbrace{
\frac12 \frac{(q)_{\infty}^{2}}{(q^{\frac12} t^2;q)_{\infty}^{2}} 
\oint \prod_{i=2}^{3}\frac{ds_{i}}{2\pi is_{i}} \prod_{i\neq j}\frac{\left(\frac{s_{i}}{s_{j}};q\right)_{\infty}}
{\left(q^{\frac12}t^2 \frac{s_{i}}{s_{j}};q\right)_{\infty}}
}_{\mathbb{II}_{\mathcal{N}'}^{\textrm{4d $U(2)$}}}
\underbrace{
\frac{1}{(q^{\frac12}t^2;q)_{\infty}}
}_{\mathbb{IV}_{\mathcal{N}'\mathcal{D}}^{\textrm{4d $U(1)$}}}
\prod_{i=2}^{3}
\frac{1}{(q^{\frac12}t^2 s_{i}^{\pm}x_{2}^{\mp};q)_{\infty}}
\underbrace{
(q^{\frac12}s_{i}^{\pm}x_{1}^{\mp};q)_{\infty}
}_{F(q^{\frac12}s_{i}x_{1}^{-1})}
\nonumber\\
&\times 
\prod_{i=2}^{3}
\underbrace{
\frac{
\left( q^{\frac34} t\frac{s_{1}}{s_{i}};q\right)_{\infty}
\left( q^{\frac34} t\frac{s_{i}}{s_{1}};q\right)_{\infty}
}
{
\left( q^{\frac14} t^{-1}\frac{s_{1}}{s_{i}};q\right)_{\infty}
\left( q^{\frac14} t^{-1}\frac{s_{i}}{s_{1}};q\right)_{\infty}
}
}_{\mathbb{I}^{\textrm{3d tHM}}\left(\frac{s_{1}}{s_{i}}\right)}. 
\end{align}

For the S-dual $\left(\begin{smallmatrix}
2&1\\ 3&1\\\end{smallmatrix}\right)$ NS5$'$-D5 junction, 
there are two 4d $\mathcal{N}=4$ $U(1)$ gauge theories in $x^2>0$ and $x^2<0$ 
with the Neumann boundary condition $\mathcal{N}'$. 
The junction has no hypermultiplet, howerver, 
it includes bosonic degrees of freedom from the broken 
$U(2)$ gauge symmetry in the upper left corner 
and the $U(3)$ gauge symmetry in the lower left corner. 
The gauge anomalies receive cancelling contributions from 
the Nahm pole and from the Dirichlet boundary conditions for the 
3d $\mathcal{N}=4$ twisted hypermultiplets. 

The quarter-index for the $\left(\begin{smallmatrix}
2&1\\ 3&1\\\end{smallmatrix}\right)$ NS5$'$-D5 junction is 
\begin{align}
\label{nd2131}
\mathbb{IV}_{\mathcal{N}'\mathcal{D}}^{\left(\begin{smallmatrix}
2&1\\3&1\\\end{smallmatrix}\right)}
&=
\underbrace{
\frac{(q)_{\infty}}{(q^{\frac12}t^2;q)_{\infty}}\oint \frac{ds_{1}}{2\pi is_{1}}
}_{\mathbb{II}_{\mathcal{N}'}^{\textrm{4d $U(1)$}}}
\underbrace{
\frac{1}{(q^{\frac12}t^2;q)_{\infty}}
}_{\mathbb{IV}_{\mathcal{N}'\mathcal{D}}^{\textrm{4d $U(1)$}}}
\frac{1}{(q^{\frac12} t^2 s_{1}^{\pm}x_{2}^{\mp};q)_{\infty}}
(q t^2 s_{1}^{\pm}x_{1}^{\mp};q)_{\infty}
\nonumber\\
&\times 
\underbrace{
\frac{(q)_{\infty}}{(q^{\frac12}t^2;q)_{\infty}}\oint \frac{ds_{2}}{2\pi is_{2}}
}_{\mathbb{II}_{\mathcal{N}'}^{\textrm{4d $U(1)$}}}
\underbrace{
\frac{1}{(q^{\frac12}t^2;q)_{\infty} (qt^4;q)_{\infty}}
}_{\mathbb{IV}_{\mathcal{N}'\textrm{Nahm}}^{\textrm{4d $U(2)$}}}
\frac{1}{(q^{\frac34}t^3 s_{2}^{\pm}x_{1}^{\mp};q)_{\infty}}
\underbrace{
(q^{\frac34} ts_{2}^{\pm}x_{2}^{\mp};q)_{\infty}
}_{\mathbb{II}_{D}^{\textrm{3d tHM}}(s_{2}x_{2}^{-1})}
\nonumber\\
&\times 
\underbrace{
\frac{
\left( q^{\frac34} t\frac{s_{1}}{s_{2}};q\right)_{\infty}
\left( q^{\frac34} t\frac{s_{2}}{s_{1}};q\right)_{\infty}
}
{
\left( q^{\frac14} t^{-1}\frac{s_{1}}{s_{2}};q\right)_{\infty}
\left( q^{\frac14} t^{-1}\frac{s_{2}}{s_{1}};q\right)_{\infty}
}
}_{\mathbb{I}^{\textrm{3d tHM}}\left(\frac{s_{1}}{s_{2}}\right)}
(q t^2x_{1}^{\pm}x_{2}^{\mp};q)_{\infty}. 
\end{align}
The contributions in the last line are the 
additional 3d $\mathcal{N}=4$ bi-fundamental twisted hypermultiplet 
and the neutral local operators associated to the 
$\left(\begin{smallmatrix}
1&0\\ 2&0\\\end{smallmatrix}\right)$ NS5$'$-D5 junction 
which we found in (\ref{nd0102}).  
It follows that 
the quarter-index (\ref{nd1123}) matches with the dual quarter-index (\ref{nd2131}).

\subsubsection{${1 \, 2 \choose 3 \, 4}$ and ${3 \, 1 \choose 4 \, 2}$}

Let us consider the $\left(\begin{smallmatrix}
1&2\\ 3&4\\\end{smallmatrix}\right)$ NS5$'$-D5 junction. 
To obtain this junction, 
we roughly combine the prescriptions for the $\left(\begin{smallmatrix}
1&2\\ 0&0\\\end{smallmatrix}\right)$ and 
$\left(\begin{smallmatrix}
0&0\\ 3&4\\\end{smallmatrix}\right)$ NS5$'$-D5 junctions. 
There is 4d $\mathcal{N}=4$ $U(1)$ gauge theory in $x^2>0$ 
and 4d $\mathcal{N}=4$ $U(3)$ gauge theory in $x^2<0$, 
with the Neumann boundary condition $\mathcal{N}'$. 
As all the numbers of D3-branes are distinct, 
there is no 3d $\mathcal{N}=4$ hypermultiplet. 

We then obtain the quarter-index for the $\left(\begin{smallmatrix}
1&2\\ 3&4\\\end{smallmatrix}\right)$ NS5$'$-D5 junction:
\begin{align}
\label{nd1234}
\mathbb{IV}_{\mathcal{N}'\mathcal{D}}^{\left(\begin{smallmatrix}
1&2\\3&4\\\end{smallmatrix}\right)}
&=
\underbrace{
\frac{(q)_{\infty}}{(q^{\frac12}t^2;q)_{\infty}}\oint \frac{ds_{1}}{2\pi is_{1}}
}_{\mathbb{II}_{\mathcal{N}'}^{\textrm{4d $U(1)$}}}
\underbrace{
\frac{1}{(q^{\frac12}t^2;q)_{\infty}}
}_{\mathbb{IV}_{\mathcal{N}'\mathcal{D}}^{\textrm{4d $U(1)$}}}
\frac{1}{(q^{\frac12}t^2 s_{1}^{\pm}x_{1}^{\mp};q)_{\infty}}
\underbrace{
(q^{\frac34}t s_{1}^{\pm}x_{2}^{\mp};q)_{\infty}
}_{\mathbb{II}_{D}^{\textrm{3d tHM}}(s_{1}x_{2}^{-1})}
\nonumber\\
&\times 
\underbrace{
\frac{1}{3!} \frac{(q)_{\infty}^{3}}{(q^{\frac12}t^2;q)_{\infty}^{3}}
\oint \prod_{i=2}^{4}\frac{ds_{i}}{2\pi is_{i}}\prod_{i\neq j}
\frac{\left(\frac{s_{i}}{s_{j}};q\right)_{\infty}}{
\left(q^{\frac12}t^2 \frac{s_{i}}{s_{j}};q\right)_{\infty}}
}_{\mathbb{II}_{\mathcal{N}'}^{\textrm{4d $U(3)$}}}
\underbrace{
\frac{1}{(q^{\frac12}t^2;q)_{\infty}}
}_{\mathbb{IV}_{\mathcal{N}'\mathcal{D}}^{\textrm{4d $U(1)$}}}
\prod_{i=2}^{4}
\frac{1}
{(q^{\frac12}t^2 s_{i}^{\pm}x_{2}^{\mp};q)_{\infty}}
\underbrace{
(q^{\frac34}ts_{i}^{\pm}x_{1}^{\mp};q)_{\infty}
}_{\mathbb{II}_{D}^{\textrm{3d tHM}}(s_{i}x_{1}^{-1})}
\nonumber\\
&\times 
\prod_{i=2}^{4}
\underbrace{
\frac{
\left(q^{\frac34}t \frac{s_{1}}{s_{i}};q\right)_{\infty}
\left(q^{\frac34}t \frac{s_{i}}{s_{1}};q\right)_{\infty}
}
{
\left(q^{\frac14}t^{-1} \frac{s_{1}}{s_{i}};q\right)_{\infty}
\left(q^{\frac14}t^{-1} \frac{s_{i}}{s_{1}};q\right)_{\infty}
}
}_{\mathbb{I}^{\textrm{3d tHM}}\left(\frac{s_{1}}{s_{i}}\right)}
(q^{\frac34}tx_{1}^{\pm}x_{2}^{\mp};q)_{\infty}. 
\end{align}

The S-dual is the $\left(\begin{smallmatrix}
3&1\\ 4&2\\\end{smallmatrix}\right)$ NS5$'$-D5 junction.
It has 4d $\mathcal{N}=4$ $U(1)$ gauge theory in $x^2>0$ 
and 4d $\mathcal{N}=4$ $U(2)$ gauge theory in $x^2<0$, 
with the Neumann boundary condition $\mathcal{N}'$. 

The resulting quarter-index for the S-dual $\left(\begin{smallmatrix}
3&1\\ 4&2\\\end{smallmatrix}\right)$ NS5$'$-D5 junction is 
\begin{align}
\label{nd3142}
\mathbb{IV}_{\mathcal{N}'\mathcal{D}}^{\left(\begin{smallmatrix}
3&1\\4&2\\\end{smallmatrix}\right)}
&=
\underbrace{
\frac{(q)_{\infty}}{(q^{\frac12}t^2;q)_{\infty}}\oint \frac{ds_{1}}{2\pi is_{1}}
}_{\mathbb{II}_{\mathcal{N}'}^{\textrm{4d $U(1)$}}}
\underbrace{
\frac{1}{(q^{\frac12}t^2;q)_{\infty} (qt^4;q)_{\infty}}
}_{\mathbb{IV}_{\mathcal{N}'\textrm{Nahm}}^{\textrm{4d $U(2)$}}}
\frac{1}{(q^{\frac34}t^3 s_{1}^{\pm}x_{2}^{\mp};q)_{\infty}}
(qt^2 s_{1}^{\pm}x_{1}^{\mp};q)_{\infty}
\nonumber\\
&\times 
\underbrace{
\frac12 \frac{(q)_{\infty}^2}{(q^{\frac12}t^2;q)_{\infty}^{2}}
\oint \prod_{i=2}^{3} \frac{ds_{i}}{2\pi is_{i}} 
\prod_{i\neq j}\frac{\left(\frac{s_{i}}{s_{j}};q\right)_{\infty}}
{\left(q^{\frac12}t^2 \frac{s_{i}}{s_{j}};q\right)_{\infty}}
}_{\mathbb{II}_{\mathcal{N}'}^{\textrm{4d $U(2)$}}}
\underbrace{
\frac{1}{(q^{\frac12}t^2;q)_{\infty} (qt^{4};q)_{\infty}}
}_{\mathbb{IV}_{\mathcal{N}'\textrm{Nahm}}^{\textrm{4d $U(2)$}}}
\prod_{i=2}^{3}
\frac{1}{(q^{\frac34}t^3 s_{i}^{\pm}x_{1}^{\mp};q)_{\infty}}
(q t^2 s_{i}^{\pm}x_{2}^{\mp};q)_{\infty}
\nonumber\\
&\times 
\prod_{i=2}^{3}
\underbrace{
\frac{
\left(q^{\frac34}t\frac{s_{1}}{s_{i}};q\right)_{\infty}
\left(q^{\frac34}t\frac{s_{i}}{s_{1}};q\right)_{\infty}
}
{
\left(q^{\frac14}t^{-1}\frac{s_{1}}{s_{i}};q\right)_{\infty}
\left(q^{\frac14}t^{-1}\frac{s_{i}}{s_{1}};q\right)_{\infty}
}
}_{\mathbb{I}^{\textrm{3d tHM}}\left(\frac{s_{1}}{s_{i}}\right)}
(q^{\frac34}tx_{1}^{\pm}x_{2}^{\mp};q)_{\infty} 
(q^{\frac54}t^3x_{1}^{\pm}x_{2}^{\mp};q)_{\infty}. 
\end{align}
The quarter-indices (\ref{nd1234}) and (\ref{nd3142}) beautifully coincide.

\subsubsection{${N \, M \choose L \, K}$ and ${L \, N \choose K \, M}$}

For general $\left(\begin{smallmatrix}
N&M\\ L&K\\\end{smallmatrix}\right)$ NS5$'$-D5 junction, 
the quarter-index will take the form
\begin{align}
\label{ndNMLK}
\mathbb{IV}_{\mathcal{N}'\mathcal{D}}^{\left(\begin{smallmatrix}
N&M\\L&K\\\end{smallmatrix}\right)}
&=
\underbrace{
\frac{1}{\min (N,M)!}\frac{(q)_{\infty}^{\min (N,M)}}
{(q^{\frac12} t^2;q)_{\infty}^{\min (N,M)}}
\oint \prod_{i=1}^{\min (N,M)}\frac{ds_{i}}{2\pi is_{i}}
\prod_{i\neq j}\frac{
\left(\frac{s_{i}}{s_{j}};q\right)_{\infty}
}
{
\left(q^{\frac12} t^2 \frac{s_{i}}{s_{j}};q\right)_{\infty}
}
}_{\mathbb{II}_{\mathcal{N}'}^{\textrm{4d $U(\min (N,M))$}}}
\nonumber\\
&\times 
\underbrace{
\prod_{k=1}^{|N-M|}\frac{1}{(q^{\frac{k}{2}}t^{2k};q)_{\infty}}
}_{\mathbb{IV}_{\mathcal{N}'\mathcal{D}/\textrm{Nahm}}^{\textrm{4d $U(|N-M|)$}}}
\prod_{i=1}^{\min (N,M)}
\frac{
\left(q^{\frac12+\frac{|L-K|}{4}} t^{|L-K|}s_{i}^{\pm}x_{2}^{\mp};q\right)_{\infty}
}{
(q^{\frac14+\frac{|N-M|}{4}}t^{1+|N-M|}s_{i}^{\pm}x_{1}^{\pm};q)_{\infty}
}
\nonumber\\
&\times 
\underbrace{
\frac{1}{\min (L,K)!}\frac{(q)_{\infty}^{\min (L,K)}}
{(q^{\frac12} t^2;q)_{\infty}^{\min (L,K)}}
\oint \prod_{i=\min (N,M)+1}^{\min (N,M)+\min (L,K)}\frac{ds_{i}}{2\pi is_{i}}
\prod_{i\neq j}\frac{
\left(\frac{s_{i}}{s_{j}};q\right)_{\infty}
}
{
\left(q^{\frac12} t^2 \frac{s_{i}}{s_{j}};q\right)_{\infty}
}
}_{\mathbb{II}_{\mathcal{N}'}^{\textrm{4d $U(\min (L,K))$}}}
\nonumber\\
&\times 
\underbrace{
\prod_{k=1}^{|L-K|}\frac{1}{(q^{\frac{k}{2}}t^{2k};q)_{\infty}}
}_{\mathbb{IV}_{\mathcal{N}'\mathcal{D}/\textrm{Nahm}}^{\textrm{4d $U(|L-K|)$}}}
\prod_{i=\min (N,M)+1}^{\min (N,M)+\min (L,K)}
\frac{
\left(q^{\frac12+\frac{|N-M|}{4}} t^{|N-M|}s_{i}^{\pm}x_{1}^{\mp};q\right)_{\infty}
}{
(q^{\frac14+\frac{|L-K|}{4}}t^{1+|L-K|}s_{i}^{\pm}x_{2}^{\mp};q)_{\infty}
}
\nonumber\\
&\times 
\prod_{i=1}^{\min (N,M)} 
\prod_{k=\min (N,M)+1}^{\min (N,M)+\min (L,K)}
\underbrace{
\frac{
\left(q^{\frac34}t \frac{s_{i}}{s_{k}};q\right)_{\infty}
\left(q^{\frac34}t \frac{s_{k}}{s_{i}};q\right)_{\infty}
}
{
\left(q^{\frac14}t^{-1} \frac{s_{i}}{s_{k}};q\right)_{\infty}
\left(q^{\frac14}t^{-1} \frac{s_{k}}{s_{i}};q\right)_{\infty}
}
}_{\mathbb{I}^{\textrm{3d tHM}}\left(\frac{s_{i}}{s_{k}}\right)}
\nonumber\\
&
\times 
\prod_{i=1}^{\min \left(|N-M|, |L-K|\right)}
\left(
q^{\frac{1+\left| |N-M|-|L-K| \right|}{4}+\frac{i}{2}} t^{\left| |N-M|-|L-K| \right|+2i-1}x_{1}^{\pm}x_{2}^{\mp};q
\right)_{\infty}^{
{\delta^{\frac{|N-M|}{N-M}}}_{\frac{|L-K|}{L-K}}
}.
\end{align}
The fifth line in (\ref{ndNMLK}) describes the 3d $\mathcal{N}=4$ bi-fundamental twisted hypermultiplets 
which are charged under the $U(\min (N,M))$ $\times$ $U(\min (L,K))$ gauge symmetry. 
From the broken gauge symmetry we may obtain the contributions from the last line in (\ref{ndNMLK}), 
which corresponds to the Nahm pole boundary conditions for 
the 3d $\mathcal{N}=4$ bi-fundamental twisted hypermultiplets 
associated to the 
$\left(\begin{smallmatrix}
\min (N,M)&0\\ \min(L,K)&0\\\end{smallmatrix}\right)$ NS5$'$-D5 junctions. 
Note that they appear only when $N>M$ and $L>K$ or $N<M$ and $L<K$. 

\begin{figure}
\begin{center}
\includegraphics[width=4cm]{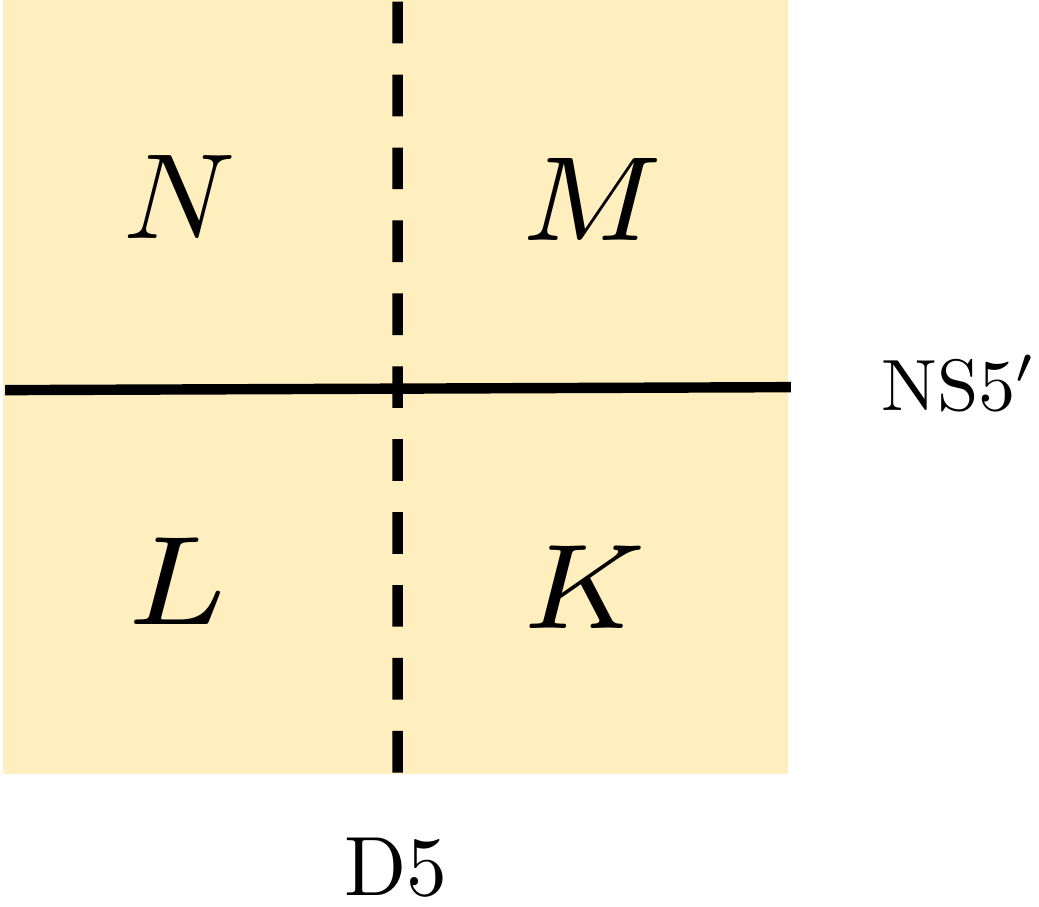}
\caption{The ${N \, M \choose L \, K}$ NS5$'$-D5 junction.}
\label{figndNMLK}
\end{center}
\end{figure}

The brane configuration is depicted in Figure \ref{figndNMLK}. 
The quarter-index (\ref{ndNMLK}) will coincide with the quarter-index for 
$\left(\begin{smallmatrix}
L&N\\ K&M\\\end{smallmatrix}\right)$ NS5$'$-D5 junction 
which will be also given by the formula (\ref{ndNMLK}).


\section{NS5-NS5$'$ and D5-D5$'$ junctions}
\label{sec_nnddjunction}
Both for NS5-NS5$'$ and D5-D5$'$ junctions we require a balancing condition: the sum of the numbers of D3-branes 
in opposite corners must be the same. 

In the D5-D5$'$ junction case, this guarantees that we can 
implement the required Nahm poles in the simplest possible way: a Nahm pole of fixed rank in the $X$ fields 
along the D5 interface and a Nahm pole of fixed rank in the $Y$ fields along the D5$'$ interface, with the two Nahm poles sitting in different commuting blocks of the largest gauge group. 

In the NS5-NS5$'$ junction case, this guarantees a cancellation of non-Abelian anomalies. Abelian anomalies 
can be cancelled by an extra ``cross-determinant'' Fermi multiplet 
which couples to the Abelian part of each of the gauge symmetries in the quadrants 
at the NS5-NS5$'$ junction as discussed in \cite{Hanany:2018hlz}. 
The following computation of quarter-indices provides us with the evidences of 
the existence of the cross-determinant Fermi multiplet 
and proposes new dualities between NS5-NS5$'$ and D5-D5$'$ junctions. 

The balancing condition is motivated in string theory by 
requiring that finite segments of fivebranes defining the common boundaries or/and corners in the effective world-volume theory of the D3-branes have the same linking numbers. 
This condition is translated into the gauge anomaly cancelation in the effective gauge theory \cite{Hanany:2018hlz}. 

Other junctions which do not satisfy the balancing condition may likely be defined in gauge theory, but 
we do not have a string theory construction which can motivate S-duality conjectures.

\subsection{${N \, N \choose 0 \, 0}$ and ${0 \, N \choose 0 \, N}$}
\label{sec_2dnnNN00}
Let us consider the $\left(\begin{smallmatrix}1&1\\0&0\\\end{smallmatrix}\right)$ NS5-NS5$'$ junction. 
It consists of two 4d $\mathcal{N}=4$ $U(1)$ gauge theories at corners with 
the boundary conditions $\mathcal{N}$ and $\mathcal{N}'$. 
It involves a 3d $\mathcal{N}=4$ bi-fundamental hypermultiplet 
obeying the Neumann boundary conditions ${\mathcal N}'$ together with the cross-determinant Fermi multiplet \cite{Hanany:2018hlz}. 

\subsubsection{${1 \, 1 \choose 0 \, 0}$ and ${0 \, 1 \choose 0 \, 1}$}
The quarter-index for the $\left(\begin{smallmatrix}
1&1\\0&0\\\end{smallmatrix}\right)$ NS5-NS5$'$ junction is 
\begin{align}
\label{nn1100}
\mathbb{IV}_{\mathcal{N}\mathcal{N}'}^{\left(\begin{smallmatrix}
1&1\\0&0\\\end{smallmatrix}\right)}
&=
\underbrace{
(q)_{\infty}\oint \frac{ds_{1}}{2\pi is_{1}}
}_{\mathbb{IV}_{\mathcal{N}\mathcal{N}'}^{\textrm{4d $U(1)$}}}
\underbrace{
(q)_{\infty}\oint \frac{ds_{2}}{2\pi is_{2}}
}_{\mathbb{IV}_{\mathcal{N}\mathcal{N}'}^{\textrm{4d $U(1)$}}}
\underbrace{
\frac{
\left(q^{\frac12}\frac{s_{1}}{s_{2}} x;q\right)_{\infty}
\left(q^{\frac12}\frac{s_{2}}{s_{1}} x^{-1};q\right)_{\infty}
}
{
\left(q^{\frac14}t\frac{s_{1}}{s_{2}} ;q\right)_{\infty}
\left(q^{\frac14}t\frac{s_{2}}{s_{1}} ;q\right)_{\infty}}}
_{\mathbb{II}_{N}^{\textrm{3d HM}}\left(\frac{s_{1}}{s_{2}}\right)\cdot 
F\left(q^{\frac12}\frac{s_{1}}{s_{2}}x\right)}. 
\end{align}

The S-dual is the $\left(\begin{smallmatrix}0&1\\0&1\\\end{smallmatrix}\right)$ D5-D5$'$ junction. 
In the absence of D5$'$-brane, 
the D5-brane requires the Dirichlet boundary condition $\mathcal{D}$ for 4d $\mathcal{N}=4$ $U(1)$ gauge theory. 
The D5$'$-brane will introduce a 3d $\mathcal{N}=4$ twisted hypermultiplet, which we 
expect to satisfy the Dirichlet boundary condition $D$ at the D5-brane. 

Therefore the quarter-index for 
the $\left(\begin{smallmatrix}0&1\\0&1\\\end{smallmatrix}\right)$ D5-D5$'$ junction takes the form 
\begin{align}
\label{dd0101}
\mathbb{IV}_{\mathcal{D}\mathcal{D}'}^{\left(\begin{smallmatrix}
0&1\\0&1\\\end{smallmatrix}\right)}
&=
\underbrace{
\frac{(q)_{\infty}}
{(q^{\frac12}t^2;q)_{\infty}}
}_{\mathbb{II}_{\mathcal{D}}^{\textrm{4d $U(1)$}}}
\underbrace{
(q^{\frac34}tx;q)_{\infty}
(q^{\frac34}tx^{-1};q)_{\infty}
}_{\mathbb{II}_{D}^{\textrm{3d tHM}}(x)}. 
\end{align}
In fact, the quarter-indices (\ref{nn1100}) and (\ref{dd0101}) coincide.

Let us draw a lesson by expanding these indices in several ways. 
We can expand the quarter-index (\ref{nn1100}) as a sum over residues at poles $\frac{s_{1}}{s_{2}}=q^{\frac14+m}t$ 
for the bi-fundamental hyper 
\begin{align}
\label{nn1100_res}
\mathbb{IV}_{\mathcal{N}\mathcal{N}'}^{\left(\begin{smallmatrix}
1&1\\0&0\\\end{smallmatrix}\right)}
&=
\underbrace{(q^{\frac34}tx;q)_{\infty}(q^{\frac14}t^{-1}x^{-1};q)_{\infty}}
_{F(q^{\frac34}tx)}
\sum_{m=0}^{\infty}
\frac{(q^{1+m};q)_{\infty}}
{(q^{\frac12+m}t^2;q)_{\infty}}q^{\frac{m}{4}} t^{-m} x^{-m}. 
\end{align}
The sum starts from the product of the half-index $\mathbb{II}_{\mathcal{N}'}^{\textrm{4d $U(1)$}}$ of the Neumann boundary condition $\mathcal{N}'$ 
for the 4d $\mathcal{N}=4$ $U(1)$ gauge theory and the Fermi index $F(q^{\frac34}tx)$. 
It is shown from the $q$-binomial theorem (\ref{q_binomial}) that the expression (\ref{nn1100_res}) is equivalent to the quarter-index (\ref{dd0101}) for 
the $\left(\begin{smallmatrix}0&1\\0&1\\\end{smallmatrix}\right)$ D5-D5$'$ junction. 

Alternatively, the quarter-index (\ref{nn1100}) can be expanded as 
\begin{align}
\label{nn1100_sum}
\mathbb{IV}_{\mathcal{N}\mathcal{N}'}^{\left(\begin{smallmatrix}
1&1\\0&0\\\end{smallmatrix}\right)}
&=
(q^{\frac14}t^{-1}x;q)_{\infty}
(q^{\frac14}t^{-1}x^{-1};q)_{\infty}
\sum_{n=0}^{\infty}
\frac{(q^{1+n};q)_{\infty}^2}
{(q^{\frac14+n}t^{-1}x;q)_{\infty} (q^{\frac14+n}t^{-1}x^{-1};q)_{\infty}}q^{\frac{n}{2}} t^{2n}.
\end{align}
The first terms in the sum can be regarded as the square of the quarter-indices $\mathbb{IV}_{\mathcal{N}\mathcal{N}'}^{\textrm{4d $U(1)$}}$. 

On the other hand, we have an expansion of the quarter-index (\ref{dd0101}) for 
the $\left(\begin{smallmatrix}0&1\\0&1\\\end{smallmatrix}\right)$ D5-D5$'$ junction:
\begin{align}
\label{dd0101_sum}
\mathbb{IV}_{\mathcal{D}\mathcal{D}'}^{\left(\begin{smallmatrix}
0&1\\0&1\\\end{smallmatrix}\right)}
&=
\frac{(q)_{\infty}}
{(q^{\frac12} t^2;q)_{\infty}}
\frac{1}{(q)_{\infty}^2}
\sum_{n=0}^{\infty}\sum_{k=0}^{n}
(q^{1+k};q)_{\infty}
(q^{1+n-k};q)_{\infty} 
(-1)^{n}
x^{2k-n}
t^{n}
q^{\frac{n^2}{2}+\frac{n}{4}+k(k-n)}. 
\end{align}
In contrast to the expansion (\ref{nn1100_res}), 
the first term in the sum is just the half-index $\mathbb{II}_{\mathcal{D}}^{\textrm{4d $U(1)$}}$ of the Dirichlet boundary condition $\mathcal{D}$ 
for 4d $\mathcal{N}=4$ $U(1)$ gauge theory. The physical interpretation of this expansion is not obvious.

\subsubsection{${2 \, 2 \choose 0 \, 0}$ and ${0 \, 2 \choose 0 \, 2}$}

As a non-Abelian example, 
consider the $\left(\begin{smallmatrix}2&2\\0&0\\\end{smallmatrix}\right)$ NS5-NS5$'$ junction. 
There are two 4d $\mathcal{N}=4$ $U(2)$ gauge theories, each defined on a quadrant, with 
boundary conditions $\mathcal{N}$ and $\mathcal{N}'$. 
In addition, there are 3d $\mathcal{N}=4$ bi-fundamental hypermultiplets 
obeying the Neumann boundary condition $N'$ and the cross-determinant Fermi multiplet 
that cancels the Abelian part of the boundary gauge anomaly. 

Then the quarter-index for the $\left(\begin{smallmatrix}
2&2\\0&0\\\end{smallmatrix}\right)$ NS5-NS5$'$ junction is given by
\begin{align}
\label{nn2200}
\mathbb{IV}_{\mathcal{N}\mathcal{N}'}^{\left(\begin{smallmatrix}
2&2\\0&0\\\end{smallmatrix}\right)}
&=
\underbrace{
\frac{1}{2} (q)_{\infty}^2 
\oint \prod_{i=1}^{2}\frac{ds_{i}}{2\pi is_{i}}
\prod_{i\neq j}\left(\frac{s_{i}}{s_{j}};q\right)_{\infty}
}_{\mathbb{IV}_{\mathcal{N}\mathcal{N}'}^{\textrm{4d $U(2)$}}}
\underbrace{
\frac{1}{2} (q)_{\infty}^2 
\oint \prod_{i=3}^{4}\frac{ds_{i}}{2\pi is_{i}}
\prod_{i\neq j}\left(\frac{s_{i}}{s_{j}};q\right)_{\infty}
}_{\mathbb{IV}_{\mathcal{N}\mathcal{N}'}^{\textrm{4d $U(2)$}}}
\nonumber\\
&\times 
\prod_{i=1}^{2}\prod_{k=3}^{4}
\underbrace{
\frac{
(q^{\frac12}\frac{s_{1}s_{2}}{s_{3}s_{4}}x;q)_{\infty}
(q^{\frac12}\frac{s_{3}s_{4}}{s_{1}s_{2}}x^{-1};q)_{\infty}
}
{
(q^{\frac14} t \frac{s_{i}}{s_{k}};q)_{\infty}
(q^{\frac14} t \frac{s_{k}}{s_{i}};q)_{\infty}
}
}_
{\mathbb{II}_{N}^{\textrm{3d HM}}\left(\frac{s_{i}}{s_{k}}\right)
\cdot F\left(q^{\frac12} \frac{\prod_{j=1}^{2}s_{j}}{\prod_{l=3}^{4}s_{l}} x\right)
}. 
\end{align}

For the S-dual $\left(\begin{smallmatrix}0&2\\0&2\\\end{smallmatrix}\right)$ D5-D5$'$ junction, 
the D5-brane should introduce the Nahm pole boundary condition 
specified by a homomorphism $\rho:$ $\mathfrak{su}(2)$ $\rightarrow$ $\mathfrak{u}(2)$. 
The additional D5$'$-brane will introduce a 3d $\mathcal{N}=4$ twisted hypermultiplet, 
which should satisfy the boundary condition associated to the Nahm pole boundary condition, 
which will be different from the Dirichlet boundary condition $D$. 

We find that 
the quarter-index (\ref{nn2200}) for the $\left(\begin{smallmatrix}
2&2\\0&0\\\end{smallmatrix}\right)$ NS5-NS5$'$ junction coincides with the following quarter-index for 
$\left(\begin{smallmatrix}0&2\\0&2\\\end{smallmatrix}\right)$ D5-D5$'$ junction:
\begin{align}
\label{dd0202}
\mathbb{IV}_{\mathcal{D}\mathcal{D}'}^{\left(\begin{smallmatrix}
0&2\\0&2\\\end{smallmatrix}\right)}
&=
\underbrace{
\frac{(q)_{\infty} (q^{\frac32}t^2;q)_{\infty}}
{(q^{\frac12}t^2;q)_{\infty} (q t^4;q)_{\infty}}
}_{\mathbb{II}_{\textrm{Nahm}}^{\textrm{4d $U(2)$}}} 
(q t^2x;q)_{\infty}
(q t^2x^{-1};q)_{\infty}. 
\end{align}
We see that 
the contributions from the 3d $\mathcal{N}=4$ fundamental twisted hypermultiplet 
are different from the Dirichlet boundary condition $D$. They can be obtained by the by-now standard Higgsing procedure:
specializing Dirichlet fugacities to describe the RG flow to a Nahm pole would give 
$(q^{\frac12} x;q)_{\infty}(q^{\frac12}x^{-1};q)_{\infty}(q t^2x;q)_{\infty}(q t^2x^{-1};q)_{\infty}$,
but we strip off the free Fermi contribution.

We can expand the quarter-index (\ref{dd0202}) as
\begin{align}
\label{dd0202_sum}
\mathbb{IV}_{\mathcal{D}\mathcal{D}'}^{\left(\begin{smallmatrix}
0&2\\0&2\\\end{smallmatrix}\right)}
&=
\frac{(q)_{\infty} (q^{\frac32} t^2;q)_{\infty}}
{(q^{\frac12} t^2;q)_{\infty} (qt^4;q)_{\infty}}
\frac{1}{(q)_{\infty}^2}
\sum_{n=0}^{\infty}\sum_{k=0}^{n}
(q^{1+k};q)_{\infty}
(q^{1+n-k};q)_{\infty} 
(-1)^{n}
x^{2k-n}
t^{2n}
q^{\frac{n^2+n}{2}+k(k-n)}
\end{align}
which begins with the half-index $\mathbb{II}_{\textrm{Nahm}}^{\textrm{4d $U(2)$}}$ of 
the Nahm pole boundary condition for 4d $\mathcal{N}=4$ $U(2)$ SYM theory.

\subsubsection{${3 \, 3 \choose 0 \, 0}$ and ${0 \, 3 \choose 0 \, 3}$}

Let us examine the $\left(\begin{smallmatrix}
3&3\\0&0\\\end{smallmatrix}\right)$ NS5-NS5$'$ junction. 
This has two 4d $\mathcal{N}=4$ $U(3)$ gauge theories, each defined on a quadrant, with 
boundary conditions $\mathcal{N}$ and $\mathcal{N}'$. 
The matter content consists of the 3d $\mathcal{N}=4$ bi-fundamental hypermultiplets 
obeying the Neumann boundary condition $N'$ and the cross-determinant Fermi multiplet 
which cancels the Abelian part of the boundary gauge anomaly. 

We obtain the quarter-index for $\left(\begin{smallmatrix}
3&3\\0&0\\\end{smallmatrix}\right)$ NS5-NS5$'$ junction
\begin{align}
\label{nn3300}
\mathbb{IV}_{\mathcal{N}\mathcal{N}'}^{\left(\begin{smallmatrix}
3&3\\0&0\\\end{smallmatrix}\right)}
&=
\underbrace{
\frac{1}{3!} (q)_{\infty}^3 
\oint \prod_{i=1}^{3}\frac{ds_{i}}{2\pi is_{i}}
\prod_{i\neq j}\left(\frac{s_{i}}{s_{j}};q\right)_{\infty}
}_{\mathbb{IV}_{\mathcal{N}\mathcal{N}'}^{\textrm{4d $U(3)$}}}
\underbrace{
\frac{1}{3!} (q)_{\infty}^3 
\oint \prod_{i=4}^{6}\frac{ds_{i}}{2\pi is_{i}}
\prod_{i\neq j}\left(\frac{s_{i}}{s_{j}};q\right)_{\infty}
}_{\mathbb{IV}_{\mathcal{N}\mathcal{N}'}^{\textrm{4d $U(3)$}}}
\nonumber\\
&\times 
\prod_{i=1}^{3}\prod_{k=4}^{6}
\underbrace{
\frac{
\left(q^{\frac12}\frac{\prod_{j=1}^{3}s_{j}}{\prod_{l=4}^{6}s_{l}}x;q\right)_{\infty}
\left(q^{\frac12}\frac{\prod_{l=4}^{6}s_{l}}{\prod_{j=1}^{3}s_{j}}x^{-1};q\right)_{\infty}
}
{
\left(q^{\frac14} t \frac{s_{i}}{s_{k}};q\right)_{\infty}
\left(q^{\frac14} t \frac{s_{k}}{s_{i}};q\right)_{\infty}
}
}_
{\mathbb{II}_{N}^{\textrm{3d HM}}\left(\frac{s_{i}}{s_{k}}\right)
\cdot F\left(q^{\frac12} \frac{\prod_{j=1}^{3}s_{j}}{\prod_{l=4}^{6}s_{l}} x\right)
}. 
\end{align}

The S-dual $\left(\begin{smallmatrix}0&3\\0&3\\\end{smallmatrix}\right)$ D5-D5$'$ junction 
involves the Nahm pole boundary condition 
with an embedding $\rho:$ $\mathfrak{su}(2)$ $\rightarrow$ $\mathfrak{u}(3)$. 
The 3d $\mathcal{N}=4$ twisted hypermultiplet arising from D3-D5$'$ strings 
should satisfy the boundary condition associated to this Nahm pole boundary condition.

The quarter-index (\ref{nn3300}) for the $\left(\begin{smallmatrix} 
3&3\\0&0\\\end{smallmatrix}\right)$ NS5-NS5$'$ junction turns out to be 
coincide with the following simple expression:
\begin{align}
\label{dd0303}
\mathbb{IV}_{\mathcal{D}\mathcal{D}'}^{\left(\begin{smallmatrix}
0&3\\0&3\\\end{smallmatrix}\right)}
&=
\underbrace{
\frac{
(q)_{\infty}
(q^{\frac32}t^2;q)_{\infty}
(q^2 t^4;q)_{\infty}
}
{
(q^{\frac12}t^2;q)_{\infty}
(q t^4;q)_{\infty}
(q^{\frac32}t^6;q)_{\infty}
}
}_{\mathbb{II}_{\textrm{Nahm}}^{\textrm{4d $U(3)$}}} 
(q^{\frac54}t^3 x;q)_{\infty} 
(q^{\frac54}t^3 x^{-1};q)_{\infty}. 
\end{align}
We can view the factor 
$(q^{\frac54}t^3 x;q)_{\infty}$ $(q^{\frac54}t^3 x^{-1};q)_{\infty}$ 
as the contributions from the 3d $\mathcal{N}=4$ twisted hypermultiplet 
obeying the boundary condition associated to the Nahm pole boundary condition with 
an embedding $\rho:$ $\mathfrak{su}(2)$ $\rightarrow$ $\mathfrak{u}(3)$. 

They can be obtained by the Higgsing procedure:
specializing Dirichlet fugacities to describe the RG flow to a Nahm pole would give 
$(q^{\frac14}t^{-1} x;q)_{\infty}$ $(q^{\frac14}t^{-1} x^{-1};q)_{\infty}(q^{\frac34}t x;q)_{\infty}$ $(q^{\frac34}t x^{-1};q)_{\infty}(q^{\frac54}t^3 x;q)_{\infty}$ $(q^{\frac54}t^3 x^{-1};q)_{\infty}$,
but we strip off the free Fermi contribution. 

We can expand the quarter-index as a sum over residues:
\begin{align}
\label{dd0303_sum}
\mathbb{IV}_{\mathcal{D}\mathcal{D}'}^{\left(\begin{smallmatrix}
0&3\\0&3\\\end{smallmatrix}\right)}
&=
\frac{(q)_{\infty} (q^{\frac32} t^2;q)_{\infty} (q^2 t^4;q)_{\infty}}
{(q^{\frac12} t^2;q)_{\infty} (qt^4;q)_{\infty} (q^{\frac32} t^6;q)_{\infty}}
\nonumber\\
&\times 
\frac{1}{(q)_{\infty}^2}
\sum_{n=0}^{\infty}\sum_{k=0}^{n}
(q^{1+k};q)_{\infty}
(q^{1+n-k};q)_{\infty} 
(-1)^{n}x^{2k-n}
t^{3n}
q^{\frac{n^2}{2}+\frac{3n}{4}+k(k-n)},
\end{align}
which begins with the half-index $\mathbb{II}_{\textrm{Nahm}}^{\textrm{4d $U(3)$}}$ of the Nahm pole boundary condition 
for 4d $\mathcal{N}=4$ $U(3)$ gauge theory.

\subsubsection{${N \, N \choose 0 \, 0}$ and ${0 \, N \choose 0 \, N}$}
We would like to propose the generalization of 
the duality between the $\left(\begin{smallmatrix}
N&N\\0&0\\\end{smallmatrix}\right)$ NS5-NS5$'$ junction 
and $\left(\begin{smallmatrix}
0&N\\0&N\\\end{smallmatrix}\right)$ D5-D5$'$ junction. 

For the $\left(\begin{smallmatrix}
N&N\\0&0\\\end{smallmatrix}\right)$ NS5-NS5$'$ junction, 
there are two 4d $\mathcal{N}=4$ $U(N)$ gauge theories, each defined in a quadrant, with 
boundary conditions $\mathcal{N}$ and $\mathcal{N}'$. 
There are the 3d $\mathcal{N}=4$ bi-fundamental hypermultiplets 
with the Neumann boundary condition $N'$ and the cross-determinant Fermi multiplet. 

The quarter-index for $\left(\begin{smallmatrix}
N&N\\0&0\\\end{smallmatrix}\right)$ NS5-NS5$'$ junction takes the form
\begin{align}
\label{nnNN00}
\mathbb{IV}_{\mathcal{N}\mathcal{N}'}^{\left(\begin{smallmatrix}
N&N\\0&0\\\end{smallmatrix}\right)}
&=
\underbrace{
\frac{1}{N!}(q)_{\infty}^N 
\oint \prod_{i=1}^{N}
\frac{ds_{i}}{2\pi is_{i}}
\prod_{i\neq j}\left(\frac{s_{i}}{s_{j}};q\right)_{\infty}
}_{\mathbb{IV}_{\mathcal{N}\mathcal{N}'}^{\textrm{4d $U(N)$}}}
\underbrace{
\frac{1}{N!}(q)_{\infty}^N 
\oint \prod_{i=1}^{N}
\frac{ds_{i}}{2\pi is_{i}}
\prod_{i\neq j}\left(\frac{s_{i}}{s_{j}};q\right)_{\infty}
}_{\mathbb{IV}_{\mathcal{N}\mathcal{N}'}^{\textrm{4d $U(N)$}}}
\nonumber\\
&\times 
\prod_{i=1}^{N}\prod_{k=N+1}^{2N}
\underbrace{
\frac{
\left(
q^{\frac12}\frac{\prod_{j=1}^{N}s_{j}}{\prod_{l=N+1}^{2N}s_{l}}x;q
\right)_{\infty}
\left(
q^{\frac12}\frac{\prod_{l=N+1}^{2N}s_{l}}{\prod_{j=1}^{N}s_{j}}x^{-1};q
\right)_{\infty}
}
{
\left(q^{\frac14}t \frac{s_{i}}{s_{k}};q\right)_{\infty}
\left(q^{\frac14}t \frac{s_{k}}{s_{i}};q\right)_{\infty}
}
}_{\mathbb{II}_{N}^{\textrm{3d HM}}\left(\frac{s_{i}}{s_{k}}\right)
\cdot F\left(q^{\frac12}\frac{\prod_{j=1}^{N}s_{j}}{\prod_{l=N+1}^{2N}s_{l}}x\right)}. 
\end{align}

\begin{figure}
\begin{center}
\includegraphics[width=10cm]{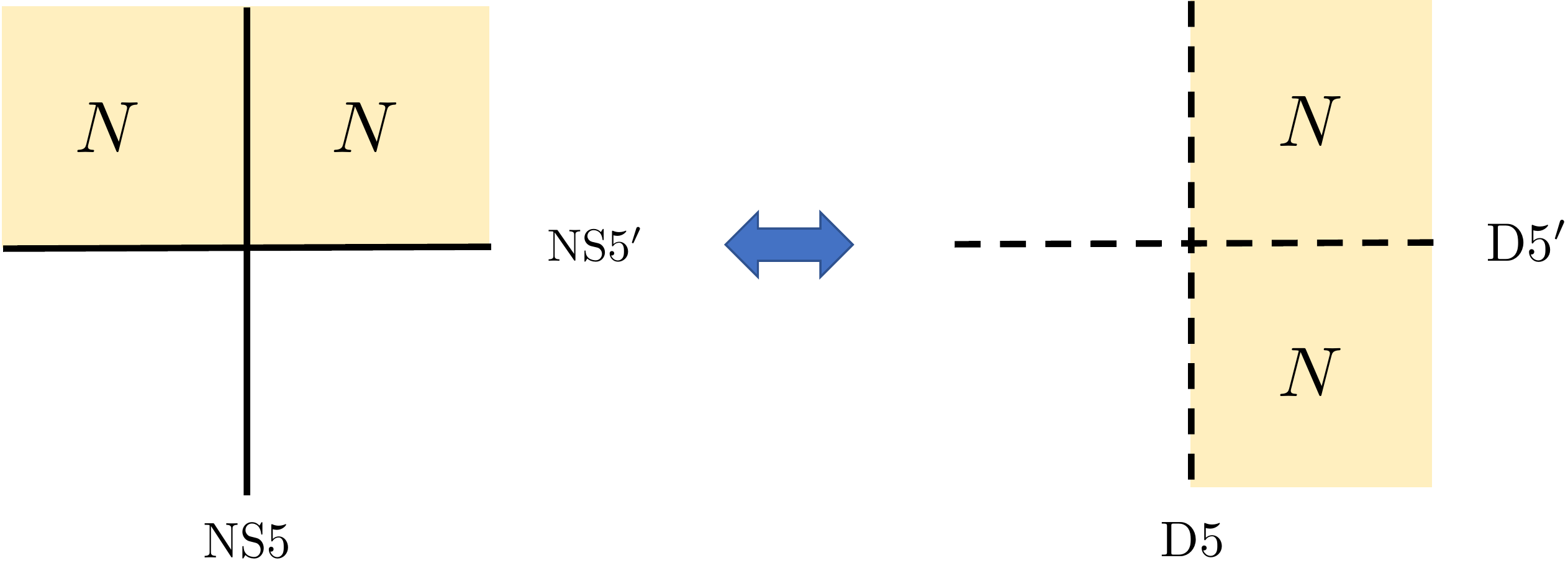}
\caption{${N \, N \choose 0 \, 0}$ NS5-NS5$'$-junction and ${0 \, N \choose 0 \, N}$ D5-D5$'$ junction.}
\label{fignnddjunction1}
\end{center}
\end{figure}

The S-dual $\left(\begin{smallmatrix}0&N\\0&N\\\end{smallmatrix}\right)$ D5-D5$'$ junction 
is characterized by the Nahm pole boundary condition 
with an embedding $\rho:$ $\mathfrak{su}(2)$ $\rightarrow$ $\mathfrak{u}(N)$.  
We expect that 
the quarter-index for $\left(\begin{smallmatrix}
N&N\\0&0\\\end{smallmatrix}\right)$ NS5-NS5$'$ junction is equal to 
the following quarter-index for $\left(\begin{smallmatrix}0&N\\0&N\\\end{smallmatrix}\right)$ D5-D5$'$ junction:
\begin{align}
\label{dd0N0N}
\mathbb{IV}_{\mathcal{D}\mathcal{D}'}^{\left(\begin{smallmatrix}
0&N\\0&N\\\end{smallmatrix}\right)}
&=
\underbrace{
\prod_{k=1}^{N}
\frac{
(q^{\frac{k+1}{2}}t^{2(k-1)};q)_{\infty}
}
{
(q^{\frac{k}{2}}t^{2k};q)_{\infty}
}
}_{\mathbb{II}_{\textrm{Nahm}}^{\textrm{4d $U(N)$}}}
\left(q^{\frac34+\frac{N-1}{4}}t^{1+(N-1)}x;q\right)_{\infty}
\left(q^{\frac34+\frac{N-1}{4}}t^{1+(N-1)}x^{-1};q\right)_{\infty}. 
\end{align}
The brane configuration is shown in Figure \ref{fignnddjunction1}

We can expand the quarter-index as a sum over residues:
\begin{align}
\label{dd0N0N_sum}
\mathbb{IV}_{\mathcal{N}\mathcal{N}'}^{\left(\begin{smallmatrix}
N&N\\0&0\\\end{smallmatrix}\right)}
&=
\mathbb{IV}_{\mathcal{D}\mathcal{D}'}^{\left(\begin{smallmatrix}
0&N\\0&N\\\end{smallmatrix}\right)}
\nonumber\\
&=
\prod_{k=1}^{N}
\frac{
\left(
q^{\frac{k+1}{2}}t^{2(k-1)};q
\right)_{\infty}
}
{
\left(
q^{\frac{k}{2}}t^{2k};q
\right)_{\infty}
}
\frac{1}{(q)_{\infty}^2}
\sum_{n=0}\sum_{k=0}^{n}
(q^{1+k};q)_{\infty}
(q^{1+n-k};q)_{\infty} 
(-1)^{n}
x^{2k-n}
t^{Nn}
q^{\frac{n^2}{2}+\frac{Nn}{4}+k(k-n)}. 
\end{align}
The first term in the series expression (\ref{dd0N0N}) appears as the half-index $\mathbb{II}_{\textrm{Nahm}}^{\textrm{4d $U(N)$}}$ of the Nahm pole boundary condition 
for the 4d $\mathcal{N}=4$ $U(N)$ SYM theory.

%
%
%
%
%

\subsection{${N \, N \choose N \, N}$}
\label{sec_2dnnNNNN}
\subsubsection{${1 \, 1 \choose 1 \, 1}$ and ${1 \, 1 \choose 1 \, 1}$}
\begin{figure}
\begin{center}
\includegraphics[width=10cm]{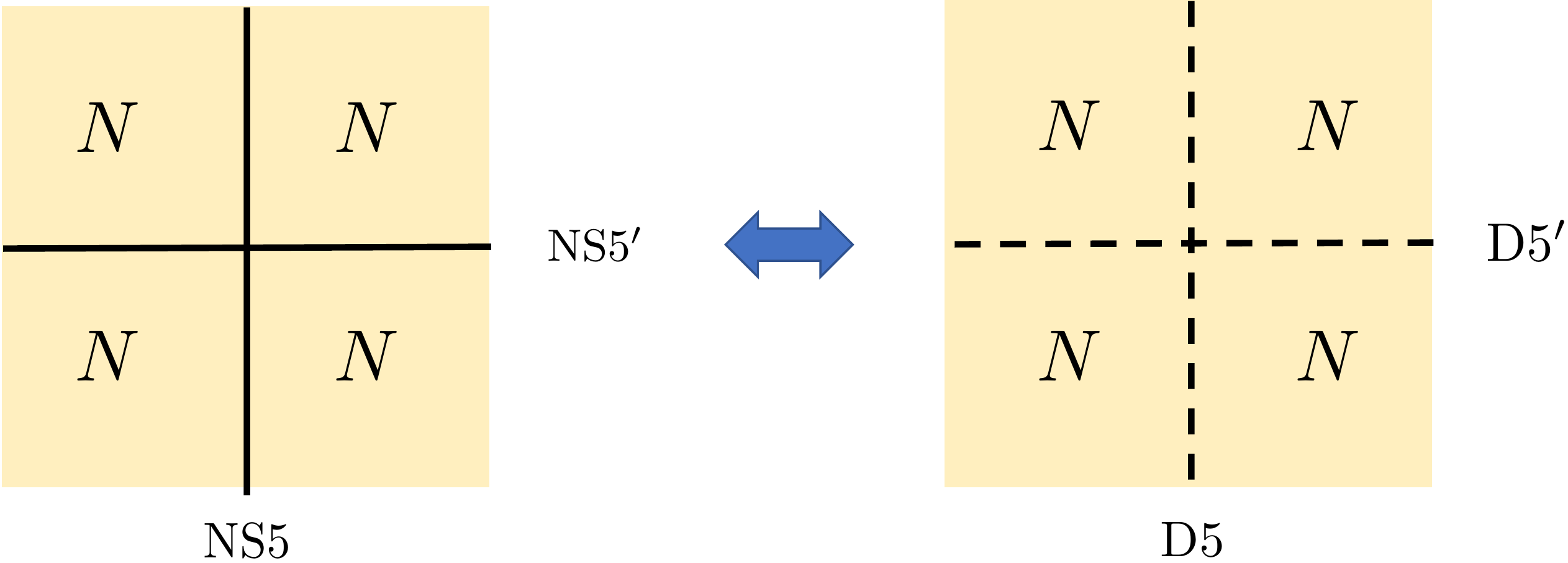}
\caption{
The ${N \, N \choose N \, N}$ NS5-NS5$'$ junction and ${N \, N \choose N \, N}$ D5-D5$'$ junction.
}
\label{fignnddNNNN}
\end{center}
\end{figure}

Let us consider the $\left(\begin{smallmatrix}
1&1\\1&1\\\end{smallmatrix}\right)$ NS5-NS5$'$ junction. 
There are four
4d $\mathcal{N}=4$ $U(1)$ gauge theories, each defined in a quadrant, with 
boundary conditions $\mathcal{N}$ and $\mathcal{N}'$. 
We denote by $U(1)_{1}$, $U(1)_{2}$, $U(1)_{3}$ and $U(1)_{4}$ 
the upper left, lower left, upper right and lower right quadrants respectively. 
The brane configuration is depicted in Figure \ref{fignnddNNNN}

The matter content can be read off from the brane box analysis in \cite{Hanany:2018hlz}. 
It involves  two 3d $\mathcal{N}=4$ bi-fundamental hypers $\mathbb{H}^{(1)}$, $\mathbb{H}^{(2)}$, 
two 3d $\mathcal{N}=4$ bi-fundamental twisted hypers $\mathbb{T}^{(1)}$, $\mathbb{T}^{(2)}$, 
two bi-fundamental Fermi multiplets $\Gamma^{(1)}$, $\Gamma^{(2)}$ 
and cross-determinant Fermi multiplet $\Xi$. 
They have charges as follows:
\begin{align}
\label{1111_ch}
\begin{array}{c|c|c|c|c}
&U(1)_{1}&U(1)_{2}&U(1)_{3}&U(1)_{4} \\ \hline 
\textrm{3d $\mathcal{N}=4$ hyper $\mathbb{H}^{(1)}$}&+&0&-&0 \\
\textrm{3d $\mathcal{N}=4$ hyper $\mathbb{H}^{(2)}$}&+&0&0&- \\ \hline
\textrm{3d $\mathcal{N}=4$ twisted hyper $\mathbb{T}^{(1)}$}&+&-&0&0 \\
\textrm{3d $\mathcal{N}=4$ twisted hyper $\mathbb{T}^{(2)}$}&0&0&+&- \\  \hline
\textrm{2d Fermi multiplet $\Gamma^{(1)}$}&+&0&0&- \\
\textrm{2d Fermi multiplet $\Gamma^{(2)}$}&0&-&+&0 \\ \hline
\textrm{2d Fermi multiplet $\Xi$}&+&-&-&+ \\
\end{array}
\end{align}
According to the NS5$'$- and NS5-branes, 
the 3d $\mathcal{N}=4$ hyper and twisted hypermultiplets 
receive the Neumann boundary conditions. 

For consistency of the Neumann boundary condition for a gauge fields, 
the boundary gauge anomaly must be canceled out. 
Let ${\bf s}_{i}$ be the field strength of the $U(1)_{i}$ gauge field with $i=1,\cdots,4$. 
The boundary gauge anomaly polynomial is computed as 
\begin{align}
\label{NN1111_AN}
\mathcal{I}_{\mathcal{N}\mathcal{N}'}^{
\left(
\begin{smallmatrix}
1&1\\
1&1\\
\end{smallmatrix}
\right)}
&=
\underbrace{
-\frac12 ({\bf s}_{1}-{\bf s}_{3})^{2}
-\frac12 (-{\bf s}_{1}+{\bf s}_{3})^{2}
}_{\textrm{$N'$ of $\mathbb{H}^{(1)}$}}
\underbrace{
-\frac12 ({\bf s}_{2}-{\bf s}_{4})^{2}
-\frac12 (-{\bf s}_{2}+{\bf s}_{4})^{2}
}_{\textrm{$N'$ of $\mathbb{H}^{(2)}$}}
\nonumber\\
&
\underbrace{
-\frac12 ({\bf s}_{1}-{\bf s}_{2})^{2}
-\frac12 (-{\bf s}_{1}+{\bf s}_{2})^{2}}
_{\textrm{$N$ of $\mathbb{T}^{(1)}$}}
\underbrace{
-\frac12 ({\bf s}_{3}-{\bf s}_{4})^{2}
-\frac12 (-{\bf s}_{3}+{\bf s}_{4})^{2}
}_{\textrm{$N$ of $\mathbb{T}^{(2)}$}}
\nonumber\\
&
+\underbrace{({\bf s}_{1}-{\bf s}_{4})^{2}}_{\Gamma^{(1)}}
+
\underbrace{
({\bf s}_{3}-{\bf s}_{2})^{2}}_{\Gamma^{(2)}}
+
\underbrace{
({\bf s}_{1}+{\bf s}_{4}-{\bf s}_{2}-{\bf s}_{3})^{2}
}_{\Xi}
\nonumber\\
&=0. 
\end{align}

The quarter-index for the 
$\left(\begin{smallmatrix}
1&1\\1&1\\\end{smallmatrix}\right)$ NS5-NS5$'$ junction is given by
\begin{align}
\label{NN1111_index}
&\mathbb{IV}_{\mathcal{N}\mathcal{N}'}^{
\left(
\begin{smallmatrix}
1&1\\
1&1\\
\end{smallmatrix}
\right)}
=
\underbrace{
(q)_{\infty}\oint \frac{ds_{1}}{2\pi is_{1}}
}_{\mathbb{IV}_{\mathcal{N}\mathcal{N}'}^{\textrm{4d $U(1)$}}}
\underbrace{
(q)_{\infty}\oint \frac{ds_{2}}{2\pi is_{2}}
}_{\mathbb{IV}_{\mathcal{N}\mathcal{N}'}^{\textrm{4d $U(1)$}}}
\underbrace{
(q)_{\infty}\oint \frac{ds_{3}}{2\pi is_{3}}
}_{\mathbb{IV}_{\mathcal{N}\mathcal{N}'}^{\textrm{4d $U(1)$}}}
\underbrace{
(q)_{\infty}\oint \frac{ds_{4}}{2\pi is_{4}}
}_{\mathbb{IV}_{\mathcal{N}\mathcal{N}'}^{\textrm{4d $U(1)$}}}
\nonumber\\
&\times 
\underbrace{
\frac{
\left(q^{\frac12} s_{1}^{\pm}s_{4}^{\mp} ;q\right)_{\infty}
\cdot 
\left(q^{\frac12} s_{2}^{\pm}s_{3}^{\mp};q\right)_{\infty}
\cdot 
\left(q^{\frac12} s_{1}^{\pm}s_{4}^{\pm}s_{2}^{\mp}s_{3}^{\mp}x^{\pm 2};q\right)_{\infty}
}
{
\left(q^{\frac14}t s_{1}^{\pm}s_{3}^{\mp};q\right)_{\infty}
\cdot 
\left(q^{\frac14}t s_{2}^{\pm}s_{4}^{\mp};q\right)_{\infty}
\cdot 
\left(q^{\frac14}t^{-1} s_{1}^{\pm}s_{2}^{\mp};q\right)_{\infty}
\cdot 
\left(q^{\frac14}t^{-1} s_{3}^{\pm}s_{4}^{\mp};q\right)_{\infty}
}
}_
{
\mathbb{II}_{N}^{\textrm{3d HM}}\left(\frac{s_{1}}{s_{3}}\right)\cdot 
\mathbb{II}_{N}^{\textrm{3d HM}}\left(\frac{s_{2}}{s_{4}}\right)\cdot 
\mathbb{II}_{N}^{\textrm{3d tHM}}\left(\frac{s_{1}}{s_{2}}\right)\cdot 
\mathbb{II}_{N}^{\textrm{3d tHM}}\left(\frac{s_{3}}{s_{4}}\right)\cdot 
F\left(q^{\frac12}\frac{s_{1}}{s_{4}}\right)\cdot 
F\left(q^{\frac12}\frac{s_{2}}{s_{3}}\right)\cdot 
F\left(q^{\frac12} \frac{s_{1}s_{4}}{s_{2}s_{3}}x^2\right)
}
\end{align}
where the cross-determinant Fermi multiplet has flvebrane charges associated to the fugacity $x$.

The S-dual is the $\left(\begin{smallmatrix}11\\11\\\end{smallmatrix}\right)$ D5-D5$'$ junction. 
We have a 4d $\mathcal{N}=4$ $U(1)$ gauge theory defined on the whole plane, 
coupled to the 3d hypermultiplet arising from D3-D5 string and 
the 3d twisted hypermultiplet arising from D3-D5$'$ string. 
In addition, there will be the neutral Fermi multiplet arising from D5-D5$'$ string \cite{Hanany:2018hlz},
possibly with a cubic coupling to the hypers and twisted hypers.

Then the quarter-index for the $\left(\begin{smallmatrix}11\\11\\\end{smallmatrix}\right)$ D5-D5$'$ junction takes the form 
\begin{align}
\label{DD1111_index}
\mathbb{IV}_{\mathcal{D}\mathcal{D}'}^{
\left(
\begin{smallmatrix}
1&1\\
1&1\\
\end{smallmatrix}
\right)}
&=
\underbrace{
\frac{(q)_{\infty}^2}
{(q^{\frac12}t^2;q)_{\infty}
(q^{\frac12}t^{-2};q)_{\infty}}
\oint \frac{ds}{2\pi is}
}_{\mathbb{I}^{\textrm{4d $U(1)$}}}
\underbrace{
(q^{\frac12}x^2;q)_{\infty}
(q^{\frac12}x^{-2};q)_{\infty}
}_{F(q^{\frac12}x^2)}
\nonumber\\
&\times 
\underbrace{
\frac{
\left(q^{\frac34}t^{-1}sx;q\right)_{\infty}
\left(q^{\frac34}t^{-1}s^{-1}x^{-1};q\right)_{\infty}
}
{
\left(q^{\frac14}t sx;q\right)_{\infty}
\left(q^{\frac14}t s^{-1}x^{-1};q\right)_{\infty}
}}_{\mathbb{I}^{\textrm{3d HM}}(sx)}
\underbrace{
\frac{
\left(q^{\frac34}tsx^{-1};q\right)_{\infty}
\left(q^{\frac34}ts^{-1}x;q\right)_{\infty}
}
{
\left(q^{\frac14}t^{-1} sx^{-1};q\right)_{\infty}
\left(q^{\frac14}t^{-1} s^{-1}x;q\right)_{\infty}
}}_{\mathbb{I}^{\textrm{3d tHM}}(sx^{-1})}. 
\end{align}
We find that 
the quarter-indices (\ref{NN1111_index}) and (\ref{DD1111_index}) beautifully coincide. 

The contour integral in (\ref{DD1111_index}) is evaluated by taking the sum over residues 
associated to two sets of poles where 
one is hypermultiplet poles at $s=q^{\frac14+m}tx^{-1}$ and 
the other is twisted hypermultiplet poles at $s=q^{\frac14+m}t^{-1}x$. 

When we take the sum over the residues at both poles, we find that 
\begin{align}
\label{NN1111_series}
\mathbb{IV}_{\mathcal{N}\mathcal{N}'}^{
\left(
\begin{smallmatrix}
1&1\\
1&1\\
\end{smallmatrix}
\right)}
&=\mathbb{IV}_{\mathcal{D}\mathcal{D}'}^{
\left(
\begin{smallmatrix}
1&1\\
1&1\\
\end{smallmatrix}
\right)}
\nonumber\\
&=
\Biggl[
\frac{
(q^{\frac12}x^2;q)_{\infty}^2 (q^{\frac12}x^{-2};q)_{\infty}^2
}
{(t^{-2}x^{2};q)_{\infty}(qt^{2}x^{-2};q)_{\infty}}
\sum_{n=0}^{\infty}
\frac{(q^{1+n};q)_{\infty}^{2} (q^{1+n}t^{2}x^{-2};q)_{\infty}^{2}}
{(q^{\frac12 +n}x^{-2};q)_{\infty}^2 (q^{\frac12 +n}t^{2};q)_{\infty}^{2}}q^{n}
\nonumber\\
&+
\frac{
(q^{\frac12}x^2;q)_{\infty}^2 (q^{\frac12}x^{-2};q)_{\infty}^2
}{(t^{2}x^{-2};q)_{\infty}(qt^{-2}x^{2};q)_{\infty}}
\sum_{n=0}^{\infty}
\frac{
(q^{1+n};q)_{\infty}^2
(q^{1+n}t^{-2}x^{2};q)_{\infty}^{2}
}
{(q^{\frac12+n}x^{2}:q)_{\infty}^2 (q^{\frac12+n}t^{-2};q)_{\infty}^2}q^{n}
\Biggr]
\end{align}
where the first term is the residue sum of hypermultiplet poles 
while the second is that of twisted hypermultiplet poles. 

\begin{figure}
\begin{center}
\includegraphics[width=14.5cm]{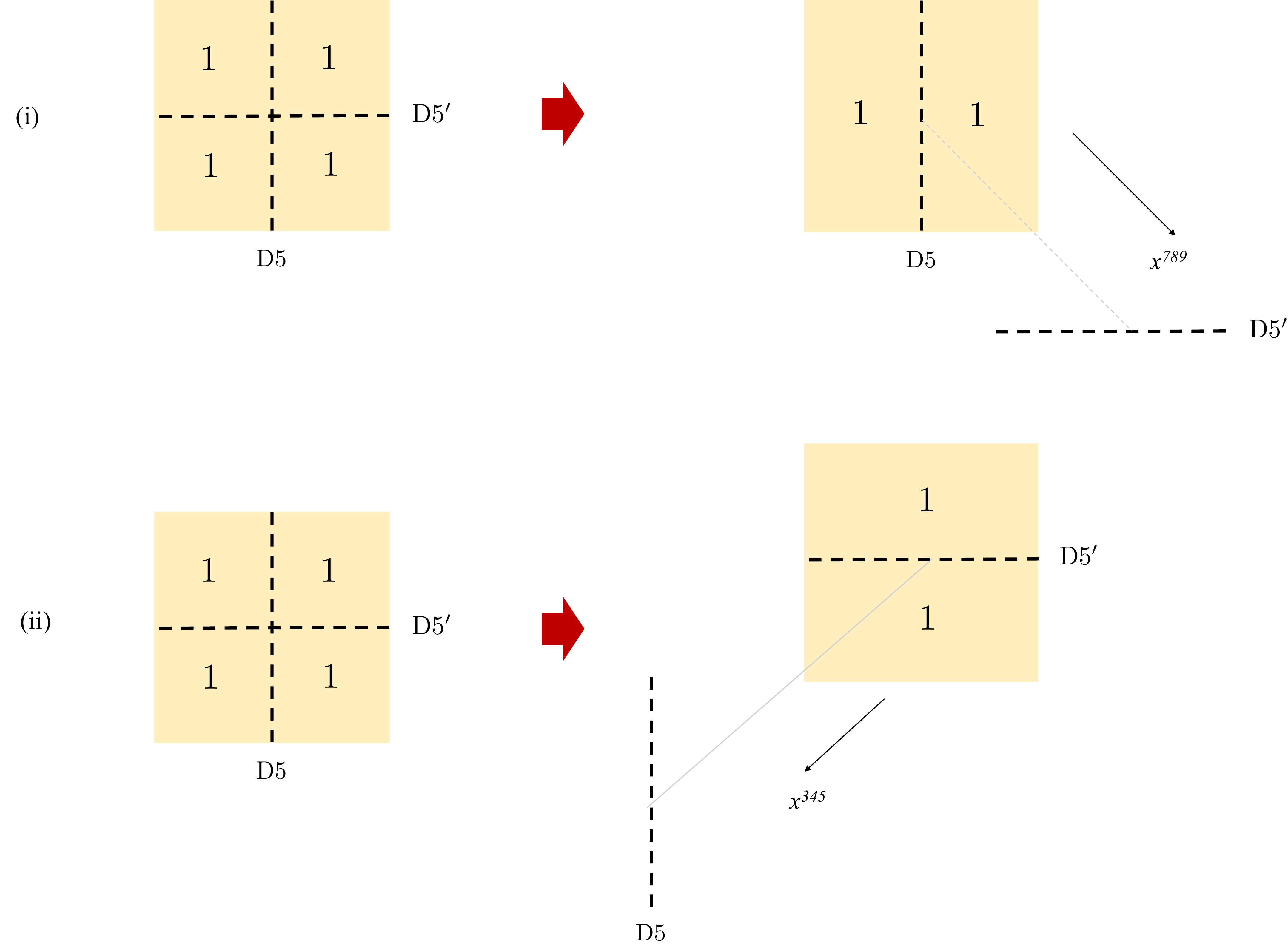}
\caption{Higgsing procedures of the ${1 \, 1 \choose 1 \, 1}$ D5-D5$'$ junction.  
(i) A separation of the D5-brane. (ii) A separation of the D5$'$-brane. }
\label{fighiggsing3a}
\end{center}
\end{figure}

We can also expand the quarter-index (\ref{DD1111_index}) as
\begin{align}
\label{DD1111_indexsum1}
\mathbb{IV}_{\mathcal{D}\mathcal{D}'}^{
\left(
\begin{smallmatrix}
1&1\\
1&1\\
\end{smallmatrix}
\right)}
&=
\frac{(q^{\frac12}t^2;q)_{\infty}^2 (q^{\frac12}x^{\pm2};q)_{\infty}}
{(q)_{\infty}^2}
\sum_{n=0}^{\infty}\sum_{m=0}^{\infty}\sum_{k=0}^{n}
\frac{(q^{1+k};q)_{\infty} (q^{1+n-k};q)_{\infty} (q^{1+m};q)_{\infty}^2}
{(q^{\frac12+k}t^{2};q)_{\infty} (q^{\frac12+n-k}t^{2};q)_{\infty} (q^{\frac12+m}t^2;q)_{\infty}^2}
\nonumber\\
&\times 
q^{\frac{n}{2}+\frac{m}{2}-\frac{k}{2}+m(n-2k)}t^{-2k-2m} x^{-2n+4k}. 
\end{align}
The sum begins with the square of half-index $\mathbb{II}_{\mathcal{D}}^{\textrm{4d $U(1)$}}$ and the extra Fermi index. 
The associated Higgsing procedure would be the separation of the D5$'$-brane (see Figure \ref{fighiggsing3a}). 

Alternatively, we also have 
\begin{align}
\label{DD1111_indexsum2}
\mathbb{IV}_{\mathcal{D}\mathcal{D}'}^{
\left(
\begin{smallmatrix}
1&1\\
1&1\\
\end{smallmatrix}
\right)}
&=\frac{(q^{\frac12}t^{-2};q)_{\infty} (q^{\frac12}x^{\pm};q)_{\infty}}
{(q)_{\infty}^2}
\sum_{n=0}^{\infty}\sum_{m=0}^{\infty}\sum_{k=0}^{n}
\frac{(q^{1+k};q)_{\infty} (q^{1+n-k};q)_{\infty} (q^{1+m};q)_{\infty}^2}
{(q^{\frac12+k}t^{-2};q)_{\infty} (q^{\frac12+n-k} t^{-2};q)_{\infty} (q^{\frac12+m}t^{-2};q)_{\infty}^2}
\nonumber\\
&\times 
q^{\frac{n}{2}+\frac{m}{2}-\frac{k}{2}+m(n-2k)} 
t^{2k+2m} x^{2n-4k}.
\end{align} 
The sum begins with the square of half-index $\mathbb{II}_{\mathcal{D}'}^{\textrm{4d $U(1)$}}$ and the extra Fermi index. 
This would be associated to the Higgsing process that separates the D5-brane (see Figure \ref{fighiggsing3a}). 

\subsubsection{${2 \, 2 \choose 2 \, 2}$ and ${2 \, 2 \choose 2 \, 2}$}
The next step is the $\left(\begin{smallmatrix}
2&2\\2&2\\\end{smallmatrix}\right)$ NS5-NS5$'$ junction. 
The $\left(\begin{smallmatrix}
2&2\\2&2\\\end{smallmatrix}\right)$ NS5-NS5$'$ 
junction has $U(2)_{1}$ $\times$ $U(2)_{2}$ $\times$ $U(2)_{3}$ $\times$ $U(2)_{4}$ gauge symmetry 
with $U(2)_{1}$, $U(2)_{2}$, $U(2)_{3}$ and $U(2)_{4}$ being 
the gauge groups corresponding to the upper left, lower left, upper right and lower right quadrant D3-branes respectively. 
The matter fields are the two 3d $\mathcal{N}=4$ bi-fundamental hypers   
transforming as $({\bf 2}, {\bf 1},\overline{\bf 2}, {\bf1})$ $\oplus$ $(\overline{\bf 2}, {\bf 1},{\bf 2}, {\bf1})$ 
and as $({\bf 1}, {\bf 2},{\bf 1}, \overline{\bf 2})$ $\oplus$ $({\bf 1}, \overline{\bf 2},{\bf 1}, {\bf 2})$, 
two 3d $\mathcal{N}=4$ bi-fundamental twisted hypers 
transforming as $({\bf 2}, \overline{\bf 2},{\bf 1}, {\bf1})$ $\oplus$ $(\overline{\bf 2}, {\bf 2},{\bf 1}, {\bf1})$ 
and as $({\bf 1}, {\bf 1},{\bf 2}, \overline{\bf 2})$ $\oplus$ $({\bf 1}, {\bf 1},\overline{\bf 2}, {\bf 2})$, 
two bi-fundamental Fermi multiplets  
transforming as $({\bf 2}, {\bf 1},{\bf 1}, \overline{\bf 2})$ $\oplus$ $(\overline{\bf 2}, {\bf 1},{\bf 1}, {\bf 2})$ 
and as $({\bf 1}, {\bf 2},\overline{\bf 2}, {\bf1})$ $\oplus$ $({\bf 1}, \overline{\bf 2},{\bf 2}, {\bf1})$ 
and cross-determinant Fermi multiplet 
transforming as $(\det$, $\det^{-1}$, $\det^{-1}$, $\det)$. 
The cross-determinant Fermi multiplet has the fivebrane charge $+2$.

The quarter-index for the $\left(\begin{smallmatrix}
2&2\\2&2\\\end{smallmatrix}\right)$ NS5-NS5$'$ junction is evaluated as 
\begin{align}
\label{NN2222}
&\mathbb{IV}_{\mathcal{N}\mathcal{N}'}^{
\left(
\begin{smallmatrix}
2&2\\
2&2\\
\end{smallmatrix}
\right)}
=
\underbrace{
\frac12 (q)_{\infty}^2 \oint \prod_{i=1}^{2}\frac{ds_{i}}{2\pi is_{i}}
\prod_{i\neq j}\left(\frac{s_{i}}{s_{j}};q\right)_{\infty}
}_{\mathbb{IV}_{\mathcal{N}\mathcal{N}'}^{\textrm{4d $U(2)$}}}
\underbrace{
\frac12 (q)_{\infty}^2 \oint \prod_{i=3}^{4}\frac{ds_{i}}{2\pi is_{i}}
\prod_{i\neq j}\left(\frac{s_{i}}{s_{j}};q\right)_{\infty}
}_{\mathbb{IV}_{\mathcal{N}\mathcal{N}'}^{\textrm{4d $U(2)$}}}
\nonumber\\
&\times 
\underbrace{
\frac12 (q)_{\infty}^2 \oint \prod_{i=5}^{6}\frac{ds_{i}}{2\pi is_{i}}
\prod_{i\neq j}\left(\frac{s_{i}}{s_{j}};q\right)_{\infty}
}_{\mathbb{IV}_{\mathcal{N}\mathcal{N}'}^{\textrm{4d $U(2)$}}}
\underbrace{
\frac12 (q)_{\infty}^2 \oint \prod_{i=7}^{8}\frac{ds_{i}}{2\pi is_{i}}
\prod_{i\neq j}\left(\frac{s_{i}}{s_{j}};q\right)_{\infty}
}_{\mathbb{IV}_{\mathcal{N}\mathcal{N}'}^{\textrm{4d $U(2)$}}}
\nonumber\\
&\times 
\underbrace{
\prod_{i=1}^{2}\prod_{j=3}^{4}\prod_{k=5}^{6}\prod_{l=7}^{8}
\frac{
\left(q^{\frac12} s_{i}^{\pm}s_{l}^{\mp} ;q\right)_{\infty}
\cdot 
\left(q^{\frac12} s_{k}^{\pm}s_{j}^{\mp};q\right)_{\infty}
\cdot 
\left(q^{\frac12} s_{1}^{\pm}s_{2}^{\pm}s_{7}^{\pm}s_{8}^{\pm} 
s_{3}^{\mp}s_{4}^{\mp}s_{5}^{\mp}s_{6}^{\mp}x^{\pm 2};q\right)_{\infty}
}
{
\left(q^{\frac14}t s_{i}^{\pm}s_{k}^{\mp};q\right)_{\infty}
\cdot 
\left(q^{\frac14}t s_{j}^{\pm}s_{l}^{\mp};q\right)_{\infty}
\cdot 
\left(q^{\frac14}t^{-1} s_{i}^{\pm}s_{j}^{\mp};q\right)_{\infty}
\cdot 
\left(q^{\frac14}t^{-1} s_{k}^{\pm}s_{l}^{\mp};q\right)_{\infty}
}
}_
{
\mathbb{II}_{N}^{\textrm{3d HM}}\left(\frac{s_{i}}{s_{k}}\right)\cdot 
\mathbb{II}_{N}^{\textrm{3d HM}}\left(\frac{s_{j}}{s_{l}}\right)\cdot 
\mathbb{II}_{N}^{\textrm{3d tHM}}\left(\frac{s_{i}}{s_{j}}\right)\cdot 
\mathbb{II}_{N}^{\textrm{3d tHM}}\left(\frac{s_{k}}{s_{l}}\right)\cdot 
F\left(q^{\frac12}\frac{s_{i}}{s_{l}}\right)\cdot 
F\left(q^{\frac12}\frac{s_{k}}{s_{j}}\right)\cdot 
F\left(q^{\frac12} \frac{s_{1}s_{2}s_{7}s_{8}}{s_{3}s_{4}s_{5}s_{6}}x^2\right)
}
\nonumber\\
&=
1+2(t^{-2}+t^{2})q^{\frac12}
+(1+5t^{-4}+5t^4)q
\nonumber\\
&+(8t^{-6}+8t^6-x^{-2}-x^2-t^{-4}x^{-2}-t^{-4}x^2-t^4x^{-2}-t^4x^2)q^{\frac32}+\cdots
\end{align}

Under S-duality 
the $\left(\begin{smallmatrix}
2&2\\2&2\\\end{smallmatrix}\right)$ NS5-NS5$'$ junction maps to 
the $\left(\begin{smallmatrix}
2&2\\2&2\\\end{smallmatrix}\right)$ D5-D5$'$ junction. 
There is a whole 4d $\mathcal{N}=4$ $U(2)$ gauge theory which 
couples to the 3d $\mathcal{N}=4$ fundamental hypermultiplet 
and to the 3d $\mathcal{N}=4$ fundamental twisted hypermultiplet. 
The junction involves the neutral Fermi multiplet arising from the D5-D5$'$ string. 

The quarter-index for the $\left(\begin{smallmatrix}22\\22\\\end{smallmatrix}\right)$ D5-D5$'$ junction is
\begin{align}
\label{DD2222}
\mathbb{IV}_{\mathcal{D}\mathcal{D}'}^{
\left(
\begin{smallmatrix}
2&2\\
2&2\\
\end{smallmatrix}
\right)}
&=
\underbrace{
\frac12 \frac{(q)_{\infty}^4}
{(q^{\frac12}t^2;q)_{\infty}^2
(q^{\frac12}t^{-2};q)_{\infty}^2}
\oint \prod_{i=1}^{2}\frac{ds_{i}}{2\pi is_{i}}
\prod_{i\neq j}\frac{\left(\frac{s_{i}}{s_{j}};q\right)_{\infty} \left(q\frac{s_{i}}{s_{j}};q\right)_{\infty}}
{
\left(q^{\frac12}t^2\frac{s_{i}}{s_{j}};q\right)_{\infty}
\left(q^{\frac12}t^{-2}\frac{s_{i}}{s_{j}};q\right)_{\infty}
}
}_{\mathbb{I}^{\textrm{4d $U(2)$}}}
\underbrace{
(q^{\frac12}x^2;q)_{\infty}
(q^{\frac12}x^{-2};q)_{\infty}
}_{F(q^{\frac12}x^2)}
\nonumber\\
&\times 
\prod_{i=1}^{2}
\underbrace{
\frac{
\left(q^{\frac34}t^{-1}s_{i}x;q\right)_{\infty}
\left(q^{\frac34}t^{-1}s_{i}^{-1}x^{-1};q\right)_{\infty}
}
{
\left(q^{\frac14}t s_{i}x;q\right)_{\infty}
\left(q^{\frac14}t s_{i}^{-1}x^{-1};q\right)_{\infty}
}}_{\mathbb{I}^{\textrm{3d HM}}(s_{i}x)}
\underbrace{
\frac{
\left(q^{\frac34}ts_{i}x^{-1};q\right)_{\infty}
\left(q^{\frac34}ts_{i}^{-1}x;q\right)_{\infty}
}
{
\left(q^{\frac14}t^{-1} s_{i}x^{-1};q\right)_{\infty}
\left(q^{\frac14}t^{-1} s_{i}^{-1}x;q\right)_{\infty}
}}_{\mathbb{I}^{\textrm{3d tHM}}(s_{i}x^{-1})}
\nonumber\\
&=
1+2(t^{-2}+t^{2})q^{\frac12}
+(1+5t^{-4}+5t^4)q
\nonumber\\
&+(8t^{-6}+8t^6-x^{-2}-x^2-t^{-4}x^{-2}-t^{-4}x^2-t^4x^{-2}-t^4x^2)q^{\frac32}+\cdots
\end{align}

It appears that 
the quarter-indices (\ref{NN2222}) and (\ref{DD2222}) coincide.

\subsubsection{${N \, N \choose N \,N}$ and ${N \, N \choose N \, N}$}
The generalization to the 
$\left(\begin{smallmatrix}
N&N\\N&N\\\end{smallmatrix}\right)$ NS5-NS5$'$ junction 
and 
$\left(\begin{smallmatrix}
N&N\\N&N\\\end{smallmatrix}\right)$ D5-D5$'$ junction 
is straightforward. 
For the $\left(\begin{smallmatrix}
N&N\\N&N\\\end{smallmatrix}\right)$ NS5-NS5$'$ 
the gauge symmetry is 
$U(N)_{1}$ $\times$ $U(N)_{2}$ $\times$ $U(N)_{3}$ $\times$ $U(N)_{4}$ 
where $U(N)_{1}$, $U(N)_{2}$, $U(N)_{3}$ and $U(N)_{4}$ respectively 
corresponds to the upper left, lower left, upper right and lower right D3-branes. 
The junction involves the two 3d $\mathcal{N}=4$ bi-fundamental hypers   
transforming as $({\bf N}, {\bf 1},\overline{\bf N}, {\bf1})$ $\oplus$ $(\overline{\bf N}, {\bf 1},{\bf N}, {\bf1})$ 
and as $({\bf 1}, {\bf N},{\bf 1}, \overline{\bf N})$ $\oplus$ $({\bf 1}, \overline{\bf N},{\bf 1}, {\bf N})$, 
two 3d $\mathcal{N}=4$ bi-fundamental twisted hypers 
transforming as $({\bf N}, \overline{\bf N},{\bf 1}, {\bf1})$ $\oplus$ $(\overline{\bf N}, {\bf N},{\bf 1}, {\bf1})$ 
and as $({\bf 1}, {\bf 1},{\bf N}, \overline{\bf N})$ $\oplus$ $({\bf 1}, {\bf 1},\overline{\bf N}, {\bf N})$, 
two bi-fundamental Fermi multiplets  
transforming as $({\bf N}, {\bf 1},{\bf 1}, \overline{\bf N})$ $\oplus$ $(\overline{\bf N}, {\bf 1},{\bf 1}, {\bf N})$ 
and as $({\bf 1}, {\bf N},\overline{\bf N}, {\bf1})$ $\oplus$ $({\bf 1}, \overline{\bf N},{\bf N}, {\bf1})$ 
and cross-determinant Fermi multiplet  
transforming as $(\det$, $\det^{-1}$, $\det^{-1}$, $\det)$ with the fivebrane charge $+2$.

The quarter-index for the $\left(\begin{smallmatrix}
N&N\\N&N\\\end{smallmatrix}\right)$ NS5-NS5$'$ junction takes the form 
\begin{align}
\label{NNNNNN_index}
&\mathbb{IV}_{\mathcal{N}\mathcal{N}'}^{
\left(
\begin{smallmatrix}
N&N\\
N&N\\
\end{smallmatrix}
\right)}
=
\underbrace{
\frac{1}{N!} (q)_{\infty}^N \oint \prod_{i=1}^{N}\frac{ds_{i}}{2\pi is_{i}}
\prod_{i\neq j}\left(\frac{s_{i}}{s_{j}};q\right)_{\infty}
}_{\mathbb{IV}_{\mathcal{N}\mathcal{N}'}^{\textrm{4d $U(N)$}}}
\underbrace{
\frac{1}{N!} (q)_{\infty}^N \oint \prod_{i=N+1}^{2N}\frac{ds_{i}}{2\pi is_{i}}
\prod_{i\neq j}\left(\frac{s_{i}}{s_{j}};q\right)_{\infty}
}_{\mathbb{IV}_{\mathcal{N}\mathcal{N}'}^{\textrm{4d $U(N)$}}}
\nonumber\\
&\times 
\underbrace{
\frac{1}{N!} (q)_{\infty}^N \oint \prod_{i=2N+1}^{3N}\frac{ds_{i}}{2\pi is_{i}}
\prod_{i\neq j}\left(\frac{s_{i}}{s_{j}};q\right)_{\infty}
}_{\mathbb{IV}_{\mathcal{N}\mathcal{N}'}^{\textrm{4d $U(N)$}}}
\underbrace{
\frac{1}{N!} (q)_{\infty}^N \oint \prod_{i=3N+1}^{4N}\frac{ds_{i}}{2\pi is_{i}}
\prod_{i\neq j}\left(\frac{s_{i}}{s_{j}};q\right)_{\infty}
}_{\mathbb{IV}_{\mathcal{N}\mathcal{N}'}^{\textrm{4d $U(N)$}}}
\nonumber\\
&\times 
\underbrace{
\prod_{i=1}^{N}\prod_{j=N+1}^{2N}\prod_{k=2N+1}^{3N}\prod_{l=3N+1}^{4N}
\frac{
\left(q^{\frac12} s_{i}^{\pm}s_{l}^{\mp} ;q\right)_{\infty}
\cdot 
\left(q^{\frac12} s_{k}^{\pm}s_{j}^{\mp};q\right)_{\infty}
\cdot 
\left(q^{\frac12} 
\prod_{i}s_{i}^{\pm} \prod_{l}s_{l}^{\pm} 
\prod_{j}s_{j}^{\mp} \prod_{k}s_{k}^{\mp}
x^{\pm 2};q\right)_{\infty}
}
{
\left(q^{\frac14}t s_{i}^{\pm}s_{k}^{\mp};q\right)_{\infty}
\cdot 
\left(q^{\frac14}t s_{j}^{\pm}s_{l}^{\mp};q\right)_{\infty}
\cdot 
\left(q^{\frac14}t^{-1} s_{i}^{\pm}s_{j}^{\mp};q\right)_{\infty}
\cdot 
\left(q^{\frac14}t^{-1} s_{k}^{\pm}s_{l}^{\mp};q\right)_{\infty}
}
}_
{
\mathbb{II}_{N}^{\textrm{3d HM}}\left(\frac{s_{i}}{s_{k}}\right)\cdot 
\mathbb{II}_{N}^{\textrm{3d HM}}\left(\frac{s_{j}}{s_{l}}\right)\cdot 
\mathbb{II}_{N}^{\textrm{3d tHM}}\left(\frac{s_{i}}{s_{j}}\right)\cdot 
\mathbb{II}_{N}^{\textrm{3d tHM}}\left(\frac{s_{k}}{s_{l}}\right)\cdot 
F\left(q^{\frac12}\frac{s_{i}}{s_{l}}\right)\cdot 
F\left(q^{\frac12}\frac{s_{k}}{s_{j}}\right)\cdot 
F\left(q^{\frac12} \frac{\prod s_{i}\prod s_{l}}{\prod s_{j}\prod s_{k}}x^2\right)
}. 
\end{align}

The S-dual $\left(\begin{smallmatrix}
N&N\\N&N\\\end{smallmatrix}\right)$ D5-D5$'$ junction is described by  
the 4d $\mathcal{N}=4$ $U(N)$ gauge theory coupled to the 3d $\mathcal{N}=4$ fundamental hypermultiplet 
and to the 3d $\mathcal{N}=4$ fundamental twisted hypermultiplet 
as well as the neutral Fermi multiplet arising from the D5-D5$'$ string. 

The quarter-index for the $\left(\begin{smallmatrix}NN\\NN\\\end{smallmatrix}\right)$ D5-D5$'$ junction is
\begin{align}
\label{DDNNNN_index}
\mathbb{IV}_{\mathcal{D}\mathcal{D}'}^{
\left(
\begin{smallmatrix}
N&N\\
N&N\\
\end{smallmatrix}
\right)}
&=
\underbrace{
\frac{1}{N!} \frac{(q)_{\infty}^{2N}}
{(q^{\frac12}t^2;q)_{\infty}^N
(q^{\frac12}t^{-2};q)_{\infty}^N}
\oint \prod_{i=1}^{N}\frac{ds_{i}}{2\pi is_{i}}
\prod_{i\neq j}\frac{\left(\frac{s_{i}}{s_{j}};q\right)_{\infty} \left(q\frac{s_{i}}{s_{j}};q\right)_{\infty}}
{
\left(q^{\frac12}t^2\frac{s_{i}}{s_{j}};q\right)_{\infty}
\left(q^{\frac12}t^{-2}\frac{s_{i}}{s_{j}};q\right)_{\infty}
}
}_{\mathbb{I}^{\textrm{4d $U(N)$}}}
\nonumber\\
&\times 
\underbrace{
(q^{\frac12}x^2;q)_{\infty}
(q^{\frac12}x^{-2};q)_{\infty}
}_{F(q^{\frac12}x^2)}
\nonumber\\
&\times 
\prod_{i=1}^{N}
\underbrace{
\frac{
\left(q^{\frac34}t^{-1}s_{i}x;q\right)_{\infty}
\left(q^{\frac34}t^{-1}s_{i}^{-1}x^{-1};q\right)_{\infty}
}
{
\left(q^{\frac14}t s_{i}x;q\right)_{\infty}
\left(q^{\frac14}t s_{i}^{-1}x^{-1};q\right)_{\infty}
}}_{\mathbb{I}^{\textrm{3d HM}}(s_{i}x)}
\underbrace{
\frac{
\left(q^{\frac34}ts_{i}x^{-1};q\right)_{\infty}
\left(q^{\frac34}ts_{i}^{-1}x;q\right)_{\infty}
}
{
\left(q^{\frac14}t^{-1} s_{i}x^{-1};q\right)_{\infty}
\left(q^{\frac14}t^{-1} s_{i}^{-1}x;q\right)_{\infty}
}}_{\mathbb{I}^{\textrm{3d tHM}}(s_{i}x^{-1})}. 
\end{align}

The quarter-indices (\ref{NNNNNN_index}) and (\ref{DDNNNN_index}) are expected to give the same answer.

\begin{figure}
\begin{center}
\includegraphics[width=10.5cm]{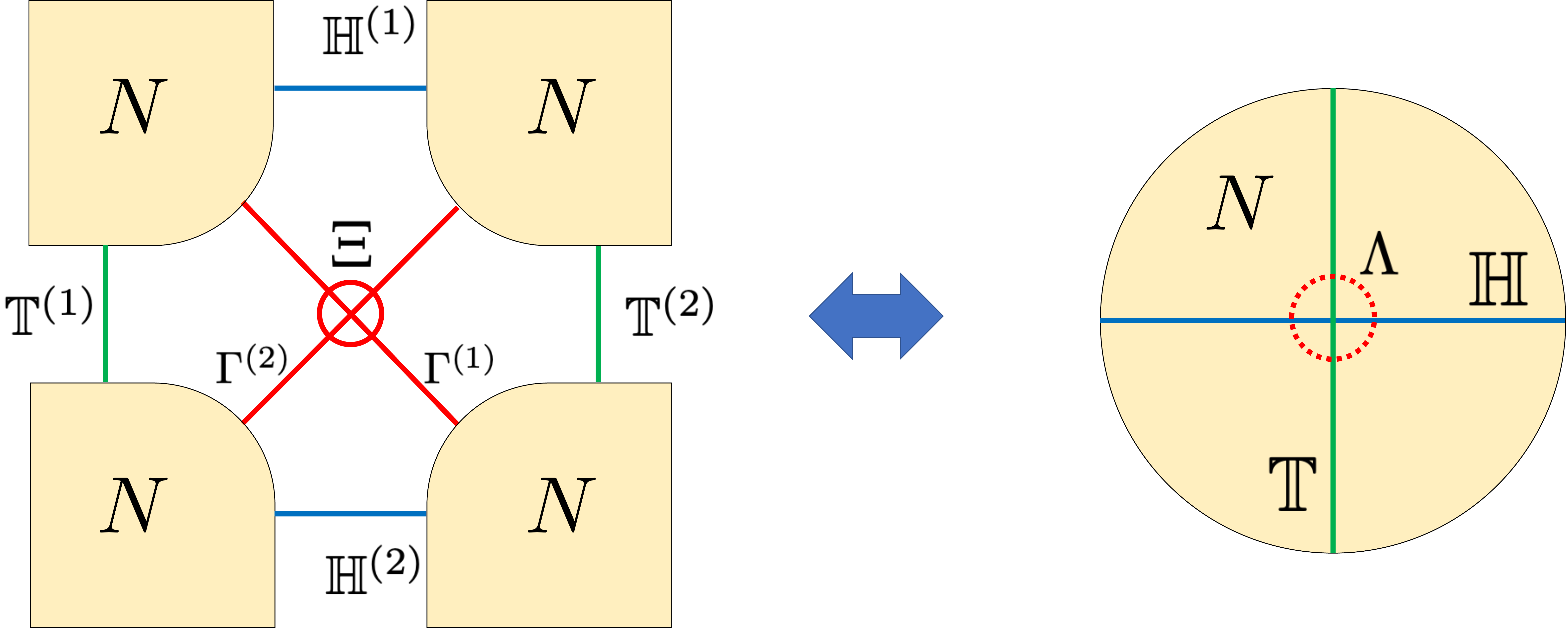}
\caption{
The quiver diagrams for the dual 4d-3d-2d systems resulting from the ${N \, N \choose N \, N}$  NS5-NS5$'$ and D5-D5$'$ junctions. 
}
\label{fignnddquiver}
\end{center}
\end{figure}

The resulting duality conjecture in the 4d-3d-2d systems from the $\left(\begin{smallmatrix}
N&N\\N&N\\\end{smallmatrix}\right)$ NS5-NS5$'$ junction 
and 
$\left(\begin{smallmatrix}
N&N\\N&N\\\end{smallmatrix}\right)$ D5-D5$'$ junction is summarized as follows:
\begin{align}
\label{NNNN_dual}
&\begin{cases}
\textrm{4d $\mathcal{N}=4$ $U(N)_{1}\times U(N)_{2}\times U(N)_{3}\times U(N)_{4}$ SYM w/ b.c. $(\mathcal{N},\mathcal{N}')$}\cr
+\textrm{3d $\mathcal{N}=4$ bi-fund. hypers $\mathbb{H}^{(1)}$ as $({\bf N}, {\bf 1},\overline{\bf N}, {\bf1})$ $\oplus$ $(\overline{\bf N}, {\bf 1},{\bf N}, {\bf1})$ w/ b.c. $N'$}\cr
+\textrm{3d $\mathcal{N}=4$ bi-fund. hypers $\mathbb{H}^{(2)}$ as $({\bf 1}, {\bf N},{\bf 1}, \overline{\bf N})$ $\oplus$ $({\bf 1}, \overline{\bf N},{\bf 1}, {\bf N})$ w/ b.c. $N'$}\cr
+\textrm{3d $\mathcal{N}=4$ bi-fund. twisted hypers $\mathbb{T}^{(1)}$ as $({\bf N}, \overline{\bf N},{\bf 1}, {\bf1})$ $\oplus$ $(\overline{\bf N}, {\bf N},{\bf 1}, {\bf1})$ w/ b.c. $N$}\cr
+\textrm{3d $\mathcal{N}=4$ bi-fund. twisted hypers $\mathbb{T}^{(2)}$ as $({\bf 1}, {\bf 1},{\bf N}, \overline{\bf N})$ $\oplus$ $({\bf 1}, {\bf 1},\overline{\bf N}, {\bf N})$ w/ b.c. $N$}\cr
+\textrm{2d bi-fund. Fermi's $\Gamma^{(1)}$ as $({\bf N}, {\bf 1},{\bf 1}, \overline{\bf N})$ $\oplus$ $(\overline{\bf N}, {\bf 1},{\bf 1}, {\bf N})$}\cr
+\textrm{2d bi-fund. Fermi's $\Gamma^{(2)}$ as $({\bf 1}, {\bf N},\overline{\bf N}, {\bf1})$ $\oplus$ $({\bf 1}, \overline{\bf N},{\bf N}, {\bf1})$}\cr
+\textrm{2d cross-det. Fermi $\Xi$ as $(\det$, $\det^{-1}$, $\det^{-1}$, $\det)$}\cr
\end{cases}
\nonumber\\
\leftrightarrow 
&\begin{cases}
\textrm{4d $\mathcal{N}=4$ $U(N)$ SYM}\cr
+\textrm{3d $\mathcal{N}=4$ fund. hypers $\mathbb{H}$}\cr
+\textrm{3d $\mathcal{N}=4$ fund. twisted hypers $\mathbb{T}$}\cr
+\textrm{2d neutral Fermi $\Lambda$}\cr
\end{cases}
\end{align}
The corresponding quiver diagram is depicted in Figure \ref{fignnddquiver}. 
\footnote{See also \cite{Hanany:2018hlz} for the notation. }
The orange objects represent 4d $\mathcal{N}=4$ SYM theories.  
The blue and green lines stand for 3d $\mathcal{N}=4$ hypers and twisted hypers respectively. 
The red bold and dotted lines describe 2d charged and neutral Fermi multiplets respectively. 
%
%
%
%
%

\subsection{${N \, 2N \choose 0 \, \, N}$ and ${\, \, 0 \, \, \, N \choose 2N \, N}$}
\label{sec_2dnnNN00}

\subsubsection{${1 \, 2 \choose 0 \, 1}$ and ${0 \, 1 \choose 1 \, 2}$}
Let us consider the NS5-NS5$'$ and D5-D5$'$ junction 
for which three of four quadrants are filled by the D3-branes. 
In this case there are three 4d $\mathcal{N}=4$ gauge theories, each defined in the appropriate quadrant.

As an example, we examine the $\left(\begin{smallmatrix}
1&2\\0&1\\\end{smallmatrix}\right)$ NS5-NS5$'$ junction. 
This includes three corners of 4d $\mathcal{N}=4$ $U(1)_{1}$, $U(2)$ and $U(1)_{2}$ gauge theories 
where the $U(1)_{1}$ and $U(1)_{2}$ are associated to the upper left and lower right D3-branes. 
There are the 3d $\mathcal{N}=4$ bi-fundamental hyper, 
3d $\mathcal{N}=4$ bi-fundamental twisted hyper, 
the diagonal Fermi multiplet, 
and the cross-determinant Fermi multiplet. 
They have the charges as follows:
\begin{align}
 \label{121_ch}
\begin{array}{c|c|c|c}
&U(1)_{1}&U(1)_{2}&U(2) \\ \hline 
\textrm{3d $\mathcal{N}=4$ hyper $\mathbb{H}$}&+&0&\overline{\bf 2} \\ \hline
\textrm{3d $\mathcal{N}=4$ twisted hyper $\mathbb{T}$}&0&+&\overline{\bf 2} \\ \hline
\textrm{2d Fermi multiplet $\Gamma$}&+&-&0 \\ \hline 
\textrm{2d Fermi multiplet $\Xi$}&+&+&\mathrm{det}^{-1} \\
\end{array}
\end{align}

The quarter-index for the $\left(\begin{smallmatrix}
1&2\\0&1\\\end{smallmatrix}\right)$ NS5-NS5$'$ junction is evaluated as  
\begin{align}
\label{NN1201_index}
\mathbb{IV}_{\mathcal{N}\mathcal{N}'}^{
\left(
\begin{smallmatrix}
1&2\\
0&1\\
\end{smallmatrix}
\right)}
&=
\underbrace{
(q)_{\infty}\oint \frac{ds_{1}}{2\pi is_{1}}
}_{\mathbb{IV}_{\mathcal{N}\mathcal{N}'}^{\textrm{4d $U(1)$}}}
\underbrace{
\frac{1}{2}(q)_{\infty}^2 \oint \prod_{i=2}^{3}\frac{ds_{i}}{2\pi is_{i}}
\prod_{i\neq j}\left(\frac{s_{i}}{s_{j}};q\right)_{\infty}
}_{\mathbb{IV}_{\mathcal{N}\mathcal{N}'}^{\textrm{4d $U(2)$}}}
\underbrace{
(q)_{\infty}\oint \frac{ds_{4}}{2\pi is_{4}}
}_{\mathbb{IV}_{\mathcal{N}\mathcal{N}'}^{\textrm{4d $U(1)$}}}
\nonumber\\
&\times 
\prod_{i=2}^{3}
\underbrace{
\frac{
\left(q^{\frac12}\frac{s_{1}}{s_{4}};q\right)_{\infty}
\left(q^{\frac12}\frac{s_{4}}{s_{1}};q\right)_{\infty}\cdot 
\left(q^{\frac12}\frac{s_{1}s_{4}}{s_{2}s_{3}} x;q \right)_{\infty}
\left(q^{\frac12}\frac{s_{2}s_{3}}{s_{1}s_{4}} x^{-1};q \right)_{\infty}
}
{
\left(q^{\frac14}t \frac{s_{1}}{s_{i}}\right)_{\infty}
\left(q^{\frac14}t \frac{s_{i}}{s_{1}}\right)_{\infty}\cdot 
\left(q^{\frac14}t^{-1} \frac{s_{i}}{s_{4}}\right)_{\infty}
\left(q^{\frac14}t^{-1} \frac{s_{4}}{s_{i}}\right)_{\infty}
}
}_{
\mathbb{II}_{N}^{\textrm{3d HM}}\left(\frac{s_{1}}{s_{i}}\right)\cdot 
\mathbb{II}_{N}^{\textrm{3d tHM}}\left(\frac{s_{i}}{s_{4}}\right) \cdot 
F\left(q^{\frac12} \frac{s_{1}}{s_{4}}\right)\cdot 
F\left(q^{\frac12}\frac{s_{1}s_{4}}{s_{2}s_{3}}x\right)
}. 
\end{align}

Under S-duality we obtain the $\left(\begin{smallmatrix}
0&1\\1&2\\\end{smallmatrix}\right)$ D5-D5$'$ junction. 
This has no surviving gauge symmetry at the junction. 
The D5- and D5$'$-branes will impose the Dirichlet boundary condition $\mathcal{D}$ and $\mathcal{D}'$ 
for 4d $\mathcal{N}=4$ $U(1)$ gauge theories in $x^6>0$ and in $x^2<0$ respectively. 

In fact we find that 
the quarter-index (\ref{NN1201_index}) for the $\left(\begin{smallmatrix}
1&2\\0&1\\\end{smallmatrix}\right)$ NS5-NS5$'$ junction 
matches with the follwing quarter-index for 
the mirror $\left(\begin{smallmatrix}
0&1\\1&2\\\end{smallmatrix}\right)$ D5-D5$'$ junction:
\begin{align}
\label{DD0112_index}
\mathbb{IV}_{\mathcal{D}\mathcal{D}'}^{
\left(
\begin{smallmatrix}
0&1\\
1&2\\
\end{smallmatrix}
\right)}
&=
\underbrace{
\frac{(q)_{\infty}}
{(q^{\frac12}t^2;q)_{\infty}}
}_{\mathbb{II}_{\mathcal{D}}^{\textrm{4d $U(1)$}}}
\underbrace{
\frac{(q)_{\infty}}
{(q^{\frac12}t^{-2};q)_{\infty}}
}_{\mathbb{II}_{\mathcal{D}'}^{\textrm{4d $U(1)$}}}
(qx;q)_{\infty}
(qx^{-1};q)_{\infty}. 
\end{align}

\subsubsection{${2 \, 4 \choose 0 \, 2}$ and ${0 \, 2 \choose 2 \, 4}$}

Consider the $\left(\begin{smallmatrix}
2&4\\0&2\\\end{smallmatrix}\right)$ NS5-NS5$'$ junction and 
$\left(\begin{smallmatrix}
0&2\\2&4\\\end{smallmatrix}\right)$ D5-D5$'$ junction. 
For $\left(\begin{smallmatrix}
2&4\\0&2\\\end{smallmatrix}\right)$ NS5-NS5$'$ junction, 
there are three corners of 4d $\mathcal{N}=4$ $U(2)_{1}$, $U(4)$ and $U(2)_{2}$. 
There are the 3d $\mathcal{N}=4$ bi-fundamental hypermultiplet transforming as 
$({\bf 2},\overline{\bf 4},{\bf 1})$ $\oplus$ $(\overline{\bf 2},{\bf 4},{\bf 1})$, 
3d $\mathcal{N}=4$ bi-fundamental twisted hypermultiplet transforming as 
$({\bf 1},\overline{\bf 4},{\bf 2})$ $\oplus$ $({\bf 1},{\bf 4},\overline{\bf 2})$, 
the diagonal Fermi transforming as 
$({\bf 2},{\bf 1},\overline{\bf 2})$ $\oplus$ $(\overline{\bf 2},{\bf 1},{\bf 2})$
and the cross-determinant Fermi multiplet transforming as $(\det, \det^{-1},\det)$. 

We can then compute the quarter-index for $\left(\begin{smallmatrix}
2&4\\0&2\\\end{smallmatrix}\right)$ NS5-NS5$'$ junction as
\begin{align}
\label{NN2402_index}
\mathbb{IV}_{\mathcal{N}\mathcal{N}'}^{
\left(
\begin{smallmatrix}
2&4\\
0&2\\
\end{smallmatrix}
\right)}
&=
\underbrace{
\frac12 (q)_{\infty}^{2}\oint \prod_{i=1}^{2}\frac{ds_{i}}{2\pi is_{i}}\prod_{i\neq j}\left(\frac{s_{i}}{s_{j}};q\right)_{\infty}
}_{\mathbb{IV}_{\mathcal{N}\mathcal{N}'}^{\textrm{4d $U(2)$}}}
\underbrace{
\frac{1}{4!} (q)_{\infty}^{4}\oint \prod_{i=3}^{6}\frac{ds_{i}}{2\pi is_{i}}\prod_{i\neq j}\left(\frac{s_{i}}{s_{j}};q\right)_{\infty}
}_{\mathbb{IV}_{\mathcal{N}\mathcal{N}'}^{\textrm{4d $U(4)$}}}
\underbrace{
\frac12 (q)_{\infty}^{2}\oint \prod_{i=7}^{8}\frac{ds_{i}}{2\pi is_{i}}\prod_{i\neq j}\left(\frac{s_{i}}{s_{j}};q\right)_{\infty}
}_{\mathbb{IV}_{\mathcal{N}\mathcal{N}'}^{\textrm{4d $U(2)$}}}
\nonumber\\
&\times 
\prod_{i=1}^{2}
\prod_{k=3}^{6}
\prod_{j=7}^{8}
\underbrace{
\frac{
\left(q^{\frac12}\frac{s_{i}}{s_{j}};q\right)_{\infty}
\left(q^{\frac12}\frac{s_{j}}{s_{i}};q\right)_{\infty}\cdot 
\left(q^{\frac12}\frac{s_{1}s_{2}s_{7}s_{8}}{s_{3}s_{4}s_{5}s_{6}} x;q \right)_{\infty}
\left(q^{\frac12}\frac{s_{3}s_{4}s_{5}s_{6}}{s_{1}s_{2}s_{7}s_{8}} x^{-1};q \right)_{\infty}
}
{
\left(q^{\frac14}t \frac{s_{i}}{s_{k}}\right)_{\infty}
\left(q^{\frac14}t \frac{s_{k}}{s_{i}}\right)_{\infty}\cdot 
\left(q^{\frac14}t^{-1} \frac{s_{k}}{s_{j}}\right)_{\infty}
\left(q^{\frac14}t^{-1} \frac{s_{j}}{s_{k}}\right)_{\infty}
}
}_{
\mathbb{II}_{N}^{\textrm{3d HM}}\left(\frac{s_{i}}{s_{k}}\right)\cdot 
\mathbb{II}_{N}^{\textrm{3d tHM}}\left(\frac{s_{k}}{s_{j}}\right) \cdot 
F\left(q^{\frac12} \frac{s_{i}}{s_{j}}\right)\cdot 
F\left(q^{\frac12}\frac{s_{1}s_{s}s_{7}s_{8}}{s_{3}s_{4}s_{5}s_{6}}x\right)
}. 
\end{align}

The S-dual $\left(\begin{smallmatrix}
0&2\\2&4\\\end{smallmatrix}\right)$ D5-D5$'$ junction has no gauge symmetry. 
This configuration includes 4d $\mathcal{N}=4$ $U(2)$ gauge theory in $x^2<0$ 
and 4d $\mathcal{N}=4$ $U(2)$ gauge theory in $x^6>0$ 
obeying the Nahm pole boundary conditions.

The quarter-index (\ref{NN2402_index}) appears to coincide with the 
following quarter-index for the $\left(\begin{smallmatrix}
0&2\\2&4\\\end{smallmatrix}\right)$ D5-D5$'$ junction:
\begin{align}
\label{DD0224_index}
\mathbb{IV}_{\mathcal{D}\mathcal{D}'}^{
\left(
\begin{smallmatrix}
0&2\\
2&4\\
\end{smallmatrix}
\right)}
&=
\underbrace{
\frac{(q)_{\infty} (q^{\frac32}t^2;q)_{\infty}}
{(q^{\frac12}t^2;q)_{\infty}(qt^4;q)_{\infty}}
}_{\mathbb{II}_{\textrm{Nahm}}^{\textrm{4d $U(2)$}}}
\underbrace{
\frac{(q)_{\infty} (q^{\frac32}t^{-2};q)_{\infty}}
{(q^{\frac12}t^{-2};q)_{\infty} (qt^{-4};q)_{\infty}}
}_{\mathbb{II}_{\textrm{Nahm}'}^{\textrm{4d $U(2)$}}}
(q^{\frac32}x;q)_{\infty}
(q^{\frac32}x^{-1};q)_{\infty}. 
\end{align}

\subsubsection{${N \, 2N \choose 0 \, N}$ and ${0 \, N \choose N \, 2N}$}

Now we would like to give the generalization for the duality between the 
$\left(\begin{smallmatrix}
N&2N\\0&N\\\end{smallmatrix}\right)$ NS5-NS5$'$ junction and 
$\left(\begin{smallmatrix}
0&N\\N&2N\\\end{smallmatrix}\right)$ D5-D5$'$ junction. 

The $\left(\begin{smallmatrix}
N&2N\\0&N\\\end{smallmatrix}\right)$ NS5-NS5$'$ junction 
has three gauge symmetry factors $U(N)_{1}$, $U(2N)$ and $U(N)_{2}$ 
which come from the three corners of 4d $\mathcal{N}=4$ gauge theories. 
It has the 3d $\mathcal{N}=4$ bi-fundamental hypermultiplet transforming as 
$({\bf N},\overline{\bf 2N},{\bf 1})$ $\oplus$ $(\overline{\bf N},{\bf 2N},{\bf 1})$, 
the 3d $\mathcal{N}=4$ bi-fundamental twisted hypermultiplet transforming as 
$({\bf 1},\overline{\bf 2N},{\bf N})$ $\oplus$ $({\bf 1},{\bf 2N},\overline{\bf N})$, 
the diagonal Fermi multiplet transforming as 
$({\bf N},{\bf 1},\overline{\bf N})$ $\oplus$ $(\overline{\bf N},{\bf 1},{\bf N})$ 
and the cross-determinant Fermi multiplet transforming as $(\det,\det^{-1},\det)$.

The quarter-index for $\left(\begin{smallmatrix}
N&2N\\0&N\\\end{smallmatrix}\right)$ NS5-NS5$'$ junction takes the form 
\begin{align}
\label{NNN2N0N_index}
&\mathbb{IV}_{\mathcal{N}\mathcal{N}'}^{
\left(
\begin{smallmatrix}
N&2N\\
0&N\\
\end{smallmatrix}
\right)}
=
\underbrace{
\frac{1}{N!} (q)_{\infty}^{N}\oint \prod_{i=1}^{N}\frac{ds_{i}}{2\pi is_{i}}\prod_{i\neq j}\left(\frac{s_{i}}{s_{j}};q\right)_{\infty}
}_{\mathbb{IV}_{\mathcal{N}\mathcal{N}'}^{\textrm{4d $U(N)$}}}
\underbrace{
\frac{1}{(2N)!} (q)_{\infty}^{2N}\oint \prod_{i=N+1}^{3N}\frac{ds_{i}}{2\pi is_{i}}\prod_{i\neq j}\left(\frac{s_{i}}{s_{j}};q\right)_{\infty}
}_{\mathbb{IV}_{\mathcal{N}\mathcal{N}'}^{\textrm{4d $U(2N)$}}}
\nonumber\\
&\times 
\underbrace{
\frac{1}{N!} (q)_{\infty}^{N}\oint \prod_{i=3N+1}^{4N}\frac{ds_{i}}{2\pi is_{i}}\prod_{i\neq j}\left(\frac{s_{i}}{s_{j}};q\right)_{\infty}
}_{\mathbb{IV}_{\mathcal{N}\mathcal{N}'}^{\textrm{4d $U(N)$}}}
\nonumber\\
&\times 
\prod_{i=1}^{N}
\prod_{k=N+1}^{3N}
\prod_{j=3N+1}^{4N}
\underbrace{
\frac{
\left(q^{\frac12}\frac{s_{i}}{s_{j}};q\right)_{\infty}
\left(q^{\frac12}\frac{s_{j}}{s_{i}};q\right)_{\infty}\cdot 
\left(q^{\frac12}\frac{\prod_{i=1}^{N}s_{i}\prod_{j=3N+1}^{4N}s_{j}}{\prod_{k=N+1}^{3N}s_{k}} x;q \right)_{\infty}
\left(q^{\frac12}\frac{\prod_{k=N+1}^{3N}s_{k}}{\prod_{i=1}^{N}s_{i}\prod_{j=3N+1}^{4N}s_{j}} x^{-1};q \right)_{\infty}
}
{
\left(q^{\frac14}t \frac{s_{i}}{s_{k}}\right)_{\infty}
\left(q^{\frac14}t \frac{s_{k}}{s_{i}}\right)_{\infty}\cdot 
\left(q^{\frac14}t^{-1} \frac{s_{k}}{s_{j}}\right)_{\infty}
\left(q^{\frac14}t^{-1} \frac{s_{j}}{s_{k}}\right)_{\infty}
}
}_{
\mathbb{II}_{N}^{\textrm{3d HM}}\left(\frac{s_{i}}{s_{k}}\right)\cdot 
\mathbb{II}_{N}^{\textrm{3d tHM}}\left(\frac{s_{k}}{s_{j}}\right) \cdot 
F\left(q^{\frac12} \frac{s_{i}}{s_{j}}\right)\cdot 
F\left(q^{\frac12}\frac{\prod_{i=1}^{N}s_{i}\prod_{j=3N+1}^{4N}s_{j} }{\prod_{k=N+1}^{3N}s_{k} }x\right)
}. 
\end{align}

Although the expression (\ref{NNN2N0N_index}) for 
the $\left(\begin{smallmatrix}
N&2N\\0&N\\\end{smallmatrix}\right)$ NS5-NS5$'$ junction is complicated 
as it is given by the contour integral over $4N$ gauge fugacities, 
the quarter-index for the S-dual $\left(\begin{smallmatrix}
0&N\\N&2N\\\end{smallmatrix}\right)$ D5-D5$'$ junction should be extremely simple 
as there is no gauge symmetry. 
There will be two types of Nahm pole boundary conditions 
required by the D5- and D5$'$-branes 
which are specified by embeddings 
$\rho:$ $\mathfrak{su}(2)$ $\rightarrow$ $\mathfrak{u}(N)$ 
and $\rho':$ $\mathfrak{su}(2)$ $\rightarrow$ $\mathfrak{u}(N)$. 

\begin{figure}
\begin{center}
\includegraphics[width=10cm]{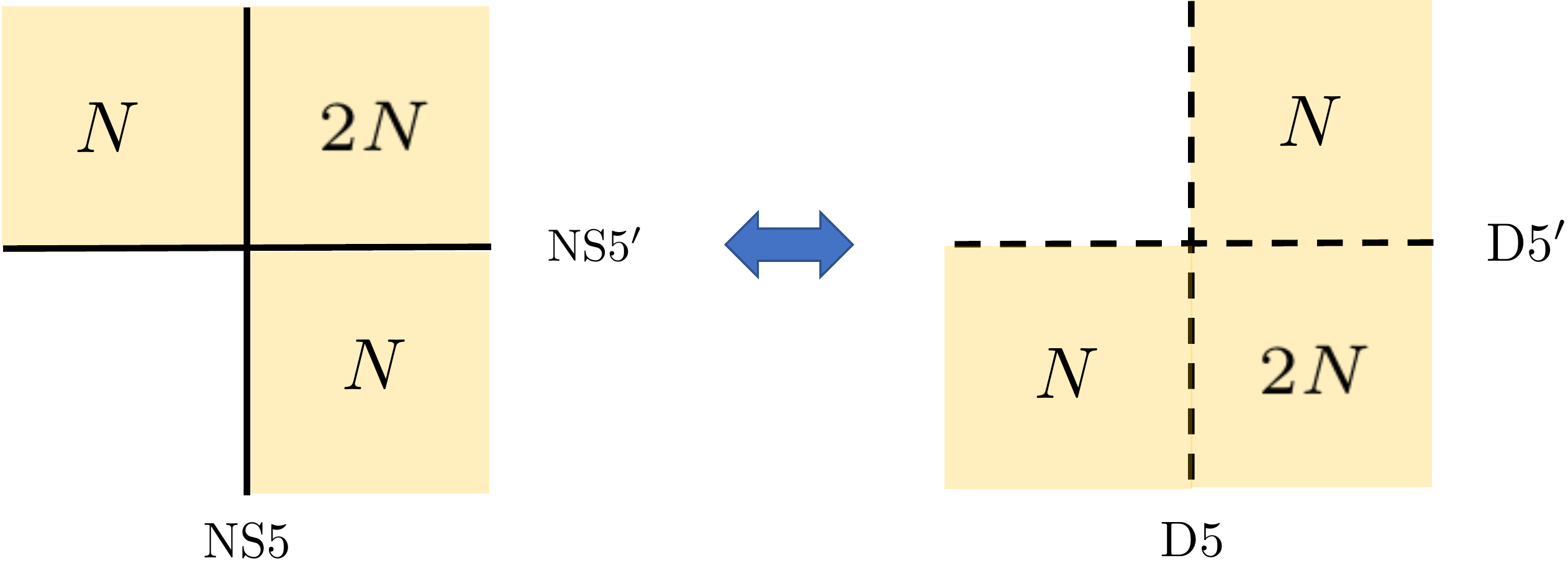}
\caption{
The ${N \, 2N \choose 0 \, \, N}$ NS5-NS5$'$ junction
and ${\, \, 0 \, \, \, N \choose 2N \, N}$ D5-D5$'$ junction 
with three quadrants of D3-branes.
}
\label{fignn0N2NN}
\end{center}
\end{figure}

We expect that 
the quarter-index for the $\left(\begin{smallmatrix}
0&N\\N&2N\\\end{smallmatrix}\right)$ D5-D5$'$ junction is given by 
$\left(\begin{smallmatrix}
0&N\\N&2N\\\end{smallmatrix}\right)$ D5-D5$'$ junction:
\begin{align}
\label{DD0NN2N_index}
\mathbb{IV}_{\mathcal{D}\mathcal{D}'}^{
\left(
\begin{smallmatrix}
0&N\\
N&2N\\
\end{smallmatrix}
\right)}
&=
\underbrace{
\prod_{k=1}^{N}
\frac{\left(q^{\frac{k-1}{2}}t^{2(k-1)};q\right)_{\infty}}
{\left(q^{\frac{k}{2}}t^{2k};q\right)_{\infty}}
}_{\mathbb{II}_{\textrm{Nahm}}^{\textrm{4d $U(N)$}}}
\underbrace{
\prod_{k=1}^{N}
\frac{\left(q^{\frac{k-1}{2}}t^{-2(k-1)};q\right)_{\infty}}
{\left(q^{\frac{k}{2}}t^{-2k};q\right)_{\infty}}
}_{\mathbb{II}_{\textrm{Nahm}'}^{\textrm{4d $U(N)$}}}
(q^{\frac{N+1}{2}}x;q)_{\infty}
(q^{\frac{N+1}{2}}x^{-1};q)_{\infty}. 
\end{align}
and this will be equal to the quarter-index (\ref{NNN2N0N_index}) for 
the $\left(\begin{smallmatrix}
N&2N\\0&N\\\end{smallmatrix}\right)$ NS5-NS5$'$ junction. 
The brane configuration is shown in Figure \ref{fignn0N2NN}
%
%
%
%
%

\subsection{${N \, M \choose N \, M}$ and ${N \, N \choose M \, M}$}
\label{sec_2dnn1212}

\subsubsection{${1 \, 2 \choose 1 \, 2}$ and ${1 \, 1 \choose 2 \, 2}$}
Let us examine the case with  
all the quadrants filled by some different numbers of D3-branes. 

Consider the $\left(\begin{smallmatrix}
1&2\\1&2\\\end{smallmatrix}\right)$ NS5-NS5$'$ junction. 
This has four 
4d $\mathcal{N}=4$ $U(1)_{1}$, $U(1)_{2}$, $U(2)_{3}$ and $U(2)_{4}$ gauge theories 
where subscripts $1,2,3,4$ indicate upper left, lower left, upper right and lower right quadrants respectively. 
Each of them obeys a pair of boundary conditions $\mathcal{N}$ and $\mathcal{N}'$. 
The matter content consists of 
the two 3d $\mathcal{N}=4$ bi-fundamental hypermultiplets 
$\mathbb{H}^{(1)}$, $\mathbb{H}^{(2)}$ 
with the Neumann boundary condition $N'$, 
two 3d $\mathcal{N}=4$ bi-fundamental twisted hypermultiplets 
$\mathbb{T}^{(1)}$, $\mathbb{T}^{(2)}$ 
with the Neumann boundary condition $N$, 
the two diagonal Fermi multiplets $\Gamma^{(1)}$, $\Gamma^{(2)}$ 
and the cross-determinant Fermi multiplet $\Xi$. 
They have charges as follows:
\begin{align}
\label{1212_ch}
\begin{array}{c|c|c|c|c}
&U(1)_{1}&U(1)_{2}&U(2)_{3}&U(2)_{4} \\ \hline 
\textrm{3d $\mathcal{N}=4$ hyper $\mathbb{H}^{(1)}$}&+&0&\overline{\bf 2}&0 \\
\textrm{3d $\mathcal{N}=4$ hyper $\mathbb{H}^{(2)}$}&+&0&0&\overline{\bf 2} \\ \hline
\textrm{3d $\mathcal{N}=4$ twisted hyper $\mathbb{T}^{(1)}$}&+&-&{\bf 1}&{\bf 1} \\
\textrm{3d $\mathcal{N}=4$ twisted hyper $\mathbb{T}^{(2)}$}&0&0&{\bf 2}&\overline{\bf 2} \\  \hline
\textrm{2d Fermi multiplet $\Gamma^{(1)}$}&+&0&0&\overline{\bf 2} \\
\textrm{2d Fermi multiplet $\Gamma^{(2)}$}&0&-&{\bf 2}&0 \\ \hline
\textrm{2d Fermi multiplet $\Xi$}&+&-&\det^{-1}&\det \\
\end{array}
\end{align}

The quarter-index for the $\left(\begin{smallmatrix}
1&2\\1&2\\\end{smallmatrix}\right)$ NS5-NS5$'$ junction is then expresses as 
\begin{align}
\label{NN1212_index}
\mathbb{IV}_{\mathcal{N}\mathcal{N}'}^{
\left(
\begin{smallmatrix}
1&2\\
1&2\\
\end{smallmatrix}
\right)}
&=
\underbrace{
(q)_{\infty}\oint \frac{ds_{1}}{2\pi is_{1}}
}_{\mathbb{IV}_{\mathcal{N}\mathcal{N}'}^{\textrm{4d $U(1)$}}}
\underbrace{
(q)_{\infty}\oint \frac{ds_{2}}{2\pi is_{2}}
}_{\mathbb{IV}_{\mathcal{N}\mathcal{N}'}^{\textrm{4d $U(1)$}}}
%
\underbrace{
\frac12 (q)_{\infty}^2\oint\prod_{i=3}^{4} \frac{ds_{i}}{2\pi is_{i}}
\prod_{i\neq j}\left(\frac{s_{i}}{s_{j}};q\right)_{\infty}
}_{\mathbb{IV}_{\mathcal{N}\mathcal{N}'}^{\textrm{4d $U(2)$}}}
\underbrace{
\frac12 (q)_{\infty}^2\oint\prod_{i=5}^{6} \frac{ds_{i}}{2\pi is_{i}}
\prod_{i\neq j}\left(\frac{s_{i}}{s_{j}};q\right)_{\infty}
}_{\mathbb{IV}_{\mathcal{N}\mathcal{N}'}^{\textrm{4d $U(2)$}}}
\nonumber\\
&\times 
\prod_{i=3}^{4}
\prod_{k=5}^{6}
\underbrace{
\frac{1}
{
\left(q^{\frac14}t s_{1}^{\pm}s_{i}^{\mp};q\right)_{\infty}
}
}_{\mathbb{II}_{N}^{\textrm{3d HM}}\left(\frac{s_{1}}{s_{i}}\right)}
\underbrace{
\frac{1}
{
\left(q^{\frac14}t s_{2}^{\pm}s_{k}^{\mp};q\right)_{\infty}
}
}_{\mathbb{II}_{N}^{\textrm{3d HM}}\left(\frac{s_{2}}{s_{k}}\right)}
\underbrace{
\frac{1}
{
\left(q^{\frac14}t^{-1} s_{1}^{\pm}s_{2}^{\mp};q\right)_{\infty}
}
}_{\mathbb{II}_{N}^{\textrm{3d tHM}}\left(\frac{s_{1}}{s_{2}}\right)}
\underbrace{
\frac{1}
{
\left(q^{\frac14}t^{-1} s_{i}^{\pm}s_{k}^{\mp};q\right)_{\infty}
}
}_{\mathbb{II}_{N}^{\textrm{3d tHM}}\left(\frac{s_{i}}{s_{k}}\right)}
\nonumber\\
&\times 
\prod_{i=3}^{4}\prod_{k=5}^{6}
\underbrace{
\left(q^{\frac12}s_{1}^{\pm}s_{k}^{\mp};q\right)_{\infty}
}_{F\left(q^{\frac12}\frac{s_{1}}{s_{k}}\right)}
\underbrace{
\left(q^{\frac12}s_{2}^{\pm}s_{i}^{\mp};q\right)_{\infty}
}_{F\left(q^{\frac12}\frac{s_{2}}{s_{i}}\right)}
\underbrace{
\left(q^{\frac12}s_{1}^{\pm}s_{5}^{\pm}s_{6}^{\pm} s_{2}^{\mp}s_{3}^{\mp}s_{4}^{\mp}x^{\pm};q\right)_{\infty}
}_{F\left(q^{\frac12}\frac{s_{1}s_{5}s_{6}}{s_{2}s_{3}s_{4}}x\right)}
\nonumber\\
&=1+2(t^{-2}+t^2)q^{\frac12}
+(-1+4t^{-4}+3t^{4})q
-t^{-3}x(1+t^{4})(1+x^2)q^{\frac54}
\nonumber\\
&
+t^{-6}(6-2t^{4}-t^{8}+4t^{12})q^{\frac32}
-t^{-5}x(2+t^4+2t^8)(1+x^2)q^{\frac{7}{4}}+\cdots
\end{align}

The S-dual is the $\left(\begin{smallmatrix}
1&1\\2&2\\\end{smallmatrix}\right)$ D5-D5$'$ junction. 
The initial 4d $\mathcal{N}=4$ $U(2)$ gauge symmetry in the half-space $x^2<0$ is broken to $U(1)$ at the boundary and identified with the
$U(1)$ gauge symmetry in the half-space $x^2>0$. Similarly, we take one of the two 3d $\mathcal{N}=4$ hypermultiplets
introduced by the D5 interface at $x^2<0$ and identify it with the single hypermultiplet at $x^2>0$. The other hypermultiplet 
gets Dirichlet boundary conditions. 

Then the quarter-index for the dual $\left(\begin{smallmatrix}
1&1\\2&2\\\end{smallmatrix}\right)$ D5-D5$'$ junction reads 
\begin{align}
\label{DD1122_index}
\mathbb{IV}_{\mathcal{D}\mathcal{D}'}^{
\left(
\begin{smallmatrix}
1&1\\
2&2\\
\end{smallmatrix}
\right)}
&=
\underbrace{
\frac{(q)_{\infty}^2}
{
\left(q^{\frac12}t^2;q\right)_{\infty}
\left(q^{\frac12}t^{-2};q\right)_{\infty}
}
\oint \frac{ds}{2\pi is}
}_{\mathbb{I}^{\textrm{4d $U(1)$}}}
\underbrace{
\frac{(q)_{\infty}}{(q^{\frac12}t^{-2};q)_{\infty}}
}_{\mathbb{II}_{\mathcal{D}'}^{\textrm{4d $U(1)$}}}
\underbrace{
(q^{\frac34} t^{-1}x;q)_{\infty}
(q^{\frac34} t^{-1}x^{-1};q)_{\infty}
}_{\mathbb{II}_{D}^{\textrm{3d HM}}(x)}
\nonumber\\
&\times 
\underbrace{
\frac{
(q^{\frac34} t^{-1}s;q)_{\infty}
(q^{\frac34} t^{-1}s^{-1};q)_{\infty}
}
{
(q^{\frac14} ts;q)_{\infty}
(q^{\frac14} ts^{-1};q)_{\infty}
}
}_{\mathbb{I}^{\textrm{3d HM}}}
\frac{
\left(q sx^{-1};q\right)_{\infty}
\left(q s^{-1}x;q\right)_{\infty}
}
{
\left(q^{\frac12}t^{-2} sx^{-1};q\right)_{\infty}
\left(q^{\frac12}t^{-2} s^{-1}x;q\right)_{\infty}
}
\nonumber\\
&=1+2(t^{-2}+t^2)q^{\frac12}
+(-1+4t^{-4}+3t^{4})q
-t^{-3}x(1+t^{4})(1+x^2)q^{\frac54}
\nonumber\\
&
+t^{-6}(6-2t^{4}-t^{8}+4t^{12})q^{\frac32}
-t^{-5}x(2+t^4+2t^8)(1+x^2)q^{\frac{7}{4}}+\cdots
\end{align}

We see that the quarter-index (\ref{NN1212_index}) for the $\left(\begin{smallmatrix} 
1&2\\1&2\\\end{smallmatrix}\right)$ NS5-NS5$'$ junction 
beautifully matches with the quarter-index (\ref{DD1122_index}) for the dual $\left(\begin{smallmatrix}
1&1\\2&2\\\end{smallmatrix}\right)$ D5-D5$'$ junction.

The quarter-index (\ref{DD1122_index}) can be evaluated 
as the sum of residues at poles $s=q^{\frac14+n}t$ and $s=q^{\frac12+n}t^{-2}x$
\begin{align}
\label{dd1122_sum}
&\mathbb{IV}_{\mathcal{D}\mathcal{D}'}^{\left(\begin{smallmatrix}
1&1\\2&2\\\end{smallmatrix}\right)}
=
\frac{(q)_{\infty}}{(q^{\frac12} t^{-2};q)_{\infty}}
(q^{\frac34}t^{-1}x^{\pm};q)_{\infty}
\cdot 
\underbrace{
(q^{\frac34}t^{-1} x;q)_{\infty}
(q^{\frac14}t x^{-1};q)_{\infty}
}_{F(q^{\frac34}t^{-1} x)}
\nonumber\\
&\times 
\Biggl[
\frac{1}
{
(q^{\frac14}t^{-3}x;q)_{\infty}
(q^{\frac34}t^3x^{-1};q)_{\infty}
}
\sum_{m=0}^{\infty}
\frac{
(q^{1+m};q)_{\infty}^2
(q^{\frac54+m}t x^{-1};q)_{\infty}
(q^{\frac34+m}t^3 x^{-1};q)_{\infty}
}
{
(q^{\frac12+m}t^2;q)_{\infty}^2
(q^{\frac34+m}t^{-1}x^{-1};q)_{\infty}
(q^{\frac14+m}tx^{-1};q)_{\infty}
}
q^{m}
\nonumber\\
&+
\frac{1}
{(q^{\frac54}t^{-3}x;q)_{\infty}
(q^{-\frac14}t^3x^{-1};q)_{\infty}}
\sum_{m=0}^{\infty}
\frac{
(q^{1+m};q)_{\infty}
(q^{\frac32+m}t^{-2};q)_{\infty}
(q^{\frac54+m}t^{-3}x;q)_{\infty}^{2}
}
{
(q^{\frac12+m}t^{-2};q)_{\infty}
(q^{1+m}t^{-4};q)_{\infty}
(q^{\frac34+m}t^{-1}x;q)_{\infty}^2
}
q^{m}
\Biggr]. 
\end{align}

\subsubsection{${1 \, 3 \choose 1 \, 3}$ and ${1 \, 1 \choose 3 \, 3}$}

The next example is the $\left(\begin{smallmatrix}
1&3\\1&3\\\end{smallmatrix}\right)$ NS5-NS5$'$ junction 
and $\left(\begin{smallmatrix}
1&1\\3&3\\\end{smallmatrix}\right)$ D5-D5$'$ junction. 
The $\left(\begin{smallmatrix}
1&3\\1&3\\\end{smallmatrix}\right)$ NS5-NS5$'$ junction has 
four 
4d $\mathcal{N}=4$ $U(1)_{1}$, $U(1)_{2}$, $U(3)_{3}$ and $U(3)_{4}$ gauge theories 
which live in upper left, lower left, upper right and lower right quadrants respectively. 
The junction contains the two 3d $\mathcal{N}=4$ bi-fundamental hypermultiplets 
with the Neumann boundary condition $N'$, 
two 3d $\mathcal{N}=4$ bi-fundamental twisted hypermultiplets 
with the Neumann boundary condition $N$, 
the two diagonal Fermi multiplets 
and the cross-determinant Fermi multiplet. 

The quarter-index of the $\left(\begin{smallmatrix}
1&3\\1&3\\\end{smallmatrix}\right)$ NS5-NS5$'$ junction is expressed as
\begin{align}
\label{NN1313}
\mathbb{IV}_{\mathcal{N}\mathcal{N}'}^{
\left(
\begin{smallmatrix}
1&3\\
1&3\\
\end{smallmatrix}
\right)}
&=
\underbrace{
(q)_{\infty}\oint \frac{ds_{1}}{2\pi is_{1}}
}_{\mathbb{IV}_{\mathcal{N}\mathcal{N}'}^{\textrm{4d $U(1)$}}}
\underbrace{
(q)_{\infty}\oint \frac{ds_{2}}{2\pi is_{2}}
}_{\mathbb{IV}_{\mathcal{N}\mathcal{N}'}^{\textrm{4d $U(1)$}}}
\underbrace{
\frac{1}{3!} (q)_{\infty}^3\oint\prod_{i=3}^{5} \frac{ds_{i}}{2\pi is_{i}}
\prod_{i\neq j}\left(\frac{s_{i}}{s_{j}};q\right)_{\infty}
}_{\mathbb{IV}_{\mathcal{N}\mathcal{N}'}^{\textrm{4d $U(3)$}}}
\underbrace{
\frac{1}{3!} (q)_{\infty}^3\oint\prod_{i=6}^{8} \frac{ds_{i}}{2\pi is_{i}}
\prod_{i\neq j}\left(\frac{s_{i}}{s_{j}};q\right)_{\infty}
}_{\mathbb{IV}_{\mathcal{N}\mathcal{N}'}^{\textrm{4d $U(3)$}}}
\nonumber\\
&\times 
\prod_{i=3}^{5}
\prod_{k=6}^{8}
\underbrace{
\frac{1}
{
\left(q^{\frac14}t s_{1}^{\pm}s_{i}^{\mp};q\right)_{\infty}
}
}_{\mathbb{II}_{N}^{\textrm{3d HM}}\left(\frac{s_{1}}{s_{i}}\right)}
\underbrace{
\frac{1}
{
\left(q^{\frac14}t s_{2}^{\pm}s_{k}^{\mp};q\right)_{\infty}
}
}_{\mathbb{II}_{N}^{\textrm{3d HM}}\left(\frac{s_{2}}{s_{k}}\right)}
\underbrace{
\frac{1}
{
\left(q^{\frac14}t^{-1} s_{1}^{\pm}s_{2}^{\mp};q\right)_{\infty}
}
}_{\mathbb{II}_{N}^{\textrm{3d tHM}}\left(\frac{s_{1}}{s_{2}}\right)}
\underbrace{
\frac{1}
{
\left(q^{\frac14}t^{-1} s_{i}^{\pm}s_{k}^{\mp};q\right)_{\infty}
}
}_{\mathbb{II}_{N}^{\textrm{3d tHM}}\left(\frac{s_{i}}{s_{k}}\right)}
\nonumber\\
&\times 
\prod_{i=3}^{5}\prod_{k=6}^{8}
\underbrace{
\left(q^{\frac12}s_{1}^{\pm}s_{k}^{\mp};q\right)_{\infty}
}_{F\left(q^{\frac12}\frac{s_{1}}{s_{k}}\right)}
\underbrace{
\left(q^{\frac12}s_{2}^{\pm}s_{i}^{\mp};q\right)_{\infty}
}_{F\left(q^{\frac12}\frac{s_{2}}{s_{i}}\right)}
\underbrace{
\left(q^{\frac12} s_{1}^{\pm} \prod_{k=6}^{8}s_{k}^{\pm} s_{2}^{\mp} \prod_{i=3}^{5}s_{i}^{\mp}x^{\pm}
;q\right)_{\infty}
}_{F\left(q^{\frac12}\frac{s_{1}\prod s_{k}}{s_{2}\prod s_{i}}x\right)}
\nonumber\\
&
=1+2(t^{-2}+t^2)q^{\frac12}+(-1+4t^{-4}+3t^4)q+(7t^{-6}-t^2+4t^6-x^{-1}-x-t^{-4}x^{-1}-t^{-4}x)q^{\frac32}+\cdots
\end{align}

The S-dual is the 
$\left(\begin{smallmatrix}
1&1\\3&3\\\end{smallmatrix}\right)$ D5-D5$'$ junction. 
We can compute the quarter-index of the $\left(\begin{smallmatrix}
1&1\\3&3\\\end{smallmatrix}\right)$ D5-D5$'$ junction as 
\begin{align}
\label{DD1133_index}
\mathbb{IV}_{\mathcal{D}\mathcal{D}'}^{
\left(
\begin{smallmatrix}
1&1\\
3&3\\
\end{smallmatrix}
\right)}
&=
\underbrace{
\frac{(q)_{\infty}^2}
{
\left(q^{\frac12}t^2;q\right)_{\infty}
\left(q^{\frac12}t^{-2};q\right)_{\infty}
}
\oint \frac{ds}{2\pi is}
}_{\mathbb{I}^{\textrm{4d $U(1)$}}}
\underbrace{
\frac{(q)_{\infty} (q^{\frac32}t^{-2};q)_{\infty}}
{(q^{\frac12}t^{-2};q)_{\infty} (qt^{-4};q)_{\infty}}
}_{\mathbb{II}_{\textrm{Nahm}'}^{\textrm{4d $U(2)$}}}
(q t^{-2}x;q)_{\infty}
(q t^{-2}x^{-1};q)_{\infty}
\nonumber\\
&\times 
\underbrace{
\frac{
(q^{\frac34} t^{-1}s;q)_{\infty}
(q^{\frac34} t^{-1}s^{-1};q)_{\infty}
}
{
(q^{\frac14} ts;q)_{\infty}
(q^{\frac14} ts^{-1};q)_{\infty}
}
}_{\mathbb{I}^{\textrm{3d HM}}}
\frac{
\left(q^{\frac54} t^{-1} sx^{-1};q\right)_{\infty}
\left(q^{\frac54} t^{-1} s^{-1}x;q\right)_{\infty}
}
{
\left(q^{\frac34}t^{-3} sx^{-1};q\right)_{\infty}
\left(q^{\frac34}t^{-3} s^{-1}x;q\right)_{\infty}
}
\nonumber\\
&
=1+2(t^{-2}+t^2)q^{\frac12}+(-1+4t^{-4}+3t^4)q+(7t^{-6}-t^2+4t^6-x^{-1}-x-t^{-4}x^{-1}-t^{-4}x)q^{\frac32}+\cdots
\end{align}
We have numerically checked that the quarter-index 
(\ref{NN1313}) matches with the quarter-index (\ref{DD1133_index}), as expected.

\subsubsection{${1 \, N \choose 1 \, N}$ and ${1\  \, 1 \choose N \, N}$}
\label{sec_2dnn1N1N}
The quarter-index of the $\left(\begin{smallmatrix}
1&N\\1&N\\\end{smallmatrix}\right)$ NS5-NS5$'$ junction is 
\begin{align}
\label{NN1N1N}
\mathbb{IV}_{\mathcal{N}\mathcal{N}'}^{
\left(
\begin{smallmatrix}
1&N\\
1&N\\
\end{smallmatrix}
\right)}
&=
\underbrace{
(q)_{\infty}\oint \frac{ds_{1}}{2\pi is_{1}}
}_{\mathbb{IV}_{\mathcal{N}\mathcal{N}'}^{\textrm{4d $U(1)$}}}
\underbrace{
(q)_{\infty}\oint \frac{ds_{2}}{2\pi is_{2}}
}_{\mathbb{IV}_{\mathcal{N}\mathcal{N}'}^{\textrm{4d $U(1)$}}}
\underbrace{
\frac{1}{N!} (q)_{\infty}^N\oint\prod_{i=3}^{2+N} \frac{ds_{i}}{2\pi is_{i}}
\prod_{i\neq j}\left(\frac{s_{i}}{s_{j}};q\right)_{\infty}
}_{\mathbb{IV}_{\mathcal{N}\mathcal{N}'}^{\textrm{4d $U(N)$}}}
\underbrace{
\frac{1}{N!} (q)_{\infty}^N\oint\prod_{i=3+N}^{2+2N} \frac{ds_{i}}{2\pi is_{i}}
\prod_{i\neq j}\left(\frac{s_{i}}{s_{j}};q\right)_{\infty}
}_{\mathbb{IV}_{\mathcal{N}\mathcal{N}'}^{\textrm{4d $U(N)$}}}
\nonumber\\
&\times 
\prod_{i=3}^{2+N}
\prod_{k=3+N}^{2+2N}
\underbrace{
\frac{1}
{
\left(q^{\frac14}t s_{1}^{\pm}s_{i}^{\mp};q\right)_{\infty}
}
}_{\mathbb{II}_{N}^{\textrm{3d HM}}\left(\frac{s_{1}}{s_{i}}\right)}
\underbrace{
\frac{1}
{
\left(q^{\frac14}t s_{2}^{\pm}s_{k}^{\mp};q\right)_{\infty}
}
}_{\mathbb{II}_{N}^{\textrm{3d HM}}\left(\frac{s_{2}}{s_{k}}\right)}
\underbrace{
\frac{1}
{
\left(q^{\frac14}t^{-1} s_{1}^{\pm}s_{2}^{\mp};q\right)_{\infty}
}
}_{\mathbb{II}_{N}^{\textrm{3d tHM}}\left(\frac{s_{1}}{s_{2}}\right)}
\underbrace{
\frac{1}
{
\left(q^{\frac14}t^{-1} s_{i}^{\pm}s_{k}^{\mp};q\right)_{\infty}
}
}_{\mathbb{II}_{N}^{\textrm{3d tHM}}\left(\frac{s_{i}}{s_{k}}\right)}
\nonumber\\
&\times 
\prod_{i=3}^{2+N}\prod_{k=3+N}^{2+2N}
\underbrace{
\left(q^{\frac12}s_{1}^{\pm}s_{k}^{\mp};q\right)_{\infty}
}_{F\left(q^{\frac12}\frac{s_{1}}{s_{k}}\right)}
\underbrace{
\left(q^{\frac12}s_{2}^{\pm}s_{i}^{\mp};q\right)_{\infty}
}_{F\left(q^{\frac12}\frac{s_{2}}{s_{i}}\right)}
\underbrace{
\left(q^{\frac12} s_{1}^{\pm} \prod_{k=3+N}^{2+2N}s_{k}^{\pm} s_{2}^{\mp} \prod_{i=3}^{2+N}s_{i}^{\mp}x^{\pm}
;q\right)_{\infty}
}_{F\left(q^{\frac12}\frac{s_{1}\prod s_{k}}{s_{2}\prod s_{i}}x\right)}. 
\end{align}

The quarter-index for the S-dual 
$\left(\begin{smallmatrix}
1&1\\N&N\\\end{smallmatrix}\right)$ D5-D5$'$ junction is expressed as the contour integral form:
\begin{align}
\label{dd11NN}
\mathbb{IV}_{\mathcal{D}\mathcal{D}'}^{\left(\begin{smallmatrix}
1&1\\N&N\\\end{smallmatrix}\right)}
&=
\underbrace{
\frac{(q)_{\infty}^2}
{(q^{\frac12}t^2;q)_{\infty}
(q^{\frac12}t^{-2};q)_{\infty}}\oint \frac{ds}{2\pi is}
}_{\mathbb{I}^{\textrm{4d $U(1)$}}}
\underbrace{
\prod_{k=1}^{N-1}
\frac
{(q^{\frac{k+1}{2}} t^{-2(k-1)};q)_{\infty}}
{(q^{\frac{k}{2}} t^{-2k};q)_{\infty}}
}_{\mathbb{II}_{\textrm{Nahm}'}^{\textrm{4d $U(N-1)$}}}
\left( q^{\frac12+\frac{N-1}{4}}t^{-(N-1)}x^{\pm};q \right)_{\infty}
\nonumber\\
&\times 
\underbrace{
\frac{
(q^{\frac34}t^{-1}s;q)_{\infty}
(q^{\frac34}t^{-1}s^{-1};q)_{\infty}
}
{
(q^{\frac14}ts;q)_{\infty}
(q^{\frac14}ts^{-1};q)_{\infty}
}
}_{\mathbb{I}^{\textrm{3d HM}}(s)}
\frac{
\left(q^{\frac34+\frac{N-1}{4}}t^{1-(N-1)}s^{\pm}x^{\mp};q\right)_{\infty}
}
{
\left(q^{\frac14+\frac{N-1}{4}}t^{-1-(N-1)}s^{\pm}x^{\mp};q\right)_{\infty}
}.
\end{align}
We expect that the quarter-indices (\ref{NN1N1N}) and (\ref{dd11NN}) coincide. 
We can evaluate the integral in (\ref{dd11NN}) by taking 
the sum of the residues of the integrand at two sets of poles $s=q^{\frac14+m}t$ and $s=q^{\frac{N}{4}+m}t^{-N}$. 
We obtain
\begin{align}
\label{dd11NN_sum}
&\mathbb{IV}_{\mathcal{D}\mathcal{D}'}^{\left(\begin{smallmatrix}
1&1\\N&N\\\end{smallmatrix}\right)}
=
\prod_{k=1}^{N-1}
\frac
{(q^{\frac{k+1}{2}} t^{-2(k-1)};q)_{\infty}}
{(q^{\frac{k}{2}} t^{-2k};q)_{\infty}}
\left( q^{\frac12+\frac{N-1}{4}}t^{-(N-1)}x^{\pm};q \right)_{\infty}
\underbrace{
(q^{\frac14+\frac{N}{4}} t^{1-N}x;q )_{\infty}
( q^{\frac34-\frac{N}{4}} t^{N-1}x^{-1};q )_{\infty}
}
_{F\left(
q^{\frac14+\frac{N}{4}} t^{1-N}x
\right)}
\nonumber\\
&\times 
\Biggl[
\frac{1}
{( q^{\frac{N-1}{4}}t^{-N-1}x;q )_{\infty}
( q^{\frac{5-N}{4}}t^{N+1}x^{-1};q )_{\infty}
}
\sum_{m=0}^{\infty}
\frac{
(q^{1+m};q)_{\infty}^2 
(q^{\frac{N+3}{4}}t^{3-N}x^{-1};q)_{\infty}
(q^{\frac{5-N}{4}+m}t^{N+1}x^{-1};q)_{\infty}
}
{
(q^{\frac12+m}t^2;q)_{\infty}^2 
(q^{\frac{N+1}{4}+m} t^{1-N}x^{-1};q)_{\infty}
(q^{\frac{3-N}{4}+m}t^{N-1}x^{-1};q)_{\infty}
}
q^{m}
\nonumber\\
&+
\frac{1}
{
(q^{\frac{1-N}{4}} t^{N+1}x^{-1};q)_{\infty}
(q^{\frac{3+N}{4}} t^{-N-1}x;q)_{\infty}
}
\sum_{m=0}^{\infty}
\frac{
(q^{1+m};q)_{\infty}
(q^{\frac{N+1}{2}+m} t^{2-2N};q)_{\infty}
(q^{\frac{N+3}{4}+m}t^{-N-1}x;q)_{\infty}^2
}
{
(q^{\frac12+m}t^{-2};q)_{\infty}
(q^{\frac{N}{2}+m}t^{-2N};q)_{\infty}
(q^{\frac{N+1}{4}+m}t^{-N+1}x;q)_{\infty}^2
}
q^{m}
\Biggr]. 
\end{align}
This generalizes the formula (\ref{NN1111_series}) and (\ref{dd1122_sum}). 

Inserting a Wilson line $\mathcal{W}_{n}$ of charge $n$ modifies 
the quarter-index (\ref{dd11NN_sum}) for the $\left(\begin{smallmatrix}
1&1\\N&N\\\end{smallmatrix}\right)$ D5-D5$'$ junction as 
\begin{align}
\label{dd11NNw}
\mathbb{IV}_{\mathcal{D}\mathcal{D}'+\mathcal{W}_{n}}^{\left(\begin{smallmatrix}
1&1\\N&N\\\end{smallmatrix}\right)}
&=
\underbrace{
\frac{(q)_{\infty}^2}
{(q^{\frac12}t^2;q)_{\infty}
(q^{\frac12}t^{-2};q)_{\infty}}\oint \frac{ds}{2\pi is}
}_{\mathbb{I}^{\textrm{4d $U(1)$}}}
\underbrace{
\prod_{k=1}^{N-1}
\frac
{(q^{\frac{k+1}{2}} t^{-2(k-1)};q)_{\infty}}
{(q^{\frac{k}{2}} t^{-2k};q)_{\infty}}
}_{\mathbb{II}_{\textrm{Nahm'}}^{\textrm{4d $U(N-1)$}}}
\left( q^{\frac12+\frac{N-1}{4}}t^{-(N-1)}x^{\pm};q \right)_{\infty}
\nonumber\\
&\times 
\underbrace{
\frac{
(q^{\frac34}t^{-1}s;q)_{\infty}
(q^{\frac34}t^{-1}s^{-1};q)_{\infty}
}
{
(q^{\frac14}ts;q)_{\infty}
(q^{\frac14}ts^{-1};q)_{\infty}
}
}_{\mathbb{I}^{\textrm{3d HM}}(s)}
\frac{
\left(q^{\frac34+\frac{N-1}{4}}t^{1-(N-1)}s^{\pm}x^{\mp};q\right)_{\infty}
}
{
\left(q^{\frac14+\frac{N-1}{4}}t^{-1-(N-1)}s^{\pm}x^{\mp};q\right)_{\infty}
}
s^{n}. 
\end{align}

Again we can evaluate the contour integral (\ref{dd11NNw}) 
by summing over the residues at poles $s=q^{\frac14+m}t$ and $s=q^{\frac{N}{4}+m}t^{-N}$, 
we find that 
\begin{align}
\label{dd11NNw_sum}
&\mathbb{IV}_{\mathcal{D}\mathcal{D}'+\mathcal{W}_{n}}^{\left(\begin{smallmatrix}
1&1\\N&N\\\end{smallmatrix}\right)}
=
\prod_{k=1}^{N-1}
\frac
{(q^{\frac{k+1}{2}} t^{-2(k-1)};q)_{\infty}}
{(q^{\frac{k}{2}} t^{-2k};q)_{\infty}}
\left( q^{\frac12+\frac{N-1}{4}}t^{-(N-1)}x^{\pm};q \right)_{\infty}
\underbrace{
(q^{\frac14+\frac{N}{4}} t^{1-N}x;q )_{\infty}
( q^{\frac34-\frac{N}{4}} t^{N-1}x^{-1};q )_{\infty}
}
_{F\left(
q^{\frac14+\frac{N}{4}} t^{1-N}x
\right)}
\nonumber\\
&\times 
\sum_{m=0}^{\infty}
\Biggl[
\frac{q^{(1+n)m+\frac{n}{4}} 
t^{n}}
{( q^{\frac{N-1}{4}}t^{-N-1}x;q )_{\infty}
( q^{\frac{5-N}{4}}t^{N+1}x^{-1};q )_{\infty}
}
\frac{
(q^{1+m};q)_{\infty}^2 
(q^{\frac{N+3}{4}}t^{3-N}x^{-1};q)_{\infty}
(q^{\frac{5-N}{4}+m}t^{N+1}x^{-1};q)_{\infty}
}
{
(q^{\frac12+m}t^2;q)_{\infty}^2 
(q^{\frac{N+1}{4}+m} t^{1-N}x^{-1};q)_{\infty}
(q^{\frac{3-N}{4}+m}t^{N-1}x^{-1};q)_{\infty}
}
\nonumber\\
&+
\frac{q^{(1+n)m+\frac{Nn}{4}} 
t^{-Nn}x^{n}}
{
(q^{\frac{1-N}{4}} t^{N+1}x^{-1};q)_{\infty}
(q^{\frac{3+N}{4}} t^{-N-1}x;q)_{\infty}
}
\sum_{m=0}^{\infty}
\frac{
(q^{1+m};q)_{\infty}
(q^{\frac{N+1}{2}+m} t^{2-2N};q)_{\infty}
(q^{\frac{N+3}{4}+m}t^{-N-1}x;q)_{\infty}^2
}
{
(q^{\frac12+m}t^{-2};q)_{\infty}
(q^{\frac{N}{2}+m}t^{-2N};q)_{\infty}
(q^{\frac{N+1}{4}+m}t^{-N+1}x;q)_{\infty}^2
}
\Biggr].
\end{align}
It would be interesting to explore further dualities with the insertion of the line operators, 
but we defer the relevant problems to future work.

\subsubsection{${2 \, 3 \choose 2 \, 3}$ and ${2 \, 2 \choose 3 \, 3}$}

Next, we compare the  
$\left(\begin{smallmatrix}
2&3\\2&3\\\end{smallmatrix}\right)$ NS5-NS5$'$ junction 
and the 
$\left(\begin{smallmatrix}
2&2\\3&3\\\end{smallmatrix}\right)$ D5-D5$'$ junction. 
This is the simplest non-Abelian example. 
For the $\left(\begin{smallmatrix}
2&3\\2&3\\\end{smallmatrix}\right)$ NS5-NS5$'$ junction 
there are four
4d $\mathcal{N}=4$ $U(2)_{1}$, $U(2)_{2}$, $U(3)_{3}$ and $U(3)_{4}$ gauge theories 
at upper left, lower left, upper right and lower right quadrants respectively. 
The matter content is the two 3d $\mathcal{N}=4$ bi-fundamental hypermultiplets 
with the Neumann boundary condition $N'$, 
two 3d $\mathcal{N}=4$ bi-fundamental twisted hypermultiplets 
with the Neumann boundary condition $N$, 
the two diagonal Fermi multiplets 
and the cross-determinant Fermi multiplet. 

The quarter-index of the $\left(\begin{smallmatrix}
2&3\\2&3\\\end{smallmatrix}\right)$ NS5-NS5$'$ junction is
\begin{align}
\label{NN2323}
\mathbb{IV}_{\mathcal{N}\mathcal{N}'}^{
\left(
\begin{smallmatrix}
2&3\\
2&3\\
\end{smallmatrix}
\right)}
&=
\underbrace{
\frac{1}{2}(q)_{\infty}^{2} \oint\prod_{i=1}^{2} \frac{ds_{i}}{2\pi is_{i}} 
\prod_{i\neq j}\left(\frac{s_{i}}{s_{j}};q\right)_{\infty}
}_{\mathbb{IV}_{\mathcal{N}\mathcal{N}'}^{\textrm{4d $U(2)$}}}
\underbrace{
\frac{1}{2}(q)_{\infty}^{2} \oint\prod_{i=3}^{4} \frac{ds_{i}}{2\pi is_{i}} 
\prod_{i\neq j}\left(\frac{s_{i}}{s_{j}};q\right)_{\infty}
}_{\mathbb{IV}_{\mathcal{N}\mathcal{N}'}^{\textrm{4d $U(2)$}}}
\nonumber\\
&\times 
\underbrace{
\frac{1}{3!}(q)_{\infty}^{3} \oint\prod_{i=5}^{7} \frac{ds_{i}}{2\pi is_{i}} 
\prod_{i\neq j}\left(\frac{s_{i}}{s_{j}};q\right)_{\infty}
}_{\mathbb{IV}_{\mathcal{N}\mathcal{N}'}^{\textrm{4d $U(3)$}}}
\underbrace{
\frac{1}{3!} (q)_{\infty}^3 \oint\prod_{i=8}^{10} \frac{ds_{i}}{2\pi is_{i}}
\prod_{i\neq j}\left(\frac{s_{i}}{s_{j}};q\right)_{\infty}
}_{\mathbb{IV}_{\mathcal{N}\mathcal{N}'}^{\textrm{4d $U(3)$}}}
\nonumber\\
&\times 
\prod_{i=1}^{2}
\prod_{j=3}^{4}
\prod_{k=5}^{7}
\prod_{l=8}^{10}
\underbrace{
\frac{1}
{
\left(q^{\frac14}t s_{i}^{\pm}s_{k}^{\mp};q\right)_{\infty}
}
}_{\mathbb{II}_{N}^{\textrm{3d HM}}\left(\frac{s_{i}}{s_{k}}\right)}
\underbrace{
\frac{1}
{
\left(q^{\frac14}t s_{j}^{\pm}s_{l}^{\mp};q\right)_{\infty}
}
}_{\mathbb{II}_{N}^{\textrm{3d HM}}\left(\frac{s_{j}}{s_{l}}\right)}
\underbrace{
\frac{1}
{
\left(q^{\frac14}t^{-1} s_{i}^{\pm}s_{j}^{\mp};q\right)_{\infty}
}
}_{\mathbb{II}_{N}^{\textrm{3d tHM}}\left(\frac{s_{i}}{s_{j}}\right)}
\underbrace{
\frac{1}
{
\left(q^{\frac14}t^{-1} s_{k}^{\pm}s_{l}^{\mp};q\right)_{\infty}
}
}_{\mathbb{II}_{N}^{\textrm{3d tHM}}\left(\frac{s_{k}}{s_{l}}\right)}
\nonumber\\
&\times 
\
\prod_{i=1}^{2}
\prod_{j=3}^{4}
\prod_{k=5}^{7}
\prod_{l=8}^{10}
\underbrace{
\left(q^{\frac12}s_{i}^{\pm}s_{l}^{\mp};q\right)_{\infty}
}_{F\left(q^{\frac12}\frac{s_{i}}{s_{l}}\right)}
\underbrace{
\left(q^{\frac12}s_{j}^{\pm}s_{k}^{\mp};q\right)_{\infty}
}_{F\left(q^{\frac12}\frac{s_{j}}{s_{k}}\right)}
\underbrace{
\left(q^{\frac12} \prod s_{i}^{\pm} \prod_{k} s_{l}^{\pm} \prod s_{j}^{\mp} \prod_{k} s_{k}^{\mp}x^{\pm}
;q\right)_{\infty}
}_{F\left(q^{\frac12}\frac{\prod s_{i}\prod s_{l}}{\prod s_{j}\prod s_{k}}x\right)}
\nonumber\\
&=
1+2(t^{-2}+t^{2})q^{\frac12}
+(1+5t^{-4}+5t^4)q
+t^{-6}(9+2t^4+2t^8+8t^{12})q^{\frac32}+\cdots
\end{align}

The quarter-index for the $\left(\begin{smallmatrix}
1&1\\3&3\\\end{smallmatrix}\right)$ D5-D5$'$ junction reads
\begin{align}
\label{DD2233}
\mathbb{IV}_{\mathcal{D}\mathcal{D}'}^{
\left(
\begin{smallmatrix}
2&2\\
3&3\\
\end{smallmatrix}
\right)}
&=
\underbrace{
\frac{1}{2}
\frac{(q)_{\infty}^{4}}
{
\left(q^{\frac12}t^2;q\right)_{\infty}^{2}
\left(q^{\frac12}t^{-2};q\right)_{\infty}^{2}
}
\oint \prod_{i=1}^{2} \frac{ds_{i}}{2\pi is_{i}} 
\prod_{i\neq j} 
\frac{\left(\frac{s_{i}}{s_{j}};q\right)_{\infty} \left(q \frac{s_{i}}{s_{j}};q\right)_{\infty}}
{\left(q^{\frac12}t^2 \frac{s_{i}}{s_{j}};q\right)_{\infty}
\left(q^{\frac12}t^{-2} \frac{s_{i}}{s_{j}};q\right)_{\infty}
}
}_{\mathbb{I}^{\textrm{4d $U(2)$}}}
\nonumber\\
&\times 
\underbrace{
\frac{(q)_{\infty}}{(q^{\frac12}t^{-2};q)_{\infty}}
}_{\mathbb{II}_{\mathcal{D}'}^{\textrm{4d $U(1)$}}}
\underbrace{
\left(q^{\frac34}t^{-1}x;q\right)_{\infty}
\left(q^{\frac34}t^{-1}x^{-1};q\right)_{\infty}
}_{\mathbb{II}_{D}^{\textrm{3d HM}}(x)}
\nonumber\\
&\times 
\prod_{i=1}^{2}
\underbrace{
\frac{
(q^{\frac34} t^{-1}s_{i};q)_{\infty}
(q^{\frac34} t^{-1}s_{i}^{-1};q)_{\infty}
}
{
(q^{\frac14} ts_{i};q)_{\infty}
(q^{\frac14} ts_{i}^{-1};q)_{\infty}
}
}_{\mathbb{I}^{\textrm{3d HM}}(s_{i})}
\frac{
\left(q s_{i}x^{-1};q\right)_{\infty}
\left(q s_{i}^{-1}x;q\right)_{\infty}
}
{
\left(q^{\frac12} t^{-2} s_{i} x^{-1};q\right)_{\infty}
\left(q^{\frac12} t^{-2} s_{i}^{-1}x;q\right)_{\infty}
}
\nonumber\\
&=
1+2(t^{-2}+t^{2})q^{\frac12}
+(1+5t^{-4}+5t^4)q
+t^{-6}(9+2t^4+2t^8+8t^{12})q^{\frac32}+\cdots
\end{align}

The quarter-index (\ref{NN2323}) for the 
$\left(\begin{smallmatrix}
2&3\\2&3\\\end{smallmatrix}\right)$ NS5-NS5$'$ junction 
and the quarter-index (\ref{DD2233}) for the 
$\left(\begin{smallmatrix}
2&2\\3&3\\\end{smallmatrix}\right)$ D5-D5$'$ junction appear to coincide.

\subsubsection{${N \, M \choose N \, M}$ and ${N \, N \choose M \, M}$}
Now we would like to propose the generalization for the 
duality between the $\left(\begin{smallmatrix}
N&M\\N&M\\\end{smallmatrix}\right)$ NS5-NS5$'$ junction 
and the 
$\left(\begin{smallmatrix}
N&N\\M&M\\\end{smallmatrix}\right)$ D5-D5$'$ junction. 

\begin{figure}
\begin{center}
\includegraphics[width=10cm]{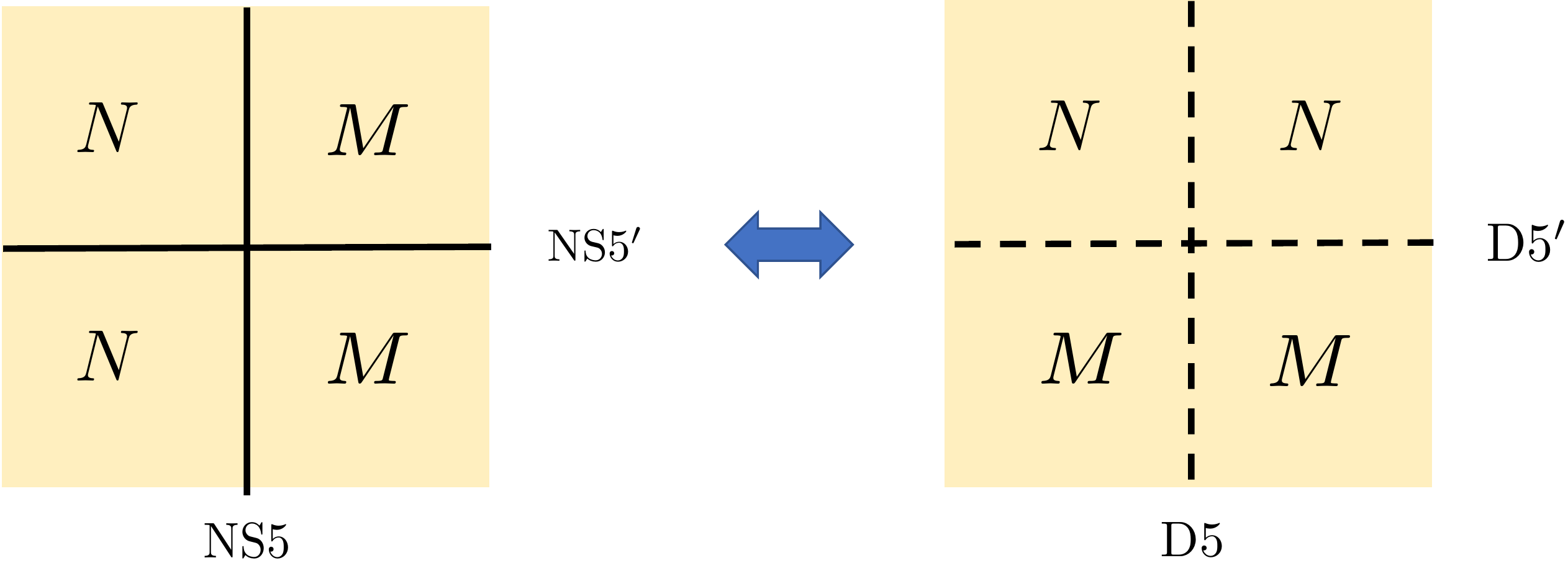}
\caption{
The ${N \, M \choose N \, M}$ NS5-NS5$'$ junction
and ${N \, N \choose M \, M}$ D5-D5$'$ junction 
with four quadrants of D3-branes.
}
\label{fignnNNMM}
\end{center}
\end{figure}

In the $\left(\begin{smallmatrix}
N&M\\N&M\\\end{smallmatrix}\right)$ NS5-NS5$'$ junction, 
we have four 
4d $\mathcal{N}=4$ $U(N)_{1}$, $U(N)_{2}$, $U(M)_{3}$ and $U(M)_{4}$ gauge theories 
with the boundary conditions $\mathcal{N}$ and $\mathcal{N}'$ 
at the upper left, lower left, upper right and lower right quadrants respectively. 
The matter content is 
the two 3d $\mathcal{N}=4$ bi-fundamental hypermultiplets 
with the Neumann boundary condition $N'$, 
two 3d $\mathcal{N}=4$ bi-fundamental twisted hypermultiplets 
with the Neumann boundary condition $N$, 
the two diagonal Fermi multiplets and the cross-determinant Fermi multiplet. 

We can compute the quarter-index for 
the $\left(\begin{smallmatrix}
N&M\\N&M\\\end{smallmatrix}\right)$ NS5-NS5$'$ junction as
\begin{align}
\label{NNNMNM}
\mathbb{IV}_{\mathcal{N}\mathcal{N}'}^{
\left(
\begin{smallmatrix}
N&M\\
N&M\\
\end{smallmatrix}
\right)}
&=
\underbrace{
\frac{1}{N!}(q)_{\infty}^{N} \oint\prod_{i=1}^{N} \frac{ds_{i}}{2\pi is_{i}} 
\prod_{i\neq j}\left(\frac{s_{i}}{s_{j}};q\right)_{\infty}
}_{\mathbb{IV}_{\mathcal{N}\mathcal{N}'}^{\textrm{4d $U(N)$}}}
\underbrace{
\frac{1}{N!}(q)_{\infty}^{N} \oint\prod_{i=N+1}^{2N} \frac{ds_{i}}{2\pi is_{i}} 
\prod_{i\neq j}\left(\frac{s_{i}}{s_{j}};q\right)_{\infty}
}_{\mathbb{IV}_{\mathcal{N}\mathcal{N}'}^{\textrm{4d $U(N)$}}}
\nonumber\\
&\times 
\underbrace{
\frac{1}{M!}(q)_{\infty}^{M} \oint\prod_{i=2N+1}^{2N+M} \frac{ds_{i}}{2\pi is_{i}} 
\prod_{i\neq j}\left(\frac{s_{i}}{s_{j}};q\right)_{\infty}
}_{\mathbb{IV}_{\mathcal{N}\mathcal{N}'}^{\textrm{4d $U(M)$}}}
\underbrace{
\frac{1}{M!} (q)_{\infty}^M \oint\prod_{i=2N+M+1}^{2N+2M} \frac{ds_{i}}{2\pi is_{i}}
\prod_{i\neq j}\left(\frac{s_{i}}{s_{j}};q\right)_{\infty}
}_{\mathbb{IV}_{\mathcal{N}\mathcal{N}'}^{\textrm{4d $U(M)$}}}
\nonumber\\
&\times 
\prod_{i=1}^{N}
\prod_{j=N+1}^{2N}
\prod_{k=2N+1}^{2N+M}
\prod_{l=2N+M+1}^{2N+2M}
\underbrace{
\frac{1}
{
\left(q^{\frac14}t s_{i}^{\pm}s_{k}^{\mp};q\right)_{\infty}
}
}_{\mathbb{II}_{N}^{\textrm{3d HM}}\left(\frac{s_{i}}{s_{k}}\right)}
\underbrace{
\frac{1}
{
\left(q^{\frac14}t s_{j}^{\pm}s_{l}^{\mp};q\right)_{\infty}
}
}_{\mathbb{II}_{N}^{\textrm{3d HM}}\left(\frac{s_{j}}{s_{l}}\right)}
\underbrace{
\frac{1}
{
\left(q^{\frac14}t^{-1} s_{i}^{\pm}s_{j}^{\mp};q\right)_{\infty}
}
}_{\mathbb{II}_{N}^{\textrm{3d tHM}}\left(\frac{s_{i}}{s_{j}}\right)}
\underbrace{
\frac{1}
{
\left(q^{\frac14}t^{-1} s_{k}^{\pm}s_{l}^{\mp};q\right)_{\infty}
}
}_{\mathbb{II}_{N}^{\textrm{3d tHM}}\left(\frac{s_{k}}{s_{l}}\right)}
\nonumber\\
&\times 
\
\prod_{i=1}^{N}
\prod_{j=N+1}^{2N}
\prod_{k=2N+1}^{2N+M}
\prod_{l=2N+M+1}^{2N+2M}
\underbrace{
\left(q^{\frac12}s_{i}^{\pm}s_{l}^{\mp};q\right)_{\infty}
}_{F\left(q^{\frac12}\frac{s_{i}}{s_{l}}\right)}
\underbrace{
\left(q^{\frac12}s_{j}^{\pm}s_{k}^{\mp};q\right)_{\infty}
}_{F\left(q^{\frac12}\frac{s_{j}}{s_{k}}\right)}
\underbrace{
\left(q^{\frac12} \prod s_{i}^{\pm} \prod_{k} s_{l}^{\pm} \prod s_{j}^{\mp} \prod_{k} s_{k}^{\mp}x^{\pm}
;q\right)_{\infty}
}_{F\left(q^{\frac12}\frac{\prod s_{i}\prod s_{l}}{\prod s_{j}\prod s_{k}}x\right)}. 
\end{align}

According to the presence of the D5- and D5$'$-branes, 
the S-dual $\left(\begin{smallmatrix}
N&N\\M&M\\\end{smallmatrix}\right)$ D5-D5$'$ junction has a smaller $U(\min (N,M))$ gauge group surviving at the junction. 
When the difference of $N$ and $M$ is equal to one, 
the gauge theory with higher rank obeys the Dirichlet boundary condition $\mathcal{D}'$ 
and the extra 3d $\mathcal{N}=4$ hypermultiplet in $x^2<0$ receives the Dirichlet boundary condition $D'$. 
When the difference is larger than one, 
the gauge theory with higher rank obeys the Nahm$'$ boundary condition 
associated to a homomorphism $\rho:$ $\mathfrak{su}(2)$ $\rightarrow$ $\mathfrak{u}(|N-M|)$. 
Correspondingly, the 3d $\mathcal{N}=4$ hypermultiplet has the Nahm pole boundary condition 
as we found in (\ref{4duNuM_hindex2}).

Then one gets the quarter-index for the 
S-dual $\left(\begin{smallmatrix}
N&N\\M&M\\\end{smallmatrix}\right)$ D5-D5$'$ junction
\begin{align}
\label{DDNNMM}
\mathbb{IV}_{\mathcal{D}\mathcal{D}'}^{
\left(
\begin{smallmatrix}
N&N\\
M&M\\
\end{smallmatrix}
\right)}
&=
\underbrace{
\frac{1}{\min (N,M)!}
\frac{(q)_{\infty}^{2\min (N,M)}}
{
\left(q^{\frac12}t^2;q\right)_{\infty}^{\min (N,M)}
\left(q^{\frac12}t^{-2};q\right)_{\infty}^{\min (N,M)}
}
\oint \prod_{i=1}^{\min (N,M)} \frac{ds_{i}}{2\pi is_{i}} 
\prod_{i\neq j} 
\frac{\left(\frac{s_{i}}{s_{j}};q\right)_{\infty} \left(q \frac{s_{i}}{s_{j}};q\right)_{\infty}}
{\left(q^{\frac12}t^2 \frac{s_{i}}{s_{j}};q\right)_{\infty}
\left(q^{\frac12}t^{-2} \frac{s_{i}}{s_{j}};q\right)_{\infty}
}
}_{\mathbb{I}^{\textrm{4d $U(\min (N,M))$}}}
\nonumber\\
&\times 
\underbrace{
\prod_{k=1}^{|N-M|} 
\frac{(q^{\frac{k+1}{2}}t^{-2(k-1)};q)_{\infty}}
{(q^{\frac{k}{2}}t^{-2k};q)_{\infty}}
}_{\mathbb{II}_{\textrm{Nahm}'}^{\textrm{4d $U(|N-M|)$}}}
\left(q^{\frac12+\frac{|N-M|}{4}}t^{-|N-M|}x;q\right)_{\infty}
\left(q^{\frac12+\frac{|N-M|}{4}}t^{-|N-M|}x^{-1};q\right)_{\infty}
\nonumber\\
&\times 
\prod_{i=1}^{\min (N,M)}
\underbrace{
\frac{
(q^{\frac34} t^{-1}s_{i};q)_{\infty}
(q^{\frac34} t^{-1}s_{i}^{-1};q)_{\infty}
}
{
(q^{\frac14} ts_{i};q)_{\infty}
(q^{\frac14} ts_{i}^{-1};q)_{\infty}
}
}_{\mathbb{I}^{\textrm{3d HM}}(s_{i})}
\frac{
\left(q^{\frac34+\frac{|N-M|}{4}} t^{1-|N-M|} s_{i}^{\pm}x^{\mp};q\right)_{\infty}
}
{
\left(q^{\frac14+\frac{|N-M|}{4}}t^{-1-|N-M|} s_{i}^{\pm}x^{\mp};q\right)_{\infty}
}. 
\end{align}
In the second line 
we have the contributions from the 3d $\mathcal{N}=4$ hypermultiplets 
obeying Nahm$'$ boundary condition, which can be compared with (\ref{dd0N0N}). 
In the last line 
we have the full index for the 3d $\mathcal{N}=4$ fundamental hypermultiplets. 
The quarter-indices (\ref{NNNMNM}) and (\ref{DDNNMM}) are expected to give the same answer. 
The brane configuration is illustrated in Figure \ref{fignnNNMM}.

\section{Y-junctions}
\label{sec_yjunction}
\begin{figure}
\begin{center}
\includegraphics[width=10cm]{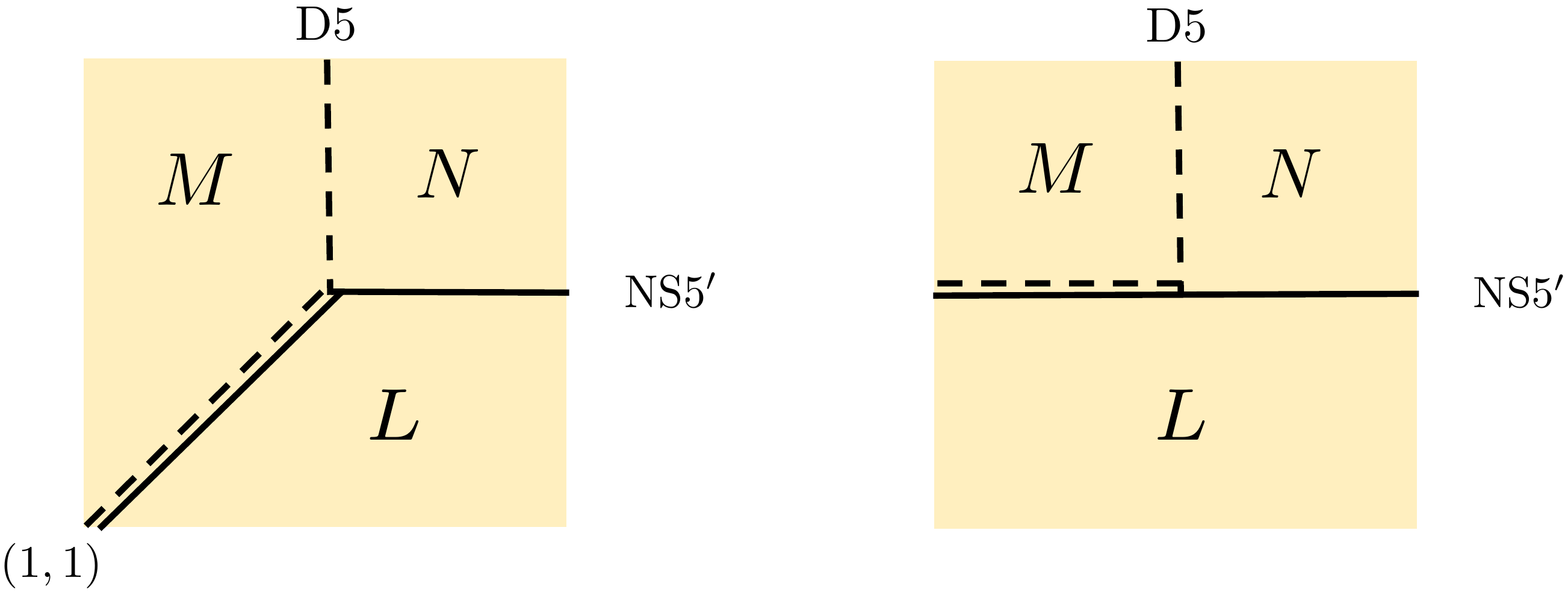}
\caption{
The Y-junctions formed by an NS5-brane, a D5-brane and a $(1,1)$ fivebrane in string theory (Left). 
The T-shaped junction as its gauge theory image (Right). 
}
\label{figyjunction}
\end{center}
\end{figure}

Now consider a trivalent junction 
of the NS5$'$-, D5- and $(1,1)$-fivebranes 
which is obtained by combining the NS5$'$-brane and D5-brane. 
The faces of the Y-junction between 
the NS5$'$-brane and $(1,1)$ fivebrane, 
between 
the $(1,1)$ fivebrane and D5-brane, 
and 
the D5-brane and NS5-brane, 
are filled by $L$, $M$ and $N$ D3-branes respectively. 
We refer to this junction as a Y-junction. 
From the perspective of low-energy effective theory on the D3-branes, 
the Y-junction is described by 
a junction of the three distinct fivebrane interfaces, 
i.e. NS5$'$-type, D5-type and $(1,1)$-type fivebrane interfaces in 4d $\mathcal{N}=4$ $U(L)$, $U(M)$ and $U(N)$ gauge theories. 
Although charge conservation in string theory requires that 
the $(1,1)$ fivebrane is constrained to have a slope in the $(x^2,x^6)$ plane \cite{Aharony:1997ju}, 
we schematically take the $(1,1)$ fivebrane to be represented by a line along $x^6$ 
so that it can be viewed as a T-shaped junction (see Figure \ref{figyjunction}). 
This is because the $(1,1)$ fivebrane introduces boundary conditions analogous to an NS5$'$-type junction, but deformed by
a unit of boundary Chern-Simons coupling. 
We represent the Y-junction shown in Figure \ref{figyjunction} 
by $\left(\begin{smallmatrix}M&|&N\\ \hline &L&\\ \end{smallmatrix}\right)$.

An important property of the Y-junction is that 
the Chern-Simons coupling introduced by the $(1,1)$ fivebrane contributes to the 2d gauge anomaly depending on 
whether the corresponding boundary condition exists in $x^2>0$ or $x^2<0$ \cite{Dimofte:2017tpi}. 
The Chern-Simons coupling has the same contribution as the 2d left-handed chiral fermion for the gauge symmetry in $x^2\ge0$. 
On the other hand, it has the same contribution as the right-handed chiral fermion for the gauge symmetry in $x^2\le0$. 
The Y-junction includes the NS5$'$-brane at $x^2=0$ which imposes the Neumann boundary condition $N'$ for the 3d $\mathcal{N}=4$ hypermultiplets and the D5-brane at $x^{6}=0$ which requires the Dirichlet boundary condition $D$ for the 3d $\mathcal{N}=4$ twisted hypermultiplets. 
The gauge anomaly at the junction is 
\begin{align}
\label{Y_AN}
\begin{array}{c|cc}
\textrm{b.c.}\setminus \textrm{location}&x^2\ge0&x^2\le0\\ \hline
\textrm{3d $\mathcal{N}=4$ fund. hyper w/ $N'$ b.c.}&-\Tr ({\bf s}_{+}^2)&\textrm{none} \\
\textrm{3d $\mathcal{N}=4$ fund. twisted hyper w/ $D$ or Nahm b.c.}&\textrm{none}&\Tr ({\bf s}_{-}^2) \\
\textrm{Chern-Simons coupling}&\Tr ({\bf s}_{+}^2)&-\Tr ({\bf s}_{-}^2) \\
\end{array}
\end{align}
where ${\bf s}_{+}$ and ${\bf s}_{-}$ are the curvatures of gauge symmetry groups in 
the upper and lower half-planes respectively. 

For the gauge symmetry in the upper half-plane, 
the Neumann boundary condition $N'$ for the 3d $\mathcal{N}=4$ hypermultiplet contributes to the gauge anomaly. 
It will be canceled by the Chern-Simons coupling for the Y-junction, 
whereas it was compensated by an additional fundamental Fermi multiplet for the NS5$'$-D5 junction. 

For the gauge symmetry in the lower half-plane, 
the Dirichlet boundary condition or Nahm pole boundary condition for the 3d $\mathcal{N}=4$ twisted hypermultiplet contributes to the gauge anomaly. When the numbers of D3-branes in the upper two quadrants are the same, the junction has an additional fundamental Fermi multiplet to cancel the gauge anomaly for the gauge symmetry in the lower half-plane. 

Starting from the Y-junctions of the GL-twisted $\mathcal{N}=4$ SYM theories \cite{Kapustin:2006pk}, 
a four parameter class of the corner VOA $Y_{L,M,N}[\Psi]$ is introduced in \cite{Gaiotto:2017euk}. 
We will find that 
the quarter-indices for the Y-junctions of 4d $\mathcal{N}=4$ SYM theories 
give enrichments of characters of the corner VOA $Y_{L,M,N}[\Psi]$. 

The action of S-duality on the gauge theory setup leads to trialities 
of the Y-junctions in 4d $\mathcal{N}=4$ SYM theories. 
We obtain various identities of the quarter-indices which 
enrich the identities between the characters of $Y_{L,M,N}[\Psi]$.

\subsection{$Y_{0,0,N}$ and $Y_{N,0,0}$}
\label{sec_y00N}

\subsubsection{$Y_{0,0,1}$ and $Y_{1,0,0}$}
Consider the $\left(\begin{smallmatrix}0&|&1\\ \hline &0&\\ \end{smallmatrix}\right)$ Y-junction. 
This is the simplest Y-junction 
where a single D3-brane fills in the upper right quadrant of the plane. 
This corresponds to the VOA $Y_{0,0,1}$ $=$ $\widehat{\mathfrak{gl}}(1)$ Kac-Moody algebra. 

The quarter-index for the $\left(\begin{smallmatrix}0&|&1\\ \hline &0&\\ \end{smallmatrix}\right)$ Y-junction is simply 
\begin{align}
\label{y001t}
\mathbb{IV}_{\mathcal{N}'\mathcal{D}}^{
\left(
\begin{smallmatrix}
0&|&1\\ \hline
&0&\\
\end{smallmatrix}
\right)}
&=
\mathbb{IV}_{\mathcal{N}'\mathcal{D}}^{\textrm{4d $U(1)$}}
=\frac{1}{(q^{\frac12} t^2;q)_{\infty}},
\end{align}
that is the quarter-index (\ref{qindex_u1n'd}) for 4d $\mathcal{N}=4$ $U(1)$ gauge theory 
with a pair of boundary conditions $\mathcal{N}'$ and $\mathcal{D}$.

%
Another inequivalent Y-junction which is obtained under S-duality is the $\left(\begin{smallmatrix}0&|&0\\ \hline &1&\\ \end{smallmatrix}\right)$ Y-junction defining the VOA $Y_{1,0,0}$. 
It has the 4d $\mathcal{N}=4$ $U(1)$ gauge theory in a half-space $x^2<0$ 
obeying the Neumann boundary condition $\mathcal{N}'$ at $x^2=0$. 
According to the $(1,1)$ fivebrane in $x^6<0$, 
the boundary condition is deformed by an unit of Chern-Simons coupling that leads to the negative contribution for the 
$U(1)$ gauge symmetry in $x^2<0$ as shown in (\ref{Y_AN}). 
Such gauge anomaly will be cancelled by the charged Fermi multiplet living at the junction.

Then the quarter-index for the $\left(\begin{smallmatrix}0&|&0\\ \hline &1&\\ \end{smallmatrix}\right)$ Y-junction 
can be evaluated as 
\begin{align}
\label{y100t}
\mathbb{IV}_{\mathcal{N}'\mathcal{D}}^{
\left(
\begin{smallmatrix}
0&|&0\\ \hline
&1&\\
\end{smallmatrix}
\right)}
&=
\underbrace{
\frac{(q)_{\infty}}{(q^{\frac12} t^2;q)_{\infty}}
\oint \frac{ds}{2\pi is}
}_{\mathbb{II}_{\mathcal{N}'}^{\textrm{4d $U(1)$}}}
\underbrace{
(q^{\frac12}s;q)_{\infty}(q^{\frac12}s^{-1};q)_{\infty}
}_{F(q^{\frac12}s)}
\nonumber\\
&=
\frac{(q)_{\infty}}{(q^{\frac12} t^2;q)_{\infty}}
\oint \frac{ds}{2\pi is} 
\frac{1}{(q)_{\infty}}\sum_{n\in \mathbb{Z}}(-1)^n q^{\frac{n^2}{2}}s^n 
\nonumber\\
&=
\frac{1}{(q^{\frac12} t^2;q)_{\infty}}. 
\end{align}
This agrees with the quarter-index (\ref{y001t}) for the 
$\left(\begin{smallmatrix}0&|&1\\ \hline &0&\\ \end{smallmatrix}\right)$ Y-junction.

%

\subsubsection{$Y_{0,0,2}$ and $Y_{2,0,0}$}

Let us consider the $\left(\begin{smallmatrix}0&|&2\\ \hline &0&\\ \end{smallmatrix}\right)$ Y-junction. 
This is the Y-junction in which only two D3-branes live at the upper right quadrant of the plane. 
This corresponds to the VOA $Y_{0,0,2}$ $=$ $\mathcal{W}_{2}$ algebra. 
The quarter-index for the $\left(\begin{smallmatrix}0&|&2\\ \hline &0&\\ \end{smallmatrix}\right)$ Y-junction is 
that for 4d $\mathcal{N}=4$ $U(2)$ gauge theory 
with a pair of boundary conditions $\mathcal{N}'$ and $\mathcal{D}$. 
\begin{align}
\label{y002t}
\mathbb{IV}_{\mathcal{N}'\mathcal{D}}^{
\left(
\begin{smallmatrix}
0&|&2\\ \hline
&0&\\
\end{smallmatrix}
\right)}
&=
\mathbb{IV}_{\mathcal{N}'\textrm{Nahm}}^{\textrm{4d $U(2)$}}
=
\frac{1}{(q^{\frac12} t^2;q)_{\infty}(qt^4;q)_{\infty}}. 
\end{align}
%

Under S-duality, 
the $\left(\begin{smallmatrix}0&|&2\\ \hline &0&\\ \end{smallmatrix}\right)$ Y-junction 
maps to the $\left(\begin{smallmatrix}0&|&0\\ \hline &2&\\ \end{smallmatrix}\right)$ Y-junction 
which defines the VOA $Y_{2,0,0}$. 
This has the 4d $\mathcal{N}=4$ $U(2)$ gauge theory in $x^2<0$ 
with the Neumann boundary condition $\mathcal{N}'$. 
It also contains a fundamental Fermi multiplet 
which cancels the gauge anomaly induced from the Chern-Simons coupling associated to the $(1,1)$ fivebrane.

The quarter-index for the $\left(\begin{smallmatrix}0&|&0\\ \hline &2&\\ \end{smallmatrix}\right)$ Y-junction is given by
\begin{align}
\label{y200t}
\mathbb{IV}_{\mathcal{N}'\mathcal{D}}^{
\left(
\begin{smallmatrix}
0&|&0\\ \hline
&2&\\
\end{smallmatrix}
\right)}
&=
\underbrace{
\frac12 
\frac{(q)_{\infty}^2}{(q^{\frac12} t^2;q)_{\infty}^2}
\oint\prod_{i=1}^{2} \frac{ds_{i}}{2\pi is_{i}}
\prod_{i\neq j}
\frac{
\left(\frac{s_{i}}{s_{j}};q\right)_{\infty}
}
{
\left(q^{\frac12} t^2\frac{s_{i}}{s_{j}};q\right)_{\infty}
}
}_{\mathbb{II}_{\mathcal{N}'}^{\textrm{4d $U(2)$}}}
\prod_{i=1}^2
\underbrace{
(q^{\frac12}s_{i};q)_{\infty}(q^{\frac12}s^{-1}_{i};q)_{\infty}
}_{F(q^{\frac12}s_{i})}. 
\end{align}
It turns out that 
the quarter-index (\ref{y200t}) for the $\left(\begin{smallmatrix}0&|&0\\ \hline &2&\\ \end{smallmatrix}\right)$ Y-junction 
coincides with 
the quarter-index (\ref{y002t}) for the $\left(\begin{smallmatrix}0&|&2\\ \hline &0&\\ \end{smallmatrix}\right)$ Y-junction:
\begin{align}
\label{y002t_series}
\mathbb{IV}_{\mathcal{N}'\mathcal{D}}^{
\left(
\begin{smallmatrix}
0&|&0\\ \hline
&2&\\
\end{smallmatrix}
\right)}
&=1+t^2 q^{\frac12}+2t^4 q
+(t^2+t^6)q^{\frac32}+t^4(2+3t^4)q^2
+(t^2+3t^6+3t^{10})q^{\frac52}
\nonumber\\
&+t^4(3+4t^4+4t^8)q^3
+(t^2+5t^6+5t^{10}+4t^{14})q^{\frac72}
+t^4(3+8t^4+6t^8+5t^{12})q^4
\nonumber\\
&+(t^2+7t^6+10t^{10}+7t^{14}+5t^{18})q^{\frac92}+\cdots
\end{align}

%

\subsubsection{$Y_{0,0,3}$ and $Y_{3,0,0}$}

As a further example, 
we examine the $\left(\begin{smallmatrix}0&|&3\\ \hline &0&\\ \end{smallmatrix}\right)$ Y-junction. 
This has three D3-branes at the upper right quadrant of the plane 
and it leads to the VOA $Y_{0,0,3}$ $=$ $\mathcal{W}_{3}$. 
The quarter-index for  the $\left(\begin{smallmatrix}0&|&3\\ \hline &0&\\ \end{smallmatrix}\right)$ Y-junction is 
\begin{align}
\label{y003t}
\mathbb{IV}_{\mathcal{N}'\mathcal{D}}^{
\left(
\begin{smallmatrix}
0&|&3\\ \hline
&0&\\
\end{smallmatrix}
\right)}
&=
\mathbb{IV}_{\mathcal{N}'\textrm{Nahm}}^{\textrm{4d $U(3)$}}
=
\frac{1}{(q^{\frac12} t^2;q)_{\infty}(qt^4;q)_{\infty} (q^{\frac32}t^6;q)_{\infty}}. 
\end{align}

The S-dual $\left(\begin{smallmatrix}0&|&0\\ \hline &3&\\ \end{smallmatrix}\right)$ Y-junction 
that defines the VOA $Y_{3,0,0}$ has the 4d $\mathcal{N}=4$ $U(3)$ gauge theory in $x^2<0$ 
with the Neumann boundary condition $\mathcal{N}'$ and the fundamental Fermi multiplet. 

The quarter-index reads
\begin{align}
\label{y300t}
\mathbb{IV}_{\mathcal{N}'\mathcal{D}}^{
\left(
\begin{smallmatrix}
0&|&0\\ \hline
&3&\\
\end{smallmatrix}
\right)}
&=
\underbrace{
\frac{1}{3!}
\frac{(q)_{\infty}^3}{(q^{\frac12} t^2;q)_{\infty}^3}
\oint \prod_{i=1}^{3}\frac{ds_{i}}{2\pi is_{i}}
\prod_{i\neq j}
\frac{
\left(\frac{s_{i}}{s_{j}};q\right)_{\infty}
}
{
\left(q^{\frac12} t^2\frac{s_{i}}{s_{j}};q\right)_{\infty}
}
}_{\mathbb{II}_{\mathcal{N}'}^{\textrm{4d $U(3)$}}}
\prod_{i=1}^3
\underbrace{
(q^{\frac12}s_{i};q)_{\infty}(q^{\frac12}s^{-1}_{i};q)_{\infty}
}_{F(q^{\frac12}s_{i})}. 
\end{align}

We can check that the quarter-index (\ref{y300t}) for the 
$\left(\begin{smallmatrix}0&|&0\\ \hline &3&\\ \end{smallmatrix}\right)$ Y-junction 
agrees with the quarter-index (\ref{y003t}) for the 
$\left(\begin{smallmatrix}0&|&3\\ \hline &0&\\ \end{smallmatrix}\right)$ Y-junction: 
\begin{align}
\label{y300t_series}
\mathbb{IV}_{\mathcal{N}'\mathcal{D}}^{
\left(
\begin{smallmatrix}
0&|&0\\ \hline
&3&\\
\end{smallmatrix}
\right)}
&=1+t^2 q^{\frac12}+2t^4 q
+(t^2+3t^6)q^{\frac32}+2(t^4+2t^8)q^2
+(t^2+4t^6+5t^{10})q^{\frac52}
\nonumber\\
&+t^4(3+6t^4+7t^8)q^3
+(t^2+6t^6+9t^{10}+8t^{14})q^{\frac72}
+t^4(3+11t^4+12t^8+10t^{12})q^4
\nonumber\\
&+(t^{2}+8t^6+17t^{10}+16t^{14}+12t^{18})q^{\frac92}
+\cdots
\end{align}

%

\subsubsection{$Y_{0,0,N}$ and $Y_{N,0,0}$}

Let us discuss the general $\left(\begin{smallmatrix}0&|&N\\ \hline &0&\\ \end{smallmatrix}\right)$ Y-junction. 
This is the 4d $\mathcal{N}=4$ $U(N)$ SYM theory at the corner with 
the boundary conditions $\mathcal{N}'$ and $\mathcal{D}$, 
which is realized as $N$ D3-branes filled at the upper right quadrant of the plane. 
This yields the VOA $Y_{0,0,N}$ $=$ $\mathcal{W}_{N}$ algebra. 

The quarter-index for $\left(\begin{smallmatrix}0&|&N\\ \hline &0&\\ \end{smallmatrix}\right)$ Y-junction 
which leads to the VOA $Y_{0,0,N}$ is the quarter-index (\ref{qindex_n'd})
\begin{align}
\label{y00Nt}
\mathbb{IV}_{\mathcal{N}'\mathcal{D}}^{
\left(
\begin{smallmatrix}
0&|&N\\ \hline
&0&\\
\end{smallmatrix}
\right)}
&=
\mathbb{IV}_{\mathcal{N}'\mathcal{D}}^{\textrm{4d $U(N)$}}
=\prod_{k=1}^{N}\frac{1}{(q^{\frac{k}{2}}t^{2k};q)_{\infty}}
\end{align}
for the 4d $\mathcal{N}=4$ $U(N)$ SYM theory at the corner with 
the boundary conditions $\mathcal{N}'$ and $\mathcal{D}$. 
%

The S-dual configuration is the $\left(\begin{smallmatrix}0&|&0\\ \hline &N&\\ \end{smallmatrix}\right)$ Y-junction 
associated with the VOA $Y_{N,0,0}$. 
This junction has the 4d $\mathcal{N}=4$ $U(N)$ SYM theory with the Neumann boundary condition $\mathcal{N}'$ 
together with the fundamental Fermi multiplet that 
cancels the gauge anomaly contribution from the Chern-Simons coupling. 

We can express the quarter-index for the 
$\left(\begin{smallmatrix}0&|&0\\ \hline &N&\\ \end{smallmatrix}\right)$ Y-junction as
\begin{align}
\label{yN00t}
\mathbb{IV}_{\mathcal{N}'\mathcal{D}}^{
\left(
\begin{smallmatrix}
0&|&0\\ \hline
&N&\\
\end{smallmatrix}
\right)}
&=
\underbrace{
\frac{1}{N!}
\frac{(q)_{\infty}^N}{(q^{\frac12} t^2;q)_{\infty}^N}
\oint 
\prod_{i=1}^{N}\frac{ds_{i}}{2\pi is_{i}}
\prod_{i\neq j}
\frac{
\left(\frac{s_{i}}{s_{j}};q\right)_{\infty}
}
{
\left(q^{\frac12} t^2\frac{s_{i}}{s_{j}};q\right)_{\infty}
}
}_{\mathbb{II}_{\mathcal{N}'}^{\textrm{4d $U(N)$}}}
\prod_{i=1}^N
\underbrace{
(q^{\frac12}s_{i};q)_{\infty}(q^{\frac12}s^{-1}_{i};q)_{\infty}
}_{F(q^{\frac12}s_{i})}. 
\end{align}
This will coincide with the quarter-index (\ref{y00Nt}) 
for the $\left(\begin{smallmatrix}0&|&0\\ \hline &N&\\ \end{smallmatrix}\right)$ Y-junction. 

\begin{figure}
\begin{center}
\includegraphics[width=10cm]{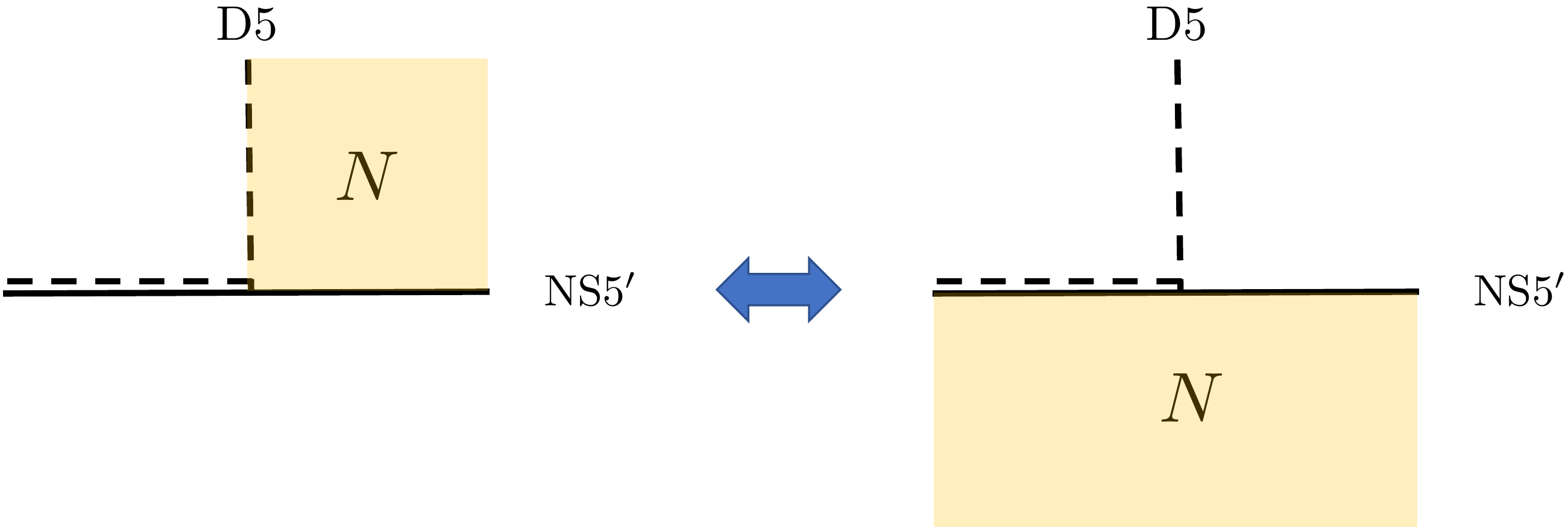}
\caption{
The Y-junctions for $Y_{0,0,N}$ and $Y_{N,0,0}$. 
}
\label{figy00n}
\end{center}
\end{figure}

The brane picture is depicted in Figure \ref{figy00n}. 
%
%
%
%
%

\subsection{$Y_{N,0,N}$ and $Y_{0,N,N}$}
\label{sec_yN0N}
\subsubsection{$Y_{1,0,1}$ and $Y_{0,1,1}$}
Let us take as the next step the Y-junction 
in which two of three faces are filled by equal numbers of D3-branes. 

The simplest example is the $\left(\begin{smallmatrix}0&|&1\\ \hline &1&\\ \end{smallmatrix}\right)$ Y-junction 
which leads to the VOA $Y_{1,0,1}$. 
This involves a 4d $\mathcal{N}=4$ $U(1)$ gauge theory in $x^2<0$ with the Neumann boundary condition $\mathcal{N}'$ at $x^2=0$ 
and another 4d $\mathcal{N}=4$ $U(1)$ gauge theory at the upper right quadrant 
with the boundary conditions $\mathcal{N}'$ and $\mathcal{D}$. 
In addition, the NS5$'$-type interface has the 3d $\mathcal{N}=4$ twisted hypermultiplet 
arising from D3-D3 strings across the NS5$'$-brane. 
At the junction it should obey the Dirichlet boundary condition $D$ required from the D5-brane. 
This cancels the gauge anomaly induced from the Chern-Simons coupling. 
Hence there would be no Fermi multiplet at the junction 
in contrast to the $\left(\begin{smallmatrix}0&|&0\\ \hline &1&\\ \end{smallmatrix}\right)$ Y-junction.

The quarter-index for the $\left(\begin{smallmatrix}0&|&1\\ \hline &1&\\ \end{smallmatrix}\right)$ Y-junction 
which leads to the VOA $Y_{1,0,1}$ is 
\begin{align}
\label{y101t}
\mathbb{IV}_{\mathcal{N}'\mathcal{D}}^{
\left(
\begin{smallmatrix}
0&|&1\\ \hline
&1&\\
\end{smallmatrix}
\right)}
&=
\underbrace{
\frac{(q)_{\infty}}{(q^{\frac12}t^2;q)_{\infty}}\oint \frac{ds}{2\pi is}
}_{\mathbb{II}_{\mathcal{N}'}^{\textrm{4d $U(1)$}}}
\underbrace{
\frac{1}{(q^{\frac12} t^2;q)_{\infty}}
}_{\mathbb{IV}_{\mathcal{N}'\mathcal{D}}^{\textrm{4d $U(1)$}}}
\underbrace{
(q^{\frac34}ts;q)_{\infty}
(q^{\frac34}ts^{-1};q)_{\infty}
}_{\mathbb{II}_{D}^{\textrm{3d tHM}}}. 
\end{align}

Under S-duality, 
the $\left(\begin{smallmatrix}0&|&1\\ \hline &1&\\ \end{smallmatrix}\right)$ Y-junction 
maps to the $\left(\begin{smallmatrix}1&|&1\\ \hline &0&\\ \end{smallmatrix}\right)$ Y-junction 
that yields the VOA $Y_{0,1,1}$. 
There is a 4d $\mathcal{N}=4$ $U(1)$ gauge theory in $x^2>0$ 
and a 3d $\mathcal{N}=4$ charged hypermultiplet arising from the D3-D5 string. 
They respectively receive the Neumann boundary condition $\mathcal{N}'$ 
and $N'$ from the NS5$'$-brane. 
Although the Neumann boundary condition $N'$ for the 3d $\mathcal{N}=4$ hypermultiplet 
yields the gauge anomaly, it will be canceled by the Chern-Simons coupling as shown in (\ref{Y_AN}).

The quarter-index for the $\left(\begin{smallmatrix}1&|&1\\ \hline &0&\\ \end{smallmatrix}\right)$ Y-junction is computed as 
\begin{align}
\label{y011t}
\mathbb{IV}_{\mathcal{N}'\mathcal{D}}^{
\left(
\begin{smallmatrix}
1&|&1\\ \hline
&0&\\
\end{smallmatrix}
\right)}
&=
\underbrace{
\frac{(q)_{\infty}}{(q^{\frac12}t^2;q)_{\infty}}\oint \frac{ds}{2\pi is}
}_{\mathbb{II}_{\mathcal{N}'}^{\textrm{4d $U(1)$}}}
\underbrace{
\frac{1}{(q^{\frac14}ts;q)_{\infty} (q^{\frac14} ts^{-1};q)_{\infty}}
}_{\mathbb{II}^{\textrm{3d HM}}_{N}}. 
\end{align}

The quarter-index (\ref{y101t}) for the $\left(\begin{smallmatrix}0&|&1\\ \hline &1&\\ \end{smallmatrix}\right)$ Y-junction 
and the quarter-index (\ref{y011t}) for the $\left(\begin{smallmatrix}1&|&1\\ \hline &0&\\ \end{smallmatrix}\right)$ Y-junction 
turn out to be equal. 
They can be written as 
\begin{align}
\label{y011tseries}
\mathbb{IV}_{\mathcal{N}'\mathcal{D}}^{
\left(
\begin{smallmatrix}
0&|&1\\ \hline
&1&\\
\end{smallmatrix}
\right)}
&=\mathbb{IV}_{\mathcal{N}'\mathcal{D}}^{
\left(
\begin{smallmatrix}
1&|&1\\ \hline
&0&\\
\end{smallmatrix}
\right)}
\nonumber\\
&=
\frac{1}{(q)_{\infty}(q^{\frac12} t^2;q)_{\infty}}
\sum_{m=0}^{\infty} (-1)^m q^{\frac{m(m+1)}{2}}
\frac{(q^{1+m};q)_{\infty}}{(q^{\frac12+m}t^2;q)_{\infty}}
\end{align}
with a clear interpretation associated to a Higgsing procedure which separates the D3-branes in the two quadrants of the upper half plane.

\subsubsection{$Y_{2,0,2}$ and $Y_{0,2,2}$}

Let us consider the $\left(\begin{smallmatrix}0&|&2\\ \hline &2&\\ \end{smallmatrix}\right)$ Y-junction 
associated to the VOA $Y_{2,0,2}$. 
This junction has a 4d $\mathcal{N}=4$ $U(2)$ gauge theory living in a lower-half plane 
with the Neumann boundary condition $\mathcal{N}'$. 
From D3-D3 strings across the NS5$'$-brane 
there appears the 3d $\mathcal{N}=4$ twisted hypermultiplet. 
As the difference of the number of D3-branes jumps from two to zero across the D5-brane, 
it would obey the Nahm pole boundary condition specified by an embedding $\rho:$ $\mathfrak{su}(2)$ $\rightarrow$ $\mathfrak{u}(2)$. 
The Nahm pole boundary condition will cancel the gauge anomaly for the $U(2)$ gauge symmetry in $x^2<0$ contributed from the Chern-Simons coupling as shown in (\ref{Y_AN}).

As a result, the quarter-index for $\left(\begin{smallmatrix}0&|&2\\ \hline &2&\\ \end{smallmatrix}\right)$ Y-junction is 
\begin{align}
\label{y202t}
\mathbb{IV}_{\mathcal{N}'\mathcal{D}}^{
\left(
\begin{smallmatrix}
0&|&2\\ \hline
&2&\\
\end{smallmatrix}
\right)}
&=
\underbrace{
\frac12 \frac{(q)_{\infty}^2}{(q^{\frac12}t^2;q)_{\infty}^2}
\oint \frac{ds_{1}}{2\pi is_{1}}\frac{ds_{2}}{2\pi is_{2}}
\frac{
\left(\frac{s_{1}}{s_{2}};q\right)_{\infty}
\left(\frac{s_{2}}{s_{1}};q\right)_{\infty}
}
{
\left(q^{\frac12} t^2\frac{s_{1}}{s_{2}};q\right)_{\infty}
\left(q^{\frac12} t^2\frac{s_{2}}{s_{1}};q\right)_{\infty}
}
}_{\mathbb{II}_{\mathcal{N}'}^{\textrm{4d $U(2)$}}}
\nonumber\\
&\times 
\underbrace{
\frac{1}{
(q^{\frac12}t^2;q)_{\infty}
(q t^4;q)_{\infty}
}
}_{\mathbb{IV}_{\mathcal{N}'\textrm{Nahm}}^{\textrm{4d $U(2)$}}}
\prod_{i=1}^{2}
(qt^2 s_{i};q)_{\infty}
(qt^2 s_{i}^{-1};q)_{\infty}
\nonumber\\
&=
1+2t^2 q^{\frac12}+(-1+5t^4)q+t^2(-1+8t^4)q^{\frac32}
+(-1+14t^8)q^2
\nonumber\\
&+2t^2 (-1+2t^4+10t^8)q^{\frac52}
+t^4(-4+11t^4+30t^8)q^3
+t^2(-1-2t^4+23t^8+40t^{12})q^{\frac72}
\nonumber\\
&
+t^4(-5+7t^4
+40t^8+55t^{12})q^4
+t^2(-1-8t^4+26t^8+64t^{12}+70t^{16})q^{\frac92}+\cdots
\end{align}
In the second line of (\ref{y202t}) we have the contributions from 
the 3d $\mathcal{N}=4$ twisted hypermultiplet obeying the Nahm pole boundary condition with a homomorphism 
$\rho:$ $\mathfrak{su}(2)$ $\rightarrow$ $\mathfrak{u}(2)$, which we have also found in (\ref{dd0202}). 

The S-dual $\left(\begin{smallmatrix}2&|&2\\ \hline &0&\\ \end{smallmatrix}\right)$ Y-junction 
defining the VOA $Y_{0,2,2}$ 
is described by a 4d $\mathcal{N}=4$ $U(2)$ SYM theory living in an upper half-space 
with the Neumann boundary condition $\mathcal{N}'$ 
and 3d $\mathcal{N}=4$ fundamental hypermultiplet with the Neumann boundary condition $N'$. 
The gauge anomaly from the Neumann boundary condition $N'$ for the 3d $\mathcal{N}=4$ hypermultiplet 
will be canceled by the Chern-Simons coupling. 

The quarter-index for the $\left(\begin{smallmatrix}2&|&2\\ \hline &0&\\ \end{smallmatrix}\right)$ Y-junction is given by
\begin{align}
\label{y022t}
\mathbb{IV}_{\mathcal{N}'\mathcal{D}}^{
\left(
\begin{smallmatrix}
2&|&2\\ \hline
&0&\\
\end{smallmatrix}
\right)}
&=
\underbrace{
\frac12 \frac{(q)_{\infty}^2}
{(q^{\frac12} t^2;q)_{\infty}^2}\oint \frac{ds_{1}}{2\pi is_{1}}\frac{ds_{2}}{2\pi is_{2}} 
\frac{
\left(\frac{s_{1}}{s_{2}};q\right)_{\infty}
\left(\frac{s_{2}}{s_{1}};q\right)_{\infty}
}
{
\left(q^{\frac12}t^2\frac{s_{1}}{s_{2}};q\right)_{\infty}
\left(q^{\frac12}t^2\frac{s_{2}}{s_{1}};q\right)_{\infty}
}
}_{\mathbb{II}_{\mathcal{N}'}^{\textrm{4d $U(2)$}}}
\nonumber\\
&\times 
\prod_{i=1}^{2}
\underbrace{
\frac{1}{
(q^{\frac14}ts_{i};q)_{\infty}
(q^{\frac14}ts_{i}^{-1};q)_{\infty}
}}
_{\mathbb{II}_{N}^{\textrm{3d HM}}(s_{i})}
\nonumber\\
&
=1+2t^2 q^{\frac12}+(-1+5t^4)q+t^2(-1+8t^4)q^{\frac32}
+(-1+14t^8)q^2
\nonumber\\
&+2t^2 (-1+2t^4+10t^8)q^{\frac52}
+t^4(-4+11t^4+30t^8)q^3
+t^2(-1-2t^4+23t^8+40t^{12})q^{\frac72}
\nonumber\\
&
+t^4(-5+7t^4
+40t^8+55t^{12})q^4
+t^2(-1-8t^4+26t^8+64t^{12}+70t^{16})q^{\frac92}+\cdots
\end{align}
It follows that 
the quarter-index (\ref{y202t}) for the $\left(\begin{smallmatrix}0&|&2\\ \hline &2&\\ \end{smallmatrix}\right)$ Y-junction 
and the quarter-index (\ref{y022t}) for the $\left(\begin{smallmatrix}2&|&2\\ \hline &0&\\ \end{smallmatrix}\right)$ Y-junction coincide.

\subsubsection{$Y_{N,0,N}$ and $Y_{0,N,N}$}
Now we propose the generalization to  
the $\left(\begin{smallmatrix}0&|&N\\ \hline &N&\\ \end{smallmatrix}\right)$ Y-junction 
and its dual $\left(\begin{smallmatrix}N&|&N\\ \hline &0&\\ \end{smallmatrix}\right)$ Y-junction.

The $\left(\begin{smallmatrix}0&|&N\\ \hline &N&\\ \end{smallmatrix}\right)$ Y-junction which gives the VOA $Y_{N,0,N}$ 
has a 4d $\mathcal{N}=4$ $U(N)$ gauge theory beying the Neumann boundary condition $\mathcal{N}'$ 
as $N$ D3-branes spanned in lower half-space $x^2<0$ end on the NS5$'$-brane at $x^2=0$. 
We have a Nahm pole boundary condition associated to an embedding $\rho:$ $\mathfrak{su}(2)$ $\rightarrow$ $\mathfrak{u}(N)$. 
These boundary conditions for the twisted hypermultiplets 
will cancel the gauge anomaly for the $U(N)$ gauge symmetry in the lower half space from the boundary Chern-Simons coupling. 

We then obtain 
the quarter-index for the $\left(\begin{smallmatrix}0&|&N\\ \hline &N&\\ \end{smallmatrix}\right)$ Y-junction
\begin{align}
\label{yN0Nt}
\mathbb{IV}_{\mathcal{N}'\mathcal{D}}^{
\left(
\begin{smallmatrix}
0&|&N\\ \hline
&N&\\
\end{smallmatrix}
\right)}
&=
\underbrace{
\frac{1}{N!}\frac{(q)_{\infty}^{N}}{(q^{\frac12}t^2;q)_{\infty}^{N}}
\oint \prod_{i=1}^{N}\frac{ds_{i}}{2\pi is_{i}}\prod_{i\neq j}\frac{\left(\frac{s_{i}}{s_{j}};q\right)_{\infty}}{\left(q^{\frac12}t^2 \frac{s_{i}}{s_{j}};q\right)_{\infty}}
}_{\mathbb{II}_{\mathcal{N}'}^{\textrm{4d $U(N)$}}}
\nonumber\\
&\times 
\underbrace{
\prod_{k=1}^{N}\frac{1}{(q^{\frac{k}{2}}t^{2k};q)_{\infty}}
}_{\mathbb{IV}_{\mathcal{N}'\mathcal{D}/\textrm{Nahm}}^{\textrm{4d $U(N)$}}}
\prod_{i=1}^{N}
\left(q^{\frac34+\frac{N-1}{4}} t^{1+(N-1)}s_{i};q\right)_{\infty}
\left(q^{\frac34+\frac{N-1}{4}} t^{1+(N-1)}s_{i}^{-1};q\right)_{\infty}. 
\end{align}
In the second line of (\ref{yN0Nt}) the index has contributions from the 3d $\mathcal{N}=4$ twisted hypermultiplet 
obeying the Dirichlet boundary condition for $N=1$ 
or the Nahm pole boundary condition associated to an embedding $\rho:$ $\mathfrak{su}(2)$ $\rightarrow$ $\mathfrak{u}(N)$ for 
$N>1$, as we have argued in (\ref{dd0N0N}). 

\begin{figure}
\begin{center}
\includegraphics[width=10cm]{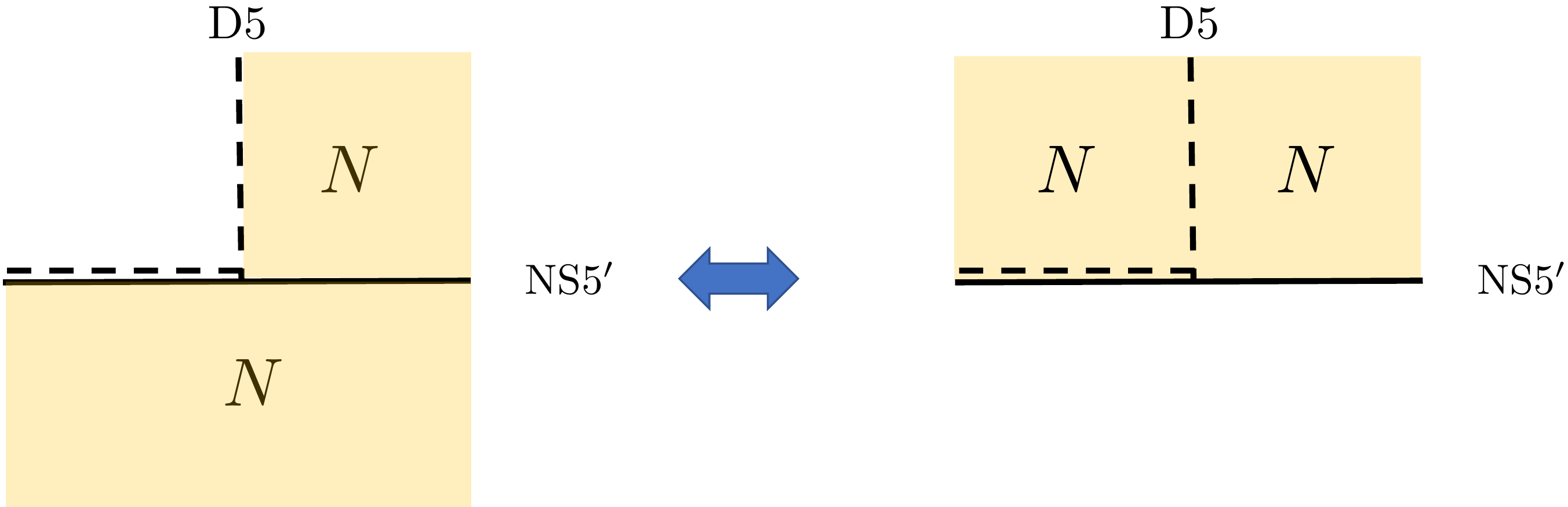}
\caption{
Y-junctions for $Y_{N,0,N}$ and $Y_{0,N,N}$. 
}
\label{figyn0n}
\end{center}
\end{figure}

The quarter-index for the $\left(\begin{smallmatrix}N&|&N\\ \hline &0&\\ \end{smallmatrix}\right)$ Y-junction reads
\begin{align}
\label{y0NNt}
\mathbb{IV}_{\mathcal{N}'\mathcal{D}}^{
\left(
\begin{smallmatrix}
N&|&N\\ \hline
&0&\\
\end{smallmatrix}
\right)}
&=
\underbrace{
\frac{1}{N!}\frac{(q)_{\infty}^{N}}{(q^{\frac12}t^2;q)_{\infty}^{N}}
\oint \prod_{i=1}^{N}\frac{ds_{i}}{2\pi is_{i}}\prod_{i\neq j}\frac{\left(\frac{s_{i}}{s_{j}};q\right)_{\infty}}{\left(q^{\frac12}t^2 \frac{s_{i}}{s_{j}};q\right)_{\infty}}
}_{\mathbb{II}_{\mathcal{N}'}^{\textrm{4d $U(N)$}}}
\prod_{i=1}^{N}
\underbrace{
\frac{1}{
(q^{\frac14}ts_{i};q)_{\infty}
(q^{\frac14}ts_{i}^{-1};q)_{\infty}
}
}_{\mathbb{II}_{N}^{\textrm{3d HM}}(s_{i})}. 
\end{align}
We expect that 
this agrees with the quarter-index (\ref{y0NNt}) for the 
$\left(\begin{smallmatrix}0&|&N\\ \hline &N&\\ \end{smallmatrix}\right)$ Y-junction. 
The brane configuration is drawn in Figure \ref{figyn0n}. 
%
%
%
%
%

\subsection{$Y_{M,0,N}$, $Y_{0,N,M}$ and $Y_{N,M,0}$}
\label{sec_yM0N}
\subsubsection{$Y_{2,0,1}$, $Y_{0,1,2}$ and $Y_{1,2,0}$}
Next consider the case with different numbers of D3-branes are filled in two of three sectors in the plane. 
In this case, S-duality provides us with three sets of Y-junctions, i.e. triality relations. 

The simplest example is a triality 
between 
the $\left(\begin{smallmatrix}0&|&1\\ \hline &2&\\ \end{smallmatrix}\right)$, 
$\left(\begin{smallmatrix}1&|&2\\ \hline &0&\\ \end{smallmatrix}\right)$ 
and $\left(\begin{smallmatrix}2&|&0\\ \hline &1&\\ \end{smallmatrix}\right)$ Y-junctions, 
which define the VOA $Y_{2,0,1}$, $Y_{0,1,2}$ and $Y_{1,2,0}$ respectively. 
For the $\left(\begin{smallmatrix}0&|&1\\ \hline &2&\\ \end{smallmatrix}\right)$ Y-junction, 
we have a 4d $\mathcal{N}=4$ $U(2)$ gauge theory in the lower half-plane 
with the Neumann boundary condition $\mathcal{N}'$ at $x^2=0$. 
In addition, there are 3d $\mathcal{N}=4$ twisted hypermultiplets arising from the D3-D3 strings across the NS5$'$-brane. 
As the number of D3-brane jumps from one to zero when crossing the D5-brane, 
they would receive a Dirichlet boundary condition $D$. 

The quarter-index for the $\left(\begin{smallmatrix}0&|&1\\ \hline &2&\\ \end{smallmatrix}\right)$ Y-junction can be evaluated as 
\begin{align}
\label{y201t}
\mathbb{IV}_{\mathcal{N}'\mathcal{D}}^{
\left(
\begin{smallmatrix}
0&|&1\\ \hline
&2&\\
\end{smallmatrix}
\right)}
&=
\underbrace{
\frac12 \frac{(q)_{\infty}^2}{(q^{\frac12}t^2;q)_{\infty}^2}
\oint \frac{ds_{1}}{2\pi is_{1}}\frac{ds_{2}}{2\pi is_{2}}
\frac{
\left(\frac{s_{1}}{s_{2}};q\right)_{\infty}
\left(\frac{s_{2}}{s_{1}};q\right)_{\infty}
}
{
\left(q^{\frac12} t^2\frac{s_{1}}{s_{2}};q\right)_{\infty}
\left(q^{\frac12} t^2\frac{s_{2}}{s_{1}};q\right)_{\infty}
}
}_{\mathbb{II}_{\mathcal{N}'}^{\textrm{4d $U(2)$}}}
\nonumber\\
&\times 
\underbrace{
\frac{1}{(q^{\frac12}t^2;q)_{\infty}}
}_{\mathbb{IV}_{\mathcal{N}'\mathcal{D}}^{\textrm{4d $U(1)$}}}
\prod_{i=1}^{2}
\underbrace{
\left(q^{\frac34} ts_{i};q\right)_{\infty}
\left(q^{\frac34} ts_{i}^{-1};q\right)_{\infty}
}_{\mathbb{II}_{D}^{\textrm{3d tHM}}(s_{i})}. 
\end{align}
%

The second junction is the $\left(\begin{smallmatrix}1&|&2\\ \hline &0&\\ \end{smallmatrix}\right)$ Y-junction. 
This closely resembles the $\left(\begin{smallmatrix}
1&2\\0&0\\\end{smallmatrix}\right)$ NS5$'$-D5 junction discussed in section \ref{sec_2dndNM00}. 
The junction has a 4d $\mathcal{N}=4$ $U(1)$ gauge theory in the upper half-plane which 
satisfies the Neumann boundary condition $\mathcal{N}'$ at $x^2=0$. 
As the number of D3-branes jumps across the D5-brane, there is no hypermultiplet. 
Unlike the $\left(\begin{smallmatrix}
1&2\\0&0\\\end{smallmatrix}\right)$ NS5$'$-D5 junction, 
there would be no charged Fermi multiplet at the junction 
since the gauge anomaly is compensated by the Chern-Simons coupling from 
the $(1,1)$ fivebrane in $x^6<0$ as in (\ref{Y_AN}). 

Thus we can express the quarter-index for the $\left(\begin{smallmatrix}1&|&2\\ \hline &0&\\ \end{smallmatrix}\right)$ Y-junction as
\begin{align}
\label{y012t}
\mathbb{IV}_{\mathcal{N}'\mathcal{D}}^{
\left(
\begin{smallmatrix}
1&|&2\\ \hline
&0&\\
\end{smallmatrix}
\right)}
&=
\underbrace{
\frac{(q)_{\infty}}{(q^{\frac12}t^2;q)_{\infty}}\oint \frac{ds}{2\pi is}
}_{\mathbb{II}_{\mathcal{N}'}^{\textrm{4d $U(1)$}}}
\underbrace{
\frac{1}{(q^{\frac12}t^2;q)_{\infty}}
}_{\mathbb{IV}_{\mathcal{N}'\mathcal{D}}^{\textrm{4d $U(1)$}}}
\frac{1}
{
(q^{\frac12} t^2s;q)_{\infty}
(q^{\frac12} t^2s^{-1};q)_{\infty}
}. 
\end{align}

The third junction is the 
$\left(\begin{smallmatrix}2&|&0\\ \hline &1&\\ \end{smallmatrix}\right)$ Y-junction. 
This junction includes a 4d $\mathcal{N}=4$ $U(1)$ gauge theory in the lower half-space 
with the Neumann boundary condition $\mathcal{N}'$. 
There is a 3d $\mathcal{N}=4$ twisted hypermultiplets 
arising from D3-D3 strings across the NS5$'$-brane 
which obey the Nahm pole boundary condition specified by a homomorphism 
$\rho:$ $\mathfrak{su}(2)$ $\rightarrow$ $\mathfrak{u}(2)$. 
The Nahm pole boundary condition will cancel the gauge anomaly for the $U(1)$ gauge symmetry 
contributed from the Chern-Simons coupling.

Then we obtain the quarter-index for the 
$\left(\begin{smallmatrix}2&|&0\\ \hline &1&\\ \end{smallmatrix}\right)$ Y-junction:
\begin{align}
\label{y120t}
\mathbb{IV}_{\mathcal{N}'\mathcal{D}}^{
\left(
\begin{smallmatrix}
2&|&0\\ \hline
&1&\\
\end{smallmatrix}
\right)}
&=
\underbrace{
\frac{(q)_{\infty}}{(q^{\frac12}t^2;q)_{\infty}}
\oint \frac{ds}{2\pi is}
}_{\mathbb{II}_{\mathcal{N}'}^{\textrm{4d $U(1)$}}} 
\underbrace{
\frac{1}{(q^{\frac12}t^2;q)_{\infty}(qt^{4};q)_{\infty}}
}_{\mathbb{IV}_{\mathcal{N}'\textrm{Nahm}}^{\textrm{4d $U(2)$}}}
(q t^2 s;q)_{\infty} (q t^2s^{-1};q)_{\infty}. 
\end{align}

We can check that the three quarter-indices 
(\ref{y201t}), (\ref{y012t}) and (\ref{y120t}) coincide 
and they can be expressed as 
\begin{align}
\label{y201tseries}
\mathbb{IV}_{\mathcal{N}'\mathcal{D}}^{
\left(
\begin{smallmatrix}
0&|&1\\ \hline
&2&\\
\end{smallmatrix}
\right)}(t;q)
&=
\mathbb{IV}_{\mathcal{N}'\mathcal{D}}^{
\left(
\begin{smallmatrix}
1&|&2\\ \hline
&0&\\
\end{smallmatrix}
\right)}(t;q)
=
\mathbb{IV}_{\mathcal{N}'\mathcal{D}}^{
\left(
\begin{smallmatrix}
2&|&0\\ \hline
&1&\\
\end{smallmatrix}
\right)}(t;q)
\nonumber\\
&=\frac{1}{(q)_{\infty}(q^{\frac12}t^2;q)_{\infty}^{2}}
\sum_{m=0}^{\infty}
\frac{(q^{1+m};q)_{\infty}}
{(q^{1+m}t^4;q)_{\infty}}
(-1)^{m}q^{\frac{m(m+1)}{2}}. 
\end{align}
In the H-twist limit, 
(\ref{y201tseries}) becomes the vacuum character of the VOA $Y_{2,0,1}$, $Y_{0,1,2}$ and $Y_{1,2,0}$ \cite{Gaiotto:2017euk}:
\begin{align}
\label{y201vch}
\mathbb{IV}_{\mathcal{N}'\mathcal{D}}^{
\left(
\begin{smallmatrix}
0&|&1\\ \hline
&2&\\
\end{smallmatrix}
\right)}(t=q^{\frac14};q)
&=
\mathbb{IV}_{\mathcal{N}'\mathcal{D}}^{
\left(
\begin{smallmatrix}
1&|&2\\ \hline
&0&\\
\end{smallmatrix}
\right)}(t=q^{\frac14};q)
=
\mathbb{IV}_{\mathcal{N}'\mathcal{D}}^{
\left(
\begin{smallmatrix}
2&|&0\\ \hline
&1&\\
\end{smallmatrix}
\right)}(t=q^{\frac14};q)
\nonumber\\
=\chi_{Y_{2,0,1}}(q)&=
\chi_{Y_{0,1,2}}(q)=
\chi_{Y_{1,2,0}}(q)
\nonumber\\
&=
\frac{1}{(q)_{\infty}^3}
\left(
1+2\sum_{m=1}^{\infty} (-1)^{m} q^{\frac{m(m+1)}{2}}
\right). 
\end{align}

\subsubsection{$Y_{3,0,1}$, $Y_{0,1,3}$ and $Y_{1,3,0}$}
To gain more insight, 
let us consider the $\left(\begin{smallmatrix}0&|&1\\ \hline &3&\\ \end{smallmatrix}\right)$ Y-junction 
associated to the VOA $Y_{3,0,1}$. 

The quarter-index for the $\left(\begin{smallmatrix}0&|&1\\ \hline &3&\\ \end{smallmatrix}\right)$ Y-junction 
can be evaluated as 
\begin{align}
\label{y301t}
\mathbb{IV}_{\mathcal{N}'\mathcal{D}}^{
\left(
\begin{smallmatrix}
0&|&1\\ \hline
&3&\\
\end{smallmatrix}
\right)}
&=
\underbrace{
\frac{1}{3!} \frac{(q)_{\infty}^3}{(q^{\frac12}t^2;q)_{\infty}^3}
\oint 
\prod_{i=1}^3
\frac{ds_{i}}{2\pi is_{i}}
\prod_{i\neq j}
\frac{
\left(\frac{s_{i}}{s_{j}};q\right)_{\infty}
}
{
\left(q^{\frac12} t^2\frac{s_{i}}{s_{j}};q\right)_{\infty}
}
}_{\mathbb{II}_{\mathcal{N}'}^{\textrm{4d $U(3)$}}}
\nonumber\\
&\times 
\underbrace{
\frac{1}{(q^{\frac12}t^2;q)_{\infty}}
}_{\mathbb{IV}_{\mathcal{N}'\mathcal{D}}^{\textrm{4d $U(1)$}}}
\prod_{i=1}^{3}
\underbrace{
\left(q^{\frac34} ts_{i};q\right)_{\infty}
\left(q^{\frac34} ts_{i}^{-1};q\right)_{\infty}
}_{\mathbb{II}_{D}^{\textrm{3d tHM}}(s_{i})}
\nonumber\\
&=
1+2t^2 q^{\frac12}+(-1+4t^4)q
+7t^6 q^{\frac32}+(-1+t^4+11t^8)q^2
+t^2(-2+5t^4+16t^8)q^{\frac52}
\nonumber\\
&+t^4(-1+11t^4+23t^8)q^3
+t^2(-2+2t^4+21t^8+31t^{12})q^{\frac72}
\nonumber\\
&+t^{4}(-4+12t^{4}+34t^{8}+41t^{12})q^{4}
+t^2(-2-3t^{4}+29t^{8}+53t^{12}+53t^{16})q^{\frac92}+\cdots
\end{align}

The action of S-duality leads to the $\left(\begin{smallmatrix}1&|&3\\ \hline &0&\\ \end{smallmatrix}\right)$ Y-junction of the VOA $Y_{0,1,3}$. 
This junction is similar to the $\left(\begin{smallmatrix}
1&3\\0&0\\\end{smallmatrix}\right)$ NS5$'$-D5 junction analyzed in section \ref{sec_2dndNM00}. 
The gauge symmetry is reduced to $U(1)$ by the D5-brane. 
The 4d $\mathcal{N}=4$ $U(3)$ gauge theory at the upper right quadrant would contribute to the index but there is no 3d $\mathcal{N}=4$ hypermultiplet. 
In contrast to the $\left(\begin{smallmatrix}
1&3\\0&0\\\end{smallmatrix}\right)$ NS5$'$-D5 junction, 
there is no Fermi multiplet at the junction as there is  now the Chern-Simons coupling in $x^6<0$. 

We can calculate 
the quarter-index for the $\left(\begin{smallmatrix}1&|&3\\ \hline &0&\\ \end{smallmatrix}\right)$ Y-junction as
\begin{align}
\label{y013t}
\mathbb{IV}_{\mathcal{N}'\mathcal{D}}^{
\left(
\begin{smallmatrix}
1&|&3\\ \hline
&0&\\
\end{smallmatrix}
\right)}
&=
\underbrace{
\frac{(q)_{\infty}}{(q^{\frac12}t^2;q)_{\infty}}\oint \frac{ds}{2\pi is}
}_{\mathbb{II}_{\mathcal{N}'}^{\textrm{4d $U(1)$}}}
\underbrace{
\frac{1}{(q^{\frac12}t^2;q)_{\infty}(qt^{4};q)_{\infty}
}
}_{\mathbb{IV}_{\mathcal{N}'\textrm{Nahm}}^{\textrm{4d $U(2)$}}}
\frac{1}
{
(q^{\frac34} t^3s;q)_{\infty}
(q^{\frac34} t^3s^{-1};q)_{\infty}
}
\nonumber\\
&=
1+2t^2 q^{\frac12}+(-1+4t^4)q
+7t^6 q^{\frac32}+(-1+t^4+11t^8)q^2
+t^2(-2+5t^4+16t^8)q^{\frac52}
\nonumber\\
&+t^4(-1+11t^4+23t^8)q^3
+t^2(-2+2t^4+21t^8+31t^{12})q^{\frac72}
\nonumber\\
&+t^{4}(-4+12t^{4}+34t^{8}+41t^{12})q^{4}
+t^2(-2-3t^{4}+29t^{8}+53t^{12}+53t^{16})q^{\frac92}+\cdots
\end{align}
where  
$(q^{\frac34}t^3 s;q)_{\infty}^{-1}$ $(q^{\frac34}t^3 s^{-1};q)_{\infty}^{-1}$ 
corresponds to the contributions which we have found in (\ref{nd0103}) for the $\left(\begin{smallmatrix}
1&3\\0&0\\\end{smallmatrix}\right)$ NS5$'$-D5 junction. 
This agrees with the quarter-index (\ref{y301t}) for the 
$\left(\begin{smallmatrix}0&|&1\\ \hline &3&\\ \end{smallmatrix}\right)$ Y-junction.

The triality gives another junction, i.e. 
the $\left(\begin{smallmatrix}3&|&0\\ \hline &1&\\ \end{smallmatrix}\right)$ Y-junction of the VOA $Y_{1,3,0}$. 
For this junction, 
there is a 4d $\mathcal{N}=4$ $U(1)$ gauge theory in the lower half-plane with 
the Neumann boundary condition $\mathcal{N}'$. 
In addition, there is a Nahm pole as the number of D3-branes jump from three to zero across the D5-brane. 
This requires the 3d $\mathcal{N}=4$ twisted hypermultiplets to have the Nahm pole boundary condition 
specified by a homomorphism $\rho:$ $\mathfrak{su}(2)$ $\rightarrow$ $\mathfrak{u}(3)$.

We can compute 
the quarter-index for the $\left(\begin{smallmatrix}3&|&0\\ \hline &1&\\ \end{smallmatrix}\right)$ Y-junction as
\begin{align} 
\label{y130t}
\mathbb{IV}_{\mathcal{N}'\mathcal{D}}^{
\left(
\begin{smallmatrix}
3&|&0\\ \hline
&1&\\
\end{smallmatrix}
\right)}
&=
\underbrace{
\frac{(q)_{\infty}}{(q^{\frac12}t^2;q)_{\infty}}\oint \frac{ds}{2\pi is}
}_{\mathbb{II}_{\mathcal{N}'}^{\textrm{4d $U(1)$}}}
\underbrace{
\frac{1}{(q^{\frac12}t^2;q)_{\infty} (qt^4;q)_{\infty} (q^{\frac32}t^6;q)_{\infty}}
}_{\mathbb{IV}_{\mathcal{N}'\textrm{Nahm}}^{\textrm{4d $U(3)$}}}
(q^{\frac54} t^3 s;q)_{\infty}
(q^{\frac54} t^3 s^{-1};q)_{\infty}
\nonumber\\
&=
1+2t^2 q^{\frac12}+(-1+4t^4)q
+7t^6 q^{\frac32}+(-1+t^4+11t^8)q^2
+t^2(-2+5t^4+16t^8)q^{\frac52}
\nonumber\\
&+t^4(-1+11t^4+23t^8)q^3
+t^2(-2+2t^4+21t^8+31t^{12})q^{\frac72}
\nonumber\\
&+t^{4}(-4+12t^{4}+34t^{8}+41t^{12})q^{4}
+t^2(-2-3t^{4}+29t^{8}+53t^{12}+53t^{16})q^{\frac92}+\cdots
\end{align}
Again the quarter-index (\ref{y130t}) agrees with the quarter-indices (\ref{y301t}) for the 
$\left(\begin{smallmatrix}0&|&1\\ \hline &3&\\ \end{smallmatrix}\right)$ Y-junction 
and the quarter-indices (\ref{y013t}) for the 
$\left(\begin{smallmatrix}1&|&3\\ \hline &0&\\ \end{smallmatrix}\right)$ Y-junction. 
The three quarter-indices (\ref{y301t}), (\ref{y013t}) and (\ref{y130t}) 
turn out to be expressed as 
\begin{align}
\label{y301t_sum}
\mathbb{IV}_{\mathcal{N}'\mathcal{D}}^{
\left(
\begin{smallmatrix}
0&|&1\\ \hline
&3&\\
\end{smallmatrix}
\right)}
&=
\mathbb{IV}_{\mathcal{N}'\mathcal{D}}^{
\left(
\begin{smallmatrix}
1&|&3\\ \hline
&0&\\
\end{smallmatrix}
\right)}
=
\mathbb{IV}_{\mathcal{N}'\mathcal{D}}^{
\left(
\begin{smallmatrix}
3&|&0\\ \hline
&1&\\
\end{smallmatrix}
\right)}
\nonumber\\
&=
\frac{1}{(q)_{\infty} (q^{\frac12} t^2;q)_{\infty}^2 (qt^4;q)_{\infty}}
\sum_{m=0}^{\infty}
\frac{
(q^{1+m};q)_{\infty}
}
{
(q^{\frac32+m}t^6;q)_{\infty}
}
(-1)^{m} q^{\frac{m^2}{2}+\frac{m}{2}}
\end{align}

\subsubsection{$Y_{N,0,1}$, $Y_{0,1,N}$ and $Y_{1,N,0}$}

For the Y-junction 
whose faces filled by 
a single D3-brane and $N$ D3-branes, we have 
\begin{align}
\label{yN01twilson}
\mathbb{IV}_{\mathcal{N}'\mathcal{D}}^{
\left(
\begin{smallmatrix}
0&|&1\\ \hline
&N&\\
\end{smallmatrix}
\right)}
&=
(-1)^{N}
\mathbb{IV}_{\mathcal{N}'\mathcal{D}}^{
\left(
\begin{smallmatrix}
1&|&N\\ \hline
&0&\\
\end{smallmatrix}
\right)}
=
(-1)^{N-1} 
\mathbb{IV}_{\mathcal{N}'\mathcal{D}}^{
\left(
\begin{smallmatrix}
N&|&0\\ \hline
&1&\\
\end{smallmatrix}
\right)}
\nonumber\\
&=
\frac{1}{(q)_{\infty} (q^{\frac12} t^{2};q)_{\infty}}
\prod_{k=1}^{N-1}\frac{1}{(q^{\frac{k}{2}} t^{2k};q)_{\infty}}
\sum_{m=0}^{\infty}
\frac{(q^{1+m};q)_{\infty}}
{(q^{\frac{N}{2}+m}t^{2N};q)_{\infty}}
(-1)^{m+N} q^{\frac{m^2}{2}+\frac{m}{2}}
\end{align}
where 
\begin{align}
\label{yN01twilson1}
\mathbb{IV}_{\mathcal{N}'\mathcal{D}}^{
\left(
\begin{smallmatrix}
0&|&1\\ \hline
&N&\\
\end{smallmatrix}
\right)}
&=
\underbrace{
\frac{1}{N!}\frac{(q)_{\infty}^N}{(q^{\frac12} t^2;q)_{\infty}^N}
\oint \prod_{i=1}^{N} 
\frac{ds_{i}}{2\pi is_{i}}
\prod_{i\neq j}
\frac{\left(\frac{s_{i}}{s_{j}};q\right)_{\infty}}
{\left(q^{\frac12} t^2 \frac{s_{i}}{s_{j}};q\right)_{\infty}}
}_{\mathbb{II}_{\mathcal{N}'}^{\textrm{4d $U(1)$}}}
\underbrace{
\frac{1}{(q^{\frac12} t^2;q)_{\infty}}
}_{\mathbb{IV}_{\mathcal{N}'\mathcal{D}}^{\textrm{4d $U(1)$}}}
\prod_{i=1}^{N}
(q^{\frac34}ts_{i}^{\pm};q)_{\infty},
\nonumber\\
\mathbb{IV}_{\mathcal{N}'\mathcal{D}}^{
\left(
\begin{smallmatrix}
1&|&N\\ \hline
&0&\\
\end{smallmatrix}
\right)}
&=
\underbrace{
\frac{(q)_{\infty}}{(q^{\frac12} t^2;q)_{\infty}}
\oint \frac{ds}{2\pi is}}
_{\mathbb{II}_{\mathcal{N}'}^{\textrm{4d $U(1)$}}}
\underbrace{
\prod_{k=1}^{N-1}\frac{1}{(q^{\frac{k}{2}}t^{2k};q)_{\infty}}
}_{\mathbb{IV}_{\mathcal{N}'\mathcal{D}/\textrm{Nahm}}^{\textrm{4d $U(N-1)$}}}
\frac{1}{(q^{\frac{N}{4}}t^{N}s;q)_{\infty} 
(q^{\frac{N}{4}}t^{N}s^{-1};q)_{\infty}
}
, 
\nonumber\\
\mathbb{IV}_{\mathcal{N}'\mathcal{D}}^{
\left(
\begin{smallmatrix}
N&|&0\\ \hline
&1&\\
\end{smallmatrix}
\right)}
&=
\underbrace{
\frac{(q)_{\infty}}{(q^{\frac12} t^2;q)_{\infty}}
\oint \frac{ds}{2\pi s}
}_{\mathbb{II}_{\mathcal{N}'}^{\textrm{4d $U(1)$}}}
\underbrace{
\prod_{k=1}^{N}\frac{1}{(q^{\frac{k}{2}}t^{2k};q)_{\infty}}
}_{\mathbb{IV}_{\mathcal{N}'\mathcal{D}/\textrm{Nahm}}^{\textrm{4d $U(N)$}}}
(q^{\frac12+\frac{N}{4}}t^{N}s;q)_{\infty}
(q^{\frac12+\frac{N}{4}}t^{N}s^{-1};q)_{\infty} . 
\end{align}

\subsubsection{$Y_{3,0,2}$, $Y_{0,2,3}$ and $Y_{2,3,0}$}

More generally, let us examine 
the $\left(\begin{smallmatrix}0&|&2\\ \hline &3&\\ \end{smallmatrix}\right)$ Y-junction 
whose multiplicities of D3-branes are larger than one. 
This is the case where each of the Y-junctions involves non-Abelian gauge symmetry. 
It has a 4d $\mathcal{N}=4$ $U(3)$ SYM theory in $x^2<0$ 
with the Neumann boundary condition $\mathcal{N}'$. 
It has the Nahm pole specified by an embedding $\rho:$ $\mathfrak{su}(2)$ $\rightarrow$ $\mathfrak{u}(2)$. 
The index has contributions from the 3d $\mathcal{N}=4$ fundamental twisted hypermultiplets 
with the Nahm pole boundary condition.

The quarter-index for the $\left(\begin{smallmatrix}0&|&2\\ \hline &3&\\ \end{smallmatrix}\right)$ Y-junction is
\begin{align}
\label{y302t}
\mathbb{IV}_{\mathcal{N}'\mathcal{D}}^{
\left(
\begin{smallmatrix}
0&|&2\\ \hline
&3&\\
\end{smallmatrix}
\right)}
&=
\underbrace{
\frac{1}{3!} \frac{(q)_{\infty}^3}{(q^{\frac12}t^2;q)_{\infty}^3}
\oint 
\prod_{i=1}^3
\frac{ds_{i}}{2\pi is_{i}}
\prod_{i\neq j}
\frac{
\left(\frac{s_{i}}{s_{j}};q\right)_{\infty}
}
{
\left(q^{\frac12} t^2\frac{s_{i}}{s_{j}};q\right)_{\infty}
}
}_{\mathbb{II}_{\mathcal{N}'}^{\textrm{4d $U(3)$}}}
\nonumber\\
&\times 
\underbrace{
\frac{1}{(q^{\frac12}t^2;q)_{\infty}(qt^4;q)_{\infty}}
}_{\mathbb{IV}_{\mathcal{N}'\textrm{Nahm}}^{\textrm{4d $U(2)$}}}
\prod_{i=1}^{3}
\left(q t^2s_{i};q\right)_{\infty}
\left(q t^2s_{i}^{-1};q\right)_{\infty}
\nonumber\\
&=
1+2t^2 q^{\frac12}+(-1+5t^4)q+t^2(-1+9t^4)q^{\frac32}+(-1-t^4+16t^8)q^2
+t^2(-2+2t^4+25t^8)q^{\frac52}
\nonumber\\
&
+t^4(-4+8t^4+39t^8)q^3
+t^2(-1-5t^4-21t^8+56t^{12})q^{\frac72}
+t^4(-4-t^4+41t^8+80t^{12})q^4
\nonumber\\
&+t^2(-1-8t^4+12t^8+74t^{12}+109t^{16})q^{\frac92}+\cdots
\end{align}

The second junction is the $\left(\begin{smallmatrix}2&|&3\\ \hline &0&\\ \end{smallmatrix}\right)$ Y-junction. 
This can be obtained from the $\left(\begin{smallmatrix}
2&3\\0&0\\\end{smallmatrix}\right)$ NS5$'$-D5 junction 
by replacing the D5-brane in $x^6<0$ by the $(1,1)$ fivebrane. 
Such deformation introduces the Chern-Simons coupling in $x^6<0$ 
so that the Fermi multiplet is not required to cancel the 2d gauge anomaly. 

For the $\left(\begin{smallmatrix}2&|&3\\ \hline &0&\\ \end{smallmatrix}\right)$ Y-junction 
the quarter-index is given by
\begin{align}
\label{y023t}
\mathbb{IV}_{\mathcal{N}'\mathcal{D}}^{
\left(
\begin{smallmatrix}
2&|&3\\ \hline
&0&\\
\end{smallmatrix}
\right)}
&=
\underbrace{
\frac12 \frac{(q)_{\infty}^2}{(q^{\frac12}t^2;q)_{\infty}^2}
\oint \frac{ds_{1}}{2\pi is_{1}}\frac{ds_{2}}{2\pi is_{2}}
\frac{
\left(\frac{s_{1}}{s_{2}};q\right)_{\infty}
\left(\frac{s_{2}}{s_{1}};q\right)_{\infty}
}
{
\left(q^{\frac12} t^2\frac{s_{1}}{s_{2}};q\right)_{\infty}
\left(q^{\frac12} t^2\frac{s_{2}}{s_{1}};q\right)_{\infty}
}}_{\mathbb{II}_{\mathcal{N}'}^{\textrm{4d $U(2)$}}}
\nonumber\\
&\times 
\underbrace{
\frac{1}{(q^{\frac12}t^2;q)_{\infty}}
}_{\mathbb{IV}_{\mathcal{N}'\mathcal{D}}^{\textrm{4d $U(1)$}}}
\prod_{i=1}^2 
\frac{1}
{
(q^{\frac12} t^2s_{i};q)_{\infty}
(q^{\frac12} t^2s_{i}^{-1};q)_{\infty}
}
\nonumber\\
&=
1+2t^2 q^{\frac12}+(-1+5t^4)q+t^2(-1+9t^4)q^{\frac32}+(-1-t^4+16t^8)q^2
+t^2(-2+2t^4+25t^8)q^{\frac52}
\nonumber\\
&
+t^4(-4+8t^4+39t^8)q^3
+t^2(-1-5t^4-21t^8+56t^{12})q^{\frac72}
+t^4(-4-t^4+41t^8+80t^{12})q^4
\nonumber\\
&+t^2(-1-8t^4+12t^8+74t^{12}+109t^{16})q^{\frac92}+\cdots
\end{align}
Note that the expression (\ref{y023t}) is obtained by 
getting rid of the contributions from Fermi multiplets in the quarter-index (\ref{nd0203}). 
The result (\ref{y023t}) coincides with the quarter-index (\ref{y302t}).

The last piece is the $\left(\begin{smallmatrix}3&|&0\\ \hline &2&\\ \end{smallmatrix}\right)$ Y-junction. 
We have a 4d $\mathcal{N}=4$ $U(2)$ gauge theory in $x^2<0$ with the Neumann boundary condition $\mathcal{N}'$. 
There is a Nahm pole associated to a homomorphism $\rho:$ $\mathfrak{su}(2)$ $\rightarrow$ $\mathfrak{u}(3)$, 
which specifies the boundary condition for the 3d $\mathcal{N}=4$ twisted hypermultiplets. 

One then finds the following quarter-index for the $\left(\begin{smallmatrix}3&|&0\\ \hline &2&\\ \end{smallmatrix}\right)$ Y-junction: 
\begin{align}
\label{y230t}
\mathbb{IV}_{\mathcal{N}'\mathcal{D}}^{
\left(
\begin{smallmatrix}
3&|&0\\ \hline
&2&\\
\end{smallmatrix}
\right)}
&=
\underbrace{
\frac12 \frac{(q)_{\infty}^{2}}
{(q^{\frac12}t^2;q)_{\infty}^2}
\oint \prod_{i=1}^{2}
\frac{ds_{i}}{2\pi is_{i}}
\prod_{i\neq j}
\frac{
\left(\frac{s_{i}}{s_{j}};q\right)_{\infty}
}
{
\left(
q^{\frac12} t^2 \frac{s_{i}}{s_{j}};q
\right)_{\infty}
}
}_{\mathbb{II}_{\mathcal{N}'}^{\textrm{4d $U(2)$}}}
\nonumber\\
&\times 
\underbrace{
\frac{1}{(q^{\frac12}t^2;q)_{\infty} (qt^{4};q)_{\infty} (q^{\frac32}t^6;q)_{\infty}}
}_{\mathbb{IV}_{\mathcal{N}'\textrm{Nahm}}^{\textrm{4d $U(3)$}}}
\prod_{i=1}^{2}
(q^{\frac54}t^3 s_{i};q)_{\infty}
(q^{\frac54}t^3 s_{i}^{-1};q)_{\infty}
\nonumber\\
&=
1+2t^2 q^{\frac12}+(-1+5t^4)q+t^2(-1+9t^4)q^{\frac32}+(-1-t^4+16t^8)q^2
+t^2(-2+2t^4+25t^8)q^{\frac52}
\nonumber\\
&
+t^4(-4+8t^4+39t^8)q^3
+t^2(-1-5t^4-21t^8+56t^{12})q^{\frac72}
+t^4(-4-t^4+41t^8+80t^{12})q^4
\nonumber\\
&+t^2(-1-8t^4+12t^8+74t^{12}+109t^{16})q^{\frac92}+\cdots
\end{align}
This again agrees with the quarter-index (\ref{y302t}) for the 
$\left(\begin{smallmatrix}0&|&2\\ \hline &3&\\ \end{smallmatrix}\right)$ Y-junction 
and the quarter-index (\ref{y023t}) for the $\left(\begin{smallmatrix}2&|&3\\ \hline &0&\\ \end{smallmatrix}\right)$ Y-junction.

\subsubsection{$Y_{4,0,2}$, $Y_{0,2,4}$ and $Y_{2,4,0}$}

As a further check of the triality between the Y-junctions including non-Abelian gauge groups, 
let us take the $\left(\begin{smallmatrix}0&|&2\\ \hline &4&\\ \end{smallmatrix}\right)$ Y-junction. 
There is a 4d $\mathcal{N}=4$ $U(4)$ SYM theory in $x^2<0$ 
with the Neumann boundary condition $\mathcal{N}'$. 
The junction has local operators from the 3d $\mathcal{N}=4$ fundamental twisted hypermultiplets 
which obey the Nahm pole boundary condition 
associated to an embedding $\rho:$ $\mathfrak{su}(2)$ $\rightarrow$ $\mathfrak{u}(2)$. 

The quarter-index for the $\left(\begin{smallmatrix}0&|&2\\ \hline &4&\\ \end{smallmatrix}\right)$ Y-junction is computed as
\begin{align}
\label{y402t}
\mathbb{IV}_{\mathcal{N}'\mathcal{D}}^{
\left(
\begin{smallmatrix}
0&|&2\\ \hline
&4&\\
\end{smallmatrix}
\right)}
&=
\underbrace{
\frac{1}{4!} \frac{(q)_{\infty}^4}{(q^{\frac12}t^2;q)_{\infty}^4}
\oint 
\prod_{i=1}^4
\frac{ds_{i}}{2\pi is_{i}}
\prod_{i\neq j}
\frac{
\left(\frac{s_{i}}{s_{j}};q\right)_{\infty}
}
{
\left(q^{\frac12} t^2\frac{s_{i}}{s_{j}};q\right)_{\infty}
}
}_{\mathbb{II}_{\mathcal{N}'}^{\textrm{4d $U(4)$}}}
\nonumber\\
&\times 
\underbrace{
\frac{1}{(q^{\frac12}t^2;q)_{\infty}(qt^4;q)_{\infty}}
}_{\mathbb{IV}_{\mathcal{N}'\textrm{Nahm}}^{\textrm{4d $U(2)$}}}
\prod_{i=1}^{3}
\left(q t^2s_{i};q\right)_{\infty}
\left(q t^2s_{i}^{-1};q\right)_{\infty}
\nonumber\\
&=
1+2t^2 q^{\frac12}
+(-1+5t^4)q+(-1+9t^4)q^{\frac32}
+(-1-t^4+17t^8)q^2
\nonumber\\
&
+t^2(-2+t^4+27t^8)q^{\frac52}
+2t^{4}
(-2+3t^4+22t^{8})q^3 
+t^2(-1-5t^4+18t^8+65 t^{12})q^{\frac72}
\nonumber\\
&
+t^4(-4-4t^4+38t^8+97t^{12})q^4
+t^2(-1-7t^4+5t^8+73t^{12}+136t^{16})q^{\frac92}+\cdots
\end{align}
%

The second $\left(\begin{smallmatrix}2&|&4\\ \hline &0&\\ \end{smallmatrix}\right)$ Y-junction can be viewed as 
a deformed $\left(\begin{smallmatrix}
2&4\\0&0\\\end{smallmatrix}\right)$ NS5$'$-D5 junction in such a way that 
the D5-brane in $x^6<0$ is replaced by the $(1,1)$ fivebrane. 
As the $(1,1)$ fivebrane introduces the additional Chern-Simons coupling in $x^6<0$, 
there is no Fermi multiplet at the junction. 

The resulting quarter-index for the $\left(\begin{smallmatrix}2&|&3\\ \hline &0&\\ \end{smallmatrix}\right)$ Y-junction is expressed as
\begin{align}
\label{y024t}
\mathbb{IV}_{\mathcal{N}'\mathcal{D}}^{
\left(
\begin{smallmatrix}
2&|&4\\ \hline
&0&\\
\end{smallmatrix}
\right)}
&=
\underbrace{
\frac12 \frac{(q)_{\infty}^2}{(q^{\frac12}t^2;q)_{\infty}^2}
\oint \frac{ds_{1}}{2\pi is_{1}}\frac{ds_{2}}{2\pi is_{2}}
\frac{
\left(\frac{s_{1}}{s_{2}};q\right)_{\infty}
\left(\frac{s_{2}}{s_{1}};q\right)_{\infty}
}
{
\left(q^{\frac12} t^2\frac{s_{1}}{s_{2}};q\right)_{\infty}
\left(q^{\frac12} t^2\frac{s_{2}}{s_{1}};q\right)_{\infty}
}}_{\mathbb{II}_{\mathcal{N}'}^{\textrm{4d $U(2)$}}}
\nonumber\\
&\times 
\underbrace{
\frac{1}{(q^{\frac12}t^2;q)_{\infty}(qt^2;q)_{\infty}}
}_{\mathbb{IV}_{\mathcal{N}'\textrm{Nahm}}^{\textrm{4d $U(2)$}}}
\prod_{i=1}^3 
\frac{1}
{
(q^{\frac34} t^3s_{i};q)_{\infty}
(q^{\frac34} t^3s_{i}^{-1};q)_{\infty}
}
\nonumber\\
&=
1+2t^2 q^{\frac12}
+(-1+5t^4)q+(-1+9t^4)q^{\frac32}
+(-1-t^4+17t^8)q^2
\nonumber\\
&
+t^2(-2+t^4+27t^8)q^{\frac52}
+2t^{4}
(-2+3t^4+22t^{8})q^3 
+t^2(-1-5t^4+18t^8+65 t^{12})q^{\frac72}
\nonumber\\
&
+t^4(-4-4t^4+38t^8+97t^{12})q^4
+t^2(-1-7t^4+5t^8+73t^{12}+136t^{16})q^{\frac92}+\cdots
\end{align}
This agrees with the quarter-index (\ref{y402t}) for the $\left(\begin{smallmatrix}0&|&2\\ \hline &4&\\ \end{smallmatrix}\right)$ Y-junction.

The third junction is the $\left(\begin{smallmatrix}4&|&0\\ \hline &2&\\ \end{smallmatrix}\right)$ Y-junction. 
It has a 4d $\mathcal{N}=4$ $U(2)$ gauge theory in $x^2<0$ with the Neumann boundary condition $\mathcal{N}'$. 
It contains a Nahm pole associated to an embedding $\rho:$ $\mathfrak{su}(2)$ $\rightarrow$ $\mathfrak{u}(4)$, 
which characterizes the boundary condition for the 3d $\mathcal{N}=4$ twisted hypermultiplets. 

We get the quarter-index for the $\left(\begin{smallmatrix}4&|&0\\ \hline &2&\\ \end{smallmatrix}\right)$ Y-junction: 
\begin{align}
\label{y240t}
\mathbb{IV}_{\mathcal{N}'\mathcal{D}}^{
\left(
\begin{smallmatrix}
4&|&0\\ \hline
&2&\\
\end{smallmatrix}
\right)}
&=
\underbrace{
\frac{1}{2} \frac{(q)_{\infty}^2}{(q^{\frac12}t^2;q)_{\infty}^2}
\oint 
\prod_{i=1}^2
\frac{ds_{i}}{2\pi is_{i}}
\prod_{i\neq j}
\frac{
\left(\frac{s_{i}}{s_{j}};q\right)_{\infty}
}
{
\left(q^{\frac12} t^2\frac{s_{i}}{s_{j}};q\right)_{\infty}
}
}_{\mathbb{II}_{\mathcal{N}'}^{\textrm{4d $U(2)$}}}
\nonumber\\
&\times 
\underbrace{
\frac{1}{(q^{\frac12}t^2;q)_{\infty} (qt^{4};q)_{\infty} (q^{\frac32}t^6;q)_{\infty} (q^2 t^8;q)_{\infty}}
}_{\mathbb{IV}_{\mathcal{N}'\mathcal{D}}^{\textrm{4d $U(4)$}}}
\prod_{i=1}^{2}
(q^{\frac32}t^4 s_{i};q)_{\infty}
(q^{\frac32}t^4 s_{i}^{-1};q)_{\infty}
\nonumber\\
&=
1+2t^2 q^{\frac12}
+(-1+5t^4)q+(-1+9t^4)q^{\frac32}
+(-1-t^4+17t^8)q^2
\nonumber\\
&
+t^2(-2+t^4+27t^8)q^{\frac52}
+2t^{4}
(-2+3t^4+22t^{8})q^3 
+t^2(-1-5t^4+18t^8+65 t^{12})q^{\frac72}
\nonumber\\
&
+t^4(-4-4t^4+38t^8+97t^{12})q^4
+t^2(-1-7t^4+5t^8+73t^{12}+136t^{16})q^{\frac92}+\cdots
\end{align}
Again this coincides with the quarter-index (\ref{y402t}) for 
the $\left(\begin{smallmatrix}4&|&0\\ \hline &2&\\ \end{smallmatrix}\right)$ Y-junction
and the quarter-index (\ref{y024t}) for 
the $\left(\begin{smallmatrix}2&|&4\\ \hline &0&\\ \end{smallmatrix}\right)$ Y-junction.

\subsubsection{$Y_{M,0,N}$, $Y_{0,N,M}$ and $Y_{N,M,0}$}

At this stage, we would like to propose the triality between 
the $\left(\begin{smallmatrix}0&|&N\\ \hline &M&\\ \end{smallmatrix}\right)$, 
$\left(\begin{smallmatrix}N&|&M\\ \hline &0&\\ \end{smallmatrix}\right)$ 
and $\left(\begin{smallmatrix}M&|&0\\ \hline &N&\\ \end{smallmatrix}\right)$ Y-junctions, 
which define the VOA $Y_{M,0,N}$, $Y_{0,N,M}$ and $Y_{N,M,0}$ respectively. 

For the first $\left(\begin{smallmatrix}0&|&N\\ \hline &M&\\ \end{smallmatrix}\right)$ Y-junction, 
there is a 4d $\mathcal{N}=4$ $U(M)$ gauge theory in the lower half-plane 
with the Neumann boundary condition $\mathcal{N}'$ at $x^2=0$ imposed by the NS5$'$-brane. 
There are local operators from the 3d $\mathcal{N}=4$ twisted hypermultiplets arising from 
D3-D3 strings across the NS5$'$-brane. 
They obey the Dirichlet boundary condition $D$ for $N=1$ 
while they satisfy the Nahm pole boundary condition associated to 
a homomorphism $\rho:$ $\mathfrak{su}(2)$ $\rightarrow$ $\mathfrak{u}(N)$ for $N>1$. 
These boundary conditions for the twisted hypermultiplets 
will cancel the $U(M)$ gauge anomaly from the Chern-Simons coupling.

The quarter-index for the $\left(\begin{smallmatrix}0&|&N\\ \hline &M&\\ \end{smallmatrix}\right)$ Y-junction takes the form  
\begin{align}
\label{yM0Nt}
\mathbb{IV}_{\mathcal{N}'\mathcal{D}}^{
\left(
\begin{smallmatrix}
0&|&N\\ \hline
&M&\\
\end{smallmatrix}
\right)}
&=
\underbrace{
\frac{1}{M!} \frac{(q)_{\infty}^M}{(q^{\frac12}t^2;q)_{\infty}^M}
\oint 
\prod_{i=1}^M
\frac{ds_{i}}{2\pi is_{i}}
\prod_{i\neq j}
\frac{
\left(\frac{s_{i}}{s_{j}};q\right)_{\infty}
}
{
\left(q^{\frac12} t^2\frac{s_{i}}{s_{j}};q\right)_{\infty}
}
}_{\mathbb{II}_{\mathcal{N}'}^{\textrm{4d $U(M)$}}}
\nonumber\\
&\times 
\underbrace{
\prod_{k=1}^{N}
\frac{1}{(q^{\frac{k}{2}}t^{2k};q)_{\infty}}
}_{\mathbb{IV}_{\mathcal{N}'\mathcal{D}/\textrm{Nahm}}^{\textrm{4d $U(N)$}}}
\prod_{i=1}^{M}
\left(q^{\frac34+\frac{N-1}{4}} t^{N}s_{i};q\right)_{\infty}
\left(q^{\frac34+\frac{N-1}{4}} t^{N}s_{i}^{-1};q\right)_{\infty}. 
\end{align}

The second $\left(\begin{smallmatrix}N&|&M\\ \hline &0&\\ \end{smallmatrix}\right)$ Y-junction 
can be obtained by deforming the $\left(\begin{smallmatrix}
N&M\\0&0\\\end{smallmatrix}\right)$ NS5$'$-D5 junction discussed in section \ref{sec_2dndNM00} 
so that the D5-brane in $x^6<0$ is replaced by the $(1,1)$ fivebrane. 
The junction has a 4d $\mathcal{N}=4$ $U(\min (N,M))$ gauge theory in the upper half-plane 
with the Neumann boundary condition $\mathcal{N}'$ at $x^2=0$. 
Also there are local operators from the scalar fields in 3d $\mathcal{N}=4$ twisted hypermultiplets. 
As opposed to the $\left(\begin{smallmatrix}
N&M\\0&0\\\end{smallmatrix}\right)$ NS5$'$-D5 junction, 
there is no charged Fermi multiplet at the junction 
since the gauge anomaly is now canceled by the Chern-Simons coupling from the $(1,1)$ fivebrane in $x^6<0$. 

Thus we can write the quarter-index for the $\left(\begin{smallmatrix}N&|&M\\ \hline &0&\\ \end{smallmatrix}\right)$ Y-junction as
\begin{align}
\label{y0NMt}
\mathbb{IV}_{\mathcal{N}'\mathcal{D}}^{
\left(
\begin{smallmatrix}
N&|&M\\ \hline
&0&\\
\end{smallmatrix}
\right)}
&=
\underbrace{
\frac{1}{\min (N,M)!} \frac{(q)_{\infty}^{\min (N,M)}}{(q^{\frac12}t^2;q)_{\infty}^{\min (N,M)}}
\oint 
\prod_{i=1}^{\min (N,M)}
\frac{ds_{i}}{2\pi is_{i}}
\prod_{i\neq j}
\frac{
\left(\frac{s_{i}}{s_{j}};q\right)_{\infty}
}
{
\left(q^{\frac12} t^2\frac{s_{i}}{s_{j}};q\right)_{\infty}
}}_{\mathbb{II}_{\mathcal{N}'}^{\textrm{4d $U(\min (N,M))$}}}
\nonumber\\
&\times 
\underbrace{
\prod_{i=1}^{|N-M|}\frac{1}{(q^{\frac{k}{2}}t^{2k};q)_{\infty}}
}_{\mathbb{IV}_{\mathcal{N}'\mathcal{D}/\textrm{Nahm}}^{\textrm{4d $U(|N-M|)$}}}
\prod_{i=1}^{\min (N,M)} 
\frac{1}
{
\left(q^{\frac14+\frac{|N-M|}{4}} t^{1+|N-M|}s_{i};q\right)_{\infty}
\left(q^{\frac14+\frac{|N-M|}{4}} t^{1+|N-M|}s_{i}^{-1};q\right)_{\infty}
}. 
\end{align}

The third $\left(\begin{smallmatrix}M&|&0\\ \hline &N&\\ \end{smallmatrix}\right)$ Y-junction 
has a 4d $\mathcal{N}=4$ $U(N)$ SYM theory in the lower half-plane 
with the Neumann boundary condition $\mathcal{N}'$ at $x^2=0$. 
When $M=1$, the 3d $\mathcal{N}=4$ fundamental twisted hypermultiplets arising from 
the D3-D3 strings across the NS5$'$-brane gets the Dirichlet boundary condition $D$. 
When $M>1$, there exists a Nahm pole associated to a homomorphism 
$\rho:$ $\mathfrak{su}(2)$ $\rightarrow$ $\mathfrak{u}(M)$ 
and therefore they have the Nahm pole boundary condition. 
These boundary conditions for the twisted hypermultiplets 
will cancel the gauge anomaly contribution from the Chern-Simons coupling.

The quarter-index for the $\left(\begin{smallmatrix}M&|&0\\ \hline &N&\\ \end{smallmatrix}\right)$ Y-junction reads
\begin{align}
\label{yNM0t}
\mathbb{IV}_{\mathcal{N}'\mathcal{D}}^{
\left(
\begin{smallmatrix}
M&|&0\\ \hline
&N&\\
\end{smallmatrix}
\right)}
&=
\underbrace{
\frac{1}{N!}\frac{(q)_{\infty}^{N}}{(q^{\frac12}t^2;q)_{\infty}^{N}}
\oint \prod_{i=1}^{N}\frac{ds_{i}}{2\pi is_{i}}\prod_{i\neq j}
\frac{\left(\frac{s_{i}}{s_{j}};q\right)_{\infty}}{\left(q^{\frac12}t^2 \frac{s_{i}}{s_{j}};q\right)_{\infty}}
}_{\mathbb{II}_{\mathcal{N}'}^{\textrm{4d $U(N)$}}}
\nonumber\\
&\times 
\underbrace{
\prod_{k=1}^{M}\frac{1}{(q^{\frac{k}{2}}t^{2k};q)_{\infty}}
}_{\mathbb{IV}_{\mathcal{N}'\mathcal{D}/\textrm{Nahm}}^{\textrm{4d $U(M)$}}}
\prod_{i=1}^{N}
\left(q^{\frac12+\frac{M}{4}}t^M s_{i};q \right)_{\infty}
\left(q^{\frac12+\frac{M}{4}}t^M s_{i}^{-1};q \right)_{\infty}. 
\end{align}

We expect that 
the three quarter-indices (\ref{yM0Nt}), (\ref{y0NMt}) and (\ref{yNM0t}) give the same answer
\begin{align}
\label{yM0Nt_identity}
\mathbb{IV}_{\mathcal{N}'\mathcal{D}}^{
\left(
\begin{smallmatrix}
0&|&N\\ \hline
&M&\\
\end{smallmatrix}
\right)}
&=
\mathbb{IV}_{\mathcal{N}'\mathcal{D}}^{
\left(
\begin{smallmatrix}
N&|&M\\ \hline
&0&\\
\end{smallmatrix}
\right)}
=\mathbb{IV}_{\mathcal{N}'\mathcal{D}}^{
\left(
\begin{smallmatrix}
M&|&0\\ \hline
&N&\\
\end{smallmatrix}
\right)}.
\end{align}

\begin{figure}
\begin{center}
\includegraphics[width=11cm]{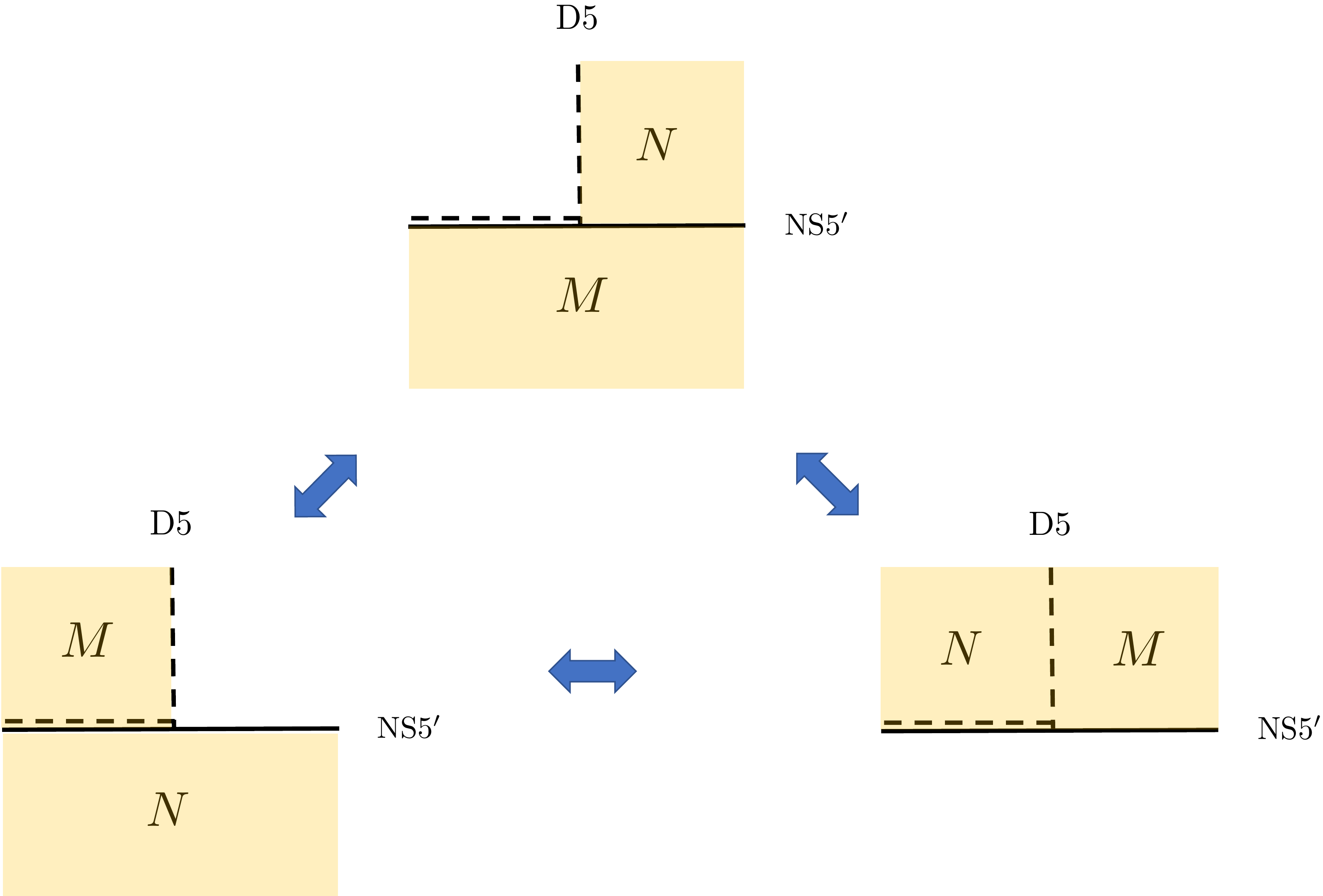}
\caption{A triality between the 
Y-junctions for $Y_{M,0,N}$, $Y_{0,N,M}$ and $Y_{N,M,0}$. 
}
\label{figym0n}
\end{center}
\end{figure}

%
Taking the H-twist limit $t\rightarrow q^{\frac14}$ of these quarter-indices, 
the relation (\ref{yM0Nt}) reduce to the identity of the vacuum characters of 
$Y_{M,0,N}$, $Y_{0,N,M}$ and $Y_{0,N,M}$ \cite{Gaiotto:2017euk}:
\begin{align}
\label{vch_Y0NM}
\chi_{Y_{M,0,N}}(q)&=\chi_{Y_{0,N,M}}(q)=\chi_{Y_{N,M,0}}(q).
\end{align}
The brane configuration is shown in Figure \ref{figym0n}. 
%
%
%
%
%

\subsection{$Y_{L,M,N}$ and triality}
\label{sec_yLMN}
Now we discuss the most general Y-junction 
in which all the three faces are filled by D3-branes.

\subsubsection{$Y_{1,1,1}$}

Consider the $\left(\begin{smallmatrix}1&|&1\\ \hline &1&\\ \end{smallmatrix}\right)$ Y-junction. 
This junction has a 4d $\mathcal{N}=4$ $U(1)$ gauge theory in $x^2>0$ and distinct one in $x^2<0$. 
Both of them have the Neumann boundary condition $\mathcal{N}'$ at $x^2=0$. 
This junction includes 
a 3d $\mathcal{N}=4$ fundamental hypermultiplet obeying Neumann boundary condition $N'$ arising from the D3-D5 string, 
and a 3d $\mathcal{N}=4$ bi-fundamental twisted hypermultiplet arising from the D3-D3 string across the NS5$'$-brane. 
As the 3d $\mathcal{N}=4$ twisted hypermultiplet persists across the junction, 
the gauge anomaly from the Chern-Simons coupling should be compensated by the charged Fermi multiplet at the junction. 

We then obtain the quarter-index for the 
$\left(\begin{smallmatrix}1&|&1\\ \hline &1&\\ \end{smallmatrix}\right)$ Y-junction
\begin{align}
\label{y111t}
\mathbb{IV}_{\mathcal{N}'\mathcal{D}}^{
\left(
\begin{smallmatrix}
1&|&1\\ \hline
&1&\\
\end{smallmatrix}
\right)}
&=
\underbrace{
\frac{(q)_{\infty}}{(q^{\frac12}t^2;q)_{\infty}}\oint \frac{ds_{1}}{2\pi is_{1}}
}_{\mathbb{II}_{\mathcal{N}'}^{\textrm{4d $U(1)$}}}
\underbrace{
\left(q^{\frac12} s_{1};q\right)_{\infty}
\left(q^{\frac12}s_{1}^{-1};q\right)_{\infty}
}_{F(q^{\frac12}s_{1})}
\nonumber\\
&\times 
\underbrace{
\frac{(q)_{\infty}}{(q^{\frac12} t^2;q)_{\infty}}\oint \frac{ds_{2}}{2\pi is_{2}}
}_{\mathbb{II}_{\mathcal{N}'}^{\textrm{4d $U(1)$}}}
\underbrace{
\frac{1}{
\left(q^{\frac14}ts_{2};q\right)_{\infty}
\left(q^{\frac14}ts_{2}^{-1};q\right)_{\infty}
}
}_{\mathbb{II}_{N}^{\textrm{3d HM}}(s_{2})}
\underbrace{
\frac{
\left(q^{\frac34}t \frac{s_{1}}{s_{2}};q\right)_{\infty}
\left(q^{\frac34}t \frac{s_{2}}{s_{1}};q\right)_{\infty}
}
{
\left(q^{\frac14}t^{-1} \frac{s_{1}}{s_{2}};q\right)_{\infty}
\left(q^{\frac14}t^{-1} \frac{s_{2}}{s_{1}};q\right)_{\infty}
}
}_{\mathbb{I}^{\textrm{3d tHM}}\left(\frac{s_{1}}{s_{2}}\right)}
\nonumber\\
&=1+\left(t^{-2}+3t^2\right)q^{\frac12}+\left(-2+t^{-4}+6t^4\right)q
+\left(t^{-6}-2t^2+10t^6\right)q^{\frac32}+\left(1+t^{-8}+t^4+15t^8\right)q^2
\nonumber\\
&+\left(t^{-10}-2t^{-2}-8t^2+7t^{6}+21t^{10}\right)q^{\frac52}
+\left(5+t^{-12}-8t^{4}+16t^{8}+28t^{12}\right)q^3+\cdots
\end{align}

\begin{figure}
\begin{center}
\includegraphics[width=12.5cm]{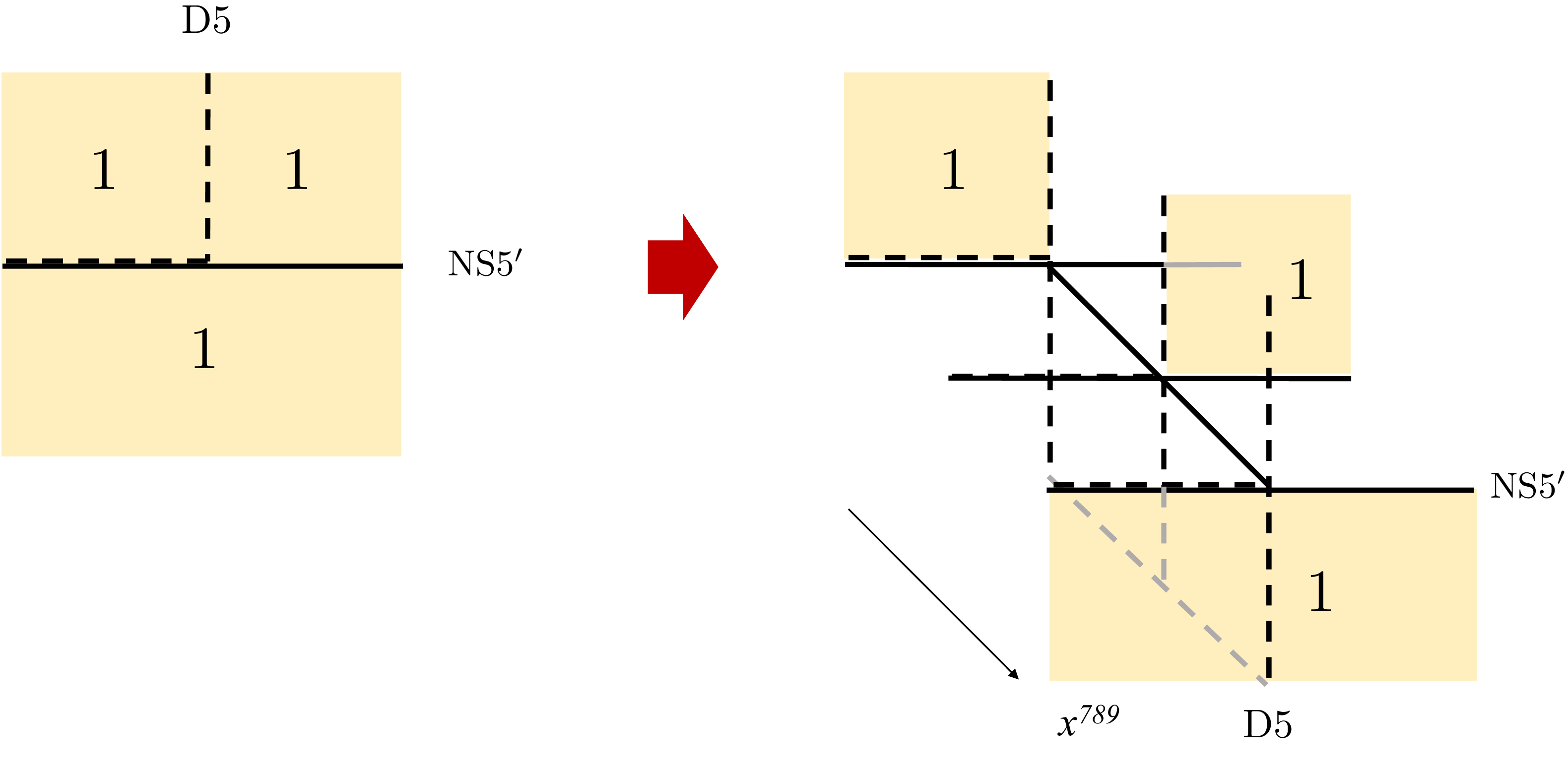}
\caption{Higgsing procedure of the Y-junction for $Y_{1,1,1}$ which splits the D3-branes into three sets along the NS5$'$- and D5-brane. }
\label{fighiggsing4}
\end{center}
\end{figure}

By picking up the residues at hypermultiplet poles $s_{2}$ $=$ $q^{\frac14+m}t$, 
we can evaluate the integral (\ref{y111t}) as
\begin{align}
\label{y111texpand}
\mathbb{IV}_{\mathcal{N}'\mathcal{D}}^{
\left(
\begin{smallmatrix}
1&|&1\\ \hline
&1&\\
\end{smallmatrix}
\right)}
&=\frac{1}{(q)_{\infty}^3}\sum_{n=0}^{\infty}\sum_{m=0}^{\infty}\sum_{k=0}^{n} 
\frac{(q^{1+k};q)_{\infty} (q^{1+n-k};q)_{\infty} (q^{1+m};q)_{\infty}}
{(q^{\frac12+k}t^2;q)_{\infty} (q^{\frac12+n-k}t^2;q)_{\infty} (q^{\frac12+m}t^2;q)_{\infty}}
\nonumber\\
&\times 
(-1)^{m-n+2k}q^{\frac{m(m+1)}{2}+m(2k-n)+\frac{(2k-n)^2}{2}+\frac{k}{2}}t^{2k-2n}. 
\end{align}
The cube of the quarter-index $\mathbb{IV}_{\mathcal{N}'\mathcal{D}}^{\textrm{4d $U(1)$}}$ appearing as the first term 
in the expansion suggests an Higgsing process 
separating D3-branes along the fivebranes (see Figure \ref{fighiggsing4}). 

In the H-twist limit $t\rightarrow q^{\frac14}$ 
the quarter-index (\ref{y111t}) gives the vacuum character of the VOA $Y_{1,1,1}$ 
\begin{align}
\label{y111}
\mathbb{IV}_{\mathcal{N}'\mathcal{D}}^{
\left(
\begin{smallmatrix}
1&|&1\\ \hline
&1&\\
\end{smallmatrix}
\right)}(t=q^{\frac14};q)
&=q+3q^2+6q^3+13q^4+24q^5+48q^6+86q^7+159q^8+279q^9+488q^{10}+\cdots
\nonumber\\
&=\chi_{Y_{1,1,1}}(q).
\end{align}
On the other hand, 
using the BRST construction, the vacuum character of the VOA $Y_{1,1,1}$ takes the form \cite{Gaiotto:2017euk}
\begin{align}
\label{y111a}
\chi_{Y_{1,1,1}}(q)&=
\oint \frac{ds_{1}}{2\pi is_{1}}\frac{ds_{2}}{2\pi is_{2}}
\frac{1}
{
\left(1-\frac{s_{1}}{s_{2}}\right)
\left(1-\frac{s_{2}}{s_{1}}\right)
}
\prod_{n=0}^{\infty}
\frac{
(1-q^{n+\frac12}s_{1})
(1-q^{n+\frac12}s_{1}^{-1})
}
{
(1-q^{n+\frac12}s_{2})
(1-q^{n+\frac12}s_{2}^{-1})
}.
\end{align}
Due to the denominator in the integrand, 
the expression (\ref{y111a}) includes an infinite sum 
so that the expression (\ref{y111}) needs an appropriate regularization. 
This happens when one considers the vacuum characters for the generic VOA $Y_{L,M,N}$ 
and a possible regularization is discussed in \cite{Prochazka:2017qum}. 
Here we get a natural regularization by specializing $t$ after the contour integral is executed.

\subsubsection{$Y_{2,1,1}$, $Y_{1,2,1}$}

Next consider the Y-junction 
where two out of three faces include the same numbers of D3-branes. 
We examine the $\left(\begin{smallmatrix}1&|&1\\ \hline &2&\\ \end{smallmatrix}\right)$ Y-junction 
which leads to the VOA $Y_{2,1,1}$. 
This has a 4d $\mathcal{N}=4$ $U(1)$ gauge theory in $x^2>0$ 
and a 4d $\mathcal{N}=4$ $U(2)$ SYM in $x^2<0$. 
Both of them obey the Neumann boundary condition $\mathcal{N}'$. 
In addition, 
there is a 3d $\mathcal{N}=4$ charged hypermultiplet in $x^2>0$ obeying the Neumann boundary condition $N'$ 
and a 3d $\mathcal{N}=4$ bi-fundamental hypermultiplets. 
As the full twisted hypermultiplets persists across the junction, 
the gauge anomaly for the $U(2)$ gauge symmetry in $x^2<0$ contributed from 
the Chern-Simons coupling must be canceled by fundametnal Fermi multiplets. 

As a result we find the quarter-index 
for the $\left(\begin{smallmatrix}1&|&1\\ \hline &2&\\ \end{smallmatrix}\right)$ Y-junction
\begin{align}
\label{y211t}
\mathbb{IV}_{\mathcal{N}'\mathcal{D}}^{
\left(
\begin{smallmatrix}
1&|&1\\ \hline
&2&\\
\end{smallmatrix}
\right)}
&=
\underbrace{
\frac{1}{2}
\frac{(q)_{\infty}^{2}}{(q^{\frac12}t^2;q)_{\infty}^{2}}\oint \prod_{i=1}^{2}\frac{ds_{i}}{2\pi is_{i}}\prod_{i\neq j}
\frac{\left(\frac{s_{i}}{s_{j}};q\right)_{\infty}}
{\left(q^{\frac12}t^2 \frac{s_{i}}{s_{j}};q\right)_{\infty}}
}_{\mathbb{II}_{\mathcal{N}'}^{\textrm{4d $U(2)$}}}
\prod_{i=1}^{2}
\left(q^{\frac12} s_{i};q\right)_{\infty}
\left(q^{\frac12}s_{i}^{-1};q\right)_{\infty}
\nonumber\\
&\times 
\underbrace{
\frac{(q)_{\infty}}{(q^{\frac12} t^2;q)_{\infty}}\oint \frac{ds_{3}}{2\pi is_{3}}
}_{\mathbb{II}_{\mathcal{N}'}^{\textrm{4d $U(1)$}}}
\frac{1}{
\left(q^{\frac14}ts_{3};q\right)_{\infty}
\left(q^{\frac14}ts_{3}^{-1};q\right)_{\infty}
}
\prod_{i=1}^{2}
\underbrace{
\frac{
\left(q^{\frac34}t \frac{s_{i}}{s_{3}};q\right)_{\infty}
\left(q^{\frac34}t \frac{s_{3}}{s_{i}};q\right)_{\infty}
}
{
\left(q^{\frac14}t^{-1} \frac{s_{i}}{s_{3}};q\right)_{\infty}
\left(q^{\frac14}t^{-1} \frac{s_{3}}{s_{i}};q\right)_{\infty}
}
}_{\mathbb{I}^{\textrm{3d tHM}}\left(\frac{s_{i}}{s_{3}}\right)}
\nonumber\\
&=
1+(t^{-2}+3t^2)q^{\frac12}
+(-1+t^{-4}+7t^{4})q
+(t^{-6}+t^{-2}-2t^{2}+13t^{6})q^{\frac32}
\nonumber\\
&
+(3+t^{-8}+t^{-4}-2t^{4}+22t^{8})q^{2}
+(t^{-10}+t^{-6}-2t^2+3t^6+34t^{10})q^{\frac52}
\nonumber\\
&
+(5+t^{-12}+t^{-8}+2t^{-4}-9t^4+13t^{8}+50t^{12})q^{3}
\nonumber\\
&
+(t^{-14}+t^{-10}+2t^{-6}+3t^{-2}+10t^2-13t^{6}+32t^{10}+70t^{14})q^{\frac72}
\nonumber\\
&
+(2+t^{-16}+t^{-12}+2t^{-8}+t^{-4}-3t^{4}-5t^{8}+60t^{12}+95t^{16})q^{4}
\nonumber\\
&
+t^{-18}
(1+t^4+2t^8+3t^{12}+4t^{16}+15t^{20}-26t^{24}+20t^{28}101t^{32}+125t^{36})q^{\frac92}
+\cdots
\end{align}

Under S-duality, 
we get the $\left(\begin{smallmatrix}2&|&1\\ \hline &1&\\ \end{smallmatrix}\right)$ Y-junction 
which leads to the VOA $Y_{1,2,1}$. 
It has two 4d $\mathcal{N}=4$ $U(1)$ gauge theories in $x^2>0$ and $x^2<0$ 
with the Neumann boundary condition $\mathcal{N}'$ at $x^2=0$. 
We can read the field content from the analysis of 
the $\left(\begin{smallmatrix}1&|&2\\ \hline &0&\\ \end{smallmatrix}\right)$ 
and $\left(\begin{smallmatrix}2&|&0\\ \hline &1&\\ \end{smallmatrix}\right)$ Y-junction. 

As in the $\left(\begin{smallmatrix}0&|&1\\ \hline &2&\\ \end{smallmatrix}\right)$ Y-junction, 
there appear the 3d $\mathcal{N}=4$ bi-fundamental twisted hypermultiplets 
arising from the D3-D3 strings across the NS5$'$-brane. 
As the number of D3-branes jumps from two to one across the D5-brane, 
they will split into two in such a way that 
one obeys the Dirichlet boundary condition $D$ and the other forms the full bi-fundamental twisted hypermultiplet. 
The Dirichlet boundary condition $D$ cancels the gauge anomaly from the Chern-Simons coupling. 

As in the $\left(\begin{smallmatrix}1&|&2\\ \hline &0&\\ \end{smallmatrix}\right)$ Y-junction, 
there is no 3d $\mathcal{N}=4$ hypermultiplet as the number of D3-branes jumps from two to one in $x^2>0$. 
Instead there are bosonic local operators from the broken $U(2)$ gauge theory. 
They will contribute to the $U(1)$ gauge anomaly in $x^2>0$ so that 
it cancels the contribution from the Chern-Simons coupling.

Then the quarter-index for the $\left(\begin{smallmatrix}2&|&1\\ \hline &1&\\ \end{smallmatrix}\right)$ Y-junction reads
\begin{align}
\label{y121t}
\mathbb{IV}_{\mathcal{N}'\mathcal{D}}^{
\left(
\begin{smallmatrix}
2&|&1\\ \hline
&1&\\
\end{smallmatrix}
\right)}
&=
\underbrace{
\frac{(q)_{\infty}}{(q^{\frac12} t^2;q)_{\infty}} \oint \frac{ds_{1}}{2\pi is_{1}}
}_{\mathbb{II}_{\mathcal{N}'}^{\textrm{4d $U(1)$}}} 
\underbrace{
\frac{1}{(q^{\frac12} t^2;q)_{\infty}}
}_{\mathbb{IV}_{\mathcal{N}'\mathcal{D}}^{\textrm{4d $U(1)$}}}
\left(q^{\frac34} t s_{1};q\right)_{\infty}
\left(q^{\frac34} t s_{1}^{-1};q\right)_{\infty}
\nonumber\\
&\times 
\underbrace{
\frac{(q)_{\infty}}{(q^{\frac12} t^2;q)_{\infty}}\oint \frac{ds_{2}}{2\pi is_{2}}
}_{\mathbb{II}_{\mathcal{N}'}^{\textrm{4d $U(1)$}}}
\frac{1}{
\left(q^{\frac12}t^2s_{2};q\right)_{\infty}
\left(q^{\frac12}t^2s_{2}^{-1};q\right)_{\infty}
}
\underbrace{
\frac{
\left(q^{\frac34}t \frac{s_{1}}{s_{2}};q\right)_{\infty}
\left(q^{\frac34}t \frac{s_{2}}{s_{1}};q\right)_{\infty}
}
{
\left(q^{\frac14}t^{-1} \frac{s_{1}}{s_{2}};q\right)_{\infty}
\left(q^{\frac14}t^{-1} \frac{s_{2}}{s_{1}};q\right)_{\infty}
}
}_{\mathbb{I}^{\textrm{3d tHM}}\left(\frac{s_{1}}{s_{2}}\right)}
\nonumber\\
&=
1+(t^{-2}+3t^2)q^{\frac12}
+(-1+t^{-4}+7t^{4})q
+(t^{-6}+t^{-2}-2t^{2}+13t^{6})q^{\frac32}
\nonumber\\
&
+(3+t^{-8}+t^{-4}-2t^{4}+22t^{8})q^{2}
+(t^{-10}+t^{-6}-2t^2+3t^6+34t^{10})q^{\frac52}
\nonumber\\
&
+(5+t^{-12}+t^{-8}+2t^{-4}-9t^4+13t^{8}+50t^{12})q^{3}
\nonumber\\
&
+(t^{-14}+t^{-10}+2t^{-6}+3t^{-2}+10t^2-13t^{6}+32t^{10}+70t^{14})q^{\frac72}
\nonumber\\
&
+(2+t^{-16}+t^{-12}+2t^{-8}+t^{-4}-3t^{4}-5t^{8}+60t^{12}+95t^{16})q^{4}
\nonumber\\
&
+t^{-18}
(1+t^4+2t^8+3t^{12}+4t^{16}+15t^{20}-26t^{24}+20t^{28}101t^{32}+125t^{36})q^{\frac92}
+\cdots
\end{align}
In fact this coincides with the quarter-index (\ref{y211t}) 
for the $\left(\begin{smallmatrix}1&|&1\\ \hline &2&\\ \end{smallmatrix}\right)$ Y-junction.

\subsubsection{$Y_{3,2,1}$, $Y_{1,3,2}$ and $Y_{2,1,3}$}

Let us consider the most general case 
with three distinct numbers of D3-branes are placed in the three sectors of the Y-junction. 

The simplest example is the set of 
the $\left(\begin{smallmatrix}2&|&1\\ \hline &3&\\ \end{smallmatrix}\right)$, 
$\left(\begin{smallmatrix}3&|&2\\ \hline &1&\\ \end{smallmatrix}\right)$ and 
$\left(\begin{smallmatrix}1&|&3\\ \hline &2&\\ \end{smallmatrix}\right)$ Y-junctions, 
which define the VOA $Y_{3,2,1}$, $Y_{1,3,2}$ and $Y_{2,1,3}$. 
The $\left(\begin{smallmatrix}2&|&1\\ \hline &3&\\ \end{smallmatrix}\right)$ Y-junction 
involves a 4d $\mathcal{N}=4$ $U(1)$ gauge theory in $x^2>0$ 
and a 4d $\mathcal{N}=4$ $U(3)$ gauge theory in $x^2<0$. 
They obey the Neumann boundary condition $\mathcal{N}'$. 
While there is no 3d $\mathcal{N}=4$ hypermultiplet 
due to the unequal numbers of D3-branes in two sides of the D5-brane, 
there are bosonic contributions to the index from the broken part of the $U(2)$ gauge theory in the upper left quadrant. 
In addition, it involves the 3d $\mathcal{N}=4$ twisted hypermultiplets 
from the D3-D3 strings across the NS5$'$-brane. 
As the number of the D3-branes jump from two to one when crossing the D5-brane, 
part of the 3d $\mathcal{N}=4$ twisted hypermultiplets should satisfy the Dirichlet boundary condition $D$.

Thus we find the quarter-index for the $\left(\begin{smallmatrix}2&|&1\\ \hline &3&\\ \end{smallmatrix}\right)$ Y-junction
\begin{align}
\label{y321t}
\mathbb{IV}_{\mathcal{N}'\mathcal{D}}^{
\left(
\begin{smallmatrix}
2&|&1\\ \hline
&3&\\
\end{smallmatrix}
\right)}
&=
\underbrace{
\frac{1}{3!}
\frac{(q)_{\infty}^{3}}{(q^{\frac12}t^2;q)_{\infty}^{3}}\oint \prod_{i=1}^{3}\frac{ds_{i}}{2\pi is_{i}}\prod_{i\neq j}
\frac{\left(\frac{s_{i}}{s_{j}};q\right)_{\infty}}
{\left(q^{\frac12}t^2 \frac{s_{i}}{s_{j}};q\right)_{\infty}}
}_{\mathbb{II}_{\mathcal{N}'}^{\textrm{4d $U(3)$}}}
\underbrace{
\frac{1}{(q^{\frac12}t^2;q)_{\infty}}
}_{\mathbb{IV}_{\mathcal{N}'\mathcal{D}}^{\textrm{4d $U(1)$}}}
\prod_{i=1}^{3}
\left(q^{\frac34} t s_{i};q\right)_{\infty}
\left(q^{\frac34} t s_{i}^{-1};q\right)_{\infty}
\nonumber\\
&\times 
\underbrace{
\frac{(q)_{\infty}}{(q^{\frac12} t^2;q)_{\infty}}\oint \frac{ds_{4}}{2\pi is_{4}}
}_{\mathbb{II}_{\mathcal{N}'}^{\textrm{4d $U(1)$}}}
\frac{1}{
\left(q^{\frac12}t^{2}s_{4};q\right)_{\infty}
\left(q^{\frac12}t^{2}s_{4}^{-1};q\right)_{\infty}
}
\prod_{i=1}^{3}
\underbrace{
\frac{
\left(q^{\frac34}t \frac{s_{i}}{s_{4}};q\right)_{\infty}
\left(q^{\frac34}t \frac{s_{4}}{s_{i}};q\right)_{\infty}
}
{
\left(q^{\frac14}t^{-1} \frac{s_{i}}{s_{4}};q\right)_{\infty}
\left(q^{\frac14}t^{-1} \frac{s_{4}}{s_{i}};q\right)_{\infty}
}
}_{\mathbb{I}^{\textrm{3d tHM}}\left(\frac{s_{i}}{s_{3}}\right)}
\nonumber\\
&=
1+(t^{-2}+3t^{2})q^{\frac12}
+(t^{-4}+8t^{4})q
+(t^{-6}+t^{-2}+17t^{6})q^{\frac32}
\nonumber\\
&
+(4+t^{-8}+t^{-4}-2t^{4}+33t^{8})q^{2}
+(t^{-10}+t^{-6}+2t^{-2}+6t^{2}-t^{6}+58t^{10})q^{\frac52}
\nonumber\\
&
+(5+t^{-12}+t^{-8}+3t^{-4}+9t^{4}+3t^{8}+97t^{12})q^{3}
\nonumber\\
&+(t^{-14}+t^{-10}+3t^{-6}+7t^{-2}+13t^{2}+6t^{6}+20t^{10}+153t^{14})q^{\frac72}
\nonumber\\
&
+(13+t^{-16}+t^{-12}+3t^{-8}+5t^{-4}+28t^{4}+3t^{8}+53t^{12}+233t^{16})q^4
\nonumber\\
&
+t^{-18}(1+t^4+3t^8+6t^{12}+11t^{16}+21t^{20}+49t^{24}+120t^{32}+342t^{36})q^{\frac92}
+\cdots
\end{align}

For the second $\left(\begin{smallmatrix}3&|&2\\ \hline &1&\\ \end{smallmatrix}\right)$ Y-junction, 
we have a 4d $\mathcal{N}=4$ $U(2)$ SYM theory in the upper half-space 
and a 4d $\mathcal{N}=4$ $U(1)$ gauge theory in the lower half-space. 
As the numbers of D3-branes changes from three to two when crossing the D5-brane, 
the broken $U(3)$ gauge theory in the upper quadrant lead to the bosonic local operators. 
Also the 3d $\mathcal{N}=4$ twisted hypermultiplets appear from the D3-D3 strings across the NS5$'$-brane. 
As the initial $U(3)$ gauge theory is broken to $U(2)$, 
part of them should obey the Dirichlet boundary condition $D$.

The quarter-index for $\left(\begin{smallmatrix}3&|&2\\ \hline &1&\\ \end{smallmatrix}\right)$ Y-junction is
\begin{align}
\label{y132t}
\mathbb{IV}_{\mathcal{N}'\mathcal{D}}^{
\left(
\begin{smallmatrix}
3&|&2\\ \hline
&1&\\
\end{smallmatrix}
\right)}
&=
\underbrace{
\frac{(q)_{\infty}}{(q^{\frac12}t^2;q)_{\infty}}\oint \frac{ds_{1}}{2\pi is_{1}}
}_{\mathbb{II}_{\mathcal{N}'}^{\textrm{4d $U(1)$}}}
\underbrace{
\frac{1}{(q^{\frac12}t^2;q)_{\infty}}
}_{\mathbb{IV}_{\mathcal{N}'\mathcal{D}}^{\textrm{4d $U(1)$}}}
\left(q^{\frac34} t s_{1};q\right)_{\infty}
\left(q^{\frac34} t s_{1}^{-1};q\right)_{\infty}
\nonumber\\
&\times 
\underbrace{
\frac12 
\frac{(q)_{\infty}^{2}}{(q^{\frac12} t^2;q)_{\infty}^{2}}\oint \prod_{i=2}^{3}\frac{ds_{i}}{2\pi is_{i}}
\prod_{i\neq j}\frac{\left(\frac{s_{i}}{s_{j}};q\right)_{\infty}}{\left(q^{\frac12}t^{2}\frac{s_{i}}{s_{j}};q\right)_{\infty}}
}_{\mathbb{II}_{\mathcal{N}'}^{\textrm{4d $U(2)$}}}
\prod_{i=2}^{3}
\frac{1}{
\left(q^{\frac12}t^{2}s_{i};q\right)_{\infty}
\left(q^{\frac12}t^{2}s_{i}^{-1};q\right)_{\infty}
}
\nonumber\\
&\times 
\prod_{i=2}^{3}
\underbrace{
\frac{
\left(q^{\frac34}t \frac{s_{i}}{s_{1}};q\right)_{\infty}
\left(q^{\frac34}t \frac{s_{1}}{s_{i}};q\right)_{\infty}
}
{
\left(q^{\frac14}t^{-1} \frac{s_{i}}{s_{1}};q\right)_{\infty}
\left(q^{\frac14}t^{-1} \frac{s_{1}}{s_{i}};q\right)_{\infty}
}
}_{\mathbb{I}^{\textrm{3d tHM}}\left(\frac{s_{i}}{s_{1}}\right)}
\nonumber\\
&=
1+(t^{-2}+3t^{2})q^{\frac12}
+(t^{-4}+8t^{4})q
+(t^{-6}+t^{-2}+17t^{6})q^{\frac32}
\nonumber\\
&
+(4+t^{-8}+t^{-4}-2t^{4}+33t^{8})q^{2}
+(t^{-10}+t^{-6}+2t^{-2}+6t^{2}-t^{6}+58t^{10})q^{\frac52}
\nonumber\\
&
+(5+t^{-12}+t^{-8}+3t^{-4}+9t^{4}+3t^{8}+97t^{12})q^{3}
\nonumber\\
&+(t^{-14}+t^{-10}+3t^{-6}+7t^{-2}+13t^{2}+6t^{6}+20t^{10}+153t^{14})q^{\frac72}
\nonumber\\
&
+(13+t^{-16}+t^{-12}+3t^{-8}+5t^{-4}+28t^{4}+3t^{8}+53t^{12}+233t^{16})q^4
\nonumber\\
&
+t^{-18}(1+t^4+3t^8+6t^{12}+11t^{16}+21t^{20}+49t^{24}+120t^{32}+342t^{36})q^{\frac92}
+\cdots
\end{align}
This agrees with the quarter-index (\ref{y321t}) for the 
$\left(\begin{smallmatrix}2&|&1\\ \hline &3&\\ \end{smallmatrix}\right)$ Y-junction.

The third one is the $\left(\begin{smallmatrix}1&|&3\\ \hline &2&\\ \end{smallmatrix}\right)$ Y-junction. 
This has a 4d $\mathcal{N}=4$ $U(1)$ gauge theory in $x^2>0$ and a 4d $\mathcal{N}=4$ $U(2)$ gauge theory in $x^2<0$. 
Unlike the other two junctions, the difference of the numbers of the D3-branes across the D5-brane is equal to two, 
which leads to a Nahm pole. 
From the broken $U(3)$ gauge theory we have the bosonic local operators 
as we found in (\ref{nd1300}) and (\ref{y013t}). 
Also the 3d $\mathcal{N}=4$ twisted hypermultiplets contribute to the index, but are affected by the Nahm pole.

\begin{figure}
\begin{center}
\includegraphics[width=11cm]{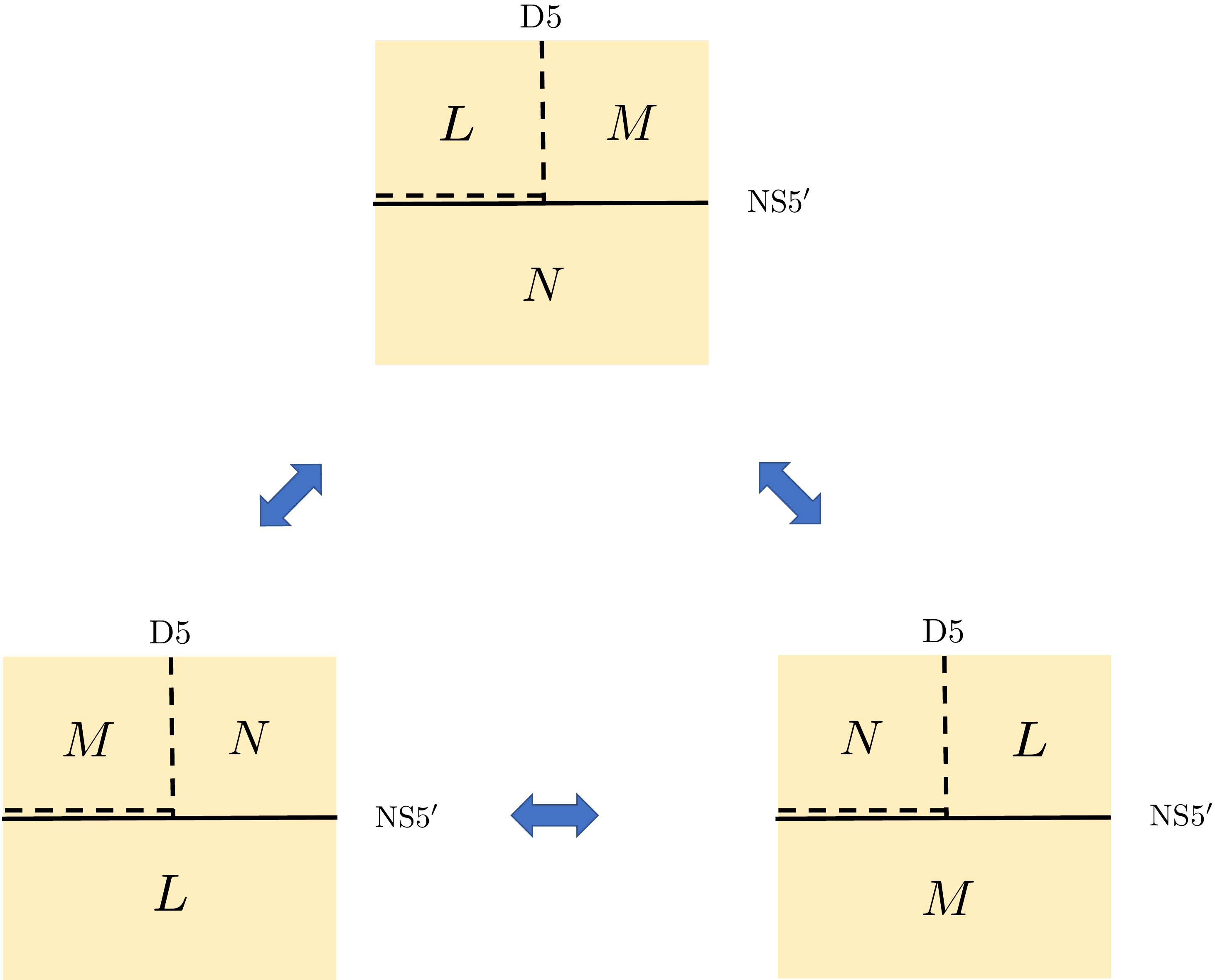}
\caption{
A triality of the Y-junctions. 
}
\label{figylmn}
\end{center}
\end{figure}

Consequently, 
we can obtain the quarter-index for the $\left(\begin{smallmatrix}1&|&3\\ \hline &2&\\ \end{smallmatrix}\right)$ Y-junction
\begin{align}
\label{y213t}
\mathbb{IV}_{\mathcal{N}'\mathcal{D}}^{
\left(
\begin{smallmatrix}
1&|&3\\ \hline
&2&\\
\end{smallmatrix}
\right)}
&=
\underbrace{
\frac12 
\frac{(q)_{\infty}^{2}}{(q^{\frac12} t^2;q)_{\infty}^{2}}\oint \prod_{i=1}^{2}\frac{ds_{i}}{2\pi is_{i}}
\prod_{i\neq j}\frac{\left(\frac{s_{i}}{s_{j}};q\right)_{\infty}}{\left(q^{\frac12}t^{2}\frac{s_{i}}{s_{j}};q\right)_{\infty}}
}_{\mathbb{II}_{\mathcal{N}'}^{\textrm{4d $U(2)$}}}
\underbrace{
\frac{1}{(q^{\frac12}t^2;q)_{\infty} (q t^{4};q)_{\infty}}
}_{\mathbb{IV}_{\mathcal{N}'\textrm{Nahm}}^{\textrm{4d $U(2)$}}}
\prod_{i=1}^{2}
\left(q t^2 s_{i};q\right)_{\infty}
\left(q t^2 s_{i}^{-1};q\right)_{\infty}
\nonumber\\
&\times 
\underbrace{
\frac{(q)_{\infty}}{(q^{\frac12} t^2;q)_{\infty}}\oint \frac{ds_{3}}{2\pi is_{3}}
}_{\mathbb{II}_{\mathcal{N}'}^{\textrm{4d $U(1)$}}}
\frac{1}{
\left(q^{\frac34}t^{3}s_{3};q\right)_{\infty}
\left(q^{\frac34}t^{3}s_{3}^{-1};q\right)_{\infty}
}
\nonumber\\
&\times 
\prod_{i=1}^{2}
\underbrace{
\frac{
\left(q^{\frac34}t \frac{s_{i}}{s_{3}};q\right)_{\infty}
\left(q^{\frac34}t \frac{s_{3}}{s_{i}};q\right)_{\infty}
}
{
\left(q^{\frac14}t^{-1} \frac{s_{i}}{s_{3}};q\right)_{\infty}
\left(q^{\frac14}t^{-1} \frac{s_{3}}{s_{i}};q\right)_{\infty}
}
}_{\mathbb{I}^{\textrm{3d tHM}}\left(\frac{s_{i}}{s_{3}}\right)}
\nonumber\\
&=
1+(t^{-2}+3t^{2})q^{\frac12}
+(t^{-4}+8t^{4})q
+(t^{-6}+t^{-2}+17t^{6})q^{\frac32}
\nonumber\\
&
+(4+t^{-8}+t^{-4}-2t^{4}+33t^{8})q^{2}
+(t^{-10}+t^{-6}+2t^{-2}+6t^{2}-t^{6}+58t^{10})q^{\frac52}
\nonumber\\
&
+(5+t^{-12}+t^{-8}+3t^{-4}+9t^{4}+3t^{8}+97t^{12})q^{3}
\nonumber\\
&+(t^{-14}+t^{-10}+3t^{-6}+7t^{-2}+13t^{2}+6t^{6}+20t^{10}+153t^{14})q^{\frac72}
\nonumber\\
&
+(13+t^{-16}+t^{-12}+3t^{-8}+5t^{-4}+28t^{4}+3t^{8}+53t^{12}+233t^{16})q^4
\nonumber\\
&
+t^{-18}(1+t^4+3t^8+6t^{12}+11t^{16}+21t^{20}+49t^{24}+120t^{32}+342t^{36})q^{\frac92}
+\cdots
\end{align}
This again coincides with 
the quarter-index (\ref{y321t}) 
for the $\left(\begin{smallmatrix}2&|&1\\ \hline &3&\\ \end{smallmatrix}\right)$ Y-junction 
and the quarter-index (\ref{y132t}) for the $\left(\begin{smallmatrix}3&|&2\\ \hline &1&\\ \end{smallmatrix}\right)$ Y-junction. 
The matching between 
three quarter-indices (\ref{y321t}), (\ref{y132t}) and (\ref{y213t}) provides us with 
a supporting evidence of the triality of the 
$\left(\begin{smallmatrix}2&|&1\\ \hline &3&\\ \end{smallmatrix}\right)$, 
$\left(\begin{smallmatrix}3&|&2\\ \hline &1&\\ \end{smallmatrix}\right)$ and 
$\left(\begin{smallmatrix}1&|&3\\ \hline &2&\\ \end{smallmatrix}\right)$ Y-junctions. 
The brane picture is shown in Figure \ref{figylmn}. 

\subsubsection{$Y_{L,M,N}$, $Y_{N,L,M}$ and $Y_{M,N,L}$}

Now we would like to provide a general formula of the 
quarter-index for the $\left(\begin{smallmatrix}M&|&N\\ \hline &L&\\ \end{smallmatrix}\right)$ Y-junction 
defining the VOA $Y_{L,M,N}$. 
The $\left(\begin{smallmatrix}M&|&N\\ \hline &L&\\ \end{smallmatrix}\right)$ Y-junction has  
a 4d $\mathcal{N}=4$ $U(\min (N,M))$ SYM theory in the upper half-space $x^2>0$ 
and a 4d $\mathcal{N}=4$ $U(L)$ SYM theory in the lower half-space $x^2<0$. 
They receive the Neumann boundary condition $\mathcal{N}'$ as required from the NS5$'$-brane. 

When $M=N$, it has the 3d $\mathcal{N}=4$ fundamental hypermultiplet arising from the D3-D5 strings. 
It should satisfy the Neumann boundary condition $N'$ according to the NS5$'$-brane. 
The resulting gauge anomaly is canceled by the Chern-Simons coupling associated to the $(1,1)$ fivebrane 
so that additional Fermi multiplet is not required at the junction. 
On the other hand, the gauge anomaly for the $U(L)$ gauge symmetry in the lower half-space 
induced from the Chern-Simons coupling will be canceled by additional fundamental Fermi multiplet at the junction. 
The junction also has the 3d $\mathcal{N}=4$ bi-fundamental twisted hypermultiplet.

For $M\neq N$ there is no 3d $\mathcal{N}=4$ fundamental hypermultiplet. 
There is the 3d $\mathcal{N}=4$ twisted hypermultiplets arising from the D3-D3 strings across the NS5$'$-brane. 
Unequal numbers of D3-branes in two sides of the D5-brane 
require that part of them, which corresponds to the broken $U(|N-M|)$ gauge symmetry, 
should obey the Dirichlet boundary condition $D$ for $|N-M|$ $=$ $1$ 
or the Nahm pole boundary condition specified by an embedding 
$\rho:$ $\mathfrak{su}(2)$ $\rightarrow$ $\mathfrak{u}(|N-M|)$ for $|N-M|$ $>$ $1$. 
Both of these boundary conditions admit the left-moving fermionic contributions 
and they will cancel the gauge anomaly for the $U(L)$ gauge symmetry in the lower half-plane 
induced from the Chern-Simons coupling. 
Hence the additional Fermi multiplet will not be required at the junction 
in contrast to the case with $M=N$. 
Note that there still remain the 3d $\mathcal{N}=4$ bi-fundamental twisted hypermultiplets.

We then obtain 
the quarter-index for the general $\left(\begin{smallmatrix}M&|&N\\ \hline &L&\\ \end{smallmatrix}\right)$ Y-junction 
\begin{align}
\label{yLMNt}
\mathbb{IV}_{\mathcal{N}'\mathcal{D}}^{
\left(
\begin{smallmatrix}
M&|&N\\ \hline
&L&\\
\end{smallmatrix}
\right)}
&=
\underbrace{
\frac{1}{L!}
\frac{(q)_{\infty}^{L}}{(q^{\frac12}t^2;q)_{\infty}^{L}}\oint \prod_{i=1}^{L}\frac{ds_{i}}{2\pi is_{i}}\prod_{i\neq j}
\frac{\left(\frac{s_{i}}{s_{j}};q\right)_{\infty}}
{\left(q^{\frac12}t^2 \frac{s_{i}}{s_{j}};q\right)_{\infty}}
}_{\mathbb{II}_{\mathcal{N}'}^{\textrm{4d $U(L)$}}}
\nonumber\\
&\times 
\underbrace{
\prod_{k=1}^{|N-M|}\frac{1}{\left(q^{\frac{k}{2}}t^{2k};q\right)_{\infty}}
}_{\mathbb{IV}_{\mathcal{N}'\mathcal{D}/\textrm{Nahm}}^{\textrm{4d $U(|N-M|)$}}}
\prod_{i=1}^{L}
\left(q^{\frac34+\frac{|N-M|-1}{4}} t^{|N-M|}s_{i}^{\pm};q\right)_{\infty}
\nonumber\\
&\times 
\underbrace{
\frac{1}{\min (N,M)!} \frac{(q)_{\infty}^{\min (N,M)}}
{(q^{\frac12}t^2;q)_{\infty}^{\min (N,M)}}\oint \prod_{i=L+1}^{L+\min (N,M)}
\frac{ds_{i}}{2\pi is_{i}}
\prod_{i\neq j}\frac{\left(\frac{s_{i}}{s_{j}};q\right)_{\infty}}{\left(q^{\frac12} t^2 \frac{s_{i}}{s_{j}}\right)_{\infty}}
}_{\mathbb{II}_{\mathcal{N}'}^{\textrm{4d $U(\min (N,M))$}}}
\nonumber\\
&\times 
\prod_{i=L+1}^{L+\min (N,M)}
\frac{1}
{
\left(q^{\frac14+\frac{|N-M|}{4}}t^{1+|N-M|}s_{i}^{\pm};q\right)_{\infty}
}
\nonumber\\
&\times 
\prod_{i=1}^{L}\prod_{k=L+1}^{L+\min (N,M)}
\underbrace{
\frac{
\left(q^{\frac34}t\frac{s_{i}}{s_{k}};q\right)_{\infty}
\left(q^{\frac34}t\frac{s_{k}}{s_{i}};q\right)_{\infty}
}
{
\left(q^{\frac14}t^{-1}\frac{s_{i}}{s_{k}};q\right)_{\infty}
\left(q^{\frac14}t^{-1}\frac{s_{k}}{s_{i}};q\right)_{\infty}
}
}_{\mathbb{I}^{\textrm{3d tHM}}\left(\frac{s_{i}}{s_{k}}\right)}. 
\end{align}

We expect that the following triality identity holds
\begin{align}
\label{yLMNt_identity}
\mathbb{IV}_{\mathcal{N}'\mathcal{D}}^{
\left(
\begin{smallmatrix}
L&|&N\\ \hline
&M&\\
\end{smallmatrix}
\right)}
&=
\mathbb{IV}_{\mathcal{N}'\mathcal{D}}^{
\left(
\begin{smallmatrix}
N&|&M\\ \hline
&L&\\
\end{smallmatrix}
\right)}
=\mathbb{IV}_{\mathcal{N}'\mathcal{D}}^{
\left(
\begin{smallmatrix}
M&|&L\\ \hline
&N&\\
\end{smallmatrix}
\right)}. 
\end{align}

In the H-twist limit $t\rightarrow q^{\frac14}$, 
the quarter-index (\ref{yLMNt}) becomes the vacuum character of the VOA $Y_{L,M,N}$
\begin{align}
\label{vch_YLNM}
\chi_{Y_{L,M,N}}&=
\frac{1}{L!}\oint \prod_{i=1}^{L}
\frac{ds_{i}}{2\pi is_{i}}
\prod_{i\neq j}
\left(1-\frac{s_{i}}{s_{j}}\right)
\left(1-\frac{s_{j}}{s_{i}}\right)
\prod_{k=1}^{|N-M|}\frac{1}{(q^{k};q)_{\infty}}
\prod_{i=1}^{L}(q^{\frac12+\frac{|N-M|}{4}}s_{i}^{\pm};q)_{\infty}
\nonumber\\
&\times 
\frac{1}{\min (N,M)!}\oint \prod_{i=L+1}^{L+\min (N,M)}
\frac{ds_{i}}{2\pi is_{i}}\prod_{i\neq j}
\left(1-\frac{s_{i}}{s_{j}}\right)
\left(1-\frac{s_{j}}{s_{i}}\right)
\nonumber\\
&\times 
\prod_{i=L+1}^{L+\min (N,M)}
\frac{1}{(q^{\frac12+\frac{|N-M|}{2}}s_{i}^{\pm};q)_{\infty}}
\prod_{i=1}^{L}\prod_{k=L+1}^{L+\min (N,M)}
\frac{1}{\left(1-\frac{s_{i}}{s_{k}}\right)\left(1-\frac{s_{k}}{s_{i}}\right) }
\end{align}
and (\ref{yLMNt_identity}) becomes the identity of the vacuum characters of 
the VOA $Y_{M,L,N}$, $Y_{L,N,M}$ and $Y_{L,N,M}$ \cite{Gaiotto:2017euk}
\begin{align}
\label{vch_YLNMid}
\chi_{Y_{M,L,N}}(q)&=\chi_{Y_{L,N,M}}(q)=\chi_{Y_{N,M,L}}(q). 
\end{align}

\subsection*{Acknowledgements}
We would like to thank Miroslav Rapcak for useful discussions and comments. 
D.G. is supported by the Perimeter Institute for Theoretical Physics.
T.O. is supported in part by Perimeter Institute for Theoretical Physics and 
JSPS Overseas Research fellowships. Research at
Perimeter Institute is supported by the Government of Canada through the Department of
Innovation, Science and Economic Development and by the Province of Ontario through the
Ministry of Research, Innovation and Science.

\appendix

\section{Notation}
\label{sec_notation}
We use the standard notation by defining $q$-shifted factorial
\begin{align}
\label{qpoch_def}
(a;q)_{0}&:=1,\qquad
(a;q)_{n}:=\prod_{k=0}^{n-1}(1-aq^{k}),\qquad 
(q)_{n}:=\prod_{k=1}^{n}(1-q^{k}),\quad 
\quad  n\ge1,
\nonumber \\
(a;q)_{\infty}&:=\prod_{k=0}^{\infty}(1-aq^{k}),\qquad 
(q)_{\infty}:=\prod_{k=1}^{\infty} (1-q^k)
\end{align}
where $a$ and $q$ are complex numbers with $|q|<1$.

\section{Formulae}
\label{sec_formula}
The following formulae are frequently used in this paper:
\begin{align}
\label{qpoch}
(x;q)_{\infty}&=\sum_{n=0}^{\infty}\frac{(-1)^{n}q^{\frac{n^2-n}{2}} x^{n}}
{(q;q)_{n}}.
\end{align}

\begin{align}
\label{1/qpoch}
\frac{1}{(x;q)_{\infty}}&=\sum_{n=0}^{\infty}\frac{x^{n}}{(q;q)_{n}},\qquad |x|<1.
\end{align}

\subsection{Jacobi's triple product identity}
Jacobi's triple product identity is given by
\begin{align}
\label{jacobi}
(q)_{\infty}(q^{\frac12}x;q)_{\infty} (q^{\frac12}x^{-1};q)_{\infty}&=
\sum_{n\in \mathbb{Z}}(-1)^n q^{\frac{n^2}{2}}x^{n},\qquad x\neq 0.
\end{align}

\subsection{$q$-binomial theorem}
The $q$-binomial theorem is 
\begin{align}
\label{q_binomial}
\sum_{n=0}^{\infty}\frac{(a;q)_{n}}{(q;q)_{n}}z^{n}
&=\frac{(az;q)_{\infty}}{(z;q)_{\infty}}.
\end{align}
This was derived 
by Cauchy \cite{cauchy1843memoire}, 
Heine \cite{MR1578577}
and other mathematicians.

\subsection{$q$-Gauss summation theorem}
The $q$-Gauss summation theorem is 
\begin{align}
\label{q_gauss}
\sum_{n=0}^{\infty}
\frac{(a;q)_{n}(b;q)_{n}}
{(c;q)_{n}(q;q)_{n}}\left(
\frac{c}{ab}
\right)^{n}
&=\frac{\left(
\frac{c}{a};q
\right)_{\infty}
\left(
\frac{c}{b};q
\right)_{\infty}
}
{
\left(c;q\right)_{\infty}
\left(
\frac{c}{ab};q
\right)_{\infty}
}.
\end{align}
This was firstly proven by Heine \cite{MR1578577} (see also \cite{MR2128719}). 

\subsection{Ramanujan's summation formula}
Ramanujan's summation formula is 
\begin{align}
\label{ramanujan_sum}
\sum_{n\in \mathbb{Z}}
\frac{(a;q)_{n}}{(b;q)_{n}}z^{n}
&=\frac{
(q)_{\infty}(ba^{-1};q)_{\infty}
(az;q)_{\infty}(q a^{-1}z^{-1};q)_{\infty}
}
{
(b;q)_{\infty}
(qa^{-1};q)_{\infty}
(z;q)_{\infty}
(ba^{-1}z^{-1};q)_{\infty}
},
\qquad |b/a|<|z|<1.
\end{align}
This was firstly given by Ramanujan \cite{MR0004860}. 
Later it was proven by 
Andrews \cite{MR241703}, 
Hahn \cite{MR30647}, 
M. Jackson \cite{MR0036882}, 
Ismail \cite{MR508183} 
and Andrews and Askey \cite{MR522519}.

\subsection{Minimal mirror identity}
Minimal mirror identity is given by
\begin{align}
\label{identity_dual}
&
\frac{(q^{\frac12} sx;q)_{\infty} (q^{\frac12}s^{-1}x^{-1};q)_{\infty}}
{(q^{\frac14}ts;q)_{\infty} (q^{\frac14}ts^{-1};q)_{\infty}}
\nonumber\\
=&\frac{1}{(q)_{\infty}(q^{\frac12} t^{2};q)_{\infty}}
\sum_{m\in \mathbb{Z}}
q^{\frac{m^{2}}{2}}
(-1)^{m}s^{m}x^{m}(q^{\frac34+m}tx;q)_{\infty}(q^{\frac34-m}tx^{-1};q)_{\infty}. 
\end{align}
This identity is physically conjectured in \cite{Okazaki:2019bok} from mirror symmetry of 
$\mathcal{N}=(0,4)$ half-BPS boundary conditions for 3d $\mathcal{N}=4$ Abelian gauge theories.

\section{Series expansions}
\label{sec_expansion}
For the examples in section \ref{sec_3bdy}, \ref{sec_ndjunction} and \ref{sec_nnddjunction}, 
we show the orders of terms in the $q$-series expansions which the indices agree up to 
as well as the first several terms in the expansions.

\subsection{Half-indices of interfaces in $\mathcal{N}=4$ SYM and S-dualities}
\label{sec_ex_3bdy}
\begin{align}
\label{3bdy_exp}
\begin{array}{c|c|c}
\textrm{interfaces}&\textrm{expansions}&\textrm{up to orders}\\ \hline
U(2)|U(2)
&1+\left(\frac{1}{t^2}+2t^2\right)q^{\frac12}+\frac{2+5t^8}{t^4}q
+\left(\frac{2}{t^6}-t^2+8t^6\right)q^{\frac32}+\cdots
& \mathcal{O}(q^5)
\\
U(3)|U(3)
&1+\left(\frac{1}{t^2}+2t^2\right)q^{\frac12}+\frac{2+5t^8}{t^4}q
+\frac{3+2t^4+t^8+10t^{12}}{t^6}q^{\frac32}+\cdots
& \mathcal{O}(q^2)
\\
U(2)|U(1)
&1+\left(\frac{1}{t^2}+2t^2 \right)q^{\frac12}
+\left(-1+\frac{1}{t^4}+4t^4\right)q
+\left(\frac{1}{t^6}-2t^2+6t^6\right)q^{\frac32}
+\cdots
& \mathcal{O}(q^5)
\\
U(3)|U(1)
&1+\left(\frac{1}{t^2}+2t^2 \right)q^{\frac12}
+\left(-1+\frac{1}{t^4}+4t^4\right)q
+\left(\frac{1}{t^6}-t^2+7t^6\right)q^{\frac32}
+\left(1+\frac{1}{t^8}-4t^4+11t^8\right)q^2
+\cdots
& \mathcal{O}(q^5)
\\
U(4)|U(1)
&1+\left(\frac{1}{t^2}+2t^2 \right)q^{\frac12}
+\left(-1+\frac{1}{t^4}+4t^4\right)q
+\left(\frac{1}{t^6}-t^2+7t^6\right)q^{\frac32}
+\left(1+\frac{1}{t^8}-3t^4+12t^{8}\right)q^2
+\cdots
& \mathcal{O}(q^5)
\\
U(2)|U(3)
&1+\left(\frac{1}{t^2}+2t^2 \right)q^{\frac12}
+\frac{2+5t^{8}}{t^4}q
+\frac{2+t^4+9t^{12}}{t^6}q^{\frac32}
+\left(1+\frac{3}{t^8}-3t^4+16t^8\right)q^2
+\cdots
& \mathcal{O}(q^5)
\\
U(2)|U(4)
&1+\left(\frac{1}{t^2}+2t^2 \right)q^{\frac12}
+\frac{2+5t^{8}}{t^4}q
+\frac{2+t^4+9t^{12}}{t^6}q^{\frac32}
+\left(2+\frac{3}{t^8}-2t^4+17t^8\right)q^2
+\cdots
& \mathcal{O}(q^5)
\\
\end{array}
\end{align}
%

\subsection{NS5$'$-D5 junction}
\label{sec_ex_ndjunction}
\begin{align}
\label{ndjunction_exp}
\begin{array}{c|c|c}
\textrm{NS5$'$-D5 jct.}&\textrm{expansions}&\textrm{up to orders}\\ \hline 
\left(\begin{smallmatrix}
2&2\\
0&0\\
\end{smallmatrix}\right)
&
1+2t^2q^{\frac12}
-\frac{t(1+x^2)}{x}q^{\frac34}
+5t^4 q
-\frac{3(t^3(1+x^2))}{x}q^{\frac54}
+t^283+8t^4)q^{\frac32}+\cdots
&
\mathcal{O}(q^5)
 \\
\left(\begin{smallmatrix}
3&3\\
0&0\\
\end{smallmatrix}\right)
&
1+t^2 q^{\frac12}
-\frac{t(1+x^2)}{x}q^{\frac34}
+3t^4q
-\frac{2(t^3(1+x^2))}{x}q^{\frac54}
+t^2(2+5t^4)q^{\frac32}
+\cdots
&
\mathcal{O}(q^5)
 \\
\left(\begin{smallmatrix}
1&2\\
0&0\\
\end{smallmatrix}\right)
&
1+2t^2 q^{\frac12}
+t^2\left(4t^2-\frac{1}{x}-x \right)q
+\left(2t^2+6t^6-\frac{2t^4(1+x^2)}{x}\right)q^{\frac32}+\cdots
&
\mathcal{O}(q^5)
 \\
\left(\begin{smallmatrix}
1&3\\
0&0\\
\end{smallmatrix}\right)
&
1+2t^2 q^{\frac12}
+4t^4q
-\frac{t^3(1+x^2)}{x}q^{\frac54}
+t^2(2+7t^4)q^{\frac32}+\cdots
&
\mathcal{O}(q^5)
 \\
\left(\begin{smallmatrix}
1&4\\
0&0\\
\end{smallmatrix}\right)
&
1+2t^2 q^{\frac12}
+4t^4q
+\left(2t^2+7t^6-\frac{t^4(1+x^2)}{x}\right)q^{\frac32}+\cdots
&
\mathcal{O}(q^5)
 \\
\left(\begin{smallmatrix}
2&3\\
0&0\\
\end{smallmatrix}\right)
&
1+2t^2q^{\frac12}
+t^2\left(5t^2-\frac{1}{x}-x\right)q
+\left(2t^2+9t^6-\frac{3t^4(1+x^2)}{x} \right)q^{\frac32}
+\cdots
&
\mathcal{O}(q^5)
 \\
\left(\begin{smallmatrix}
2&4\\
0&0\\
\end{smallmatrix}\right)
&
1+2t^2 q^{\frac12}
+5t^4q
-\frac{t^3(1+x^2)}{x}q^{\frac54}
+t^2(2+9t^4)q^{\frac32}
+\cdots
&
\mathcal{O}(q^5)
 \\
\left(\begin{smallmatrix}
3&4\\
0&0\\
\end{smallmatrix}\right)
&
1+2t^2 q^{\frac12}
+t^2\left(5t^2-\frac{1}{x}-x\right)q
+\left(2t^2+10t^6 -\frac{3t^4(1+x^2)}{x} \right)q^{\frac32}
+\cdots
&
\mathcal{O}(q^5)
 \\
\left(\begin{smallmatrix}
1&1\\
1&1\\
\end{smallmatrix}\right)
&
1+\left(\frac{1}{t^2}+4t^2 \right)q^{\frac12}
-\frac{t(1+x^4)}{x^2}q^{\frac34}
+\left(-2+\frac{1}{t^4}+10t^4 \right)q
-\frac{5(t^3(1+x^4))}{x^2}q^{\frac54}
+\cdots
&
\mathcal{O}(q^5)
 \\
\left(\begin{smallmatrix}
1&1\\
2&0\\
\end{smallmatrix}\right)
&
1+3t^2 q^{\frac12}
+(-1+7t^4)q
-\frac{t^3(1+x^2)}{x}q^{\frac54}
+(t^2+13t^6)q^{\frac32}
+\cdots
&
\mathcal{O}(q^5)
 \\
\left(\begin{smallmatrix}
2&1\\
3&0\\
\end{smallmatrix}\right)
&
1+3t^2 q^{\frac12}
+(-1+8t^4)q
-\frac{t^3(x_{1}^2+x_{2}^2)}{x_{1}x_{2}}q^{\frac54}
+17t^6 q^{\frac32}
+\cdots
&
\mathcal{O}(q^3)
 \\
\left(\begin{smallmatrix}
3&1\\
2&0\\
\end{smallmatrix}\right)
&
1+3t^2 q^{\frac12}
-\frac{t(x_{1}^2+x_{2}^2)}{x_{1}x_{2}}q^{\frac34}
+(-1+8t^4)q
-\frac{4(t^3(x_{1}^2+x_{2}^2))}{x_{1}x_{2}}q^{\frac54}
+\cdots
&
\mathcal{O}(q^3)
 \\
\left(\begin{smallmatrix}
1&2\\
1&2\\
\end{smallmatrix}\right)
&
1
+\left(\frac{1}{t^2}+4t^2 \right)q^{\frac12}
-\frac{t (x_{1}+x_{2}^2)}{x_{1}x_{2}}q^{\frac34}
+\left(\frac{1}{t^4}+12t^{4} \right)q
-\frac{2(1+5t^4)}{t}q^{\frac54}
+\cdots
&
\mathcal{O}(q^5)
 \\
\left(\begin{smallmatrix}
1&1\\
2&3\\
\end{smallmatrix}\right)
&
1
+\left(\frac{1}{t^2}+4t^2 \right)q^{\frac12}
+\left(\frac{1}{t^4}+12t^{4}-\frac{t^2(x_{1}+x_{2}^2)}{x_{1}x_{2}} \right)q
+\cdots
&
\mathcal{O}(q^5)
 \\
\left(\begin{smallmatrix}
1&2\\
3&4\\
\end{smallmatrix}\right)
&
1
+\left(\frac{1}{t^2}+4t^2 \right)q^{\frac12}
-\frac{t(x_{1}+x_{2}^2)}{x_{1}x_{2}}q^{\frac34}
+\left(1+\frac{1}{t^4}+13t^4 \right)q
-\frac{(1+5t^4)(x_{1}^2+x_{2}^2)}{tx_{1}x_{2}}q^{\frac54}
+\cdots
&
\mathcal{O}(q^3)
 \\
\end{array}
\end{align}

\subsection{NS5$'$-NS5 and D5-D5$'$ junctions}
\label{sec_ex_nnddjunction}

\begin{align}
\label{nnddjunction_exp}
\begin{array}{c|c|c}
\textrm{NS5$'$-NS5 jct.}&\textrm{expansions}&\textrm{up to orders}\\ \hline
\left(\begin{smallmatrix}
2&2\\
0&0\\
\end{smallmatrix}\right)
&
1
+t^2 q^{\frac12}
+\left(-1+2t^4-\frac{t^2(1+x^2)}{x} \right)q
+\left(-t^2+2t^6-\frac{t^4(1+x^2)}{x} \right)q^{\frac32}
+\cdots
&
\mathcal{O}(q^5)
 \\
\left(\begin{smallmatrix}
3&3\\
0&0\\
\end{smallmatrix}\right)
&
1
+t^2 q^{\frac12}
+(-1+2t^4)q
+\left(-\frac{t^3}{x}-t^3x \right)q^{\frac54}
+(-t^2+3t^6)q^{\frac32}
+\cdots
&
\mathcal{O}(q^2)
 \\
\left(\begin{smallmatrix}
1&1\\
1&1\\
\end{smallmatrix}\right)
&
1
+\frac{2(1+t^4)}{t^2}q^{\frac12}
+\left(
-3+\frac{3}{t^4}+3t^4
-\frac{1+x^4}{t^2x^2}
-\frac{t^2(1+x^4)}{x^2}
\right)q
+\cdots
&
\mathcal{O}(q^5)
 \\
\left(\begin{smallmatrix}
2&2\\
2&2\\
\end{smallmatrix}\right)
&
1
+\frac{2(1+t^4)}{t^2}q^{\frac12}
+\left(1+\frac{5}{t^4}+5t^4 \right)q
+\left(
\frac{8}{t^6}+8t^6
-\frac{1+x^4}{x^2}
-\frac{1+x^4}{t^4 x^2}
-\frac{t^4(1+x^4)}{x^2}
\right)q^{\frac32}
+\cdots
&
\mathcal{O}(q^2)
 \\
\left(\begin{smallmatrix}
1&2\\
0&1\\
\end{smallmatrix}\right)
&
1
+\left(\frac{1}{t^2}+t^2 \right)q^{\frac12}
+\left(-1+\frac{1}{t^4}+t^4-\frac{1}{x}-x \right)q
+\cdots
&
\mathcal{O}(q^3)
 \\
\left(\begin{smallmatrix}
2&4\\
0&2\\
\end{smallmatrix}\right)
&
1
+\left(\frac{1}{t^2}+t^2 \right)q^{\frac12}
+\left(-1+\frac{2}{t^4}+2t^4 \right)q
+\left(\frac{2}{t^6}+2t^6-\frac{1}{x}-x \right)q^{\frac32}
+\cdots
&
\mathcal{O}(q^2)
 \\
\left(\begin{smallmatrix}
1&2\\
1&2\\
\end{smallmatrix}\right)
&
1
+\frac{2(1+t^4))}{t^2}q^{\frac12}
+\left(-6+\frac14 \left(20+\frac{16}{t^4}+12t^4 \right) \right)q
+\frac14 \left(-\frac{8}{t^3}-8t \right)q^{\frac54}
+\cdots
&
\mathcal{O}(q^5)
 \\
\left(\begin{smallmatrix}
1&3\\
1&3\\
\end{smallmatrix}\right)
&
1
+\frac{2(1+t^4)}{t^2}q^{\frac12}
+\left(-1+\frac{4}{t^4}+3t^4 \right)q
+\left(\frac{7}{t^6}-t^2+4t^6-\frac{1+x^2}{x}-\frac{1+x^2}{t^4 x} \right)q^{\frac32}
+\cdots
&
\mathcal{O}(q^3)
 \\
\left(\begin{smallmatrix}
2&3\\
2&3\\
\end{smallmatrix}\right)
&
1
+\frac{2(1+t^4)}{t^2}q^{\frac12}
+\left(1+\frac{5}{t^4}+5t^4 \right)q
+\frac{9+2t^4+2t^8+8t^{12}}{t^6}q^{\frac32}
+\frac{(1+t^4+t^8)(1+x^2)}{t^5 x}q^{\frac74}
+\cdots
&
\mathcal{O}(q^3)
 \\
\end{array}
\end{align}

\bibliographystyle{utphys}
\bibliography{ref}

\end{document}